\RequirePackage{silence} 
\documentclass[ twoside,openright,titlepage,numbers=noenddot,
                headinclude,cleardoublepage=empty,abstract=on,
                BCOR=5mm,paper=A4,fontsize=11pt,
                ]{scrreprt}

\PassOptionsToPackage{utf8}{inputenc}
  \usepackage{inputenc}

\PassOptionsToPackage{T1}{fontenc} 
  \usepackage{fontenc}

\PassOptionsToPackage{
  drafting=false,    
  tocaligned=false, 
  dottedtoc=false,  
  eulerchapternumbers=true, 
  linedheaders=false,       
  floatperchapter=true,     
  eulermath=false,  
  beramono=true,    
  palatino=true,    
  style=classicthesis 
}{classicthesis}

\newcommand{\myTitle}{Nonlocal transport signatures of{\linebreak}Andreev bound states\xspace}
\newcommand{\mySubtitle}{PhD Thesis\xspace}

\newcommand{\myName}{Andreas P{\"o}schl\xspace}

\newcommand{\myFaculty}{Niels Bohr Institute\xspace}
\newcommand{\myDepartment}{Center for Quantum Devices\xspace}
\newcommand{\myUni}{University of Copenhagen\xspace}

\providecommand{\mLyX}{L\kern-.1667em\lower.25em\hbox{Y}\kern-.125emX\@}

\PassOptionsToPackage{ngerman,american}{babel} 
    \usepackage{babel}

\usepackage{csquotes}
\PassOptionsToPackage{%
  backend=bibtex8,bibencoding=ascii,%
  language=auto,%
  style=phys,
  eprint = true,
  sorting=nyt, 
  maxbibnames=10, 
  natbib=true 
}{biblatex}
    \usepackage{biblatex}

\PassOptionsToPackage{fleqn}{amsmath}       
  \usepackage{amsmath}

\usepackage{graphicx} %
\usepackage{scrhack} 
\usepackage{xspace} 
\PassOptionsToPackage{nohyperlinks}{acronym}
  \usepackage{acronym} 

\usepackage{tabularx} 
\usepackage{subfig}

\usepackage{listings}
\usepackage{classicthesis}

\hypersetup{%
  colorlinks=true, linktocpage=true, pdfstartpage=3, pdfstartview=FitV,%
  breaklinks=true, pageanchor=true,%
  pdfpagemode=UseNone, %
  plainpages=false, bookmarksnumbered, bookmarksopen=true, bookmarksopenlevel=1,%
  hypertexnames=true, pdfhighlight=/O,
  urlcolor=CTurl, linkcolor=CTlink, citecolor=CTcitation, 
  pdftitle={\myTitle},%
  pdfauthor={\textcopyright\ \myName, \myUni, \myFaculty},%
  pdfsubject={},%
  pdfkeywords={},%
  pdfcreator={pdfLaTeX},%
  pdfproducer={LaTeX with hyperref and classicthesis}%
}

\makeatletter
\@ifpackageloaded{babel}%
  {%
    \addto\extrasamerican{%
    }%
    \addto\extrasngerman{%
    }%
    }{\relax}
\makeatother

\areaset[current]{400pt}{700pt} 

\usepackage{pdfpages}
\usepackage{bm}
\usepackage{siunitx}
\usepackage{amsmath}
\usepackage{amssymb}
\usepackage{physics}
\usepackage{pifont}
\begin{document}
\frenchspacing
\raggedbottom
\selectlanguage{american} 
\pagenumbering{roman}
\pagestyle{plain}
\includepdf[pages=-]{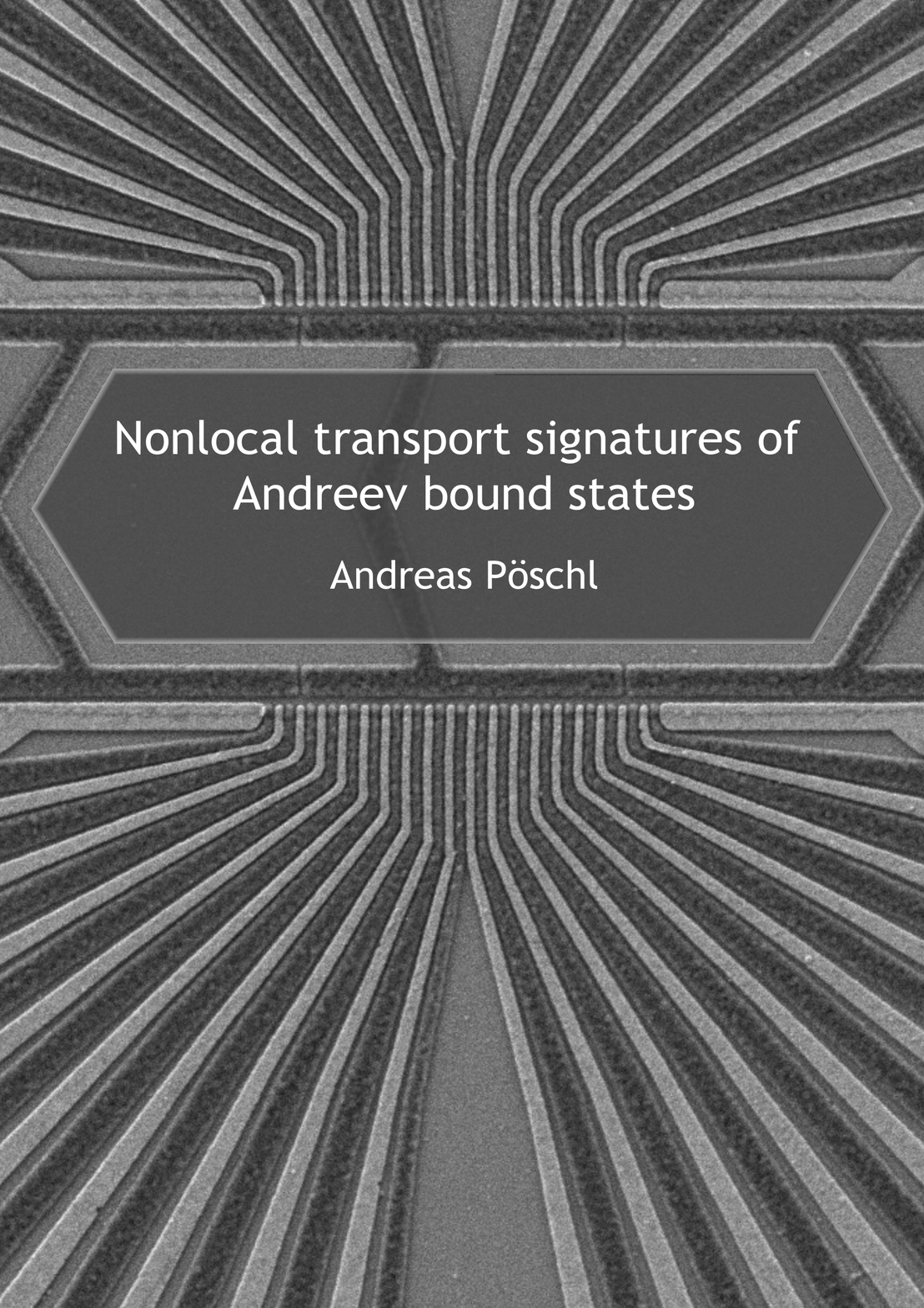}
\cleardoublepage
\begin{titlepage}
    \begin{addmargin}[-1cm]{-3cm}
    \begin{center}
        \large

        \hfill

        \vfill

        \begingroup
            \color{CTtitle}\spacedallcaps{\myTitle} \\ \bigskip
        \endgroup

        \spacedlowsmallcaps{\myName}

        \vfill

        \mySubtitle \\ \medskip
        \myDepartment \\
        \myFaculty \\
        \myUni \\ \bigskip
\vfill
        March 2022
\vfill

\begin{minipage}{0.4\textwidth}
\begin{center}
 \underline{Academic advisor:}\\
 Prof. Charles M. Marcus\newline\newline
 \end{center}
\end{minipage}%
\begin{minipage}{0.4\textwidth}
\begin{center}
\underline{Assessment committee:}\\
Prof. Karsten Flensberg\\
Dr. Ram\'{o}n Aguado\\
Dr. Silvano de Franceschi
\end{center}
\end{minipage}%
\hfill

  \vfill
  \vfill
  \vfill
  \vfill
 
\begin{minipage}{0.4\textwidth}
\begin{flushleft}
\includegraphics[width=5cm]{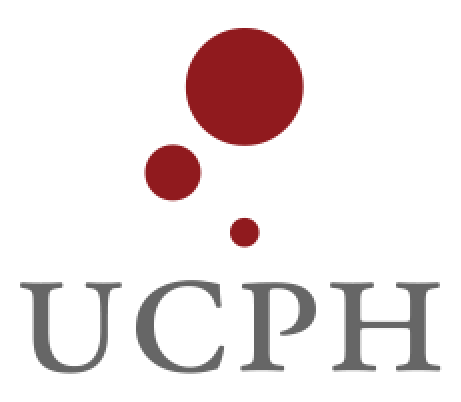}
\end{flushleft}
\end{minipage}%
\hfill
\begin{minipage}{0.4\textwidth}
\begin{flushright}
\includegraphics[width=6cm]{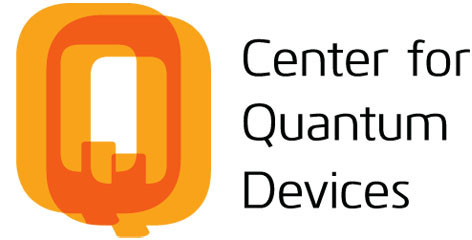} \\ \medskip
\end{flushright}
\end{minipage}%

    \end{center}
  \end{addmargin}
\end{titlepage}

\cleardoublepage
\pdfbookmark[1]{Abstract}{Abstract}
\begingroup
\let\clearpage\relax
\let\cleardoublepage\relax
\let\cleardoublepage\relax

\chapter*{Abstract}

In this PhD thesis, quantum devices based on molecular-beam epitaxy grown InAs semiconductor with an in-situ grown epitaxial Al film were investigated. Novel device geometries were realized that allow for the study of bound states that arise at low temperatures in semiconducting nanowires (NWs) due to the presence of spin-orbit coupling, Zeeman effect, and superconducting proximity effect. \\

Sideprobe devices allow for tunneling conductance measurements at discrete points of NWs that are gate-defined in a two-dimensional Al-InAs hybrid two-dimensional heterostructure. Extended Andreev bound states (ABSs) appear as correlated subgap resonances in tunneling conductance measured at neighboring probes that are $\lesssim\SI{0.8}{\micro\meter} $ apart from each other. The correlations were found to be robust with respect to magnetic field and gate voltage. The ABSs were hybridized with a local quantum dot in one of the tunnel barriers. The effect of this hybridization was observed both locally at the position of the quantum dot resonance and at the neighboring tunnel probe, which reveals nonlocal signatures of hybridization. 

The charge character of the ABSs was determined by measuring nonlocal conductance. The ABSs move as a function of gate voltage, with a modification of their charge character in agreement with theoretical predictions for the total Bardeen-Cooper-Schrieffer charge of the ABSs. 

For NWs of length $\gtrsim\SI{1.4}{\micro\meter}$ bound states no longer extend over the whole NW resulting in a lack of end-to-end correlations in the measured spectra at different locations along the NW. Accidental correlations could be produced by fine-tuning local gates. These correlations were compared to the correlations stemming from extended bound states.\\

A three-terminal device geometry based on vapor-liquid-solid grown InAs NWs with a full-shell of in-situ grown Al was realized. The devices were used to study the subgap states that appear in tunneling conductance at each NW end when the superconducting shell of the NW is threaded by a single flux quantum. States that extend over the whole length of the NW were ruled out experimentally. The subgap states lead to similar spectra at both NW ends, however, which suggests that the cross-sectional shape of the NW has a strong influence on the resulting subgap spectrum.

\vfill

\newpage

\begin{otherlanguage}{danish}
\pdfbookmark[1]{Resume}{Resume}
\chapter*{Resumé}
I denne afhandling blev kvantekredsløb baseret på molekylær-stråle-epitaxi-groede InAs halvledere med in-situ deponerede epitaksiale Al film undersøgt. Nye kredsløbsgeometrier blev realiseret. Disse muliggjorde undersøgelse af bundne Andreev tilstande (BAT) som opstår ved lave temperaturer i halvleder nano-ledninger (NL) på grund af tilstedeværelsen af spin-bane-koblingen, Zeeman effekten og den superledende nærhedseffekt. \\

Kredsløb med sideprober muliggør tunnel-konduktans målinger ved diskrete punkter langs NL som er gate-spænding defineret i en todimensionelle elektrongas. Udstrakte BAT opstår som korrelerede undergab resonanser i tunnelkonduktans, målt ved nabo-prober $\lesssim \SI{0.8}{\micro\meter}$ fra hinanden. De observerede korrelationer var robuste med henhold til det anvendte magnetiske felt samt gate-spænding. BAT blev hybridiseret med lokale kvante-punkt-resonanser i en af tunnelbarriererne. Effekten af denne hybridisering blev observeret både lokalt ved kvante-punkt-resonansen samt ved nabo-proben som viste ikke-lokale signaturer af hybridiseringen. 

Ladningskarateren af BAT blev fundet ved måling af ikke-lokal ledningsevne. BAT bevæger sig som funktion af gate-spænding, med en ændring af deres ladningskarakter, i overensstemmelse med teoretiske forudsigelser for den totale Bardeen-Cooper-Schrieffer-ladning af BAT.

For NL med længde $\gtrsim\SI{1.4}{\micro\meter}$ strakte de bundne tilstande sig ikke over hele NL, hvilket resulterede i fraværet af ende til ende  korrelationer i de målte spektre  ved forskellige lokationer langs NL. Andre korrelationer kan forekomme ved tilfælde eller ved finjustering af lokale gate-spændinger. Disse blev sammenlignet med korrelationer fra udstrakte bundne tilstande.\\

En tre-terminals kredsløbsgeometri blev realiseret, baseret på vapor-liquid-solid InAs NL med fuldt dækkende in-situ groet Al overflade. Kredsløbene blev brugt til at studere undergabstilstande, som opstår i tunnelkonduktans, målt ved hver ende af NL når den omkringliggende superledende Al cylinder er trådet af en kvanteflux. Tilstande som udstrækker sig langs hele NL blev eksperimentelt udelukket. Dog fører undergabs tilstandene til ens  spektre i begge ender, hvilket antyder at NL’s  tværsnitsform stærkt influerer det endelig spektrum. 

\end{otherlanguage}

\endgroup

\vfill

\cleardoublepage
\pdfbookmark[1]{Publications}{publications}
\chapter*{Publications}

This thesis presents results from the following publications and manuscripts:

\begin{enumerate}

\item \textbf{Nonlocal signatures of hybridization between quantum dot and Andreev bound states}\\
\emph{A. P{\"{o}}schl}, A. Danilenko, D. Sabonis, K. Kristjuhan, T. Lindemann, C. Thomas, M.J. Manfra, C.M. Marcus.\\\href{https://doi.org/10.1103/PhysRevB.106.L161301}{Phys. Rev. B 106, L161301 (2022).}\hfill

\item \textbf{Nonlocal conductance spectroscopy of Andreev bound states in gate-defined InAs/Al nanowires}\\
\emph{A. P{\"{o}}schl}, A. Danilenko, D. Sabonis, K. Kristjuhan, T. Lindemann, C. Thomas, M.J. Manfra, C.M. Marcus.\\\href{https://doi.org/10.1103/PhysRevB.106.L241301}{Phys. Rev. B 106, L241301 (2022).}\hfill

\item \textbf{Measurement circuit effects in three-terminal electrical transport measurements}\\
E.A. Martinez, \emph{A. P{\"{o}}schl}, E.B. Hansen, M.A.Y. van de Poll, S. Vaitiekėnas, A.P. Higginbotham, L. Casparis.\\
In review. \href{https://doi.org/10.48550/arXiv.2104.02671}{arXiv:2104.02671 (2021)}.\hfill

\item \textbf{Closing of the induced gap in a hybrid superconductor-semiconductor nanowire}\\
D. Puglia, E.A. Martinez, G.C. M{\'{e}}nard, \emph{A. P{\"{o}}schl}, S. Gronin, G.C. Gardner, R. Kallaher, M.J. Manfra, C.M. Marcus, A.P. Higginbotham, L. Casparis.\\
\href{https://doi.org/10.1103/PhysRevB.103.235201}{Phys. Rev. B 103, 23520 (2021).}\hfill

\end{enumerate}
\cleardoublepage
\pdfbookmark[1]{Acknowledgments}{acknowledgments}


\bigskip

\begingroup
\let\clearpage\relax
\let\cleardoublepage\relax
\let\cleardoublepage\relax
\chapter*{Acknowledgments}

Foremost, I have to thank \textbf{Charlie} for giving me the opportunity, resources, and supervision to reach the achievements during my PhD at QDev. Your training, love for clarity and simplicity, and curiosity has had the most positive influence on me and my work.

Furthermore, I would like to thank \textbf{Andrew, Ferdinand,} and \textbf{Lucas} for their guidance, helpful comments, and lessons about experimental physics. I could not have asked for better teachers.

I want to express my gratitude to my committee \textbf{Ram\'{o}n Aguado}, \textbf{Silvano de Franceschi}, and \textbf{Karsten Flensberg}, who invested their time in evaluating my work, achievements, and thesis.

Fruitful collaborations are the basis of any scientific achievement. I had the privilege to work together with the great material scientists \textbf{Candice, Geoff, Sara, Sergei, Teng, Tyler, Mike,} and \textbf{Peter}. Your materials enable major breakthroughs in this branch of physics.

I want to thank \textbf{Davydas, Denise, Esben, Esteban,} and \textbf{Sole} for a lot of great time inside and outside the lab. You definitely helped me to sustain a steep learning curve while I was working on SAG and VLS nanowire projects.

A special thanks goes to \textbf{Dima, Farhad, Georg,} and \textbf{Torsten} who were always happy to discuss about the most recent progress and the latest data. \textbf{Andrea, Max, Waldemar,} and \textbf{Karsten} were very helpful in sharing their theoretical insights. 

I had the pleasure of being introduced to nanofabrication by \textbf{ Aga, Karolis, Maren, Martin, Paschalis, Sachin, Sangeeth, Shiv} and \textbf{Smitha}. Your efforts to keep the cleanroom in perfect shape certainly helps us succeed in QDev.

I had a lot of great predecessors on the 2DEG team, who figured out difficult questions and created the foundation for my work. I want to thank \textbf{Abhishek, Alexander, Antonio, Asbj{\o}rn,} and \textbf{Felix} for teaching me a lot and getting me up to speed with my own experiments. 
I also want to thank the individuals with whom I have worked closely together during the last three years. These are \textbf{Ahnaf, Frederik, Daniel, Kaur, Magnus,} and \textbf{Serwan}. It is a great pleasure to work in such a nice team and reach great results. In particular, I want to express my gratitude to \textbf{Alisa} and \textbf{Deividas}. I was lucky enough to work with the two of you, share a lot of fun moments in the lab, and press on when it comes to our experiments.

I also want to thank my amazing `4\textsuperscript{th} floor colleagues' \textbf{Albert, Anasua, Anders, Dags, Eoin, Fabio, Fabrizio, Federico, Jos\'{e}, Lukas, Michaela, Natalie, Oscar,} as well as \textbf{Gunjan, Harry, Martin, Steffen} who were always ready to chat about physics, experiments, or trivia outside the lab over a cup of coffee.

A great thanks goes to my family, friends, and especially my partner \textbf{Johanna} for their love and support.

\endgroup

\cleardoublepage
\pagestyle{scrheadings}
\pdfbookmark[1]{\contentsname}{tableofcontents}
\setcounter{tocdepth}{1} 
\setcounter{secnumdepth}{3} 
\manualmark
\markboth{\spacedlowsmallcaps{\contentsname}}{\spacedlowsmallcaps{\contentsname}}
\tableofcontents
\automark[section]{chapter}
\renewcommand{\chaptermark}[1]{\markboth{\spacedlowsmallcaps{#1}}{\spacedlowsmallcaps{#1}}}
\renewcommand{\sectionmark}[1]{\markright{\textsc{\thesection}\enspace\spacedlowsmallcaps{#1}}}
\clearpage
\begingroup
    \let\clearpage\relax
    \let\cleardoublepage\relax
    \pdfbookmark[1]{\listfigurename}{lof}
    \listoffigures

    \vspace{8ex}





    \pdfbookmark[1]{Acronyms}{acronyms}
    \markboth{\spacedlowsmallcaps{Acronyms}}{\spacedlowsmallcaps{Acronyms}}
    \chapter*{Acronyms}
    \begin{acronym}[2DEG]
        \acro{2DEG}[2DEG]{two-dimensional electron gas}
        \acro{ABS}[ABS]{Andreev bound state}
        \acro{ALD}[ALD]{atomic layer deposition}
        \acro{EBL}[EBL]{electron-beam lithography}   
        \acro{QD}[QD]{quantum dot}
        \acro{QP}[QP]{quasiparticle}
        \acroplural{QDs}[QDs]{quantum dots}
        \acro{MBE}[MBE]{molecular-beam epitaxy}
        \acro{MZM}[MZM]{Majorana zero mode}
        \acro{NW}[NW]{nanowire}
        \acro{VLS}[VLS]{vapor-liquid-solid}
        \acro{ZBP}[ZBP]{zero-bias conductance peak}
    \end{acronym}

\endgroup

\cleardoublepage
\pagestyle{scrheadings}
\pagenumbering{arabic}
\cleardoublepage
\part{Introduction and Motivation}\label{pt:intro}
\chapter{Semiconducting Indium Arsenide and the proximity effect}
\label{ch:InAs}
Hybrid devices consisting of semiconducting InAs proximitized by superconducting Al are the subject of this work. InAs is a semiconductor of interest due to its sizeable electron $g$ factor and its spin-orbit coupling. It is furthermore possible to create low-dimensional nanostructures and versatile device geometries from it. \\

In a three-dimensional crystal of InAs, electrons can be described by Bloch waves propagating in all spatial directions. InAs can be incorporated in a heterostructure in between layers of larger bandgap, which confines the electron motion to a two-dimensional plane defined by an InAs quantum well \cite{BeenakkerNanostructures}. In one-dimensional nanowires (NWs) the electrons are further constrained to move only along the axis of the NW. The relation between the length scale to which the nanostructure confines the electrons and the Fermi wavelength determines how many modes are occupied along the respective spatial dimension. If electrons are confined in all three spatial directions, for instance, by laterally gating a two-dimensional electron gas (2DEG), a quantum dot is formed. It manifests itself in the form of discrete energy levels, analogous to the discrete energy levels of electronic orbitals in atoms \cite{Kouwenhoven1997}. For electrons confined in an InAs nanostructure additional peculiarities arise due to the band structure of InAs. \\

An important property of InAs is that its charge carriers are subjected to strong spin-orbit interaction. Spin-orbit interaction arises in semiconductors that lack inversion symmetry \cite{DRESSLHAUS, RASHBA, focusRasbha}. This may either be due to the underlying crystal structure (bulk inversion asymmetry) or due to an electrostatic potential with a finite gradient (structural inversion asymmetry). The finite gradient of the electrostatic potential leads to an electric field $\mathbf{E}$ and can contain contributions stemming both from electrostatic gating and bandgap mismatches within a heterostructure. An intuitive explanation of spin-orbit interaction suggests that the movement of an electron in a constant electric field translates - after a Lorentz transformation - into a constant magnetic field in the rest frame of the electron \cite{SOI_lorentz1, SOI_lorentz2, SOI_lorentz3, SOI_lorentz4}. While this toy model visualization of spin-orbit interaction gives the right qualitative picture it fails to describe the phenomenon quantitatively as it neglects the effect of the periodic electrostatic potential originating from the crystal lattice on the conduction electrons \cite{Winkler2003}. A quantitative description can be given within the framework of $\bm{k}\cdot \bm{p}$ theory. The contribution to spin-orbit interaction due to structural inversion asymmetry gives rise to the Rasbha spin-orbit interaction for electrons with spin $\bm{\sigma}$ and wave vector $\bm{k}$ given by the Hamiltonian \cite{RASHBA, RASHBA2}

\begin{equation}
H_\mathrm{SO}=\hbar \bm{\alpha_\mathrm{R}}\cdot(\bm{\sigma}\times\bm{k}).
\end{equation}
The magnitude of this interaction term is determined by $\bm{\alpha_\mathrm{R}}$ which is proportional to the electric field $\mathbf{E}$. 

For confinement in the two-dimensional $xy$-plane and an electric field in $z$-direction, the term simplifies to
\begin{equation}
H_{\mathrm{SO}}=\hbar \alpha_\mathrm{R} (\sigma_x k_y-\sigma_y k_x).
\end{equation}
If the electron motion is further constrained to the $x$-direction in a one-dimensional channel one arrives at 

\begin{equation}
H_{\mathrm{SO}}=\hbar \alpha_\mathrm{R} (\sigma_y k_x).
\end{equation}
Spin-orbit coupling lifts the spin degeneracy of electrons. For a one-dimensional mode, the effect of spin-orbit interaction is a shift of the spin-up and spin-down bands away from $k=0$ in opposite directions.

The contribution to spin-orbit interaction due to bulk inversion asymmetry is also present in InAs due to its crystal structure being Wurzite or Zincblende, both of which lack inversion symmetry. It is, however, smaller than the contribution coming from sturctural inversion symmetry in nanostructures like NWs and two-dimensional heterostructures \cite{rashba_vs_dressel_inas}.\\

An additional property of the InAs band structure is revealed under an applied magnetic field $\mathbf{B}$. Due to the Zeeman effect, free electrons with spin $\bm{\sigma}$ acquire an energy described by
\begin{equation}
H_{Z} = \frac{g \mu_\mathrm{B}}{\hbar} \bm{\sigma} \cdot\bm{B}
\end{equation}
where $\mu_B$ is the Bohr magneton and $g$ is the the Land\'e $g$ factor. Instead of the free electron Land\'e $g$ factor $g=2$ one finds an effective $g$ factor $g^*=-15$ for InAs bulk crystals \cite{InAsBulkG,EsrInAsG}. The large, negative $g$ factor results from the coupling of the electron spin with the orbital motion \cite{ROTH_spinorbit}. The effect is further enhanced by the small effective mass of InAs. For two-dimensional electron gases the confinement leads to a different spacing of the subbands which slightly compensates the enhancement of the effective Land\'e $g$ factor \cite{reducedGInWell,reductionGaAs,reductionWiresDots}. On the other hand, confinement in a quasi-one-dimensional NW was found to enhance spin-orbit coupling by this enhancing the effective electron $g$ factor even more \cite{soleGFactor,GeorgWireSOI}. \\



Material properties of InAs can be further enriched by interfacing it with a superconductor. Conventional superconductors above their critical temperature can be described as Fermi liquid with a continuum of excitations above the ground state \cite{fermiliquid}. This normal state of electrons exhibits an instability at the critical temperature, leading to the condensation of the electrons into the superconducting state \cite{zagoskin}. The ground-state of the superconductor is a condensate of Cooper pairs. Its excitations are superpositions of particle and hole with a finite excitation gap around the Fermi energy \cite{BCS}. 
By combining it with semiconducting InAs, one hopes to realize a novel phase of matter that inherits the spin-orbit coupling and effective $g$ factor from the semiconductor and a finite electron-hole symmetric pairing gap from the superconductor. The approach of combining unique materials of this kind to engineer novel quantum materials goes back to multiple theoretical proposals \cite{FUKANE,OREG,LUTCHYN}. Experimentally, it was found that Al can be grown by molecular-beam epitaxy on InAs with an epitaxially matched interface, a surprising fact considering the lattice mismatch between the two materials \cite{krogstrup_epitaxially}. Al undergoes a phase transition to a conventional $s$-wave superconducting phase when cooled below its critical temperature. For bulk Al the critical temperature  is $T_\mathrm{c}=\SI{1.2}{\kelvin}$ and the superconducting pairing gap is $\Delta \approx \SI{0.18}{\milli\eV}$. These values are slightly larger for thin Al films \cite{AD_AO, OA69}. In a magnetic field, the superconductivity of Al films with a thickness below the London penetration depth is destroyed by Pauli para-magnetism because orbital motion of the electrons can be neglected \cite{FuldeReview,fieldEnhancementTinkham,thinAlfilms}. This is true for magnetic fields parallel to the thin film. Consequently, Al thin films can show a high critical magnetic field at which a first order phase transition to the normal state occurs \cite{thinAl1storder}. 

The transfer of the pairing correlations of electrons from the superconductor to the electrons in the InAs is facilitated via the superconducting proximity effect, an effect discovered by Walther Mei{\ss}ner \cite{meissnerProximity}. It can be understood by considering an electron in the InAs at the Fermi energy impinging on the semiconductor-superconductor interface and undergoing Andreev reflection \cite{ANDREEV,andreevReflectionCourtois}. Due to the finite pairing gap there is no fermionic excitations at the Fermi energy in the superconductor. The electron can therefore not penetrate the superconductor. Instead, it is retro-reflected as a hole with opposite spin, while a Cooper pair is transferred to the superconductor. During this process, phase coherence of the electron and the retro-reflected hole is maintained, which leads to a build-up of the superconducting pairing of electrons in the semiconductor. This pairing is  described by the pairing term
\begin{equation}
H_\mathrm{SC} = \Delta \tau_x
\end{equation}
where $\tau_x$ is a Pauli matrix in particle-hole space. \\

When combining semiconducting InAs with superconducting Al, one hopes to find a new material system that combines large spin-orbit interaction, sizable $g$ factor, and a superconducting pairing gap. Experimental observations and advancements in the theoretical understanding suggest, however, that properties of the hybrid-material are not a simple sum of semiconducting and superconducting properties. Instead, a renormalization of the semiconductor properties arises due to the presence of the superconductor \cite{stanescuRenormalizatin, SDS_substrateRenormalization, AkhmerovRenormalization, GfactorRenormalizationAntipov}. This effect has been described as metallization of the semiconductor, as it leads to a decrease in spin-orbit coupling strength, $g$ factor, and Fermi wavelength \cite{metallizationLoss}. This results from the semiconductor bands being bent at the interface to the superconductor due to a work function mismatch. This leads to an accumulation layer in the semiconductor close to the superconductor \cite{Mikkelsen2018,unifiedNumericsWinkler, manfra_cyclotron}. An improved model for the proximity effect furthermore takes into account a strong coupling of the semiconducting modes to the superconductor. This model is applicable in the case of a highly transparent semiconductor-superconductor interface and tries to explains the renormalization of the semiconductor properties \cite{Stanescu_strongCoupling}. \\

A brief summary of the different material platforms that are commonly used to engineer superconductor-semiconductor hybrids is given in the following, before looking at the different quantum states that can emerge in proximitized semiconductors

\chapter{Promising material platforms}

To date, several promising material platforms exist that allow the fabrication of devices based on InAs semiconductor proximitized by superconducting Al.

\section{Nanowires grown by vapor-liquid-solid method}
Nanowire (NW) growth by molecular-beam epitaxy (MBE) has become an established technique. The growth of these NWs can be initiated by gold catalyst-particles on a substrate, leading to nanowire growth perpendicular to the substrate plane under the right growth conditions. The growth is understood by the vapor-liquid-solid (VLS) mechanism and can in principle yield stacking fault free single-crystalline InAs NWs. 
The possibility of growing Al epitaxially matched on InAs nanowires was first demonstrated with VLS nanowires \cite{krogstrup_epitaxially}. Recent improvements of the growth method enable NW crosses \cite{sabbir_crosses}, NWs with shadow junctions \cite{martin_shadow, sabbir_shadow}, and NWs that are fully covered by a shell of Al \cite{sole_fullshell, trivial_fullshell}. 

InAs NWs typically show a hexagonal cross-section with diameter around $\SI{100}{\nano\meter}$. It has to be assumed that several subbands are occupied in these NWs \cite{akhmerov_electrostatics, GfactorRenormalizationAntipov}. Hence, they have to be considered quasi-one-dimensional. The Al is grown \emph{in-situ} on the InAs surface with an epitaxially matched, highly transparent interface. The modes in the semiconductor are therefore strongly coupled to the superconductor resulting in an induced pairing gap in the semiconductor that is close to the one of superconducting Al film \cite{chang_hardgap}.

\section{Selective area grown InAs nanowire networks}
To allow for the scalable growth of InAs NW networks, selective area growth (SAG) techniques can be used \cite{krizek_SAG, sole_sag,palm_sag,martin_sag}. SAG predefines the desired shape of a NW network by a dielectric mask on a growth substrate (e.g., InP, GaAs). Using MBE growth the InAs network can be grown, followed by the \emph{in-situ} deposition of an epitaxially matched layer of Al. SAG allows for an engineered heterostructure instead of homogeneous NWs. This way, a buffer layer can be incorporated between the InAs channel and the substrate with the purpose of relaxing the strain in the crystal lattice. A semiconducting material with larger energy band gap than InAs can be used in between the InAs and the Al to tune the coupling between the superconductor and the modes in the semiconductor. Despite these promising features it is not possible to grow SAG InAs NWs with high crystal quality. The first studies pointed out the presence of stacking faults in particular at the crossing of NWs \cite{krizek_SAG} - a strong indication of the low quality of SAG NWs. Despite intense research efforts it remains a challenge to grow InAs SAG NWs with high selectivity and low density of defects. 

\section{InAs two-dimensional electron gases}
MBE growth can be used to grow InAs within a heterostructure of large bandgap materials. Charge carriers are confined in a quantum well forming a two-dimensional electron gas (2DEG) \cite{Beenakker1991}. InAs 2DEGs that are deep under the surface of the material can be produced with high electron mobility. Using more shallow InAs 2DEGs leads to a compromise in electron mobility as the electrons undergo scattering at the surface. On the other hand, shallow InAs 2DEGs allow for superconducting proximity effect of the InAs quantum well by an epitaxially matched layer of Al on the surface of the 2DEG wafer \cite{javad_2deg, pauka_repairingInAs}. Naturally, this approach allows for a material in between the InAs and the Al as part of the heterostructure. A tunable hybridization between the superconductor and the semiconductor is therefore possible. 

Material of this type offers the highest material quality for superconductor-semiconductor hybrid devices today in terms of electron mobility \cite{SDS_mobility}. It is furthermore possible to access the electron mobility of the semiconductor via a Hall-bar measurement. An electron peak mobility of $\SI{25000}{\square\cm\per\volt\s}$ can be reached reproducibly \cite{javad_2deg, pauka_repairingInAs, AD_AO, higgin_pip, manfra_cyclotron}. Top-down fabrication techniques allow for the Al to be shaped arbitrarily in a scalable way. Lithographically defined electrostatic gates for lateral confinement of electrons in the 2DEG enable numerous hybrid devices for the study of quantum states in proximitized semiconducting devices.

\chapter{Quantum states in proximitized Indium Arsenide}

Devices based on InAs nanostructures proximitized by superconducting Al allow for the exploration of phenomena that arise due to spin-orbit coupling, Zeeman energy, and a finite pairing gap. These three physical phenomena combined with the possibility to tune the electron density in the semiconductor by gate electric fields opens the door to an entire jungle of quantum states. 

Andreev bound states (ABSs) are a broad range of states that emerge in proximitized semiconductors, leading to a finite density of states below energies of the superconducting gap. Several physical mechanisms can give rise to these subgap states. Semiconducting quantum dots (QDs) that are coupled to a superconductor are an example, that has been realized in various experiments with VLS NWs \cite{lee_singletdoublet,Binsensitive,kasper_tuningYSR,kasper_two_impurity,coulombic_states}. In this case, discrete energy levels of the QD couple to the superconductor. Depending on the size of the superconducting pairing, the strength of the coupling to the superconductor, the spin of the electrons on QD, and the charging energy of the QD, the bound states show different characteristics \cite{Pillet2013,QD_kondo_regime,lee_scaling}. Theoretically, these states have been treated in the single impurity Anderson model, where a single spin couples to the superconducting electrons around it \cite{gorm_doubledot, RUSINOV_YSR, YU_YSR, SHIBA_YSR}. Experimental realizations in hybrid devices allow for tuning of the couplings to the leads, the spin state of the QD, and the number of electrons confined in the QD \cite{kasper_ysr, kasper_two_impurity, kasper_tuningYSR, chang_hardgap} . 

Additional richness is added to the problem by a magnetic field, that affects the spinful energy levels of the QD and decreases the size of the superconducting gap \cite{lee_singletdoublet, lee_scaling}. Several experiments added a phase difference to the problem by embedding a QD in a superconducting loop \cite{pillet_CNT,alex_QDinLoop,chang_YSR, vanDam}. A natural extension of these experiments are Josephson junctions, in which a supercurrent in a weak link is carried by ABSs \cite{nichele_ABS_Ic}. Here, the formation of ABSs can be understood as consequence of multiple resonant Andreev reflection at the two interfaces of the superconductor. 2DEG based devices were able to relate the supercurrent to the ABS spectrum in the Josephson junction. Circuit quantum electrodynamics experiments were furthermore able to exploit ABSs in Josephson junctions as nonlinear elements \cite{Hays_ABS_dynamic, Lange_NW_cqed, thorvald_NW_cqed,Anders_fullshell_cqed, Anders_leaded_cqed}.

NWs that are proximitized by a superconductor that covers their whole length can also host ABSs \cite{henri_lead,Binsensitive}. The amount of disorder, crystal defects, confining potential at the NW ends, and coupling strength between the modes in the semiconducting NW and the superconductor have a strong influence on the nature of the ABSs. For NWs with a sufficiently low level of disorder the ABSs are expected to evolve into well seperated Majorana zero modes localized at the ends of the NW under certain conditions.

\section{A route to topological states of matter}\label{ch:introduction}
It turns out, that low-dimensional, proximitized semiconductors with spin-orbit interaction in a magnetic field may give rise  to a topological phase of matter that possesses a bulk with a finite topological energy gap and edge modes at its ends. This connection is revealed by looking at a simple toy model before making the transition to a real world material system.

\subsection{A simple toy model}
The Kitaev chain, named after its creator Alexei Kitaev, is a one-dimensional tight binding model of spinless electrons with nearest neighbor interaction \cite{KITAEV}. The Hamiltonian is given by  
\begin{align}
\label{eq:kitaevham1}
H=\sum_{j}{-t(c_j c^\dagger_{j+1} + c_{j+1} c^\dagger_j)} -\mu (c_j^\dagger c_j+\frac{1}{2}) + \Delta c_j c_{j+1} + \Delta^* c_{j+1}^{\dagger} c_j^\dagger
\end{align} 
where $t$ determines the hopping strength on the lattice, $\mu$ is the chemical potential, and $\Delta$ is a complex pairing amplitude. $c_j^{\dagger}$ ($c_j$) is the creation (annihilation) operator of an electron at site $j$. The last two terms in the Hamiltonian are of the form of a superconducting pairing between spinless electrons of neighboring sites. This corresponds to $p$-wave pairing.

The model can be simplified by rewriting the complex electron creation and annihilation operators using real valued Majorana operators. The phase $\theta$ of the complex pairing amplitude $\Delta = |\Delta|  \mathrm{e}^{\mathrm{i}\theta}$ is included in the definition of the Majorana operators
\begin{align}
\gamma_{2j-1} = \mathrm{e}^{\mathrm{i} \theta/2} c_j + \mathrm{e}^{-\mathrm{i} \theta/2} c_j^\dagger && \gamma_{2j} = -\mathrm{i}\mathrm{e}^{\mathrm{i} \theta/2} c_j + \mathrm{i} \mathrm{e}^{-\mathrm{i} \theta/2} c_j^\dagger 
\end{align}

For a range in chemical potential near $\mu=0$, Majorana operators of neighboring sites are coupled and contribute to the energy of the system; the Majorana operators $\gamma_1, \: \gamma_{2N}$ at the ends of the chain drop out of the Hamiltonian. The  fermion, formed by the unpaired Majorana operators at the ends, can therefore be occupied without any cost in energy and is called Majorana fermion. This leads to a two-fold degenerate ground state at zero-energy. \\

\begin{figure}[h]
\begin{center}
\includegraphics[scale=0.9]{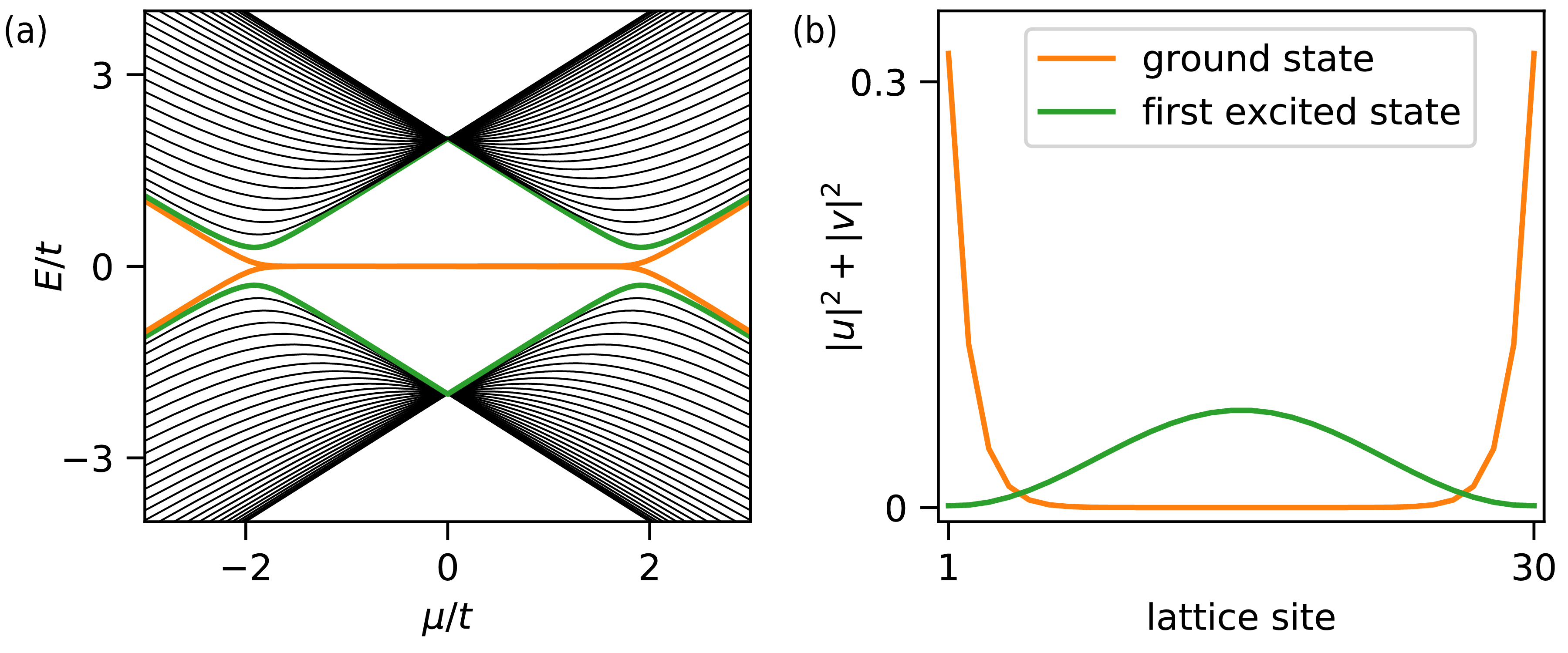}
\end{center}
\caption[Spectrum and wavefunction of the Kitaev chain]{\label{fig:kitaev_spec}Tight binding model for a Kitaev chain with $N=30$ sites and $|\Delta|=t$. (a) spectrum of eigenenergies. For a extended range of chemical potential a zero-energy state is formed by the ground state (orange). The first excited state is at finite energy. (b) Squared wavefunction amplitude of the zero-energy ground state and the first excited state along the chain for $\mu=1.2t$, calculated from the particle (hole) components $u$ ($v$) of the states.}
\end{figure}

In order to illustrate the spectrum of the system as a function of the chemical potential $\mu$ in units of the hopping strength $t$, the respective Bogoliubov-de Gennes Hamiltonian of eq. \ref{eq:kitaevham1} is diagonalized numerically on a lattice. The spectrum of eigenenergies is depicted in Fig.~\ref{fig:kitaev_spec}(a). For $|\mu|<2t$ there is a zero-energy eigenvalue. Notably, the energy of the first excited state reaches minima at the points $|\mu| = 2t$ leading to the appearance of a closing and reopening of the spectral gap. The wavefunctions of the groundstate together with the first excited state for $\mu=1.2t$ are plotted in Fig.~\ref{fig:kitaev_spec}(b). The ground state wavefunction is localized at the chain ends with an exponential decay into the bulk of the chain. These are the two Majorana zero modes (MZMs) localized at edges of the system. The first excited state extends through the whole system with vanishing weight at the boundaries. The MZMs at the two ends of the chain together form a Majorana fermion. These MZMs follow non-Abelian particle statistics.\\

In the early days of quantum mechanics in the 1930s, it was argued that there are two kinds of particles: fermions and bosons \cite{fierz, pauli}. The distinction between the two lies in their spin. Fermions have half-integer valued spin, whereas bosons have integer valued spin. This fundamental argument got refined over time and lies at the heart of the spin-statistic theorem \cite{schwinger}. Bose statistics predicts a wavefunction $\ket{\Phi\Psi}_\mathrm{bose}$ that is symmetric under exchange of two particles whereas Fermi statistics predicts a wavefunction $\ket{\Phi\Psi}_\mathrm{fermi}$ that is anti-symmetric under particle exchange

\begin{align*}
\ket{\Phi\Psi}_\mathrm{bose} = \ket{\Psi\Phi}_\mathrm{bose} && \ket{\Phi\Psi}_\mathrm{fermi} = -\ket{\Psi\Phi}_\mathrm{fermi}
\end{align*}

This is the underlying cause for the the Pauli exclusion principle.

For particles in three dimensions, this strict classification as fermions and bosons has passed numerous experimental tests and is unchallenged until today. When it comes to low-dimensional systems nature has a surprise in store. Namely, particles that are neither bosons nor fermions may emerge in one- and two-dimensional systems \cite{Leinaas}. These exotic particles, called \emph{any}ons \cite{wielszek}, can have \emph{any} fractional value of spin $s$ and the two particle wavefunction $\ket{\Psi\Phi}_\mathrm{any}$ will follow the exchange statistic 
\begin{align}
\ket{\Phi\Psi}_\mathrm{any} = \exp(2 \pi \mathrm{i} s)\ket{\Psi\Phi}_\mathrm{any}.
\end{align}
This means, the exchange of two anyons results in the wavefunction acquiring a phase factor of $2\pi s$ - a process referred to as braiding. MZMs turn out to be such anyons. The wavefunction will therefore change, if two MZMs are exchanged. Moreover, for MZMs this will not only result in a phase factor but instead in a unitary operation acting on the wavefunction. The order of consecutive exchanges of MZMs will therefore affect the outcome of the operation. This makes MZMs non-Abelian anyons. The presence of MZMs at the ends of the Kitaev chain for small values of chemical potential and the absence of the same at high values of chemical potential expresses the topological difference between the two possible ground state phases. \\

A strong motivation to investigate non-Abelian anyons is their potential application in the context of quantum information processing. Information may be encoded in the phase of anyons, manipulation can then be performed by moving the anyons around each other, and the information is read-out by fusing the anyons \cite{CHETAN, FREEDMAN, kitaev_faulttolerant}. Storing quantum information this way is in theory protected from decoherence by the nonlocal character of the MZMs. The realization of a one-dimensional system that hosts MZMs would therefore not only mean a major breakthrough for condensed matter physics but also an important leap for quantum technology.

\subsection{Towards realization: Lutchyn-Oreg model}

The Kitaev chain model leads - despite its simplicity - to the rich physics of non-Abelian MZMs, which may be applied one day in a topological quantum computer. Nevertheless, the model has severe limitations as it assumes spinless electrons and a $p$-wave pairing of electrons. Yet today, there is no known materials that shows these properties conclusively. An alternative model system that supports MZMs makes use of spin-orbit interaction, Zeeman energy, and proximity effect in one dimension. It goes back to the two independent proposals by Lutchyn, Sau, Das Sarma \cite{LUTCHYN} and Oreg, Rafael, von Oppen \cite{OREG}. \\

It is convenient, to inspect this model in the Nambu space, where wavefunctions are given by four-vector spinors $\Psi (x) =  (u_\uparrow(x), u_\downarrow(x), v_\downarrow (x), -v_\uparrow (x))^T$. This automatically incorporates the particle-hole and spin degree of freedom. The Bogoliubov-de Gennes Hamiltonian $H$ is given by
\begin{align}
\label{eq:lutchynoreg}
H&= H_\mathrm{kin} + H_\mathrm{SO} + H_\mathrm{Z} + H_\mathrm{SC}\\
 &=\left(\frac{p_z^2}{2m^*}-\mu \right) \tau_z + \alpha p_z \sigma_x \tau_z + V_\mathrm{Z} \sigma_z + \Delta \tau_x
\end{align}
where the matrices $\tau_i$ and $\sigma_i$ act in particle-hole and spin space respectively. 
The first term $H_\mathrm{kin}$ is the kinetic energy of electrons of effective mass $m^*$ and chemical potential $\mu$ confined to a one-dimensional mode propagating in the $z$-direction. The second term $H_\mathrm{SO}$ describes a Rashba spin-orbit coupling with spin-orbit interaction strength $\alpha$. The term $H_\mathrm{Z}$ is a Zeeman term of strength $V_\mathrm{Z}$. The last term $H_\mathrm{SC}$ is an $s$-wave pairing with a pairing amplitude $\Delta$.  As described in Chapter \ref{ch:InAs}, all of these terms may be realized in InAs NWs proximitized by superconducting Al in a magnetic field. 

The combination of spin-orbit interaction and Zeeman energy is of particular importance. The two terms do not commute in spin space, and lead to a mixing of different spin components. Effectively, one can achieve modes that have on average zero spin, which makes the bridge to Kitaev's original spin-less model.\\


\begin{figure}[h!]
\begin{center}
\includegraphics[scale=1.1]{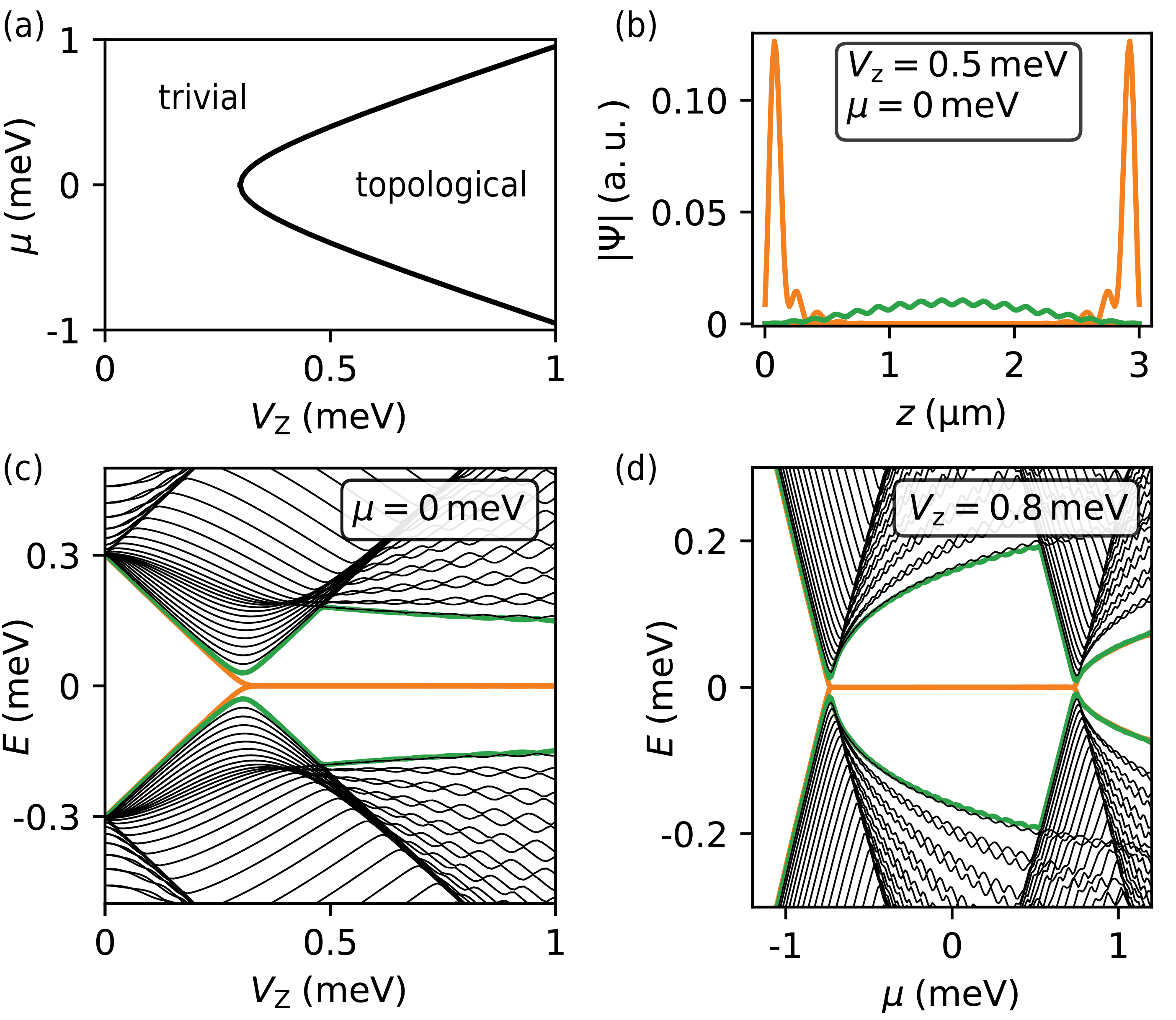}
\end{center}
\caption[Spectrum and wavefunctions of the Lutchyn-Oreg model]{\label{fig:lutchyn_oreg_panel}Tight-binding model with 200 sites for a NW with length \SI{3}{\micro\meter}, $\alpha=\SI{20}{\milli\eV\nano\meter}$, $\Delta = \SI{0.3}{\milli\eV}$, $m^*=0.023 m_\mathrm{e}$. (a) Phase diagram with parabolic phase boundary. (b) Wavefunction magnitude along the NW for the ground state (orange) and first excited state (green). (c) Eigenenergies as function of Zeeman energy $V_\mathrm{Z}$ reveal the formation of a zero-energy state. (d) Eigenenergies as function of chemical potential $\mu$ show the stability of the zero-energy state.}
\end{figure}

The spectrum of the Hamiltonian from eq. \ref{eq:lutchynoreg} can be solved exactly \cite{OREG, LUTCHYN} and shows a closing and reopening of the spectral gap. This closing and reopening marks the topological phase transition and appears at chemical potential $\mu$ given by
\begin{align}
\mu^2=V_z^2-\Delta^2
\end{align}
The resulting phase diagram is shown in Fig.~\ref{fig:lutchyn_oreg_panel}(a). In order to illustrate the emergence of the zero-energy state in the topological phase, an example of the energy eigenvalues as a function of Zeeman energy $V_\mathrm{Z}$ is shown in Fig.~\ref{fig:lutchyn_oreg_panel}(b). The zero-energy state is present in a finite range of chemical potential [see Fig.~\ref{fig:lutchyn_oreg_panel}(d)] and disappears at the points were the spectral gap of the system closes. The wavefunction of the zero-energy state is localized in the form of two modes at the two ends of the system, whereas the first excited state wavefuntion shows a finite weight throughout the whole system  [see Fig.~\ref{fig:lutchyn_oreg_panel}(c)]. This is analogous to the Kitaev chain with the only difference of a periodic modulation, which arises due to the spin and particle-hole structure of the states in the presence of spin-orbit coupling and magnetic field \cite{phas_transition_fate, elsa_nonlocality, MZM_polarization}.

\subsection{Signatures of Majorana zero modes}

Inspired by the Lutchyn-Oreg model, multiple experiments attempted to realize a topological phase hosting MZMs in semiconducting NWs proximitized by a superconductor. Semiconducting InAs in combination with superconducting Al has been among the used material combinations. In such systems, multiple signatures of MZMs have been predicted by theory, most of which have been observed in experiments. Based on the current status quo of this research field, none of these experiments could convincingly distinguish between MZMs and trivial ABSs, as both can give rise to the same experimental signatures. \\

One of the key signatures of MZMs is the observation of the zero-energy state in tunneling spectroscopy as a conductance peak at zero source-drain bias voltage. It should reach a quantized value of $2\,e^2/h$ in differential conductance \cite{quant_theory_1, Akhmerov2011, wimmer_qpc, nichele_scaling, frolov_quantized,mourik, mingtang_science, sole_fullshell}. However, ABSs due to disorder, QDs at the end of the NW, or quasi-Majoranas (ABSs that form due to a soft confining potential at the NW end) can give rise to zero-bias states that are stable with respect to experimental parameters \cite{vuik_quasiMBS, hess_nl_quasimajo, SDS_against_hao,dassarma_ABSvsMBSspectroscopy, dassarma_goodbadugly, dassarma_against_mingtang, lee_scaling, lee_singletdoublet, elsa_ABSvsMZM, Altland_Dpeak}. The zero-bias state should furthermore appear correlated at both NW ends after a closing and reopening of the energy gap in the bulk, which has never been observed \cite{SDS_nl_conductance, das_sarma_correlation, dassarma_gapclosing}. Experiments in this thesis try to address this issue.

Further signatures expected for MZMs are phase-coherent transport of $1e$ quasiparticles \cite{alex_interfero, Fu2010}, oscillations around zero-bias for shorter NW as a function of chemical potential and magnetic field \cite{dassama_meta, albrecht, dassarma_against_sven}, $4\pi$ Josephson effect \cite{FUKANE2, leo_4pi}, and topological Kondo effect \cite{topo_kondo_cooper}. In time domain measurements, MZMs should furthermore show distinct behavior due to fusion and coherent coupling \nolinebreak{of MZMs \cite{AASEN, sabonis_fullshell}.} \\

A significant body of theoretical and experimental work has identified several hurdles associated with the realization of MZMs in proximitized semiconducting NWs. These include multiple bands in the NW \cite{aguado_multiband,potter_multiband, jelena_realistic_model}, renormalization of the semiconductor properties due to the strong coupling to the superconductor \cite{metallizationLoss, stanescuRenormalizatin,SDS_substrateRenormalization, Antipov2018, akhmerov_electrostatics, AkhmerovRenormalization}, closing of the superconducting pairing gap at a magnetic field value below the topological phase transition, and  strong disorder in the used materials \cite{Lobos2012, SDS_impurity, SDS_mobility}.

\chapter{Novel device geometries in this work}

In this thesis, several novel device geometries are presented. Their use is to investigate bound states in semiconducting InAs NWs proximitized by superconducting Al with the goal of developing a better understanding of hybrid materials and their quasiparticle excitations. The work herein contributes to the development of experimental methods for the realization and characterization of a topological phase.\\

\begin{figure}[h!]
\begin{center}
\includegraphics[width=0.95\textwidth]{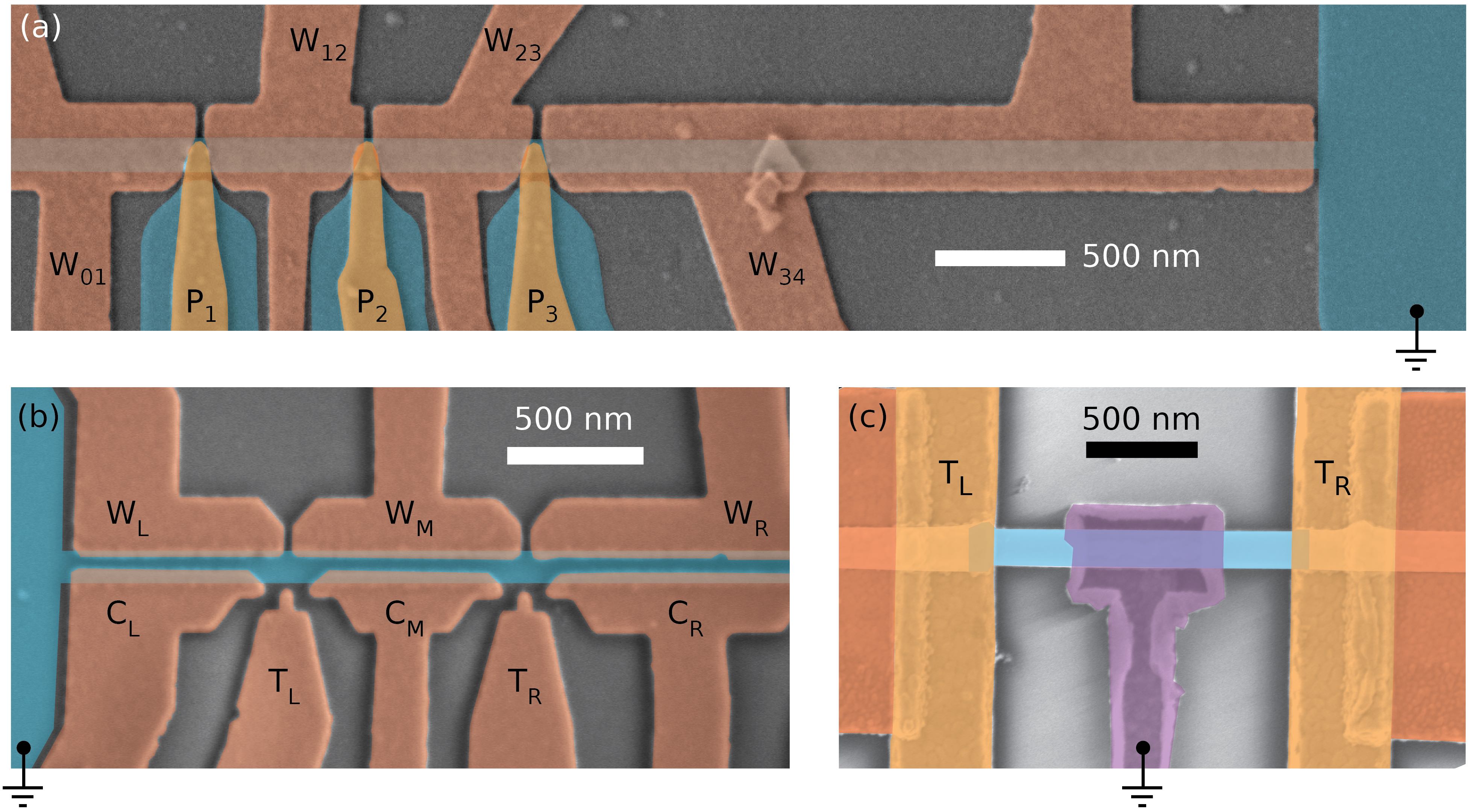}
\end{center}
\caption[Novel device geometries in this thesis]{\label{fig:dev_over}False-color micrographs of novel device geometries in this thesis. (a) NW electrostatically defined in InAs 2DEG by gate electrodes (red), proximitized by Al. Side probes from Al (blue) are coupled to the wire through tunnel barriers that can be tuned by a gate (orange). (b) gate-defined NW based on InAs 2DEG proximitized by Al. Semiconducting leads under gates $\mathrm{T_\mathrm{L,\, R}}$ are coupled from the side to the NW through a quantum point contact formed by gates $\mathrm{C_\mathrm{L,\, M,\, R}}$. (c) InAs NW with Al full-shell (blue) which is contacted with an Al lead (purple). Both NW ends are contacted by Au and have a gate tunable tunnel barrier. }
\end{figure} 

The first novel device geometry that will be presented in the Chapters \ref{ch:pradaclarke}, \ref{ch:2DEG_nonlocal}, and \ref{ch:sideprobe} are gate-defined NWs based on shallow InAs 2DEGs that are proximitized by an epitaxially matched film of superconducting Al. The superconducting Al is grounded and tunnel probes are coupled from the side to the NW. This allows for local density of states measurements at discrete points along a the NW using tunneling spectroscopy. Two device geometries are shown in Fig.~\ref{fig:dev_over}(a, b). One of them uses superconducting sideprobes that are separated by a gate tunable tunnel barrier from the NW. The second device geometry makes use of semiconducting leads as tunnel probes that are coupled through a quantum point contact to the NW. 

The control over electron density in individual segments of the NW allows for the confinement of ABSs in selected segments of the NW. In Chapter \ref{ch:pradaclarke} we present how extended ABSs appear correlated in tunneling spectroscopy on neighboring tunnel probes that are spaced apart by $\lesssim\SI{0.8}{\micro\meter}$. We furthermore hybridized a QD resonance in the tunnel barrier at one probe location with the ABSs and observed the effect on the bound state on the other probe. 

The parent superconductor of the proximitized NW is grounded in this device geometry. The wire has therefore no charging energy and bound states can be characterized by measuring local and nonlocal differential conductance. Nonlocal conductance spectroscopy of ABSs is presented in Chapter \ref{ch:2DEG_nonlocal}. At magnetic fields $\gtrsim\SI{1}{\tesla}$, intersecting ABSs create a low-energy state that oscillates around zero bias. Its charge character was determined using nonlocal conductance and agreed with its energy evolution as a function of gate voltage.

Discrete tunneling spectroscopy measurements can reveal the extent of bound states in the 2DEG based NW. Using tunneling probes at the ends and at the center of NW segments longer than $\SI{0.8}{\micro\meter}$ showed no extended bound states. The tunneling spectroscopy data of localized bound states at the ends of the NW show only accidental correlations, which is contrasted with the data from bound states in shorter NWs in Chapter \ref{ch:sideprobe}. 

Similar devices have been investigated numerically \cite{dassarma_circuits} and experimentally implemented using conventional VLS NWs \cite{grivnin_multiprobe}. The devices presented here may become a helpful tool for the characterization of topological phases in clean 2DEG NWs where a direct measurement of the bulk gap by means of a direct tunnel probe or nonlocal conductance is desirable \cite{das_sarma_qpgaps, close_or_not, das_sarma_qpgaps, andreev_rectifier, SDS_nl_conductance, hess_nl_quasimajo}. \\

Results from a device geometry based on VLS grown InAs NWs with a full-shell of Al are presented in Chapter \ref{ch:fullshell_nonlocal}. The device consists of a proximitized NW that is contacted at both ends to normal leads via gate tunable tunnel barriers [see Fig.~\ref{fig:dev_over}(c)]. The superconducting shell of the NW is grounded via a third contact. The resulting three-terminal device geometry allows for measurements of local and nonlocal conductance. The spectra of subgap states in local tunneling conductance at the two NW ends are similar in with respect to the number of conductance resonances. The overall small nonlocal conductance, lack of correlations in source-drain voltage, and hybridization with a local QD resonance rule out extended bound states as origin of the subgap states. The data point towards localized bound states at the NW ends with a spectrum dependent on the cross sectional shape of the NW.


\cleardoublepage

\part{Materials and Methods}\label{pt:matandmeth}

\chapter{Indium Arsenide/Aluminum hybrid materials}
Superconductor-semiconductor hybrid materials were used in this thesis. The semiconductor of choice was InAs, as it is known for its strong spin-orbit coupling and sizeable g-factor.

\section{Two-dimensional heterostructures}
The work in Chapters \ref{ch:pradaclarke}, \ref{ch:2DEG_nonlocal}, and \ref{ch:sideprobe} is based on heterostructures hosting a two-dimensional electron gas (2DEG). The material was provided by the Manfra group at Purdue University. Similar InAs/Al hybrid heterostructures have been developed in previous works \cite{javad_2deg, HATKE_inas, CANDICE_inas, InSbAs_Delft}.\\

\begin{figure}[h!]
\begin{center}
\includegraphics[width=0.98\textwidth]{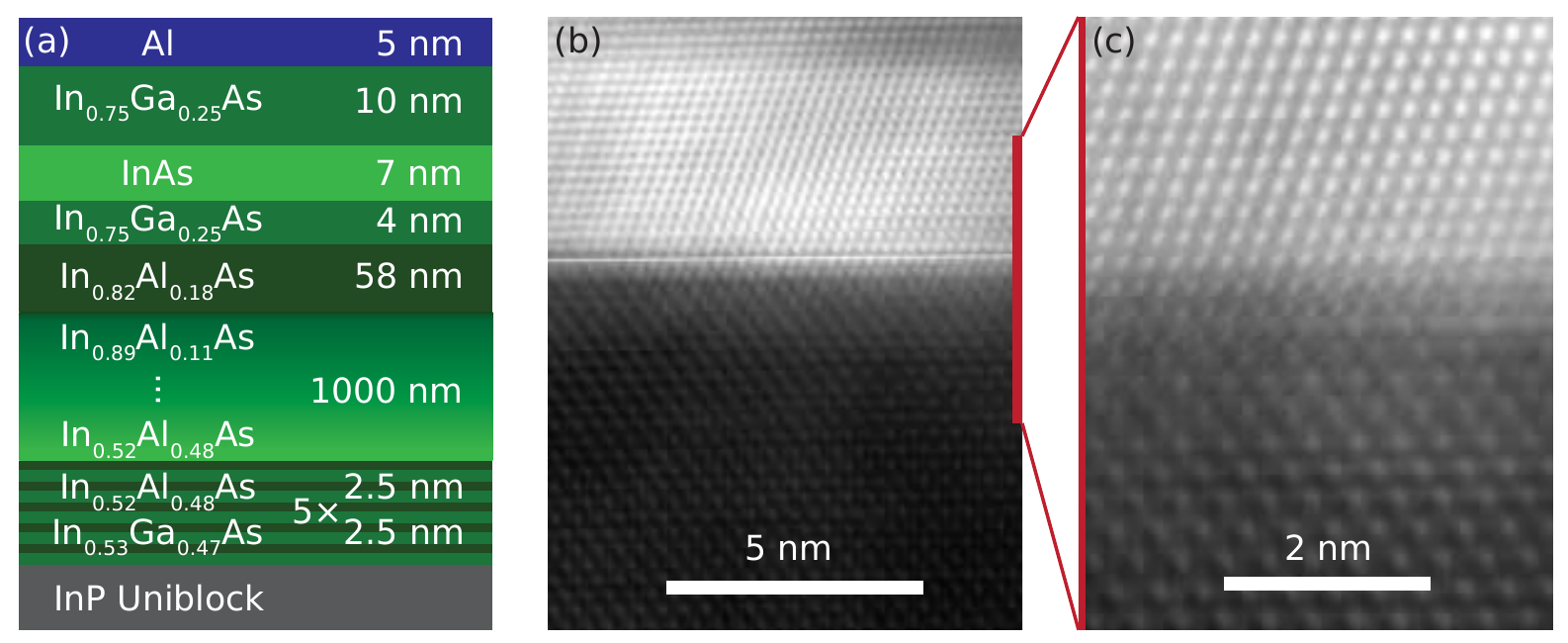}
\end{center}
\caption[Schematic and micrograph of two-dimensional superconductor-semiconductor hybrid heterostructure]{\label{fig:waferstack}InAs/Al hybrid-heterostructure. (a) Schematic overview of the used semiconductor heterostructure with a superconductor top layer. (b, c) Transmission electron micrographs showing the semiconductor-superconductor interface. Micrographs are provided by the Manfra group.}
\end{figure}

The 2-inch wafers were grown using molecular-beam epitaxy (MBE). Starting from an InP substrate, an InAlAs-InGaAs superlattice is grown, serving as trap for dislocations. Consequently, a buffer with a concentration gradient was grown to improve the lattice match. The 2DEG is confined in an InAs well in the upper layers of the heterostructure. The quantum well is confined by a top and bottom barriers with larger bandgap (e.g., InGaAs, InAlAs). A monolayer of GaAs was deposited before the \emph{in-situ} deposition of Al to prevent Al atoms from diffusing into the top barrier of the quantum well. Figure \ref{fig:waferstack} shows a schematic and transmission electron micrograph of the resulting heterostructure.\\

A successful material growth results in strain relaxation and a continuous Al film. The strain relaxation is seen from the surface morphology as hatch pattern in optical dark field microscopy \cite{MortenThesis}. Defects in the Al film are well visible in bright field optical microscopy and appear frequently for thin Al films with a thickness $\lesssim \SI{5}{\nano\meter}$. Atomic force microscopy can be used to measure the surface roughness of the material after the deposition of the superconductor. A surface roughness of $\approx\SI{1}{\nano\meter}$ root-mean-square is typically observed. Standard Hall effect measurements of the electron mobility at base temperature of a dilution refrigerator typically yield peak mobility values in the range \SIrange{20000}{90000}{\square\centi\m\per\volt\per\second}. The peak is typically reached at a electron density $n=\SI{1e16}{\per\square\meter}$. This corresponds to a Fermi wavelength $\lambda_\mathrm{F}=\SI{25}{\nano\meter}$ and mean free path $l_\mathrm{e}=\SI{0.8}{\micro\meter}$. 
The peak mobility varies from growth to growth and depends on the details of the heterostructure \cite{AD_AO,pauka_repairingInAs}. Heterostructures with a thicker barrier layer between the InAs and the Al yield a higher mobility \cite{HATKE_inas}. However, with thicker barriers the superconducting proximity effect and the size of the induced superconducting gap is reduced.

\section{Fullshell nanowires}
In Chapter \ref{ch:fullshell_nonlocal}, results from an experiment using nanowires (NWs) are presented. The NWs were provided by the Krogstrup group at the University of Copenhagen.

\begin{figure}[h!]
\begin{center}
\includegraphics[scale=0.7]{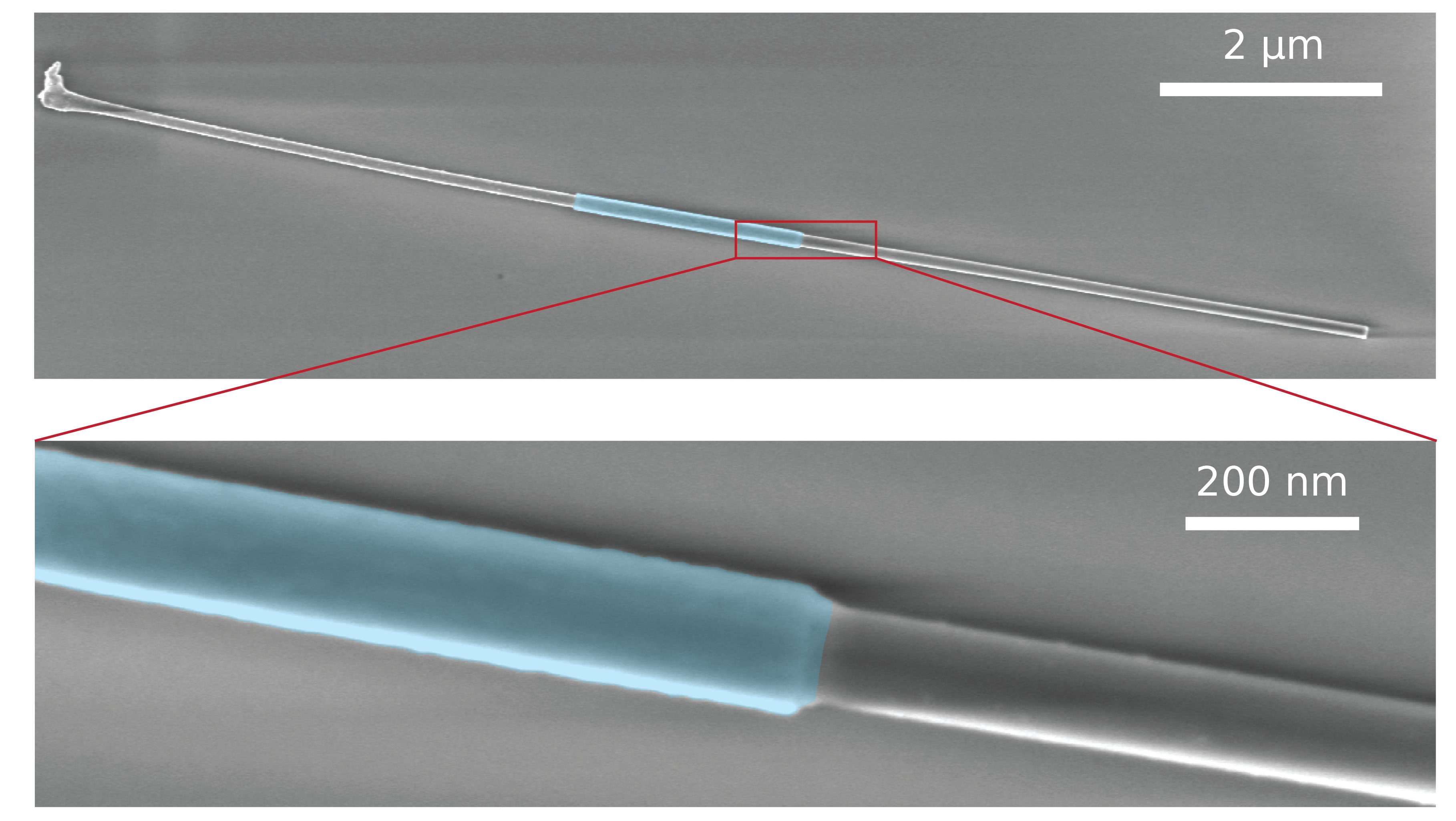}
\end{center}
\caption[Micrograph of InAs nanowire with Al full-shell.]{\label{fig:fullshell_sem}False-color micrograph of a VLS InAs NW with a full-shell of Al (blue). The Al was partially removed by wet etching.}
\end{figure}

The NWs were MBE grown on a (111)B InAs substrate using the vapor-liquid-solid (VLS) method \cite{krogstrup_epitaxially, sole_fullshell}. A Au catalyst particle was used to seed the growth; its size was a tuning parameter for the final NW diameter. The NW were grown with (0001)B wurtzite orientation at  $\SI{420}{\celsius}$. The Al was grown in-situ in the same MBE chamber at $\SI{-30}{\celsius}$ while rotating the substrate to cover all NW facets. The NWs have a typical length of \SI{10}{\micro\meter} and are transferred to a silicon substrate using a micro-manipulator proceeding device fabrication. Figure \ref{fig:fullshell_sem} shows an electron micrograph of a full-shell NW.

\chapter{Device fabrication}
Nanofabrication is the key to turn superconductor-semiconductor materials into quantum devices. The fabrication processes for different material platforms have been optimized over the past decade at the Center for Quantum Devices \cite{FedeThesis, FabioThesis,ChangThesis, MortenThesis, AlexThesis, AsbjornThesis, Razmadze2020}. The techniques applied in this thesis are discussed in this chapter. Fabrication procedures beyond the current standard process are pointed out. Detailed protocols for specific devices presented in this thesis can be found in Appendix \ref{ch:fab}. 

Nanofabrication is a process that takes place in a cleanroom environment. To minimize the risk of contamination, it is common practice to handle samples with tweezers that are always kept in the cleanroom. Moreover, a designated pair of tweezers for wet chemical processing, etching, and for the use on the hotplate may be used. Before using a beaker for a chemical, it should be rinsed with the respective chemical to remove dust particles and residues. The use of optical microscopy images helps to evaluate the success of process steps. It furthermore fulfills the purpose of documentation and enables debugging in the case of potential problems during the fabrication process. 

\section{Electron-beam lithography}
\begin{figure}[h!]
\begin{center}
\includegraphics[width=0.9\textwidth]{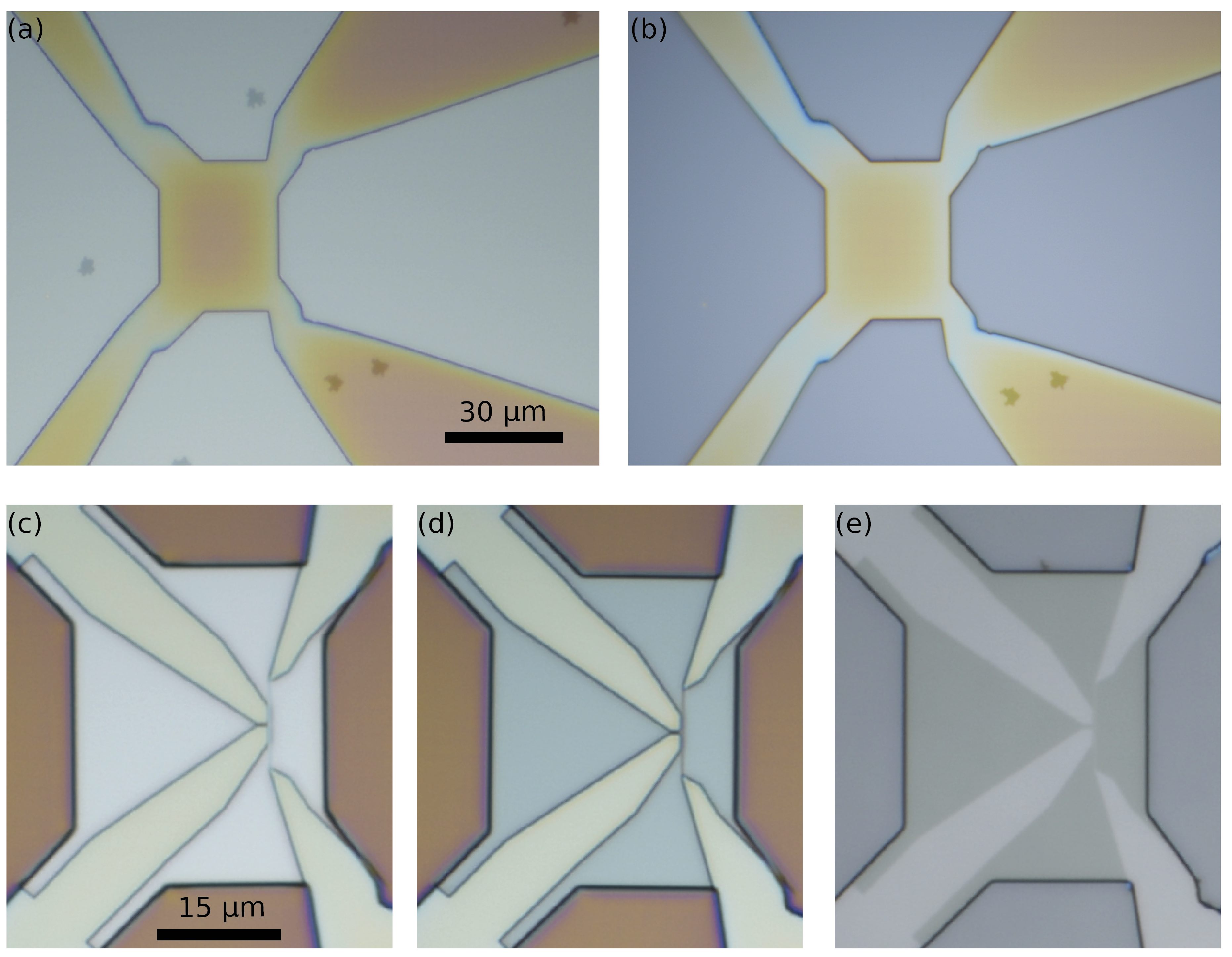}
\end{center}
\caption[Optical microscope images of at different stages of wet etching]{\label{fig:etch}Optical microscope images at different stages of the Al and mesa wet etch. (a) Exposed pattern for the mesa etch after development. Exposed Al appears light blue and shows some defects. (b) After the Al and semiconductor mesa etch, the etched semiconductor appears dark blue. (c) Exposed pattern for the Al etch of a grounded NW with two sideprobes.  (d) Bare semiconductor appears blue after the reflective Al was removed by wet etch. (e) Same as (d) after stripping of resist and ALD growth of dielectric.}
\end{figure}

All fabrication steps in this work made use of electron-beam lithography (EBL). This process consists of coating the chip with a suitable resist and exposing the desired pattern with an EBL tool (Elionix ELS-500 100kV system). After the exposure, development follows. Herein, positive resist was used exclusively which means exposed areas of resist are dissolved during development. An optical image of a developed pattern is depicted in Fig.~\ref{fig:etch}(a).\\

An optimal lithography result is achieved with the right choice of resist, exposed dose, and development. A suitable resist for most fabrication processes on 2DEGs is PMMA A4. For finer features a  higher resolution can be achieved with A3, because it is thinner. A2 was tried out as well, but did not cover the slanted walls of the mesa structure properly. 
A thicker resist stack of two layers copolymer EL9 and a single layer PMMA A4 is used for the lithography step defining the bond pads and connecting lines to gates that have to crawl up the mesa walls. 

The suitable dose for the exposure of a process is best determined in a dose test \cite{FedeThesis}. The software Beamer was used to produce machine files for lithography. The proximity effect correction of Beamer was used to account for the exposure by secondary electrons. Multipass exposure can be used to achieve reproducable lithography result for fine features (size $\lesssim \SI{50}{\nano\meter}$). A suitable development can be done in MIB:IPA solution at room temperature. Cold development at \SI{-5}{\celsius} in IPA:H20 solution has been tested \cite{FabioThesis, FedeThesis}. It did not shown a significant improvements in comparison to the room temperature development for the device types presented in this thesis.

\section{Wet etch}

Wet etch was used to selectively remove Al and to define a mesa profile in 2DEG chips. For the Al etch, the deisred pattern is defined using EBL. After development the chip undergoes oxygen plasma cleaning in an asher. The resist is reflown for \SI{2}{\min} on a hotplate at \SI{120}{\celsius}. Four beakers are required for the Al etch: One beaker containing Transene D Al etchant and two beakers containing ultra pure water. The conductivity of the water should be $\lesssim \SI{0.06}{\micro\siemens}$. All three beakers are placed in a hot bath. The level of liquid should be the same in all three and should be aligned with the liquid level of the hot bath. One of the two beakers containing water is used to measure the temperature, it should be at $\SI{50\pm0.1}{\celsius}$. The fourth beaker is filled with ultra pure water and is placed outside the hot bath. The etch is performed by dipping the chip \SI{5}{\second} in the Transene D, \SI{20}{\second} in the hot water beaker that was not used to measure the temperature, and \SI{40}{\second} in the water beaker outside the hot bath. During each step the chip should be swirled rigorously. The process ends with blow drying the chip using a nitrogen gun. 

Al strips with a width as small as \SI{100}{\nano\meter} on InAs 2DEG can be fabricated reproducibly using this procedure. Note that it is important to perform the Al etch with precise timing at the correct temperature. It is advisable to train the etch on scrap material with the correct Al film thickness, before attempting device fabrication. \\

For wet etching a mesa profile, the desired shape was defined using EBL. The top layer of Al was removed using an Al etch as described above. The mesa profile was etched using a solution $\mathrm{H_2O:C_6H_8O7:H_3PO_4:H_2O_2}$ (220:55:3:3) at room temperature. Note that $\mathrm{H_2O_2}$ has to be stored in the dark in a fridge to avoid decomposition of the $\mathrm{H_2O_2}$. The solution is mixed using a magnetic stir which is kept on during the etch process. For the etch, the chip is submerged for \SI{9}{\min} in the solution and rotated by \SI{90}{\degree} every \SI{30}{\second}. Subsequently, the chip is rinsed in ultra pure water for \SI{1}{\min} under rigorous swirling followed by blow drying. A typical etch depth of \SI{300}{\nano\meter} should result and can be checked using optical interferometric microscopy or a stylus profilometer. 

Alternatively a solution $\mathrm{H_3PO_4:H_2O_2}$ (1:1) can be used to achieve mesa walls with a more shallow slope. Note that this solution is slightly more aggressive and requires only \SI{7}{\min} etch time.

\section{Metal gate deposition}

Metallic gate electrodes were fabricated by EBL, and metal evaporation, followed by liftoff. Metal evaporation is done in an electron-beam evaporation system (AJA international). For Au gates a \SIrange{3}{5}{\nano\meter} Ti sticking layer is applied. For devices with multiple gate layers it is required to cover metallic gates with dielectric. In this case, Au gates were covered with \SI{3}{\nano\meter} Ti to promote a better dielectric growth. 

\section{Gate dielectric}

Au electrodes on InAs do not form a Shottky barrier at the metal semiconductor interface in comparison to Au on GaAs. It is therefore required to include a gate dielectric between the  superconductor-semiconductor heterostructure and the gate electrodes. Atomic layer deposition (ALD) was used to deposite $\mathrm{HfO_{x}}$ for this purpose. The used system was a Savannah S100 from Cambridge Nanotechnology. Dielectric was grown globally on the chip at \SI{90}{\celsius} by 150 pulses of TDMAH precursor each followed by a  $\mathrm{H_2O}$ pulse.

The same dielectric growth procedure can also used on top of Ti capped Au gates that are supposed to be isolated from subsequently deposited gate layers. Alternatively, one can use gates from deposited Al with native oxide. Al can be oxidized in ambient conditions at \SI{185}{\celsius} on a hotplate. The resulting native oxide serves as dielectric encapsulation of the gates \cite{FabioThesis, dotFabGuide}.  

\section{Nanofabrication of nanowire based devices}
In Chapter \ref{ch:fullshell_nonlocal}, devices based on MBE grown NWs are discussed. Prior to fabrication, individual NWs were transferred from the growth substrate  to a carrier chip using a micro manipulator. The carrier chip was made of doped silicon, which allows its use as electrostatic backgate. The carrier chip furthermore had predefined bond pads and lithography registration marks. Scanning electron microscope images after the NW deposition were used to align the desired fabrication pattern with the deposited NWs. A standard fabrication process consisting of selective removal of the Al by wet etch, contacting of the semiconductor by ohmic contacts, growth of gate dielectric, and deposition of gate electrodes followed. Each step made use of EBL. 

The devices presented in this work, are three-terminal devices, i.e., the Al shell of the NW was contacted by a third lead. A ramp of cross-linked resist ensured that the third lead made of $\SI{25}{\nano\meter}$ thin Al can crawl up the side of the NW. This method was developed in Ref. \cite{Razmadze2020} and ensures a third lead that is superconducting at high magnetic fields.

\chapter{Low-temperature electrical measurements}
In order to observe certain quantum phenomena, it is necessary to cool samples to low temperatures. This suppresses thermal excitations and enables the exploration of condensed matter phases that only exist below a certain phase transition temperature. Cryo-free dilution refrigerators are a convenient experimental tool for this purpose \cite{triton_fridge, triton_loader, triton_magnet}. The experimental techniques for electrical transport measurements are outlined in the following.

\section{Wire bonding}
\begin{figure}[h!]
\begin{center}
\includegraphics[width=0.9\textwidth]{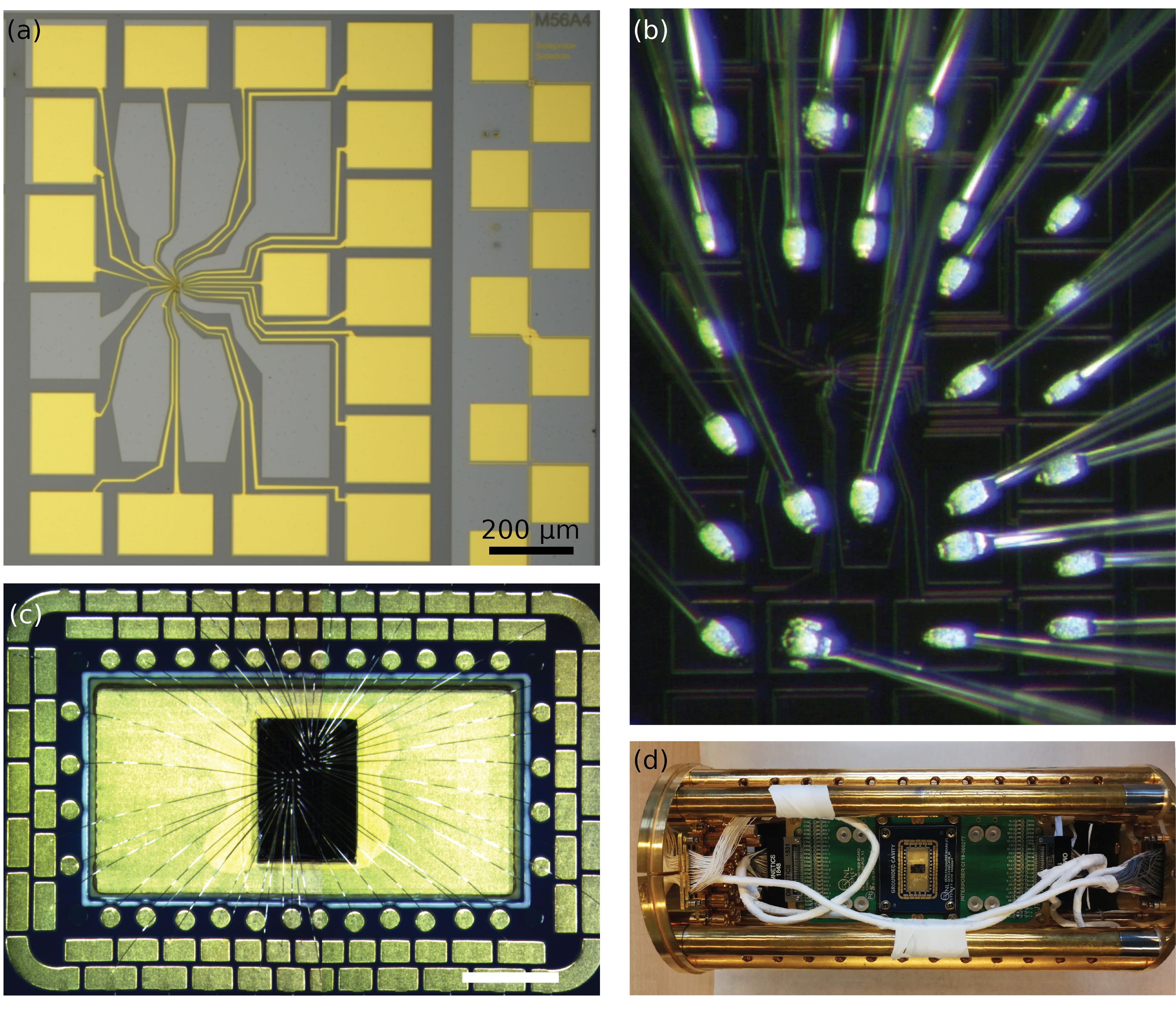}
\end{center}
\caption[Wire bonding of a device for measurements at \SI{}{mK} temperatures.]{\label{fig:bonding}(a) Optical microscope image of device after fabrication. Au bond pads with leads appear yellow. Semiconductor covered with epitaxial Al serve as ohmic contacts (gray). Dummy bond pads on the edge of the chip are used to optimize parameters for wire bonding. (b) Device after wire bonding with Al bonds. One bond pad in the bottom row of bond pads shows bite marks to its sides. (c) wire bonded chip on a chip carrier. (d) chip carrier mounted on mother board in the sample-puck of a dilution refrigerator. }
\end{figure}

For the purpose of measurements, chips were mounted on a chip carrier. The chips were glued with a drop of PMMA resist or silver paint. The samples were subsequently wire bonded with Al wires using an automatic wedge bonder (F\&S Bondtec 5630). Bondpads have a typical size of $\approx\SI{200}{\micro\meter} \times \SI{150}{\micro\meter}$ with $\approx\SI{30}{\micro\meter}$ spacing in between which allows for wire bonding given the footprint of wire bonds [see Fig.~\ref{fig:bonding}(a)]. The samples were wire bonded by bonding to Au bond pads of the gate electrodes. Semiconducting mesa covered with epitaxial Al served as ohmic contact for InAs/Al hybrid heterostructures. Wire bonding can be performed by punching through dielectric, bonding directly to the dielectric covered epitaxial Al. Note that slightly lower power has to be used in the bond parameters when bonding to Al on semiconductor instead of Au on semiconductor. Figure \ref{fig:bonding}(b, c) show a wire bonded sample.\\

Before every bonding session, it is advisable to optimize the parameters (force, power, time) on dummy bond pads [see Fig.~\ref{fig:bonding}(a)] before bonding to the devices of interest. Suitable bond parameters are found when the bonds have a slightly oval shape with no bite marks next to them. Test bonds can furthermore be slightly poked with tweezers to check for a firm connection. \\

The chip carrier was mounted on a mother board in the sample puck. The puck was cooled in a cryo-free dilution refrigerator (Oxford Instruments Triton 400 with 1-1-6 $\SI{}{\tesla}$ vector magnet). Electrical connections to room temperature electronics was made by DC lines that were filtered by cryogenic, in-house built multi-stage low pass filters.

\section{Tunneling spectroscopy}

Tunneling spectroscopy is a technique frequently applied in low-temperature solid-state physics. It relies on the measurement of a small current $I$ as a function of source-drain voltage $V_\mathrm{SD}$. The current $I$ is typically a tunneling current running through a potential barrier of the sample. A differential measurement of this current can be performed using a lock-in amplifier. For this purpose a small sinusoidal modulation of RMS amplitude $\mathrm{d}V_\mathrm{SD}$ is added to $V_\mathrm{SD}$ using the output voltage of a lock-in amplifier. By measuring the resulting tunneling current $\mathrm{d}I$ at the frequency of the modulation the differential conductance

\begin{equation}
G=\frac{\mathrm{d}I}{\mathrm{d}V_\mathrm{SD}}
\end{equation}
can be determined. This can serve as a measure of the density of states. \\

\begin{figure}[h!]
\begin{center}
\includegraphics[scale=0.9]{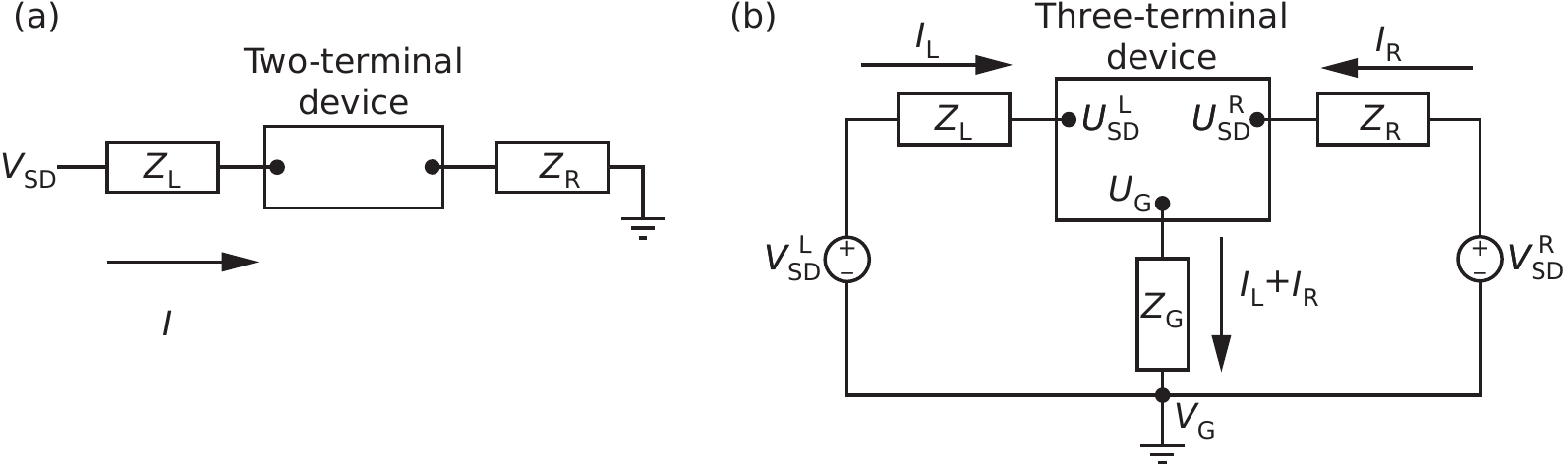}
\end{center}
\caption[Circuit of two- and three-terminal device]{\label{fig:circuit}Schematic measurement circuits for devices connected to source-drain voltage supplies $V_\mathrm{SD}$ by lines and filters with impedance $Z_\mathrm{L,\, R, \, G}$. (a) Two-terminal device. (b) Three-terminal device.}
\end{figure}

Figure \ref{fig:circuit}(a) shows the schematic circuit of such a measurement. Note that the line resistances are also measured. The contribution of the line resistance to the measured conductance can be subtracted, if the line impedances $Z_\mathrm{L},\; Z_\mathrm{R}$ are known precisely. Alternatively, two extra lines can be used in a four probe measurement to measure the voltage drop over the sample directly and use it for the calculation of the differential conductance.

\section{Three-terminal measurement scheme}
In the context of superconductor-semiconductor hybrid devices, multi-terminal devices have attracted considerable attention lately. These devices allow for local density of states measurements at both ends of a NW and enable the characterization of the bulk states by nonlocal conductance. The simplest form of a multi-terminal device is a three-terminal device. It can consists of a semiconducting NW proximitized by a superconductor that is electrically grounded and both NW ends tunnel coupled to metallic leads. A detailed description of the electrical circuit relevant for transport measurements was developed within this thesis and can be found in Ref. \cite{voltage_divider}. A summary is given in the following.\\

Electric transport measurements of three-terminal devices include the two currents $I_\mathrm{L}$, $I_\mathrm{R}$ flowing into the left and right normal metal lead as a function of the voltages $V_{\mathrm{SD}}^\mathrm{L}$, $V_{\mathrm{SD}}^\mathrm{R}$. These quantities are measured at room temperature with the sample being cooled in a dilution refrigerator. This requires finite line and filter impedances $Z_\mathrm{L}$,  $Z_\mathrm{G}$,  $Z_\mathrm{R}$ between the device and the measurement electronics. A schematic is shown in Fig.~\ref{fig:circuit}(b). When performing low frequency (DC to $\SI{100}{\Hz}$) the impedances can be taken to be dominated by their real part. The finite line impedance will result in a difference between the voltages applied at room temperature $V^\mathrm{L}_\mathrm{SD}$, $V^\mathrm{R}_\mathrm{SD}$ and the voltages $U^\mathrm{L}_\mathrm{SD}$,  $U^\mathrm{R}_\mathrm{SD}$ directly at the device terminals. The method described herein estimates the impact of the finite line impedances on the measured quantities. It includes formulas that relate the easily accessible differential conductances $G_\mathrm{ij}=\mathrm{d}I_\mathrm{i}/\mathrm{d}V_\mathrm{SD}^\mathrm{j}$ ($i,j \in \{L,R\}$) to the differential conductance directly at the device terminals $G'_\mathrm{ij}=\mathrm{d}I_\mathrm{i}/\mathrm{d}(U_\mathrm{SD}^\mathrm{j}-U_\mathrm{G})$. It furthermore shows limits where the difference between the two quantities is negligible.

We assume the potential of the superconducting terminal to be ground $U_\mathrm{G}= 0$. One can use the vectors $\mathbf{U}=(U_\mathrm{SD}^\mathrm{L}, U_\mathrm{SD}^\mathrm{R})^\mathrm{T}$ and $\mathbf{V}=(V_\mathrm{SD}^\mathrm{L}, V_\mathrm{SD}^\mathrm{R})^\mathrm{T}$ to describe the applied voltages. They are related to the currents $\mathbf{I} = (I_\mathrm{L}, I_\mathrm{R})^\mathrm{T}$ by
\begin{align}
    \label{eq:v_tilde}
    \mathbf{U} &= \mathbf{V} - Z \ \mathbf{I},
\end{align}
where $Z$ is a matrix with the line impedances:
\begin{align}
    {Z} = \left(
    \begin{matrix}
        Z_\mathrm{L} + Z_\mathrm{G} & Z_\mathrm{G}\\
        Z_\mathrm{G} & Z_\mathrm{R} + Z_\mathrm{G}
    \end{matrix}
    \right).
\end{align}
Eq.~(\ref{eq:v_tilde}) is particularly useful to calculate the DC voltage biases ${\mathbf{U}}$ at the device from the DC voltages $\mathbf{V}$ applied at the measurement terminals and the measured DC currents $\mathbf{I}$, by taking the zero-frequency component.\\

The electrical behavior of any device in a particular configuration is fully characterized by its differential conductance matrix as a function of the voltages at the terminals. The conductance matrix ${G'}$ at the device is given by:
\begin{equation}
    \label{eq:g_tilde}
    {G'}(\mathbf{U}) = \frac{\partial \mathbf{I}}{\partial \mathbf{U}} 
    = \left(
    \begin{matrix}
        \frac{\partial I_\mathrm{L}}{\partial U_\mathrm{L}} & \frac{\partial I_\mathrm{L}}{\partial U_\mathrm{R}} \\
        \frac{\partial I_\mathrm{R}}{\partial U_\mathrm{L}} & \frac{\partial I_\mathrm{R}}{\partial U_\mathrm{R}} \\
    \end{matrix} \right).
\end{equation}
This quantity can be expressed by the quantity ${G}(\mathbf{V})$ measured at room temperature by

\begin{align}
    \label{eq:g-transform}
    {G'}(\mathbf{U}) &= {G}(\mathbf{V}) ( \mathbb{1} - Z \ \mathrm{G}(\mathbf{V}) )^{-1}.
\end{align}
The difference between  ${G'}$ and ${G}$ is small if all the elements $(Z \ G(\mathbf{V}))_\mathrm{ij} \ll 1$, that is:
\begin{align}
    |Z_\mathrm{L}| &\ll |G_\mathrm{Li}|^{-1} \nonumber\\
    |Z_\mathrm{R}| &\ll |G_\mathrm{Ri}|^{-1} \nonumber\\
    |Z_\mathrm{G}| &\ll |G_\mathrm{ij}|^{-1}, \nonumber
\end{align}
for $i, j = \mathrm{L, R}$. In this case, the zeroth order the correction can be written as:
\begin{align}
    \label{eq:correction-approx}
    {G}'(\mathbf{U}) &\approx G(\mathbf{V}) + \left(
        \begin{matrix}
            (Z_\mathrm{L} + Z_\mathrm{G}) G_\mathrm{LL}^2 & Z_\mathrm{G} G_{ll} G_\mathrm{RR} \\
            Z_\mathrm{G} G_\mathrm{LL} G_\mathrm{RR} & (Z_\mathrm{R} + Z_\mathrm{G}) G_\mathrm{RR}^2 \\
        \end{matrix} \right).
\end{align}
This equation shows how the measured conductance $G$ is altered by finite line impedances in comparison to $G'$. Namely, the diagonal terms of $G$ are the same terms in $G'$ after a quadratic rescaling. The off-diagonal terms of $G$ are equal to the respective terms in $G'$ plus a spurious term proportional to the diagonal elements of $G$. \\

Equation \ref{eq:correction-approx} has two important implications. Firstly, it shows that the conductance matrix at the terminals of a three-terminal device $G'$ can be calculated from $G$ measured at room temperature if the line impedances are known. Note that this procedure leads to a decrease in signal-to-noise as two noisy quantities are combined. Secondly, this equation shows that the correction to the the off diagonal elements of $G'$ can be kept small by either grounding the superconducting terminal with a small line impedance $Z_\mathrm{G}$ or by operating the device deep in the tunneling regime where $G_\mathrm{RR}, G_\mathrm{LL}$ are small. These two principles were applied in this thesis to avoid the reduced signal-to-noise ratio. For phenomena that are only expected when the device has transparent tunneling barriers $\gtrsim1e^2/h$ the application of equation \ref{eq:correction-approx} is unavoidable. 

A further improvement to the signal-to-noise ratio can be made by bonding the ohmic contacts of a three-terminal device to lines that are routed out of the dilution refrigerator via the same Fischer cable loom. The logic is the same behind choosing a pair of braided lines for the contacts of a two-terminal devices.


\cleardoublepage
\part{Results}\label{pt:results}
\chapter{Hybridization between quantum dots and Andreev bound states}\label{ch:pradaclarke}

\section{Introduction}

Hybrid devices based on InAs two-dimensional electron gases (2DEGs) proximitized by superconducting Al \cite{javad_2deg, henri_lead,morten_sqpcn, alex_QDinLoop, eoins_dots, nichele_scaling, alex_interfero} allow exploration of various bound states in nanowires (NWs), including Yu-Shiba-Rusinov states, Andreev bound states (ABSs), and Majorana bound states \cite{alex_QDinLoop, chang_YSR, kasper_tuningYSR,lee_scaling,pillet_CNT,nichele_ABS_Ic,nichele_scaling,dassama_meta}. 

\begin{figure}[h!]
\begin{center}
\includegraphics[scale=0.85]{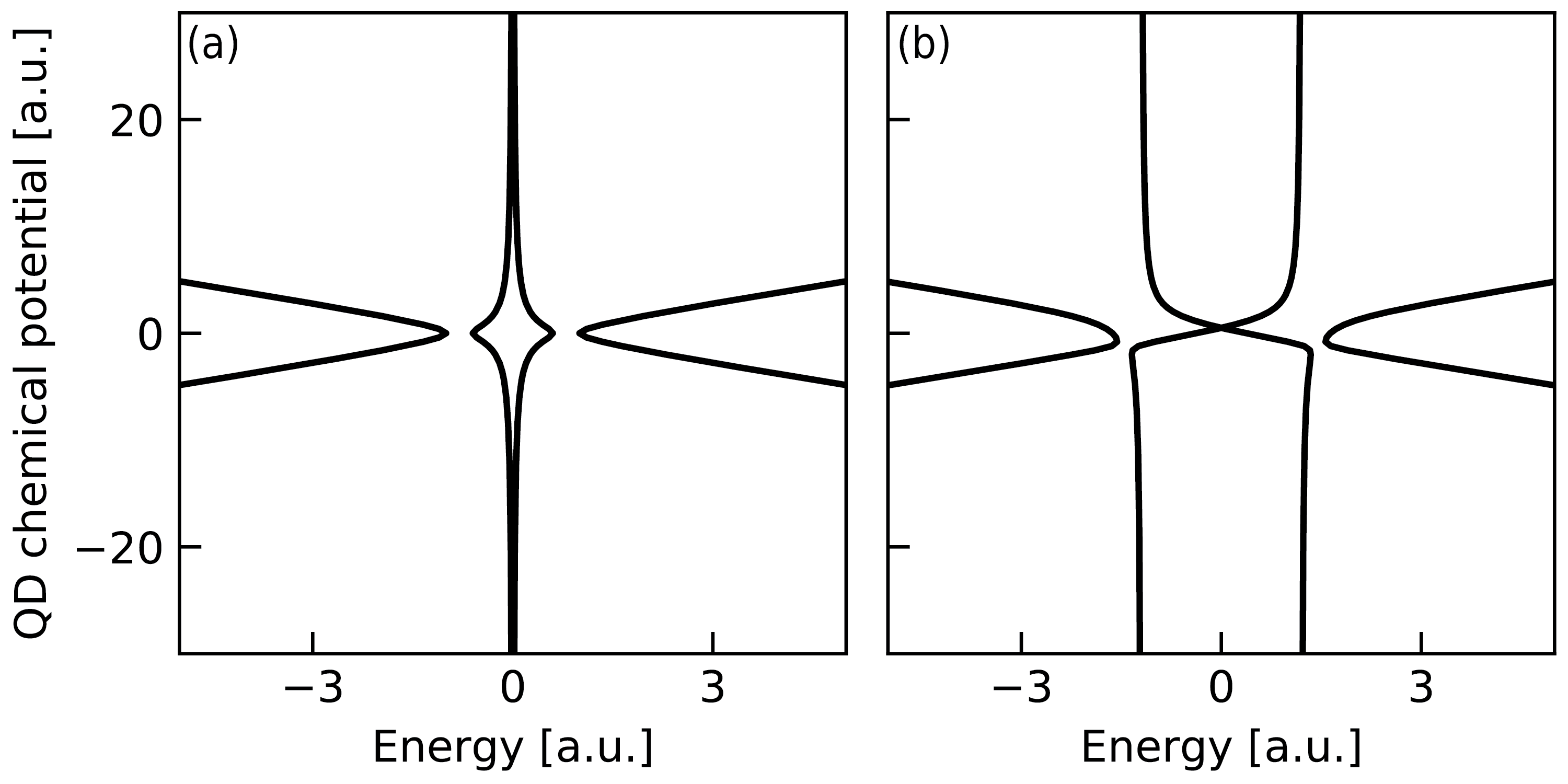}
\end{center}
\caption[Spectrum expected from theory for strongly overlapping Majorana zero modes coupled to a quantum dot]{\label{fig:pc_theory}Theoretically expected spectrum of strongly overlapping MZMs coupled to a QD. (a) Zero energy bound states split when tunnel coupled to a quantum dot level. The bound states have the shape of a `diamond'. (b) Zero crossing of finite energy bound states is observed when brought on resonance with a quantum dot level. The bound states form the `bowtie' shape.}
\end{figure}

The use of semiconductor-superconductor hybrids facilitates the realization of electrostatically controlled quantum dots (QDs) coupled to superconductors. QDs coupled to ABSs have received considerable attention from theoretical studies, including the use of the QD as a tool for measuring bound state lifetimes \cite{leijnse_dotMBSlifetime} or providing Majorana parity readout \cite{gharavi_dotMBSparity,hoffman_dotMBScomputation,karsten_dotMBS_operations,leijnse_QItransferDotMBS,plugge_boxqubits, alex_interfero}. In particular, the hybridization between a QD and a bound state leads to a shift of the bound state energy resulting in a characteristic `bowtie' or `diamond' shape \cite{clarke_quality}. The spectrum expected from theory for highly overlapping MZMs coupled to a QD is plotted in Fig.~\ref{fig:pc_theory}.  The precise shape of the resulting anti-crossing between the energy levels is determined by the nonlocality and the spin structure of the bound states \cite{elsa_nonlocality, mingtang_nonlocality, elsa_quantifying}.

In this work, we perform tunneling spectroscopy of a NW in a novel geometry that allows measurements at several side branches along the NW length using electrostatic gates patterned on an InAs/Al hybrid heterostructure. A similar configuration has been investigated theoretically \cite{dassarma_circuits}, and a related experiment has been carried out in a conventional nanowire with deposited superconductor and normal metallic side contacts \cite{grivnin_multiprobe}. In addition to ABSs due to bound states in the NW, we find conductance resonances due to accidental QDs in the tunnel barriers. We investigate hybridization of QD states with ABSs in the NW, observing signatures of hybridization both locally, that is, at the position of the accidental QD, and nonlocally, measured on another sideprobe away from the QD.

\section{Extended Andreev bound states}
\label{sec:abs_pc}
\begin{figure*}[h!]
\begin{center}
\includegraphics[scale=0.75]{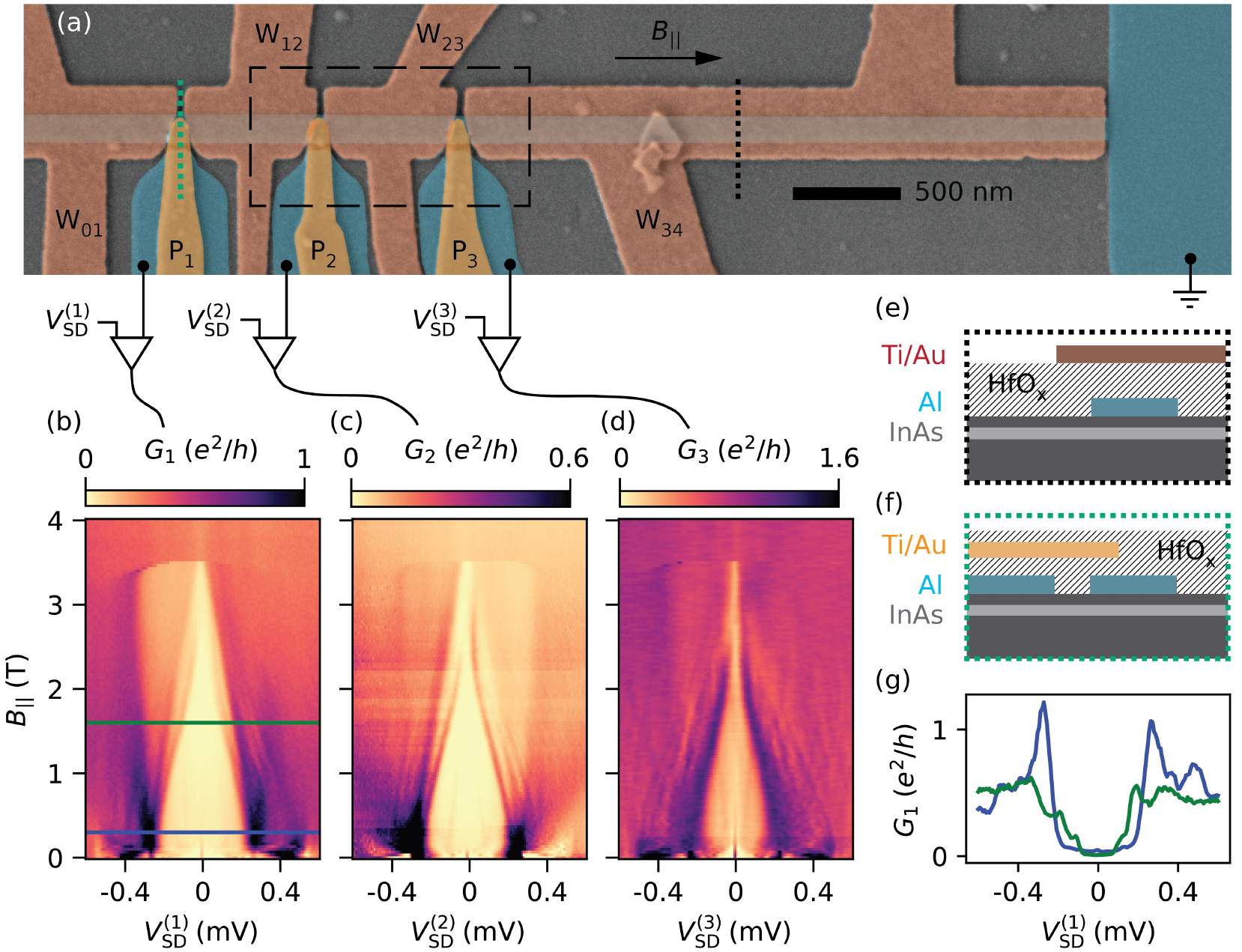}
\end{center}
\caption[2DEG device with superconducting sideprobes depleted from bound states]{\label{fig:device_pc}(a) False-color scanning electron micrograph of device 1. The device consists of patterned epitaxial Al, forming a long narrow nanowire with several tunnel probes on top of an InAs quantum well. Gates labeled $\mathrm{P}_j$ are used to tune the tunnel barrier between the tunnel probe $j$ and the wire ($j \in \{1,2,3\}$). Gates labeled $\mathrm{W}_{kl}$ with $kl \in \{01,12,23,34\}$ deplete carriers except under the Al. (b-d) Tunneling spectroscopy at three probes with all gate voltages $V_{\mathrm{W}kl}=\SI{-4.5}{\volt}$. (e, f) Schematic cross sections of the device at the positions given by the green and black dotted lines in (a). (g) Line cuts at field values $B_{||}=\SI{0.3}{\tesla}$ and $B_{||}=\SI{1.6}{\tesla}$ indicated by the blue and green line in (b).}
\end{figure*}

Figure \ref{fig:device_pc}(a) shows a micrograph of device 1, based on an InAs 2DEG with \SI{5}{\nano\meter} of epitaxial Al. The device consists of an Al strip of width \SI{100}{\nano\meter} and length \SI{5}{\micro\meter}, connected at both ends to large planes of Al that were electrically grounded. Gates labeled $\mathrm{W}_{kl}$ were Ti/Au on top of \SI{30}{\nano\meter} $\mathrm{HfO_2}$, as shown in Fig.~\ref{fig:device_pc}(e). Gates were used to deplete the semiconductor on either side of the Al wire, creating by depletion a quasi one-dimensional InAs NW self-aligned to the proximitizing Al.

Neighboring gates $\mathrm{W}_{kl}$ form a constriction that acts as a tunnel probe. The lead of the probe, away from the tunneling region, is made using the same unetched epitaxial Al. Tunneling across the bare semiconductor region between Al NW and Al lead is controlled by a probe gate,  $\mathrm{P}_j$, as shown in Fig.~\ref{fig:device_pc}(f). Details of the materials and fabrication are given in the Section \ref{app:wafer}. 

 The measurement setup is shown schematically in Fig.~\ref{fig:device_pc}(a). With the NW  grounded, individual voltage biases $V_{\mathrm{SD}}^{(j)}$ were applied on probe $j$ via current to voltage converters.
 Tunneling currents $I_j$ through the tunnel barriers were measured using lock-in detection yielding differential conductances $G_j=\mathrm{d}I_j/\mathrm{d}V^{(j)}_{\mathrm{SD}}$. Measurements were carried out in a cryo-free dilution refrigerator with a 6-1-1 \SI{}{\tesla} vector magnet at $\approx\SI{15}{\milli\kelvin}$ mixing-chamber temperature. 
 
Tunneling conductances $G_j$ as a function of magnetic field $B_{||}$ applied parallel to the NW are shown in Figs.~\ref{fig:device_pc}(b-d). For weak tunneling and in the absence of probe resonances, $G_j$ is proportional to the density of states in the NW. The superconducting gap of the Al in the leads of the probes closes at low field, $B_{||} \approx \SI{0.2}{\tesla}$ above which the probes can be regarded as normal metal, as discussed previously \cite{henri_lead, nichele_scaling}. The semiconductor under the Al in the NW was depleted by setting all $\mathrm{W}_{kl}$ gates to $\SI{-4.5}{\volt}$.  Measurements  on all three probes showed a superconducting gap closing without any subgap states crossing zero energy. For $G_1$ this is illustrated by the line cuts in Fig.~\ref{fig:device_pc}(g). Note that the measurement of $G_3$ shows finite subgap conductance, which we attribute to probe 3 being tuned to an open regime with high-bias conductance $G_3(V^{(3)}_{\mathrm{SD}}=\SI{0.4}{\milli\volt}) \gtrsim 1 \; e^2/h$.\\

\begin{figure}[h!]
\begin{center}
\includegraphics[scale=1.0]{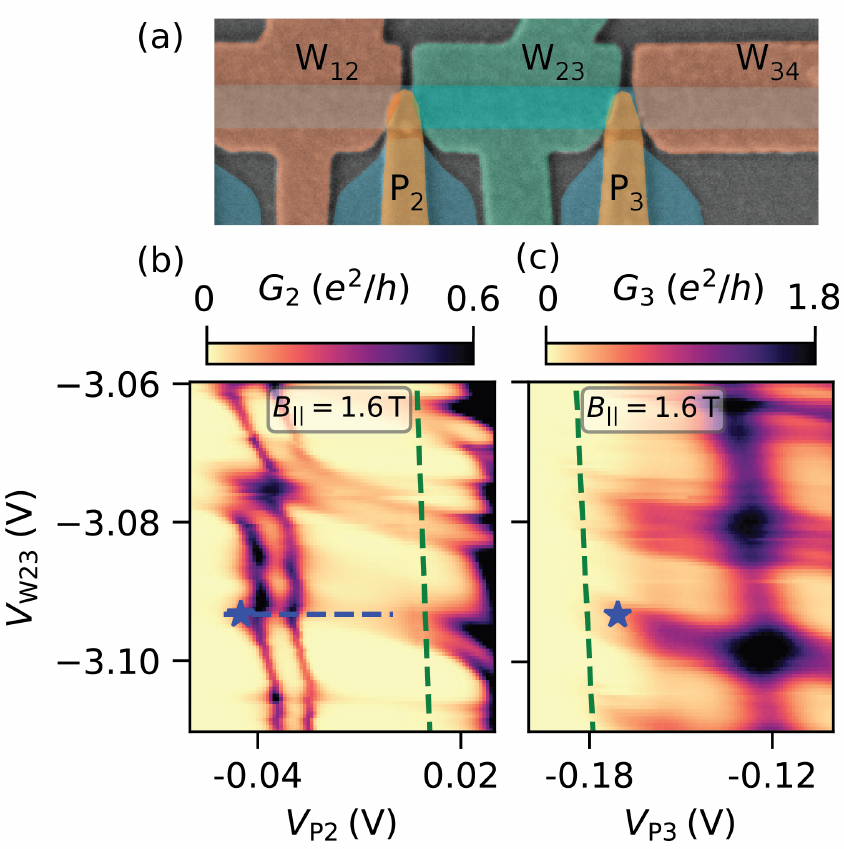}
\end{center}
\caption[Gate dependence of bound states in a \SI{0.6}{\micro\meter} long wire segment]{\label{fig:gatemap}(a) Micrograph of the NW segment under investigation. $V_\mathrm{W12}=V_\mathrm{W34}=\SI{-7.0}{\volt}$ while the voltage on gate $\mathrm{W}_{23}$ (green), is varied. (b), (c) Differential conductance at zero bias measured at the left and the right end of the NW segment. Horizontal conductance resonances appear in both maps at similar gate voltages. Vertical conductance features, strongly dependent on gates $V_\mathrm{P2}$ and $V_\mathrm{P3}$ which tune the tunnel barriers, are also visible.
}
\end{figure}

\begin{figure}[h!]
\includegraphics[scale=0.85]{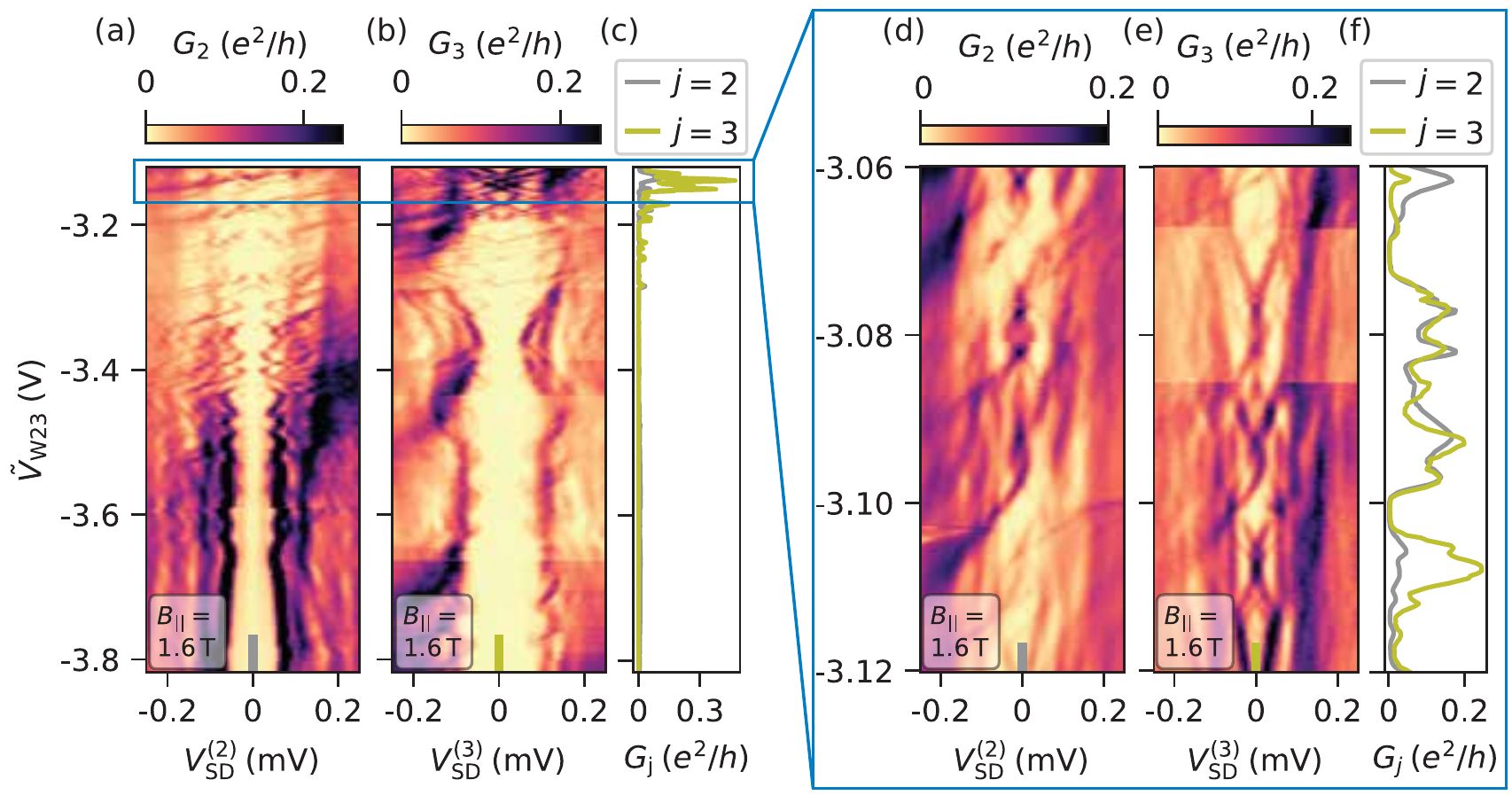}
\caption[Correlated appearance of extended bound states as a function of gate voltage]{\label{fig:evolution}(a, b) show conductance spectroscopy measurements of $G_2$ and $G_3$ over a wide range of $V_\mathrm{W23}$. The voltages on the tunnel barrier gates $V_\mathrm{P2}$ and $V_\mathrm{P3}$ were compensated according to Eqs. \ref{eq:long_plunger}. The variable on the vertical axis is therefore marked $\tilde V_\mathrm{W23}$ instead of $V_\mathrm{W23}$. No subgap state crosses zero bias for gate voltages $\tilde V_\mathrm{W23}<\SI{-3.3}{\volt}$. A line cut at zero bias for $G_2$ and $G_3$ is plotted in (c). A zoomed-in measurement of $G_2$ and $G_3$ is depicted in (d) and (e). It reveals subgap states whose peak positions show identical nontrivial gate voltage dependence in both $G_2$ and $G_3$. The data for $G_2$ and $G_3$ at zero bias is plotted in (f) for comparison.}
\end{figure}

To investigate the hybridization of a probe QD state with an ABS in the NW, we focus on the \SI{0.6}{\micro\meter} long NW segment under gate $\mathrm{W_{23}}$, see dashed box in Fig.~\ref{fig:device_pc}(a), shown in Fig.~\ref{fig:gatemap}(a). To create an ABS in this segment, the voltage on gate $\mathrm{W_{23}}$ was set less negative, in the range of $\SI{-3}{\volt}$, while voltages on neighboring gates $\mathrm{W_{12}}$ and $\mathrm{W_{34}}$ were set to $\SI{-7.0}{\volt}$. At $B_{||}=\SI{1.6}{\tesla}$ and zero source drain biases, $V_{\mathrm{SD}}^{(j)}=0$, conductances $G_2$ and $G_3$ were measured as functions of  probe-gate voltages $V_\mathrm{P2}$ and $V_\mathrm{P3}$, respectively, and wire-gate voltage $V_\mathrm{W23}$. For both tunnel junctions, two sets of conductance resonances can be distinguished in Figs.~\ref{fig:gatemap}(b) and (c) by their characteristic slope. 
The first set primarily consists of vertical features that are strongly dependent on the gate voltages $V_\mathrm{P2}$ ($V_\mathrm{P3}$), which we attribute to  QDs in the tunnel barriers. The second set are predominantly horizontal, depending more strongly on $V_\mathrm{W23}$. The latter resonances are visible in both $G_2$ and $G_3$, suggesting that they arise from ABSs that extend over the segment covered by gate $\mathrm{W}_{23}$. Figures \ref{fig:evolution}(a, b) show the evolution of $G_2$ and $G_3$ from $V_\mathrm{W23}=\SI{-3.8}{\volt}$ to $V_\mathrm{W23}=\SI{-3.1}{\volt}$. Note that while changing $V_\mathrm{W23}$, the voltages $V_\mathrm{P2}$ and $V_\mathrm{P3}$ were changed according to the relations

\begin{equation}
 \begin{aligned} \label{eq:long_plunger}
     V_\mathrm{P2}&=-\SI{10}{\milli\volt}-\frac{0.05}{0.40} \cdot(V_\mathrm{W23}+\SI{3.1}{\volt})\\
     V_\mathrm{P3}&=-\SI{135}{\milli\volt}-\frac{0.045}{0.40} \cdot (V_\mathrm{W23}+\SI{3.5}{\volt}) 
 \end{aligned}
\end{equation}
The lines given by these relations are shown in green in Figs.~\ref{fig:gatemap}(b, c). This compensates for the effect of the gate $V_\mathrm{W23}$ on the tunnel barriers and ensures that the high bias conductance stays around $\sim 0.1\; e^2/h$ throughout the complete range of $\SI{-3.8}{\volt}<V_\mathrm{W23}<\SI{-3.1}{\volt}$. To denote that more than one gate voltage was changed during the measurement, we label the variable $\tilde V_\mathrm{W23}$ instead of $V_\mathrm{W23}$.
There are no states crossing zero bias for voltage values $\tilde V_\mathrm{W23}<\SI{-3.3}{\volt}$. Above that value ABSs cross the gap. Line cuts at zero bias are displayed in Fig.~\ref{fig:evolution}(c). A measurement with higher resolution within the range $\SI{-3.06}{\volt}<\tilde V_\mathrm{W23}<\SI{-3.12}{\volt}$ in Figs.~\ref{fig:evolution}(d, e) shows that subgap states evolve with the same nontrivial dependence on $\tilde V_\mathrm{W23}$ in measurements of $G_2$ and $G_3$. At the same time, the subgap states can appear with different strength in conductance in $G_2$ and $G_3$.  A line cut taken at zero bias in Fig.~\ref{fig:evolution}(f) shows the correlated $\tilde V_\mathrm{W23}$ dependence on both sides in the form of coinciding peak positions, while the difference in conductance value by up to one order of magnitude for some of the subgap states is apparent from the difference in peak height.\\

\begin{figure}[h!]
\begin{center}
\includegraphics[scale=0.95]{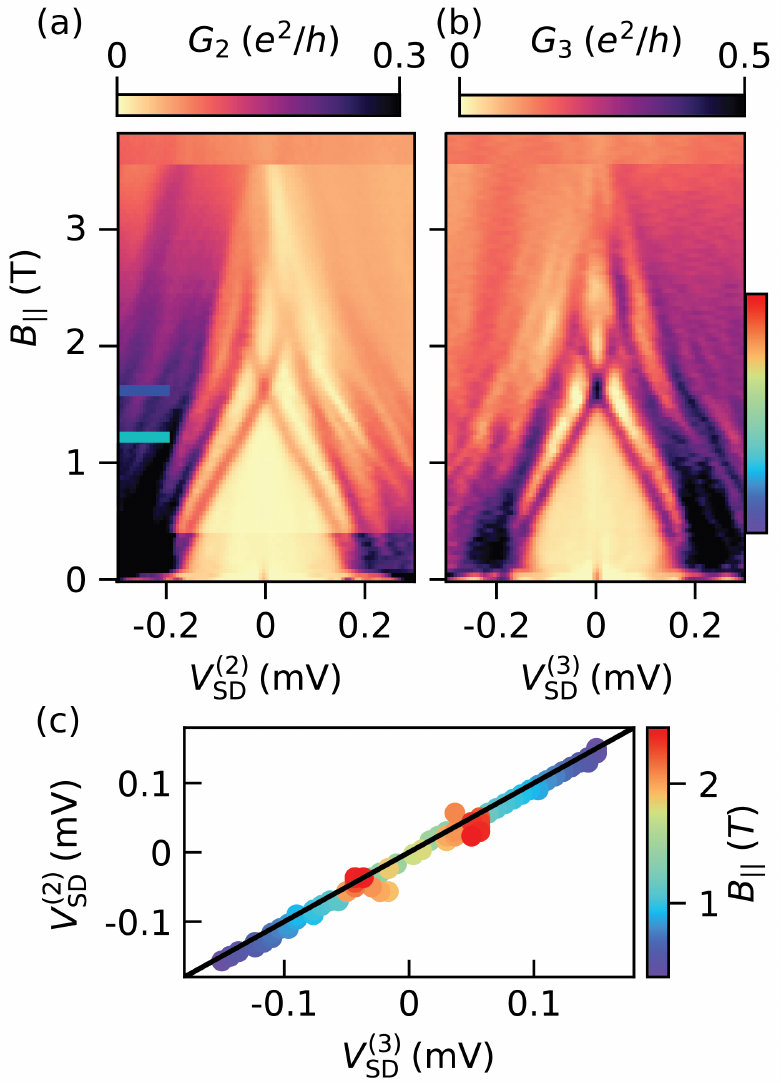}
\end{center}
\caption[Evolution of bound states as a function of magnetic field at both nanowire ends]{\label{fig:fieldscan_pc}(a, b) Tunneling spectroscopy with respect to magnetic field at the two ends of the NW with $V_\mathrm{W23}=\SI{-3.09}{\volt}$ [marked with $\star$ on Fig.~\ref{fig:gatemap}(b, c)]. Both measurements show subgap states crossing zero bias at $B_{||}=\SI{1.6}{\tesla}$ with a clear overshoot around $B_{||}=\SI{2}{\tesla}$. (c) Parametric plot of the extracted peak positions from the lowest energy subgap states in (a) and (b). 
The color of the points indicates the field value in accordance with the rainbow color bar in (b). }
\end{figure}

The blue star markers in Figs.~\ref{fig:gatemap}(b, c) at gate voltages $V_\mathrm{W23}=\SI{-3.09}{\volt}$, $V_\mathrm{P2}=\SI{-0.045}{\volt}$, and $V_\mathrm{P3}=\SI{-0.170}{\volt}$ mark ABSs that are weakly tunnel coupled to the probes. Tunneling spectroscopy of these ABSs as a function of magnetic field $B_{||}$ in Figs.~\ref{fig:fieldscan_pc}(a, b) reveals a zero-bias crossing of the ABSs at $B_{||}=\SI{1.6}{\tesla}$ followed by an overshoot at $B_{||}=\SI{2}{\tesla}$. The states appear in both tunneling conductance measurements of $G_2$ and $G_3$. We extracted the peak position in $V^{2(3)}_{\mathrm{SD}}$ of the ABS from the measurements of $G_2$ and $G_3$. 
The parametric plot of the peak positions $V^{(2/3)}_{\mathrm{SD}}$ of the ABSs in $G_2$ and $G_3$ in Fig.~\ref{fig:fieldscan_pc}(c) shows that all points lie close to the identity line, suggesting strong correlations. Details about the extraction of the peak position are outlined in the Section \ref{sec:peakextraction}.

The ABSs seen in $G_2$ and $G_3$ evolve similarly with gate voltage $V_\mathrm{W23}$ and magnetic field $B_{||}$, suggesting that they belong to the same extended quantum states. Similar experimental findings have been made previously \cite{gian_correlations,frolov_quantized}. The magnetic field dependence of the states is furthermore characteristic for ABSs in short NWs \cite{dassama_meta}.\\

\section{Nonlocal signatures of hybridization}
\label{sec:pc}
\begin{figure}[h!]
\begin{center}
\includegraphics[scale=0.95]{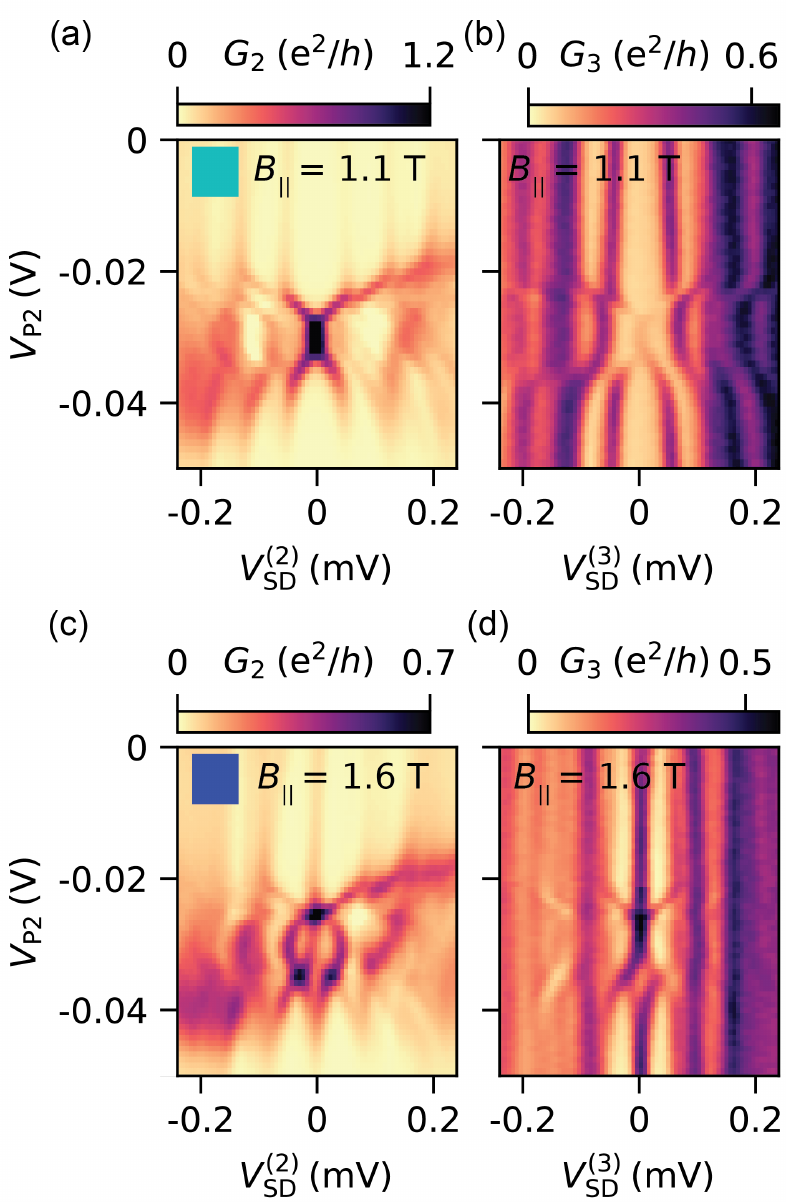}
\end{center}
\caption[Tunneling spectroscopy of nonlocal signatures of hybridization between quantum dot and bound states]{\label{fig:pc}(a) Tunneling conductance $G_2$ at the left side of the NW segment at $V_\mathrm{W23}=\SI{-3.09}{\volt}$ [marked with $\star$ and blue dashed line on Fig.~\ref{fig:gatemap}(b, c)] as a function of the gate voltage $V_\mathrm{P2}$ which tunes the tunnel barrier. 
A QD resonance is visible as an enhancement of conductance at high bias around $V_2\approx\SI{-0.030}{\volt}$. The subgap states change their energy at the point of the QD resonance, drawing a characteristic `bowtie' shape. 
(b) Tunneling spectroscopy $G_3$ at the other end of the NW. The ABS show the same change in energy as visible in the measurement of $G_2$ in (a). (c, d) same as (a, b) at higher parallel magnetic field, $B_{||}$. The ABSs split to form a `diamond' shaped energy profile at the position of the QD resonance.}
\end{figure}

Special points in the measurement in Fig.~\ref{fig:gatemap}(b) are the crossing points of the horizontal resonances with the sharp vertical resonances. At these points, an ABS in the NW is on resonance with the QD in the tunnel barrier under the gate $\mathrm{P_2}$. Tunnelling spectroscopy $G_2$ using tunnel probe 2 at a field value of $B_{||}=\SI{1.1}{\tesla}$ while sweeping $V_\mathrm{P2}$ along the values given by the blue dashed line in Fig.~\ref{fig:gatemap}(b) is shown in Fig.~\ref{fig:pc}(a). The ABSs were unaffected by the change of $V_\mathrm{P2}$ outside the range \SIrange{-0.040}{-0.020}{\volt}. Within this range, the QD resonance appears as a conductance enhancement at high bias, reflecting the fact that $G_2$ was being measured through the QD in tunnel barrier 2. As the QD went on resonance with the ABSs, the ABSs with lowest energy merged at zero bias before returning to their previous energies. This resulted in a characteristic `bowtie' shape of the resonances of the ABSs. A simultaneous measurement of $G_3$ during the sweep of $V_\mathrm{P2}$ at the other end of the NW is shown in Fig.~\ref{fig:pc}(b). The enhancement of conductance at high bias due to the QD that was present in the measurement of $G_2$ was absent in the measurement of $G_3$. The ABSs, however, showed the same `bowtie' shape around the voltage value $V_\mathrm{P2}\approx\SI{-0.030}{\volt}$ which corresponds to the resonance condition between the ABSs and the QD in tunnel barrier 2. Note that in the measurement of both $G_2$ and $G_3$ not only the lowest energy ABSs undergoes a change at the resonance condition with the QD, but also the higher excited states. In addition to the change in ABS energy, a clear change in the conductance peak height is visible when going through the resonance condition.

Around $B_{||}=\SI{1.6}{\tesla}$, the ABSs merged to yield a single conductance peak at zero bias. A measurement of $G_2$ with respect to $V_\mathrm{P2}$ in Fig.~\ref{fig:pc}(c) shows that this peak was unperturbed except for voltage values around $V_\mathrm{P2}\approx\SI{-0.030}{\volt}$ where the QD was on resonance with the ABSs. At this point, the ABS resonances split symmetrically away from zero bias, forming a `diamond' shape. The simultaneous measurement of $G_3$ in Fig.~\ref{fig:pc}(d) reveals similar $V_\mathrm{P2}$ dependence of the ABS energy. Note that higher excited states were also affected around $V_\mathrm{P2}\approx\SI{-0.030}{\volt}$. \\

\section{Discussion and conclusion}

The appearance of ABSs with `bowtie'- and `diamond'-shaped patterns while on resonance with the QD level is an indication of the QD being sufficiently tunnel coupled to the ABSs such that the two energy levels significantly hybridize, consistent with theoretical and previous experimental results \cite{elsa_nonlocality, clarke_quality, vuik_quasiMBS, mingtang_nonlocality,elsa_quantifying}. The measurement of the energy shift at both ends of the \SI{0.6}{\micro\meter} NW, while a local gate voltage at only one end is changed, is a nonlococal signature of the delocalized ABSs. If tunneling spectroscopy is measured at both ends of a NW, hybridization of bound states with a local QD can be used as a quantum mechanical tool to test whether a quantum state extends through the whole NW, similar to the analysis of cross-conductance and correlated appearance at both ends \cite{gerbold_nonlocal, gian_correlations}. This is in contrast to experiments where spectroscopy is performed at one end of a NW. In such a case, a QD in the absence of a bound states in the NW can mimic signatures of extended states inside the NW in tunneling spectroscopy \cite{dassarma_against_mingtang, lee_singletdoublet, trivial_fullshell,dassarma_ABSvsMBSspectroscopy, dassarma_goodbadugly, vuik_quasiMBS}.

In comparison to previous experiments, the present setup offers additional information about the spatial extent of the bound state, as one can perform tunneling spectroscopy at both ends of the NW segment. This also allows for the observation of the change in energy of the ABS at one position while it is being hybridized with a QD \SI{0.6}{\micro\meter} away by the means of changing a local gate. 
This nonlocal signature is a demonstration of the ABS being an extended quantum state. 

In Section \ref{sec:sup_dev_1_pc} we present data of the hybridization of an ABS in the NW segment under the gate labeled $\mathrm{W}_{12}$ with a local QD in the tunnel barrier. Results from a second device (device 2) are presented in Section \ref{sec:sup_dev_2_pc}. Device 2 is a slightly different design, with sideprobes that do not have Al leads, only the bare semiconductor. 
Device 2 showed similar results as device 1 when an ABS in a NW was brought onto resonance with a QD localized at one probe while observing the impact of the hybridization on the bound state at the other probe.


\section{\label{sec:sup_dev_1_pc}Supplementary information on device 1}

\subsection{\label{app:wafer}Wafer information and fabrication}
The material used for device 1 was an InGaAs/InAs/InAlAs heterostructure covered with an \emph{in-situ} epitaxially grown Al top layer. Electron mobility in the InAs quantum well after removal of the Al by wet etching was measured using a Hall bar to be $\SI{25000}{\centi\meter\squared\per\volt\per\second}$. Fabrication details for device 1 can be found in Appendix \ref{sec:fab_super_sideprobe}. The material for device 2 was an InAlAs/InAs/InAlAs heterostructure with a similar epitaxial Al top layer.

\subsection{\label{sec:el_setup}Measurement setup}

\begin{figure}[h!]
\begin{center}
\includegraphics[scale=0.75]{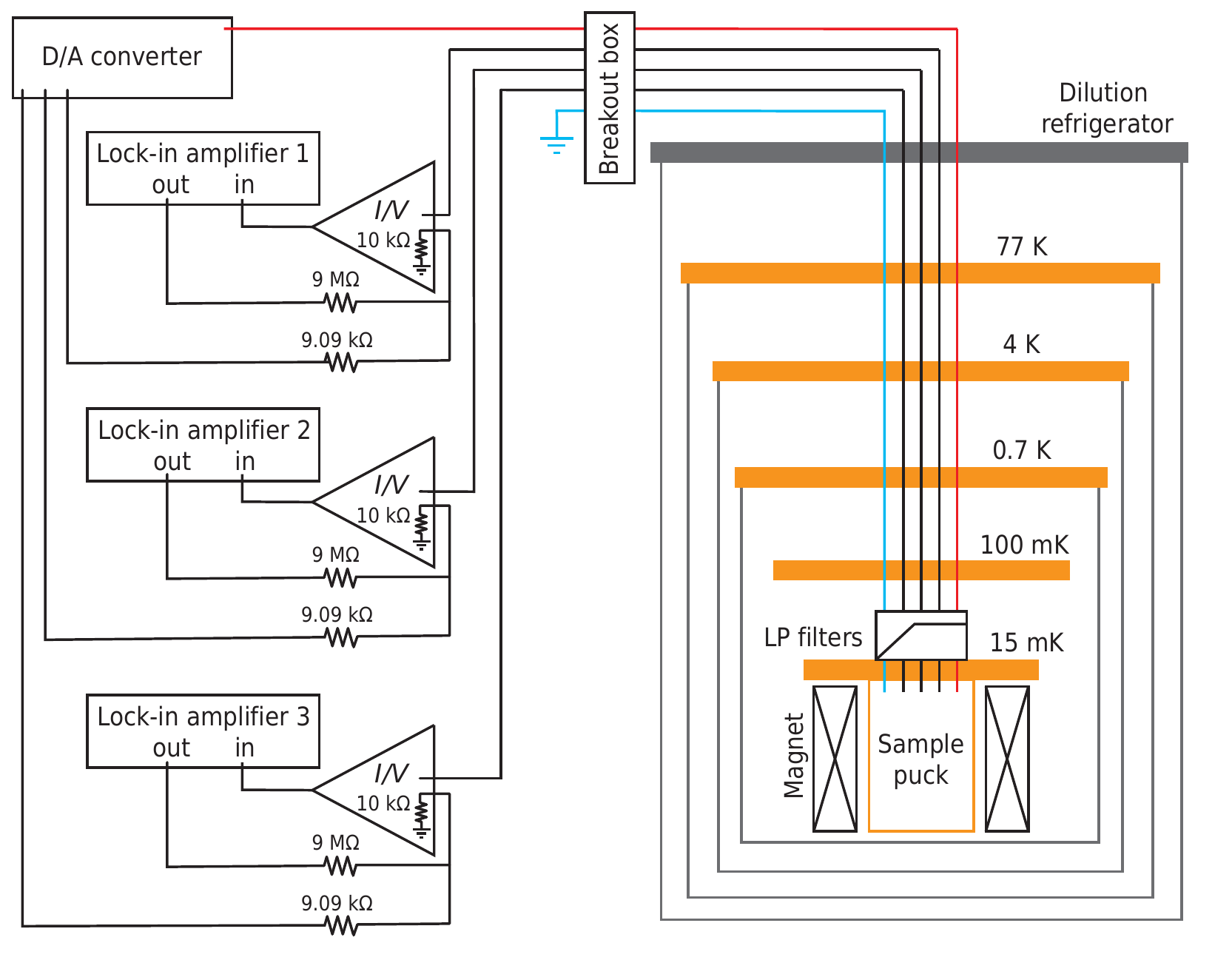}
\end{center}
\caption[Electrical measurement setup for multi-probe tunneling spectroscopy]{\label{fig:setup} Schematic overview of the experimental measurement setup. All lines that supply gate voltages are pictured by the red line. The four lines that are used to connect the ground planes of the sample to the breakout box are represented by a single blue line. The dilution refrigerator is an Oxford Instruments Triton 400. The three current-to-voltage ($I/V$) converting amplifiers are made by Basel Precision Instruments (part number SP983c). Source-drain bias voltage is supplied to the sample via the input voltage offset port of the $I/V$ converters. Suitable resistors together with the input resistance of this port form a voltage divider. The lock-in amplifiers are made by Stanford Research Systems (model SR830 and SR860).}
\end{figure}

A schematic of the experimental setup for transport measurements is depicted in Fig.~\ref{fig:setup}. The sample was mounted in a puck-loading cryo-free dilution refrigerator (Oxford Instruments, Triton 400) equipped with a 6-1-1 \SI{}{\tesla} vector magnet. Throughout the measurements, the mixing chamber of the dilution fridge was at base temperature of roughly \SI{15}{\milli\kelvin}, without active control, as measured by a $\mathrm{RuO_2}$-based thermometer. Gate voltages were generated using an in-house built digital-to-analog converter with 20-bit precision. The three tunneling currents $I_j$ were amplified using a current-to-voltage converter (Basel Precision Instruments SP983c) followed by an AC lock-in amplifier (Stanford Research Systems SR830, SR860). Bias voltages $V^{(j)}_{\mathrm{SD}}$ were applied via the offset voltage inputs of the current-to-voltage converters. The DC component of $V^{(j)}_{\mathrm{SD}}$ was supplied by the digital-to-analog (D/A) converter and an additional AC component $\mathrm{d}V^{(j)}_{\mathrm{SD}}$ at a frequency $f_j$ was supplied by the sine output of lock-in amplifier $j$. The frequencies were $f_1=\SI{42.2}{\hertz}$, $f_2=\SI{31.9}{\hertz}$, $f_3=\SI{20.4}{\hertz}$. In-house built multi-stage low-pass filters at the mixing chamber plate were used to attenuate electrical noise. The two ground planes of Al that connect to the nanowire (NW) at both ends were connected via two lines each to ground at the breakout box. 

For measurements in Figs.~\ref{fig:device_pc}(b-d), the DC bias voltage was swept simultaneously for all three tunnel probes, i.e., $V_{\mathrm{SD}}^{(1)}=V_{\mathrm{SD}}^{(2)}=V_{\mathrm{SD}}^{(3)}$. For all other tunneling spectroscopy measurements, where only two conductances are reported, the third probe was left floating. The two bias voltages were swept sequentially, with the respective other set to zero.

\subsection{\label{sec:peakextraction}Extraction of peak positions}
\begin{figure}[h!]
\begin{center}
\includegraphics[scale=0.75]{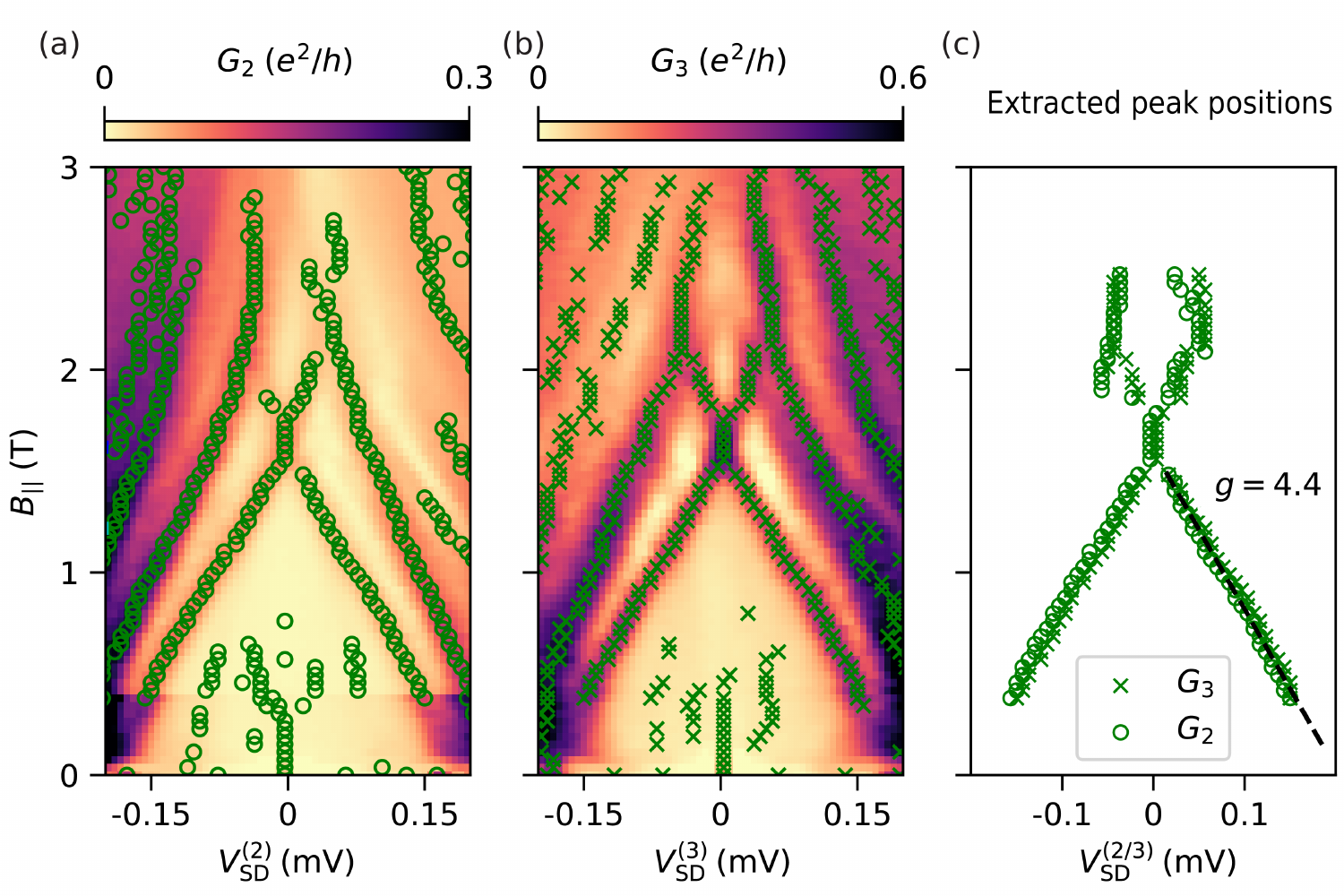}
\end{center}
\caption[]{\label{fig:peak_extraction}(a, b) Identical data to the ones in Figs.\ref{fig:fieldscan_pc}(a, b). The green markers indicate the position of identified local maxima for individual magnetic field values. They originate from subgap states, noise, and a supercurrent peak at low magnetic field values. (c) Locations of the local maxima from (a) and (b) after removing the peaks that we identify not to stem from the low energy subgap states of interest. Linear fit shown as dashed line yields a $g$ factor of 4.4.}
\end{figure}

We extracted peak positions of subgap states from  $G_2$ and $G_3$ as functions of magnetic field and $V_\mathrm{SD}^{(2)}$ and $V_\mathrm{SD}^{(3)}$ used in Figs.~\ref{fig:fieldscan_pc}(a, b) as follows: local maxima of $G_{(2/3)}$ were found for each value of $B_{||}$ within the range $\SI{0.38}{\tesla}<B_{||}<\SI{2.5}{\tesla}$, wherever the subgap state is resolved. Detected peaks are marked in the measurements of $G_2$ and $G_3$ in Figs.~\ref{fig:peak_extraction}(a, b). As a next step, the maxima that are obviously (by eye) not arising from the subgap state of interest but due to noise or higher excited states are discarded. The same applies for values of $B_{||}$ at which only one subgap peak could be detected in one of the two measurements, while other measurement showed two peaks. These post-selected peak positions associated with the subgap state of interest are plotted in Fig.~\ref{fig:peak_extraction}(c).\\

\subsection{\label{sec:control_experiment}No nonlocal signature in the absence of Andreev bound states}
In the case of a rather negative gate voltage $V_\mathrm{W23}$ around \SI{-4}{\volt} we observed no subgap state that appears in both $G_2$ and $G_3$, while the quantum dot (QD) resonance in the tunnel barrier under $\mathrm{P_2}$ was still present. In this configuration, no nonlocal signatures can be measured in tunneling spectroscopy.

\begin{figure}[h!]
\begin{center}
\includegraphics[scale=0.75]{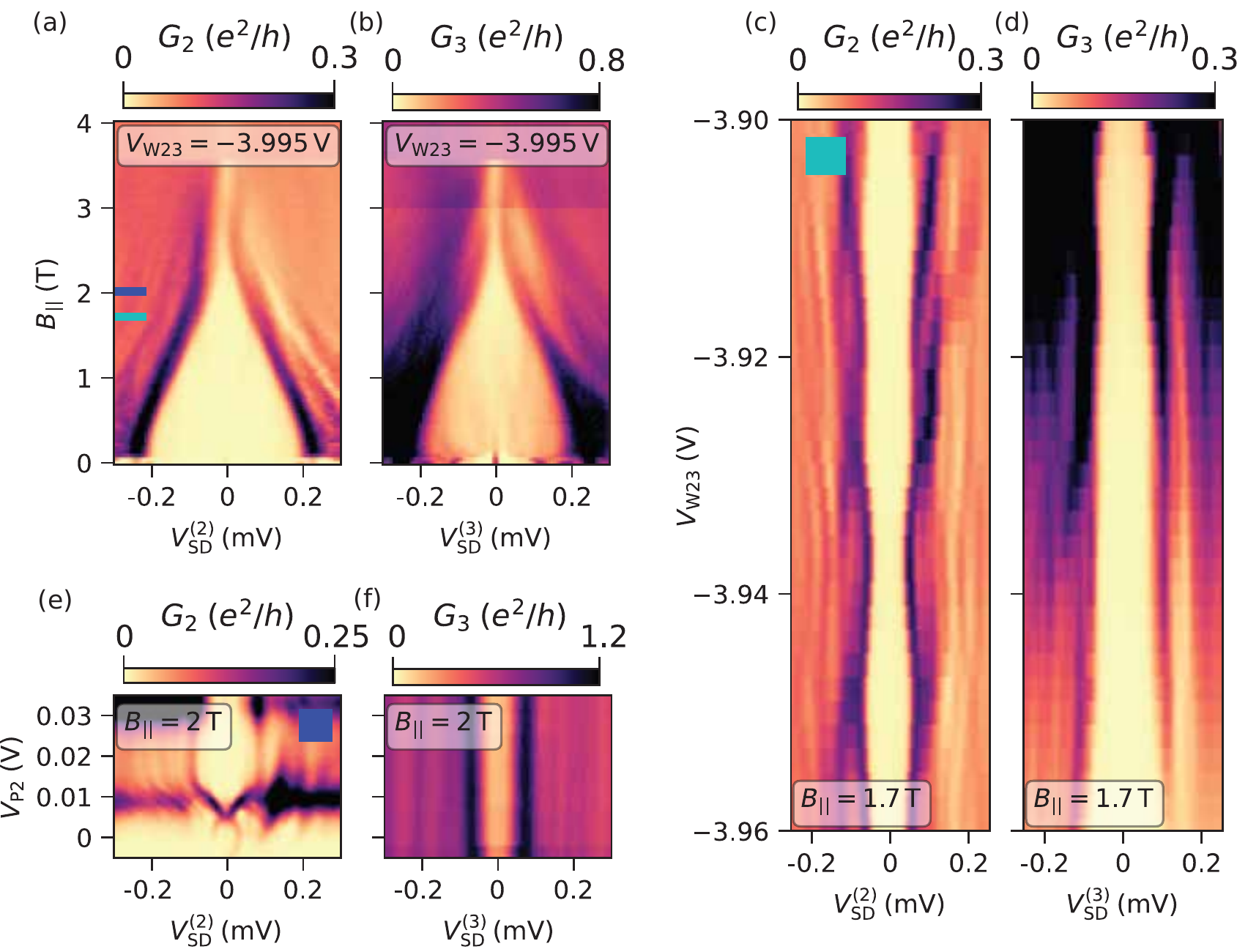}
\end{center}
\caption[No nonlocal signature in the absence of extended bound state]{\label{fig:control} (a) and (b) show tunneling spectroscopy data of $G_2$ and $G_3$ with $V_\mathrm{W12}=V_\mathrm{W34}=\SI{-7}{\volt}$ and $V_\mathrm{W23}=\SI{-3.995}{\volt}$. There is no subgap state visible. Tunneling spectroscopy with respect to changing gate voltage $V_\mathrm{W23}$ at $B_{||}=\SI{1.7}{\tesla}$ is plotted in (c) and (d). There is no subgap state present. Tunneling conductance measurement with respect to the gate voltage $V_\mathrm{P2}$ that tunes the tunnel barrier is depicted in (e) and (f). While $G_2$ shows a clear subgap state at the position of the QD resonance, the data of $G_3$ in (f) shows no measurable change at the same gate voltage.}
\end{figure}

Tunneling spectroscopy in Figs.~\ref{fig:control}(a, b) shows no subgap states crossing zero bias. A measurement of tunneling spectroscopy in a small range around $V_\mathrm{W23}=\SI{-3.93}{\volt}$ at $B_{||}=\SI{1.6}{\tesla}$ in Figs.~\ref{fig:control}(c, d) shows a gap with no subgap states persisting over the whole range of  $V_\mathrm{W23}$. In Fig.~\ref{fig:control}(e) tunneling spectroscopy as a function of the gate voltage $V_\mathrm{P2}$ reveals the QD resonance around $V_\mathrm{P2}=\SI{0.01}{\volt}$ at high bias \cite{chang_YSR, kasper_tuningYSR, lee_singletdoublet, alex_QDinLoop}.
The measurement of $G_3$ while sweeping $V_\mathrm{P2}$ over the QD resonance is shown in Fig.~\ref{fig:control}. Throughout the measurement, $G_3$ is independent of $V_\mathrm{P2}$, in particular around the value $V_\mathrm{P2}=\SI{0.01}{\volt}$ at which the QD resonance appears in $G_2$.

In summary, in the absence of a discrete subgap state extending between the two probes, no change in the density of states is measured on one probe while a QD on the other tunnel probe is brought on resonance. This shows that only an extended ABS can hybridize with a local QD and lead to a measureable nonlocal signature on a tunnel probe $\SI{0.6}{\micro\meter}$ away from the QD.

\subsection{\label{sec:S2}Andreev bound states under gate W\textsubscript{12}}
\begin{figure}[h!]
\begin{center}
\includegraphics[scale=0.75]{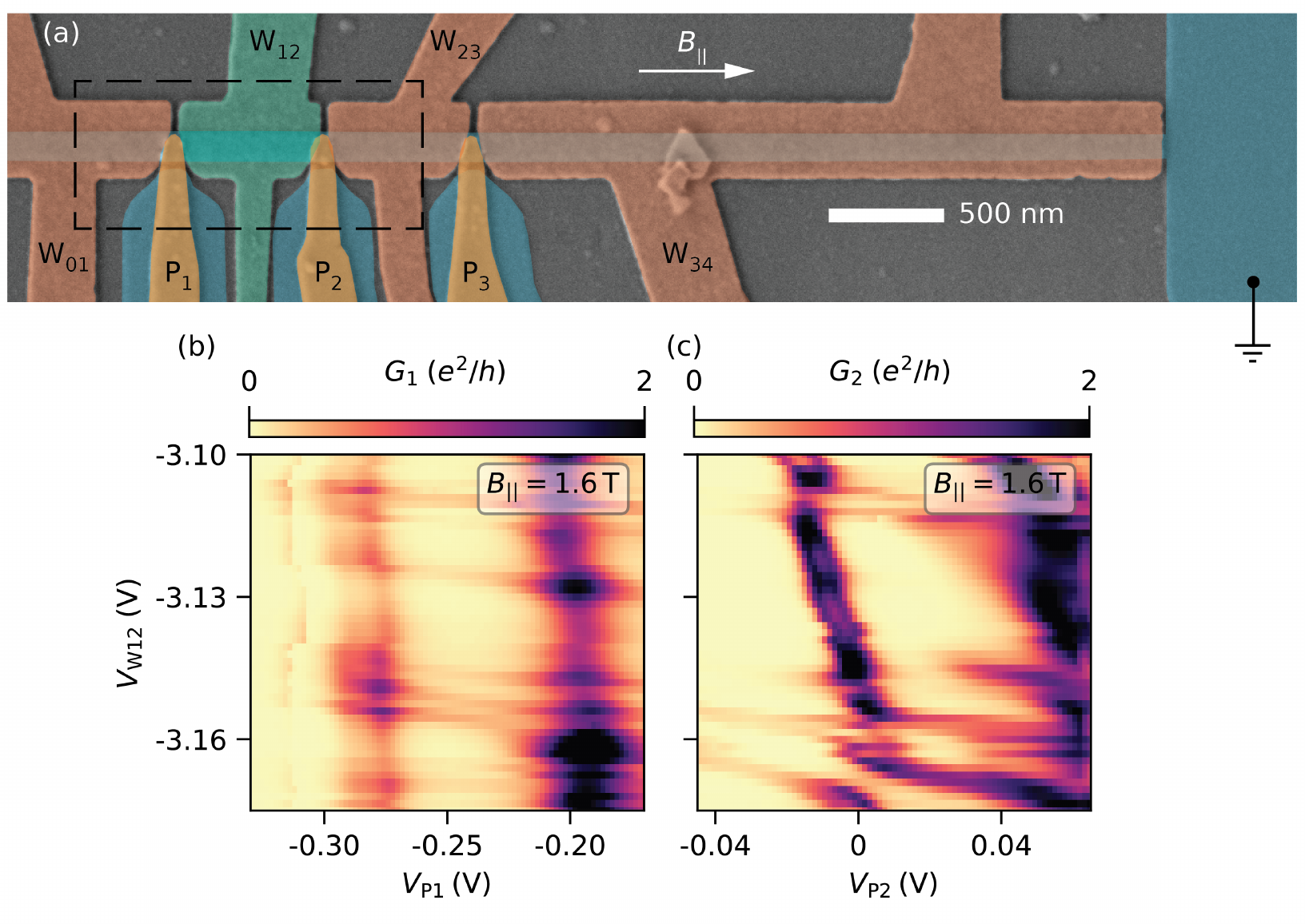}
\end{center}
\caption[]{\label{fig:sem_left}\label{fig:gatemap_left}(a) False-color scanning electron micrograph of device 1. (b) and (c) show conductance resonances measured at $B_{||}=\SI{1.6}{\tesla}$ at $V^{(j)}_{\mathrm{SD}}=\SI{0}{\volt}$. Both measurements show two distinct families of states - horizontal ones, that strongly depend on $V_\mathrm{W23}$, and vertical ones, that strongly depend on the gate voltage that tunes the respective tunnel barrier.}
\end{figure}

In the Section \ref{sec:abs_pc}, data of the hybridization between an ABS in the NW segment under the gate $\mathrm{W}_{23}$ and a local QD is shown. Here data of the comparable experiment with an ABS in the NW segment under the gate $\mathrm{W}_{12}$, which is marked green in Fig.~\ref{fig:gatemap_left}(a), is presented. 

The gate voltage dependence of conductance resonances with respect to the voltages on the gate $V_\mathrm{W12}$ that covers the NW and the two gates $\mathrm{P}_{1}$ and $\mathrm{P}_{2}$, while keeping the neighboring gate voltages at $V_\mathrm{W01}=V_\mathrm{W23}=\SI{-4.5}{\volt}$ can be seen in Fig.~\ref{fig:gatemap_left}(c, d). There are horizontal resonances in $G_2$ and $G_3$ that are associated with extended states in the NW. In addition vertical resonances, which can be attributed to localized states in the respective tunnel barrier regions, are present. 

\begin{figure}[h!]
\begin{center}
\includegraphics[scale=0.80]{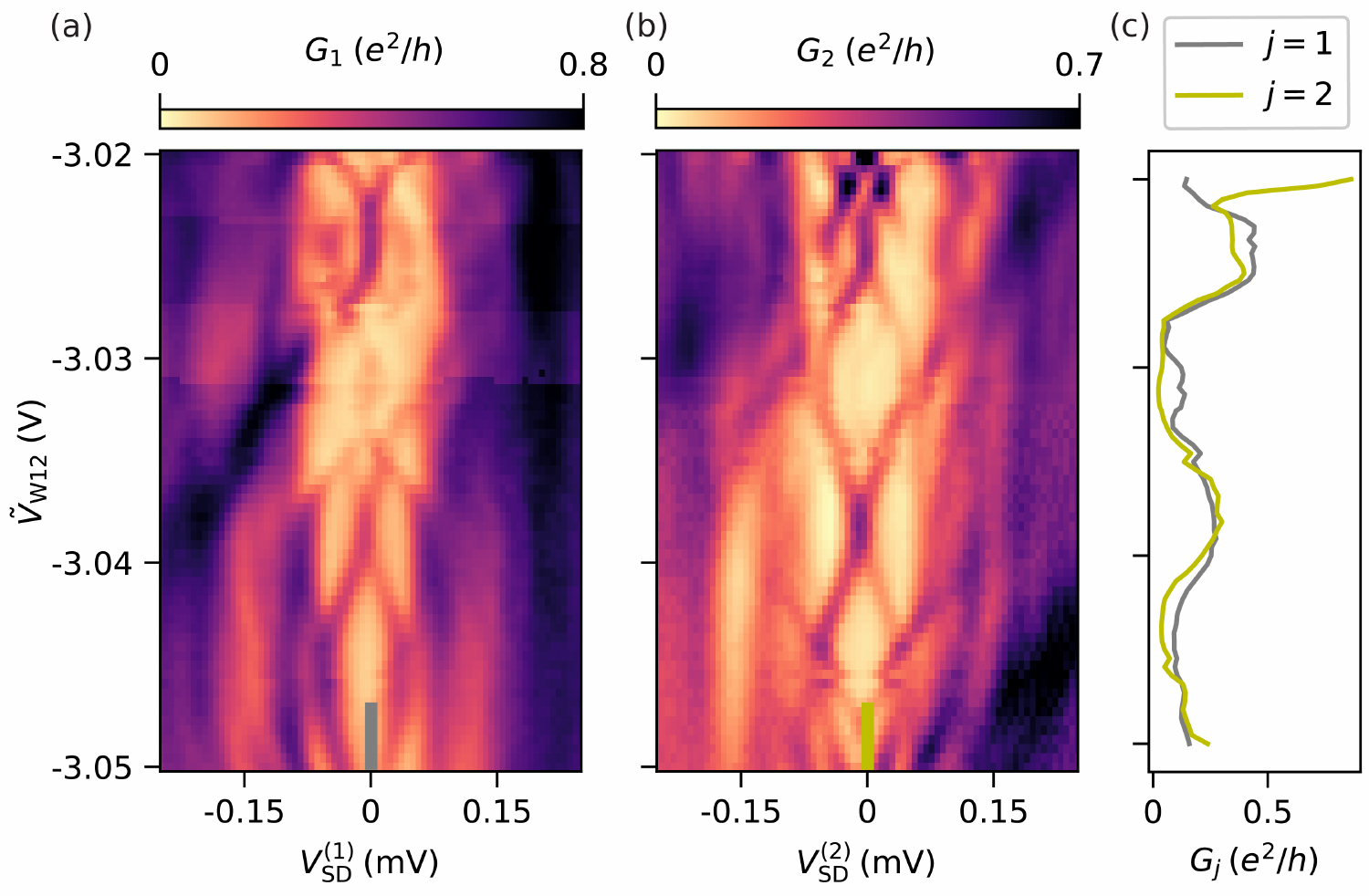}
\end{center}
\caption[]{\label{fig:plunger_scan_left}(a, b) Tunneling spectroscopy measurements of $G_2$, $G_3$ with respect to $\tilde V_\mathrm{W12}$. The voltages on the other gates were fixed at $V_\mathrm{W01}=V_\mathrm{W23}=V_\mathrm{W34}=\SI{-4.5}{\volt}$. Subgap states appear in both measurements of $G_1$ and $G_2$. The data for $G_2$ and $G_3$ at $V^{(1)}_{\mathrm{SD}}=V^{(2)}_{\mathrm{SD}}=\SI{0}{\volt}$ is plotted in (c) for comparison.}
\end{figure}

Tunneling spectroscopy data as a function of $V_\mathrm{W12}$, using tunnel probes 1 and 2, are shown in Figs.~\ref{fig:plunger_scan_left}(a, b). Voltages on the gates $\mathrm{P}_1$ and $\mathrm{P}_2$ were compensated according to:
\begin{equation}
    \begin{aligned}
     V_\mathrm{P1}&=-\SI{307}{\milli\volt}-\frac{0.01}{0.80} \cdot(V_\mathrm{W12}+\SI{3}{\volt})\\
     V_\mathrm{P2}&=-\SI{15}{\milli\volt}-\frac{0.015}{0.80} \cdot(V_\mathrm{W12}+\SI{3}{\volt}).
    \end{aligned}
\end{equation}
To denote that more than one gate voltage was changed during the measurement, the variable is labeled $\tilde V_\mathrm{W12}$ instead of $V_\mathrm{W12}$.
Tunneling conductances $G_1$ and $G_2$ in Figs.~\ref{fig:plunger_scan_left}(a, b) show subgap states as a function of $\tilde V_\mathrm{W12}$. These states show similar dependences on $\tilde V_\mathrm{W12}$ in the two measurements. This is emphasized by the line cut taken at zero bias in Fig.~\ref{fig:plunger_scan_left}(c).

\begin{figure}[h!]
\begin{center}
\includegraphics[scale=0.80]{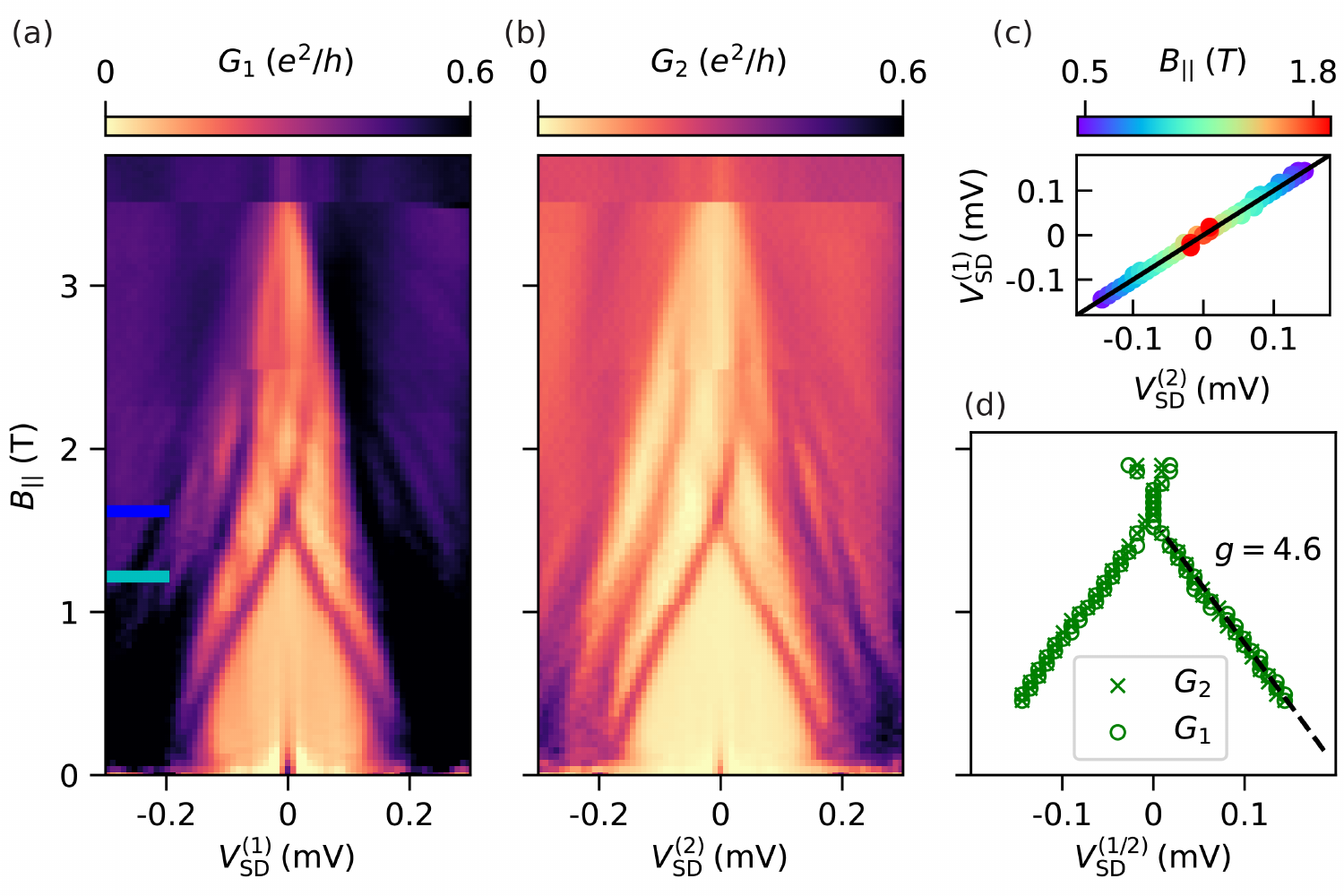}
\end{center}
\caption[]{\label{fig:field_scan_left} (a, b) Tunneling spectroscopy measurements $G_1$, $G_2$ with respect to magnetic field $B_{||}$ at $V_\mathrm{W12}=\SI{-3.02}{\volt}$. The peak positions that were extracted according to the procedure described in Section \ref{sec:peakextraction} are plotted in (d). Linear fit shown as dashed line in (d) yields a $g$ factor of 4.6. A parametric plot of the same data points as in (d) is shown in (c).}
\end{figure}
Tunneling conductance as a function of magnetic field $B_{||}$ is shown in Fig.~\ref{fig:field_scan_left}(a, b). Subgap states were found to emerge at low magnetic field from the quasiparticle continuum. At $B_{||}=\SI{1.6}{\tesla}$, subgap states cross zero bias and overshoot at $B_{||}=\SI{2}{\tesla}$. For every value of $B_{||}$ we extracted the peak position from the measured trace of $G_1(G_2)$ according to the procedure outlined in Section \ref{sec:peakextraction}. Extracted peak positions are shown in Fig.~\ref{fig:field_scan_left}(d). The same peak positions are plotted parametrically in Fig.~\ref{fig:field_scan_left}(c). In the parametric plot, all points lie on or close to the identity line demonstrating the similar behavior of the subgap states in measurements of $G_1$ and $G_2$. Due to their correlated behavior in $G_1$ and $G_2$ with respect to magnetic field and $V_\mathrm{W12}$ gate voltage, we attribute these subgap states to ABSs that extend over the NW segment under gate $V_\mathrm{W12}$.\\

\begin{figure}[h!]
\begin{center}
\includegraphics[scale=0.80]{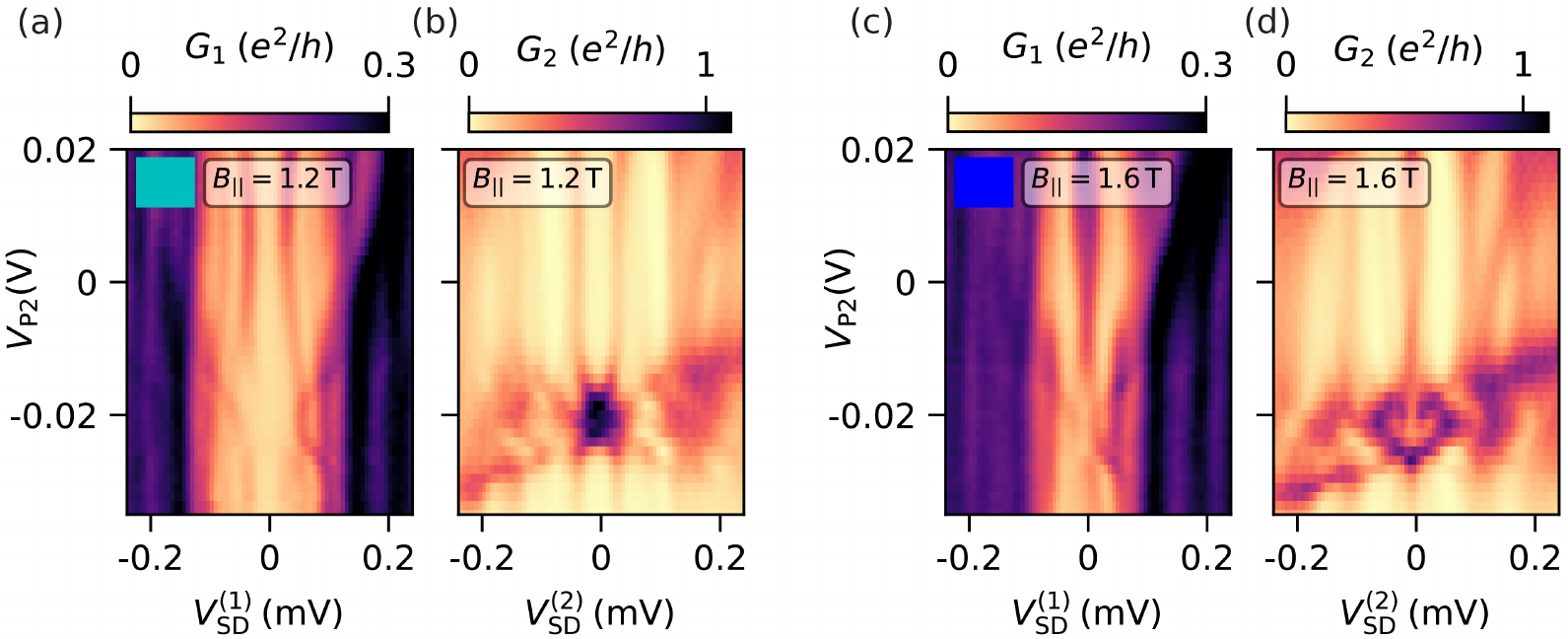}
\end{center}
\caption[]{\label{fig:pc_left}Tunneling spectroscopy data at $V_\mathrm{W12}=\SI{-3.02}{\volt}$ as a function of gate voltage $V_\mathrm{P2}$ is plotted in (a) and (b). A QD resonance is clearly visible in the measurement of $G_2$. The same measurement at higher magnetic field $B_{||}=\SI{1.6}{\tesla}$ is shown in (c) and (d).}
\end{figure}

To investigate the effect on the subgap states as the ABS is brought on resonance with the QD in the tunnel barrier under $\mathrm{P}_2$, we performed tunneling spectroscopy with respect to the voltage on gate $\mathrm{P}_2$. The measurement of $G_1$ and $G_2$ at $B_{||}=\SI{1.1}{\tesla}$ is plotted in Figs.~\ref{fig:pc_left}(a) and (b). A QD resonance can be seen around $V_\mathrm{P2}=\SI{-0.02}{\volt}$ as signal enhancement at high bias in the measurement of $G_2$. In the vicinity of the QD resonance, the subgap states exhibit a `bowtie' shape with a conductance enhancement at the point where the states cross zero bias in $G_2$. The simultaneous measurement of tunneling conductance $G_1$ shows that the subgap states are almost constant in energy, up to a small region around $V_\mathrm{P2}\sim\SI{-0.02}{\volt}$ around the QD resonance. In the vicinity of the QD resonance, the lowest energy subgap state decreases in intensity in $G_1$ such that it is not visible at the position of the QD resonance.

A measurement of the same type at a higher magnetic field value of $B_{||}=\SI{1.6}{\tesla}$ is shown in Figs.~\ref{fig:pc_left}(c) and (d). The quantity $G_2$, which is measured through the QD, shows the QD resonance crossing the gap at $V_\mathrm{P2}=\SI{-0.02}{\volt}$. A similar pattern to the one shown in Fig.\ref{fig:pc}(c) is visible around zero bias at the point where the QD is on resonance with the subgap state close to zero bias. The resolution is not sufficient to resolve the exact pattern and determine whether the subgap states shift in energy around the QD resonance to form a `diamond' shape. The measurement of $G_1$ shows that the conductance value measured for the subgap state at zero bias decreases abruptly around the QD resonance while a splitting of the states away from zero energy is resolved. 

For all measurements in Fig.~\ref{fig:pc_left} a clear effect on the ABSs can be observed at tunnel barrier 1 when the subgap state is brought on resonance with a QD in tunnel barrier 2, even though the precise line shape of the subgap state is not resolved in this measurement. The change in conductance value of the subgap state at the point of the QD at both ends is stronger in this measurement compared to the data presented in Section \ref{sec:pc}.

\section{\label{sec:sup_dev_2_pc}Supplementary information on device 2}

\begin{figure}[h!]
\begin{center}
\includegraphics[scale=0.80]{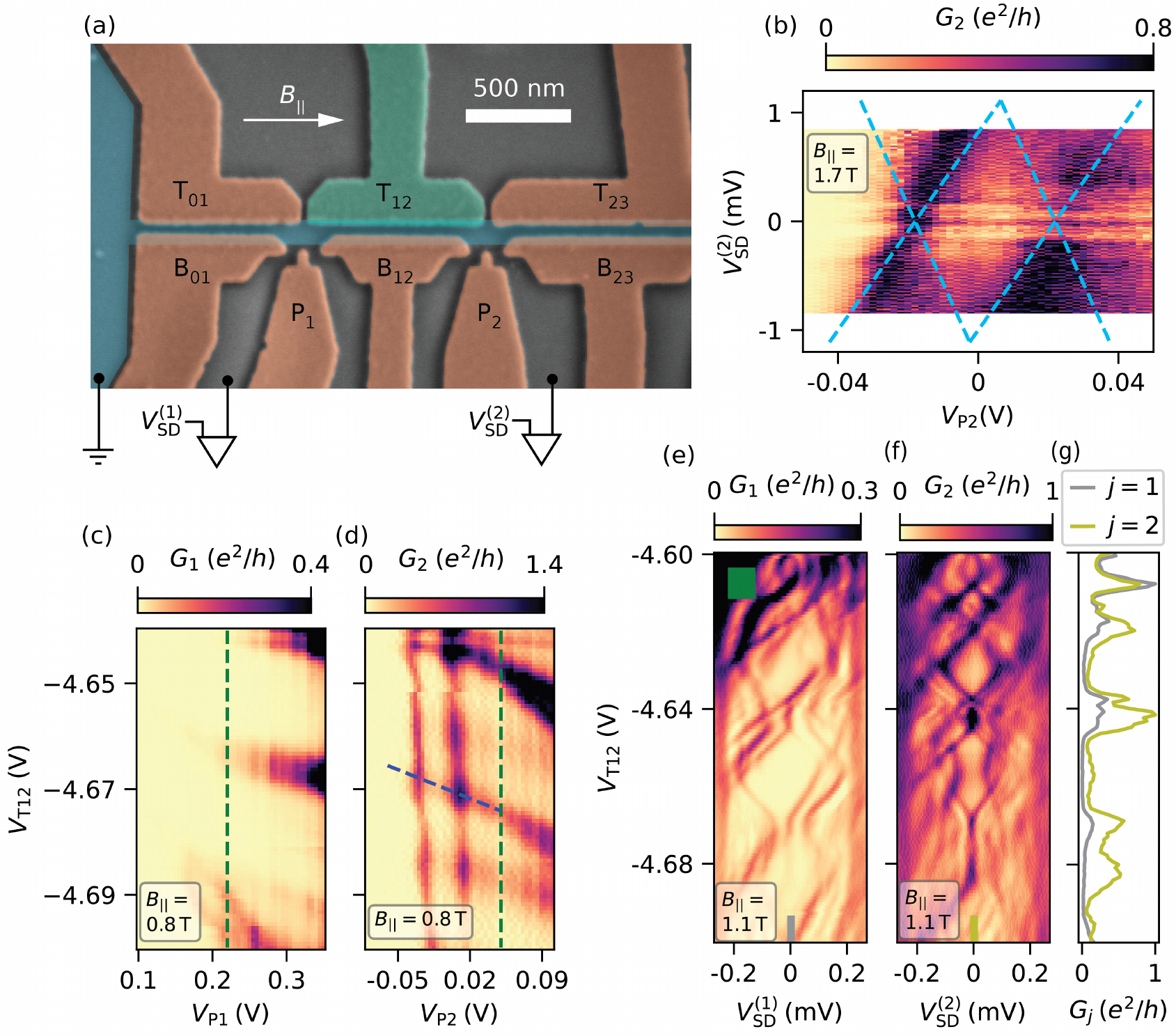}
\end{center}
\caption[]{\label{fig:gate_coupling_dev2}\label{fig:sem_dev2}\label{fig:P_scan_dev2}(a) False-color scanning electron micrograph of device 2. The device consists of a superconducting Al NW on top of a InAs quantum well. The gates labeled $V_{\mathrm{P}j}$ can be used to tune the tunnel barrier between the tunnel probe $j$ and the wire ($j \in \{1,2,3\}$). The gates labled $V_{\mathrm{B}kl}$ form the confining potential of the quasi one-dimensional NW at the bottom edge while also forming tunnel barriers. The gates labeled  $\mathrm{T}_{kl}$ confine the NW at the top edge and tune the chemical potential of individual NW segments ($kl \in \{01,12,23\}$). (c, d) Tunneling conductance measurements of $G_1$, $G_2$ with $V^{(j)}_{\mathrm{SD}}=\SI{0}{\volt}$. States that couple strongly to the gate voltage $V_\mathrm{T12}$ are visible in both measurements. Two sharp resonances that strongly depend on the gate voltage $V_\mathrm{P2}$ appear in the measurement of $G_2$. (b) Tunneling spectroscopy at $V_\mathrm{T12}=\SI{-4.67}{\volt}$. The dashed lines serve as a guide to the eye. Tunneling spectroscopy at $B_{||}=\SI{1.1}{\tesla}$, when moving the gate voltages along the green dashed lines at $V_\mathrm{P1}=\SI{0.22}{\volt}$  and at $V_\mathrm{P2}=\SI{0.05}{\volt}$ in (c, d) are plotted in (e, f). While the absolute values of tunneling conductance in $G_1$ and $G_2$ are different, the structure of the states in energy is nearly identical. A line cut at $V^{(1)}_{\mathrm{SD}}=V^{(2)}_{\mathrm{SD}}=\SI{0}{\volt}$ of the data in (e, f) is shown in (g).}
\end{figure}

We fabricated a second device with semiconducting, instead of superconducting, sideprobes, i.e, with the epitaxial Al removed in the leads of the sideprobes. Fabrication details of device 2 can be found in Appendix \ref{sec:fab_semi_sideprobe}. The device is depicted in Fig.~\ref{fig:sem_dev2}(a). It consists of a superconducting strip of Al on top of an InAs quantum well. The gates labeled $\mathrm{T}_{kl}$ are used to electrostatically define the NW on one side and tune the electron density of individual segments of the NW. The gates labeled $\mathrm{B}_{kl}$ serve the purpose of defining the wire on the other side, tuning its electron density, and forming a quantum point contact between adjacent $\mathrm{B}_{kl}$ gates. The gates $\mathrm{P}_1$ and $\mathrm{P}_2$ add additional control over the quantum point contacts. In the data presented in the following, the gate voltages $V_\mathrm{T01}=V_\mathrm{T23}=\SI{-6}{\volt}$, $V_\mathrm{B01}=V_\mathrm{B23}=\SI{-2.4}{\volt}$, $V_\mathrm{B12}=\SI{-2.5}{\volt}$ were kept at fixed values.\\ 

Measurements of conductance resonances in $G_1$ ($G_2$) at zero bias and magnetic field $B=\SI{0.8}{\tesla}$ with respect to the gate voltages $V_\mathrm{T12}$ and $V_\mathrm{P1}$ ($V_\mathrm{P2}$) are shown in Figs.~\ref{fig:gate_coupling_dev2}(c) and (d). Both measurements reveal resonances that strongly depend on $V_\mathrm{T12}$ originating from states inside the NW. The measurement of $G_2$ furthermore shows two resonances that strongly couple to the gate voltage $V_\mathrm{P2}$. These features can be associated with a QD in the tunnel barrier, which is verified by bias spectroscopy shown in Fig.~\ref{fig:gate_coupling_dev2}(b). The conductance shows a Coulomb diamond at high bias as emphasized by the cyan, dashed lines. This allows for an estimate of the charging energy to be $\sim \SI{1}{\milli\eV}$. 

Tunneling spectroscopy measurements of subgap states in $G_1$ and $G_2$ at $B_{||}=\SI{1.1}{\tesla}$ are shown in Fig.~\ref{fig:P_scan_dev2}(e) and (f). The gate voltages $V_\mathrm{P1}=\SI{0.22}{\volt}$ and $V_\mathrm{P2}=\SI{0.05}{\volt}$ were kept constant during this measurement, which corresponds to the green dashed lines in Fig.~\ref{fig:P_scan_dev2}(c, d). Similar to the results of device 1, states appear in both measurements of $G_1$ and $G_2$ with the same nontrivial dependence on the voltage $V_\mathrm{T12}$. The conductance value with which states appear in the two measurements can be very different, however. A line cut at zero bias is depicted in Fig.~\ref{fig:P_scan_dev2}(g) and captures the correlated dependence of conductance peaks with respect to $V_\mathrm{T12}$ in the form of concomitant peak positions. The difference in conductance values results in a difference in peak heights for the two different curves in Fig.~\ref{fig:P_scan_dev2}(g). \\

\begin{figure}[h!]
\begin{center}
\includegraphics[scale=0.90]{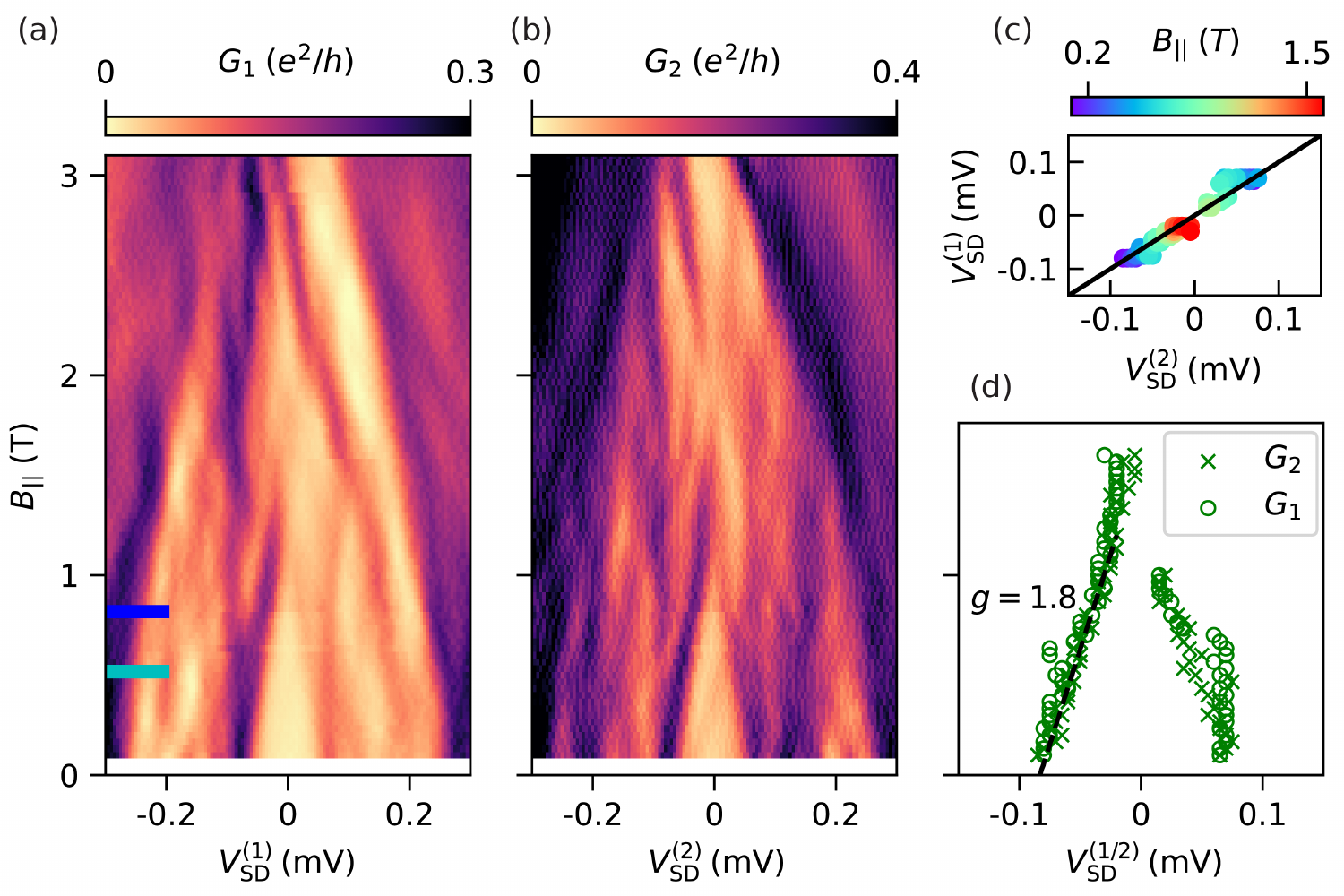}
\end{center}
\caption[]{\label{fig:fieldscan_pc_dev2}(a, b) Tunneling spectroscopy with respect to magnetic field $B_{||}$ with $V_\mathrm{T12}=\SI{-4.67}{\volt}$. The induced gap at zero field is $\sim \SI{80}{\micro\eV}$. Peak positions extracted from $G_1$ and $G_2$ associated with the lowest energy state are plotted in (d). Linear fit shown as dashed line in (d) yields a $g$ factor of 1.8. (c) Parametric plot of the peak positions from (d). }
\end{figure}

Figures~\ref{fig:fieldscan_pc_dev2}(a, b) show tunneling spectroscopy data of $G_1$ and $G_2$ for device 2 as a function of magnetic field $B_{||}$ parallel to the NW. The induced gap at low field given by the lowest energy states is $\sim \SI{80}{\micro\eV}$. Note that this device was fabricated on material where the InAs quantum well is separated by an InAlAs barrier from the superconducting Al. 

The lowest energy states in both $G_1$ and $G_2$ cross zero bias at around $B_{||}=\SI{1.5}{\tesla}$. Extracting the peak positions for each value of $B_{||}$ and plotting them parametrically reveals strong correlation as shown in Fig.~\ref{fig:fieldscan_pc_dev2}(c). Based on this strong correlation in $G_1$ and $G_2$ with respect to magnetic field and $V_\mathrm{T12}$, we attribute these subgap states to extended ABSs in the NW.\\

\begin{figure}[h!]
\begin{center}
\includegraphics[scale=0.85]{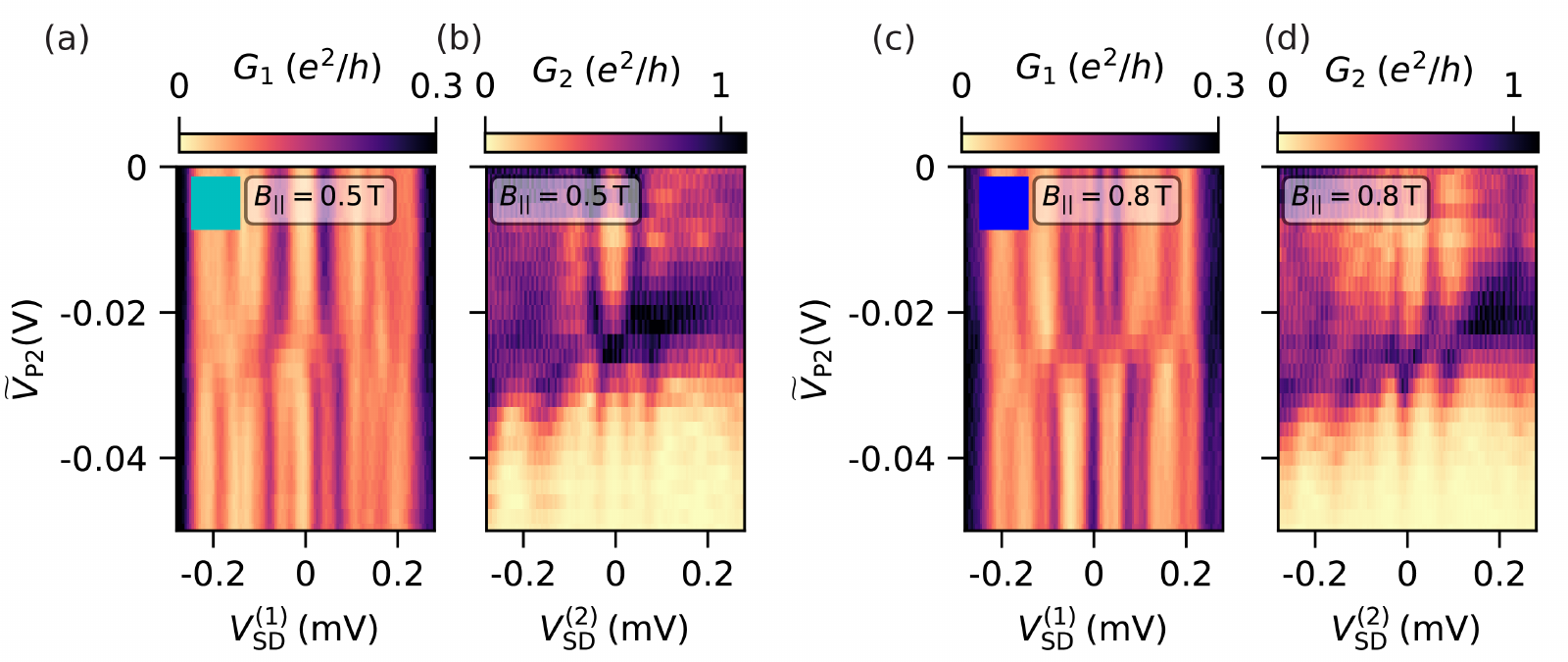}
\end{center}
\caption[]{\label{fig:pc_dev2}(a, b) Tunneling conductance measured while changing the gate voltages $V_\mathrm{P2}$ and $V_\mathrm{T12}$ along the blue dashed line according to equation \ref{eq:pradaclarke_compensation}. To indicate that more than one gate voltage was changed, the variable on the vertical axes is labeled $\tilde V_\mathrm{P2}$ instead of $V_\mathrm{P2}$. At $\tilde V_\mathrm{P2} = \SI{-0.025}{\volt}$ a QD resonance is visible in $G_2$. In both $G_1$ and $G_2$, the subgap states draw a `bowtie' shape at this value of gate voltage. (c, d) The same measurement as in (a, b) but at a higher magnetic field value of $B_{||}=\SI{0.8}{\tesla}$. In $G_1$ one can clearly see two peaks merging into a zero-bias conductance peak at the position of the QD resonance. }
\end{figure}

To investigate the behavior around places where the ABSs in the NW are on resonance with the QD in tunnel barrier 2, we measured tunneling spectroscopy with respect to $V_\mathrm{P2}$. While sweeping $V_\mathrm{P2}$, the gate voltage $V_\mathrm{T12}$ was compensated to mitigate the effect that $V_\mathrm{P2}$ has on the ABSs inside the NW. The parametric equation that was used is given by 
\begin{align}
    \label{eq:pradaclarke_compensation}
    V_\mathrm{T12}&=-\SI{4.664}{\volt}-\frac{0.014}{0.18}\cdot (V_\mathrm{P2}+\SI{0.08}{\volt})
\end{align}
and is depicted by the blue dashed line in Fig.~\ref{fig:gate_coupling_dev2}(d). To denote that more than one gate voltage was changed during the measurement, the variable is labeled $\tilde V_\mathrm{P2}$ instead of $V_\mathrm{P2}$. 
The measured tunneling spectroscopy $G_1$ and $G_2$ is depicted in Fig.~\ref{fig:pc_dev2}(a) and (b) for a magnetic field value of $B_{||}=\SI{0.5}{\tesla}$. The lowest energy subgap states in $G_2$ show a clear crossing at zero energy with the characteristic `bowtie' shape at the location of the QD resonance at $\tilde V_\mathrm{P2}=\SI{-0.02}{\volt}$. The states at higher bias are not well resolved in the measurement of $G_2$. The lowest energy subgap states in $G_1$ show a `bowtie' shape in the form of a zero bias crossing at the location of the QD resonance. For the measurement of $G_1$, the higher excited states are well resolved and a shift in energy for all higher excited states is visible at the position of the QD resonance. 

Identical measurements of $G_1$ and $G_2$ at a higher magnetic field, $B_{||}=\SI{0.8}{\tesla}$, are depicted in Figs.~\ref{fig:pc_dev2}(c, d). At the point of the QD resonance all conductance resonances shift in energy. The pair of subgap states with lowest energy merge from two separated peaks at $\tilde V_\mathrm{P2}>\SI{-0.02}{\volt}$ to a single peak at zero bias for $\tilde V_\mathrm{P2}<\SI{-0.02}{\volt}$. This pattern shows close similarity to the results presented in Ref.~\cite{mingtang_science}.

\chapter{Nonlocal conductance spectroscopy of Andreev bound states in gate-defined InAs/Al nanowires}
\chaptermark{Nonlocal conductance spectroscopy of ABS in 2DEG based nanowires}
\label{ch:2DEG_nonlocal} 

\section{Introduction}
Electrical transport measurements are a cornerstone of solid-state physics. For semiconducting nanowires (NWs) which are proximitized by a superconductor, normal metal-insulator-superconductor (N-I-S) tunneling conductance measurements reveal the spectrum of bound states. The process of Andreev reflection enables a measurable current in the presence of states below the superconducting gap. This process of quasiparticle (QP) reflection at the boundary between normal and superconducting phase bears similarity with the reflection of photons from a phase-conjugating mirror \cite{beenakker_microjunctions, beenakker_mirror, beenakker_rmt4topoNIS}. Recently, a novel device geometry has been realized that makes it possible to measure tunneling currents at two normal leads that are connected to the same proximitized, semiconducting NW with the parent superconductor being electrically grounded \cite{gerbold_nonlocal,denise_nl_gapclosing}. 

In this experimental setting the nonlocal tunneling current flowing from one normal lead to the other consists of QPs that are transmitted or crossed Andreev reflected \cite{datta_super_LBK,Takane1992, lobos_topo_pt_NSN, andreev_rectifier}. One can obtain the nonlocal conductance by a differential measurement of this current. The transport processes underlying this signal are mediated via extended quantum states that couple to both leads. This measurement technique is therefore promising for the characterization of proximitized NWs and their bulk states. Theoretical studies predict a characteristic signature in nonlocal conductance at the topological phase transition in Rashba NWs \cite{andreev_rectifier, karsten_nl_spectroscopy, SDS_nl_conductance, hess_nl_quasimajo, hao_nextstep, dima_protocol}. Experiments with InAs NWs uncovered symmetry relations between local and nonlocal conductances \cite{gerbold_nonlocal}. The closing of the induced gap measured in nonlocal conductance at finite magnetic fields has furthermore been observed \cite{denise_nl_gapclosing}. Utilizing the same transport processes, quantum dots coupled to one superconducting and two normal leads have been used to demonstrate cooper pair splitting \cite{CPS_1, CPS_2, CPS_3, CPS_4}. Nonlocal spectroscopy of subgap states induced by the quantum dot states have been performed in similar devices \cite{gramich_nl_abs, CPS_3}.\\

Here, we present local and nonlocal conductance measurements carried out on gate-confined NWs in InAs two-dimensional electron gas (2DEG) which is proximitized by superconducting Al. Tunneling probes are coupled laterally to the NW. Extended Andreev bound states (ABSs) couple to neighbouring probes and give rise to characteristic nonlocal conductance signatures. We investigated three working devices. Data from {device} 1 is presented in the following, data from {device} 2 can be found in the Section \ref{sec:dev2_nl}. {Device} 3 did not show extended ABS and is discussed in Section \ref{sec:fine_tuning}. \\

\begin{figure}[h!]
\begin{center}
\includegraphics[scale=1.0]{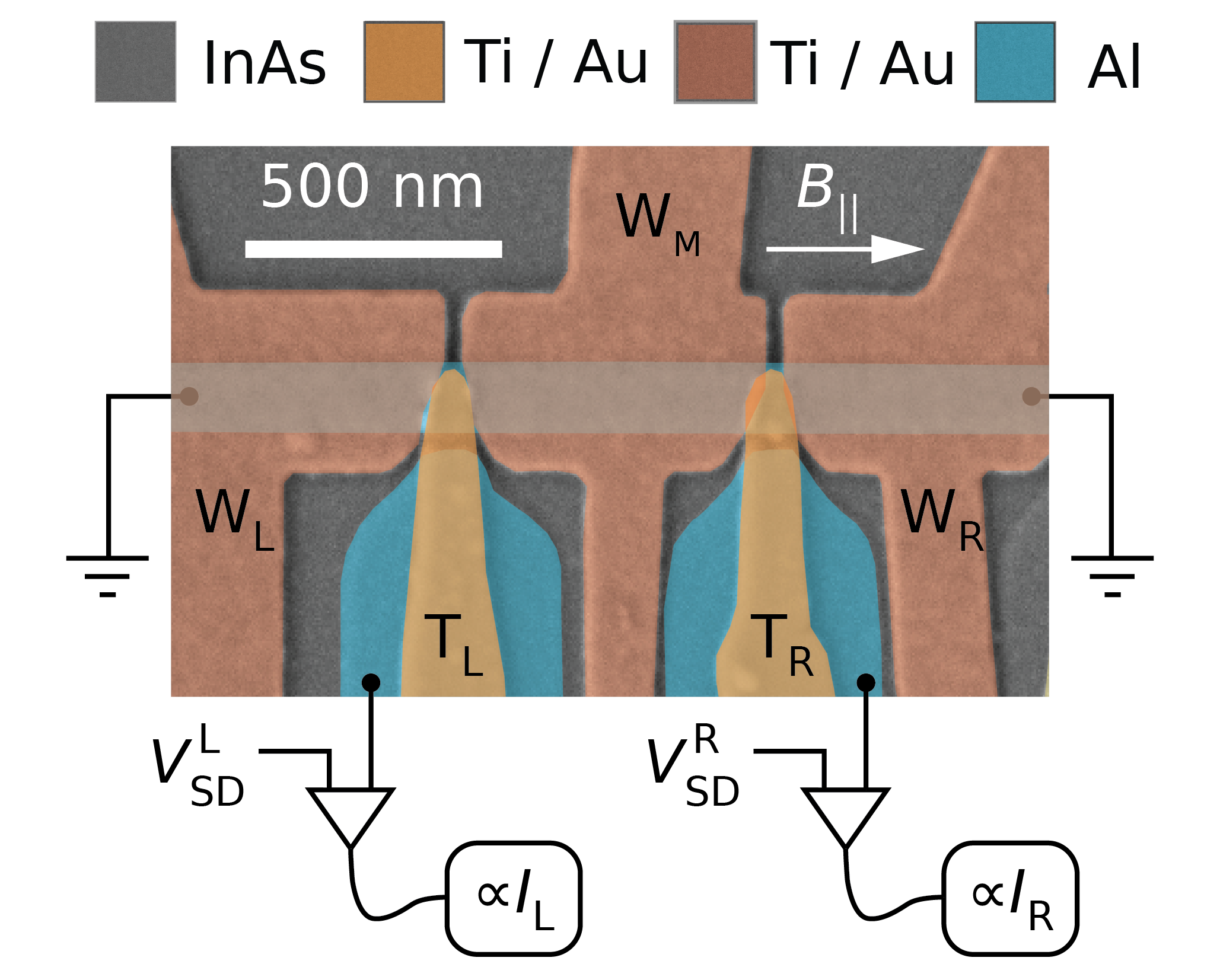}
\end{center}
\caption[Micrograph of a three-terminal device (device 1) based on proximitized 2DEG]{\label{fig:device}False-color electron micrograph of {device} 1. A proximitized quasi one-dimensional NW is formed in the InAs quantum well (gray) under the strip of superconducting Al (blue) by lateral electrostatic confinement from the gates $\mathrm{W_L,W_M,W_R}$ (red). Probes made from Al are separated by a tunnel barrier from the NW and allow for measurements of the tunneling currents $I_\mathrm{L}$, $I_\mathrm{R}$ into the NW. The gates $\mathrm{T_L}$ and $\mathrm{T_R}$ (orange) tune the tunnel barriers between the NW and the probes.}
\end{figure} 

\section{Three-terminal device and nonlocal conductance}
\label{sec:2DEG_3terminal}

{Device} 1 is shown in Fig.~\ref{fig:device}. It consists of a superconducting strip of Al on top of a shallow 2DEG formed in an InAs quantum well. Two superconducting probes made from Al were defined $\SI{50}{\nano\meter}$ away from the superconducting strip. Gates made from Ti/Au are separated by $\mathrm{HfO_x}$ gate dielectric from the semiconductor and superconductor. Details on the fabrication of device 1 are given in Appendix \ref{sec:fab_super_sideprobe}. The gates labeled $\mathrm{W_L}$, $\mathrm{W_M}$, and $\mathrm{W_R}$ cover different segments of the Al strip and electrostatically confine a quasi one-dimensional NW in the semiconductor. The gate $\mathrm{T_L}$ ($\mathrm{T_R}$) depletes the semiconductor between left (right) probe and NW forming a tunnel barrier. The ends of the superconducting strip were connected to electrically grounded planes of superconducting Al. \\

Connecting individual current-to-voltage converters to the two probes allows for the measurement of the currents $I_\mathrm{L}$ and $I_\mathrm{R}$  running into the two leads as a function of the source drain bias voltages $V^\mathrm{L}_\mathrm{SD}$ and $V_\mathrm{SD}^\mathrm{R}$. Note that positive current direction is defined as current running from the amplifier towards the device. The lock-in detection technique as described in Ref. \cite{gerbold_nonlocal} was used to measure the tunneling conductances
\begin{align}\label{eq:Gmat}
    G_\mathrm{LL}=\mathrm{d}I_\mathrm{L}/\mathrm{d}V^\mathrm{L}_\mathrm{SD} && G_\mathrm{LR} = \mathrm{d}I_\mathrm{L}/\mathrm{d}V^\mathrm{R}_\mathrm{SD}
\end{align}
and $G_\mathrm{RR}$, $G_\mathrm{RL}$ defined analogously. 
Further details on the measurement technique are outlined in the Section \ref{sec:setup_nl}. \\

\begin{figure}[h!]
\begin{center}
\includegraphics[scale=0.9]{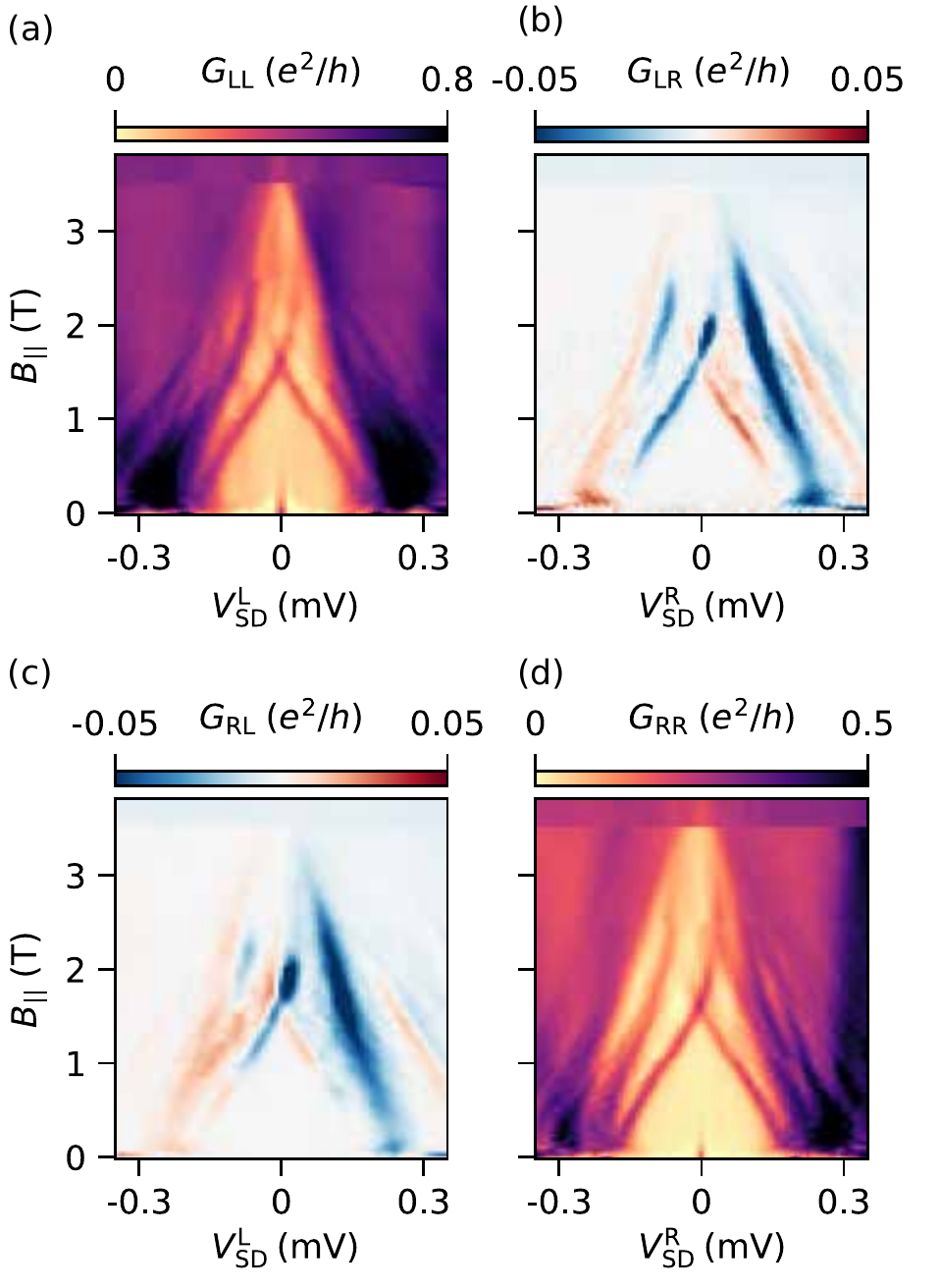}
\end{center}
\caption[Local and nonlocal conductance of ABSs as function of magnetic field]{\label{fig:fieldscan}Measurement of the four conductances as a function of magnetic field parallel to the NW $B_{||}$ at $V_\mathrm{WM}=\SI{-3.02}{\volt}$. Both local conductances in (a) and (d) reveal subgap states that cross zero bias. The subgap states also appear in the nonlocal conductances in (b) and (c).
}
\end{figure}

 In order to confine an ABS in the NW, a modulation of the electron density along the elongated NW dimension was created using the three gates $\mathrm{W_L,\,W_M,\,W_R}$. The gate voltages $V_\mathrm{WL}$ and $V_\mathrm{WR}$ were set to \SI{-4.50}{\volt}. The NW segments underneath the respective gates $\mathrm{W_L}$ and $\mathrm{W_R}$ have a hard superconducting gap with no subgap states and act as QP filters between the Al ground planes and the NW segment under the gate $\mathrm{W_M}$ \cite{gerbold_malinowski_qpfilter, delft_quasiparticletrap}. The gate voltage $V_\mathrm{WM}$ was set to $\SI{-3.02}{\volt}$; significantly more positive than the voltages on the neighboring gates $\mathrm{W_L}$, $\mathrm{W_R}$. A measurement of the four conductances (\ref{eq:Gmat}) as a function of magnetic field $B_{||}$ is depicted in Fig.~\ref{fig:fieldscan}. Note that the superconducting probes serve as normal leads for magnetic field values $B_{||}>\SI{0.2}{\tesla}$ \cite{henri_lead}. Both local tunneling conductances $G_\mathrm{LL}$, $G_\mathrm{RR}$ reveal subgap resonances which emerge at low magnetic fields from the QP continuum at high bias, and cross zero voltage bias at parallel magnetic field $B_{||}=\SI{1.6}{\tesla}$. The resonances can be attributed to an extended ABS in the \SI{0.6}{\micro\meter} long NW segment under the gate $\mathrm{W_M}$ due to the appearance in both local tunneling conductances with identical dependence on magnetic field and gate voltage $V_\mathrm{WM}$, which will be discussed in the following. Hybridization with a quantum dot resonance localized in one of the tunnel barriers was furthermore employed to ensure that we are measuring an extended ABS \cite{PoschlPC2022}.

Both nonlocal conductances $G_\mathrm{LR}$ and $G_\mathrm{RL}$ show sizeable conductance $\approx 5 \cdot 10^{-2} e^2/h$. The absolute value is only one order of magnitude smaller than the local conductances. Numerical simulations for NWs of similar length suggest that this is a sign of low or intermediate disorder \cite{SDS_nl_conductance}. The regions of strong nonlocal conductance $G_\mathrm{LR},G_\mathrm{RL}\geq 25 \cdot 10^{-3} e^2/h$ are bound at high bias values by an envelope. This boundary can be interpreted as the gap $\Delta(B_{||})$ of the parent superconductor that proximitizes the semiconductor and closes with applied magnetic field. For magnetic fields below the zero-crossing of the ABSs, there is a region of vanishing nonlocal conductance around zero bias which extends to the $V^\mathrm{L/R}_\mathrm{SD}$ values which mark the ABS energy. The ABSs are the lowest lying excited states that extend over the whole NW under the gate $\mathrm{W_M}$. Their energy therefore sets the size of the energy gap  $\Delta_{\mathrm{ind}}$ that is induced in the semiconductor by proximity effect and Zeeman energy. Exponentially suppressed nonlocal conductance is expected in the case $eV^\mathrm{L/R}_\mathrm{SD}<\Delta_\mathrm{ind}$ for NWs that are longer than the wavefunction decay length at these energies \cite{andreev_rectifier}. In the voltage range $\Delta_\mathrm{ind}< eV_\mathrm{SD}\leq\Delta$ there is finite nonlocal conductance that can be interpreted as QP transport through higher excited states.

\begin{figure}[h!]
\begin{center}
\includegraphics[scale=1.05]{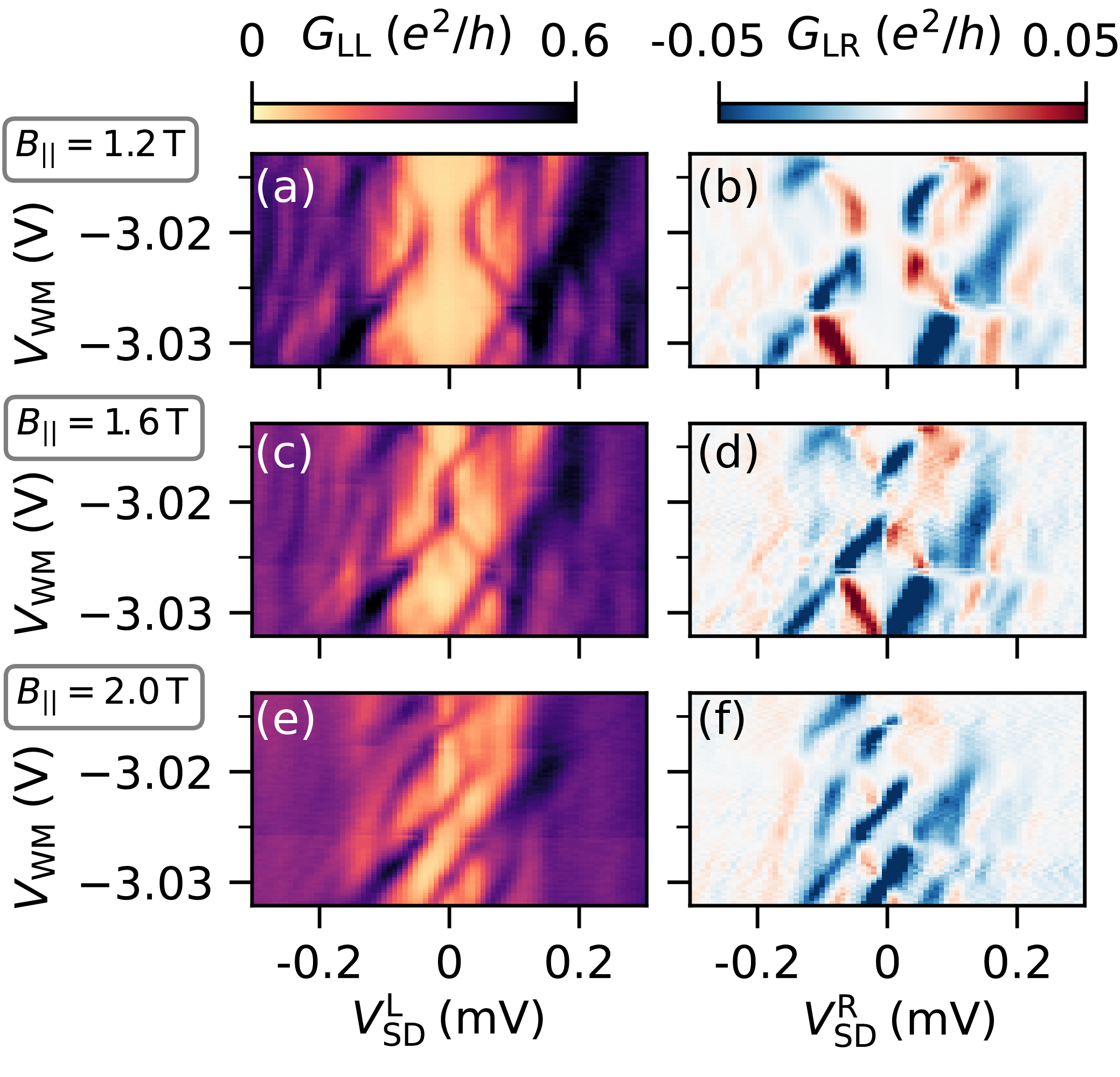}
\end{center}
\caption[Local and nonlocal conductance of ABSs as function of gate voltage]{\label{fig:plungerscans}Local conductance $G_\mathrm{LL}$ and nonlocal conductance $G_\mathrm{LR}$ as a function of $V_\mathrm{WM}$ at three different values of magnetic field $B_{||}$. (a) Subgap states appear as lobes in the superconducting gap. (b) The lowest excited state appears together with a spectrum of higher excited states in the nonlocal conductance. (c) At $B_{||}=\SI{1.6}{\tesla}$ the ABSs merge at zero bias. (d) The nonlocal conductance is suppressed around that region. (e) At $B_{||}=\SI{2}{\tesla}$ the ABSs intersect forming a low energy state that oscillates around zero bias. (f) The nonlocal conductance changes sign at the turning points of the low energy state. }
\end{figure}

\section{Evolution of Andreev bound states and their charge character}
\label{sec:2DEG_nl_gate}
At a magnetic field of $B_{||}=\SI{1.2}{\tesla}$, the ABSs trace out a pair of lobes which do not reach zero bias as a function of gate voltage $V_\mathrm{WM}$, as seen in the the local conductance $G_\mathrm{LL}$ [see Fig.~\ref{fig:plungerscans}(a)]. The corresponding nonlocal conductance  $G_\mathrm{LR}$, plotted in Fig.~\ref{fig:plungerscans}(b), has its largest value at a value $V_\mathrm{SD}^\mathrm{R}$ that tracks the position of the low energy subgap state in $G_\mathrm{LL}$. Following this state, the nonlocal conductance changes sign in two cases. The first case is a value $V_\mathrm{WM}$ at which the ABS reaches a minimum in energy. The second case are points where two ABSs cross, which leads to the energy of the lowest lying state changing its slope abruptly from positive to negative and vice versa. Note that there is a spectrum of additional excited states visible at higher bias values $V_\mathrm{SD}^\mathrm{R}$. We interpret these states as a result of the NW being sufficiently long such that the spacing between excited states is decreased \cite{Mishmash2016}. The absence of a similarly dense spectrum of excited states in nonlocal conductance measurements on proximitized quantum dots \cite{gramich_nl_abs} and shorter NWs \cite{gerbold_nonlocal} is in agreement with this interpretation. 
At a magnetic field of $B_{||}=\SI{1.6}{\tesla}$ the lowest lying ABSs merge at zero voltage bias for a small interval of $V_\mathrm{WM}$ as seen from the local conductance in Fig.~\ref{fig:plungerscans}(c). Within this range, the nonlocal conductance through the ABS is smaller compared to $V_\mathrm{WM}$ values for which the ABSs are at finite bias voltage $[$see Fig.~\ref{fig:plungerscans}(d)$]$. This can be understood as a result of the rates for crossed Andreev reflection and QP transmission being equal at this point due to particle-hole symmetry \cite{Akhmerov2011, lobos_topo_pt_NSN}. The nonlocal conductance consequently vanishes because it is proportional to the difference of these two rates \cite{Takane1992,datta_super_LBK,lobos_topo_pt_NSN}. At a magnetic field $B_{||}=\SI{2}{\tesla}$ the ABSs intersect each other, creating a low energy state that oscillates around zero bias, as seen in the measurement of $G_\mathrm{LL}$ in Fig.~\ref{fig:plungerscans}(e). \\

$G_\mathrm{LL}$ shows an asymmetry with respect to $V_\mathrm{SD}^\mathrm{L}$. Local tunneling conductances are expected to be symmetric with respect to source-drain bias for the case of two-terminal devices, but for the case of three terminal devices a finite asymmetry is expected. In a linear transport theory, the anti-symmetric part of the conductances $G_\mathrm{ij}^\mathrm{anti}(V_\mathrm{SD}^\mathrm{j})=[G_\mathrm{ij}(V_\mathrm{SD}^\mathrm{j})-G_\mathrm{ij}(-V_\mathrm{SD}^\mathrm{j})]/2$ with ${i,j}\in \{ \mathrm{L,R}\}$ fulfil the relations 
\begin{equation}
    G_\mathrm{ij}^\mathrm{anti}(V_\mathrm{SD}^\mathrm{j}) = -G_\mathrm{ii}^\mathrm{anti}(V_\mathrm{SD}^\mathrm{i})
\end{equation}
 at subgap voltages $eV_\mathrm{SD}<\Delta$ as a consequence of particle-hole symmetry and current conservation \cite{karsten_nl_spectroscopy, gerbold_nonlocal}. We find that these relations are quantitatively fulfilled for the lowest excited state, while they are violated for higher excited states. Consequently, the sum over all conductance matrix elements $G_\mathrm{sum}=\sum_\mathrm{i,j} G_\mathrm{ij}$ is symmetric up to source-drain bias voltages of the lowest energy state. A detailed analysis can be found in Section \ref{sec:antisymmetries_dev1}. Possible reasons for deviations from the symmetry relations were given in \cite{gerbold_nonlocal}. In addition, numerical studies have shown that an energy dependence of the tunnel barriers in a nonlinear transport theory can give rise to violations of the symmetry relations \cite{melo_asym}.\\

\begin{figure}[h!]
\begin{center}
\includegraphics[width=\textwidth]{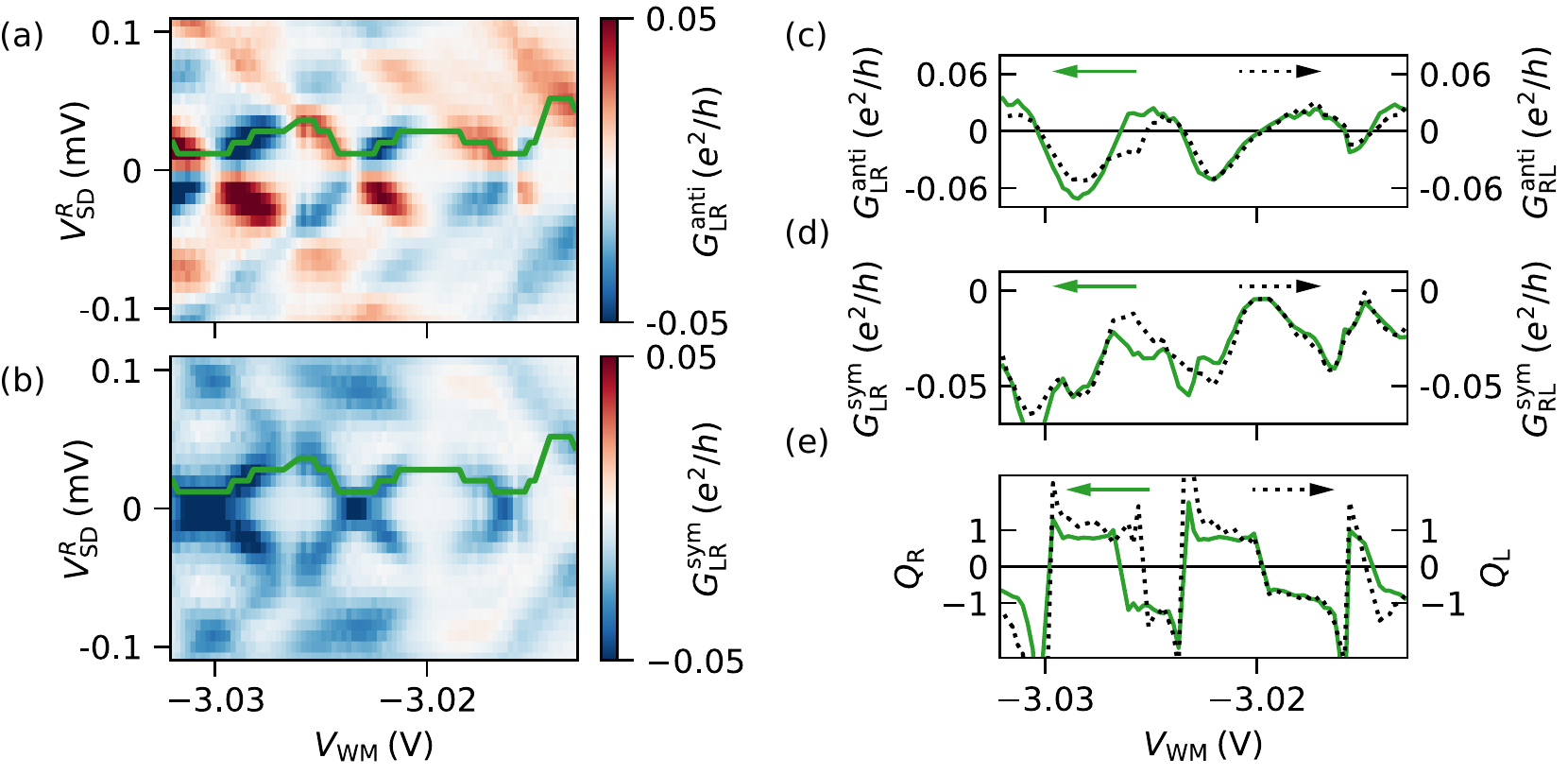}
\end{center}
\caption[Symmetric and anti-symmetric component of nonlocal conductance and extracted charge character]{\label{fig:Qextraction}(a, b) Anti-symmetric and symmetric component of the nonlocal conductance $G_\mathrm{LR}$ measured at $B_{||}=\SI{2}{\tesla}$. 
(c, d) Anti-symmetric and symmetric part of the nonlocal conductances $G_\mathrm{LR}$,\;$G_\mathrm{RL}$ extracted at the position of the lowest lying state marked by the green lines in (a, b). (e) Resulting values for $Q_\mathrm{L}$ and $Q_\mathrm{R}$ are approximately equal and show clear oscillations between +1 and -1. Positive (negative) values of $Q_\mathrm{j}$ coincide with regions of positive (negative) slope of the state energy as a function of $V_\mathrm{WM}$. }
\end{figure}

The quantity
\begin{equation}
 \begin{aligned}\label{eq:BCS_relation}
 Q_\mathrm{j}=\mathrm{sign}(V_\mathrm{SD}^\mathrm{j})\left.\frac{G^\mathrm{sym}_\mathrm{ij}(V_\mathrm{SD}^\mathrm{j})}{G^\mathrm{anti}_\mathrm{ij}(V_\mathrm{SD}^\mathrm{j})}\right\vert_{E=eV_\mathrm{SD}^\mathrm{j}}
 \end{aligned}
\end{equation} of a subgap state at energy $E=eV^\mathrm{j}_\mathrm{SD}$ can be extracted from the anti-symmetric and symmetric components of the measured nonlocal conductance $G_\mathrm{ij}(V^\mathrm{j}_\mathrm{SD})$ \cite{gerbold_nonlocal}. The symmetric and anti-symmetric components of the nonlocal conductance $G_\mathrm{LR}$ measured at a magnetic field value $B_{||}=\SI{2}{\tesla}$ are plotted as a function of source-drain voltage in Figs.~\ref{fig:Qextraction}(a, b). The values stemming from the low energy state were extracted along the positions given by the green lines in Figs.~\ref{fig:Qextraction}(c, d). The extracted values of the (anti-)symmetric part of $G_\mathrm{LR}$ ($G_\mathrm{RL}$) are shown in Figs.~\ref{fig:Qextraction}(c, d) with a solid green (dotted black) line. These values correspond to the conductance values which enter Eq.~\ref{eq:BCS_relation}. The (anti-)symmetric components $G_\mathrm{LR}^\mathrm{sym(anti)}$ and $G_\mathrm{RL}^\mathrm{sym(anti)}$ are approximately equal. The resulting values for $Q_\mathrm{L}$ and $Q_\mathrm{R}$ according to Eq.~\ref{eq:BCS_relation} are shown in Fig.~\ref{fig:Qextraction}(e). $Q_\mathrm{L}$ closely follows $Q_\mathrm{R}$.
\\

Theory suggests that $Q_\mathrm{j}$ is proportional to the local BCS charge of the bound state at the position $\mathrm{j}$ of the conductance probe \cite{Hellenes2019, karsten_nl_spectroscopy, gerbold_nonlocal}.
Here we found that the local charge character on the left and the right are approximately equal $Q_\mathrm{L}\approx Q_\mathrm{R}$. For {device} 1 and $B_{||}=\SI{2}{\tesla}$, there are extended plateaus $Q_\mathrm{j}\approx+1$ ($Q_\mathrm{j}\approx-1$) indicative of a state which is locally fully electron (hole) like. Regions of constant positive (negative) $Q_\mathrm{j}$ coincide with ranges in $V_\mathrm{WM}$ where the state energy has a positive (negative) slope with respect to $V_\mathrm{WM}$. Abrupt changes in $Q_\mathrm{j}$ appear at crossing points of states at finite and zero source-drain bias. This is in agreement with the interpretation of $Q_\mathrm{j}$ measuring the local charge of the bound state. For lower magnetic field values, at which the ABSs appear as parabolic lobes without zero energy crossings, a continuous change of  $Q_\mathrm{j}$ from -1 to 1 is found at the point of minimal ABS energy [See Fig.~\ref{fig:Q_extraction_left_1.2T}]. For {device 2}, the same behavior of $Q_\mathrm{L}\approx Q_\mathrm{R}$ is found with either abrupt changes or continuous crossover from positive to negative $Q_\mathrm{j}$.

\begin{figure}[h!]
\begin{center}
\includegraphics[scale=0.9]{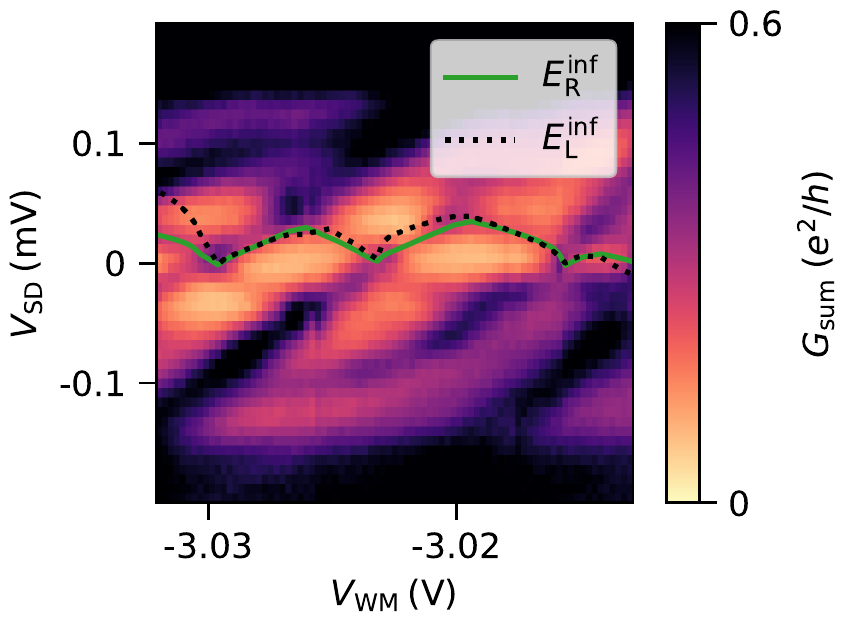}
\end{center}
\caption[Map of differential conductance at $B_{||}=\SI{2}{\tesla}$ with inferred energies $E_\mathrm{L,\:R}^\mathrm{inf}$ superimposed]{\label{fig:E_inf_2T}Map of the sum of all conductances $G_\mathrm{sum}$ at $B_{||}=\SI{2}{\tesla}$. The inferred energies $E_\mathrm{L,\:R}^\mathrm{inf}$ from the integrated $Q_\mathrm{L}$, $Q_\mathrm{R}$ are superimposed. The $E_\mathrm{L,\:R}^\mathrm{inf}$ curves roughly retrace the energy of the subgap state.}
\end{figure}

The total, integrated charge $Q$ of a bound state at energy $E$ is proportional to $\mathrm{d}E/\mathrm{d}V_\mathrm{WM}$ \cite{karsten_nl_spectroscopy}. Integrating the function $Q(V_\mathrm{WM})$ should therefore recover the energy of the subgap state as a function of $V_\mathrm{WM}$. From the experimentally determined $Q_\mathrm{j}$ the energies  
\begin{equation}
    \begin{aligned}
    E_\mathrm{j}^\mathrm{inf}=a\int Q_{j} \mathrm{d}V_\mathrm{WM} + b V_\mathrm{WM} + c.
    \end{aligned}
\end{equation}
 can be inferred, which is the numerical integrated value of $Q_\mathrm{j}$ after re-scaling by a lever arm $a$ and taking into account a linear background $b$ and integration constant $c$.  The inferred energy $E_\mathrm{L,\:R}^\mathrm{inf}$ curves are plotted in Fig.~\ref{fig:E_inf_2T} superimposed on a map showing the sum of all conductances $G_\mathrm{sum}$. The curves roughly match the evolution of the low energy subgap state over an extended range of $V_\mathrm{WM}$. This suggests that the experimentally determined $Q_\mathrm{j}$ not only reflects the local charge character of the ABS, but serves as measure for the total charge $Q$ of the bound states. Deviations from this behavior are expected for longer devices were the QP charge can vary along the spatial extent of bound states \cite{karsten_nl_spectroscopy,Hellenes2019}.
 
\section{Conclusion}
In summary, we have performed local and nonlocal conductance spectroscopy of ABSs in a 2DEG based NW as a function of magnetic field $B_{||}$ and gate voltage $V_\mathrm{WM}$. The predicted symmetry relations between the anti-symmetric components of local and nonlocal conductances are fulfilled for the lowest excited state. In addition, we find a dense spectrum of excited states that give rise to nonlocal conductance. For the lowest excited state, the extracted charge character is the same at both NW ends. This is similar to previous studies \cite{gerbold_nonlocal} despite a longer NW being used here. At high magnetic fields the charge character $Q_\mathrm{L,\:R}$ of the low energy state alternates between fully electron and hole like. The oscillations in the charge character $Q_\mathrm{L,\:R}$ are found to be in agreement with the energy evolution $E(V_\mathrm{\mathrm{WM}})$ of the subgap state which suggests that $Q_{j}$ reflects the total charge of the ABS measured. \\

In Section \ref{sec:SM_nl_dev1} we show details on the materials used, the data analysis, and additional data measured on {device} 1. Section \ref{sec:dev2_nl} contains data from a second device ({device} 2).

\section{Supplementary information on device 1}
\label{sec:SM_nl_dev1}
\subsection{Wafer information}

The material for {device} 1 was a wafer hosting an $\mathrm{In_{1-x}Ga_{x}As-InAs-In_{1-x}Ga_{x}As}$ quantum well covered by \textit{in situ} epitaxially grown Al. The InAs layer had a thickness of \SI{7}{\nano\meter} and the Al was $\approx\SI{5}{\nano\meter}$ thick. The structure was grown on an InP substrate using molecular beam epitaxy. A buffer between the quantum well and the InP was grown to filter dislocation defects and improve the lattice matching. The material for {device} 2 was based on an $\mathrm{In_{1-y}Al_{y}As-InAs-In_{1-x}Ga_{x}As}$ quantum well (InAs thickness \SI{7}{\nano\meter}) with a $\approx\SI{5}{\nano\meter}$ thick layer of Al. For both wafers the Al was passivated with a layer of native oxide formed in a controlled oxidation succeeding the growth. Fabrication details for device 1 can be found in Appendix \ref{sec:fab_super_sideprobe}.

\subsection{Electrical measurement details}
\label{sec:setup_nl}
\begin{figure}[h!]
\begin{center}
\includegraphics[scale=0.75]{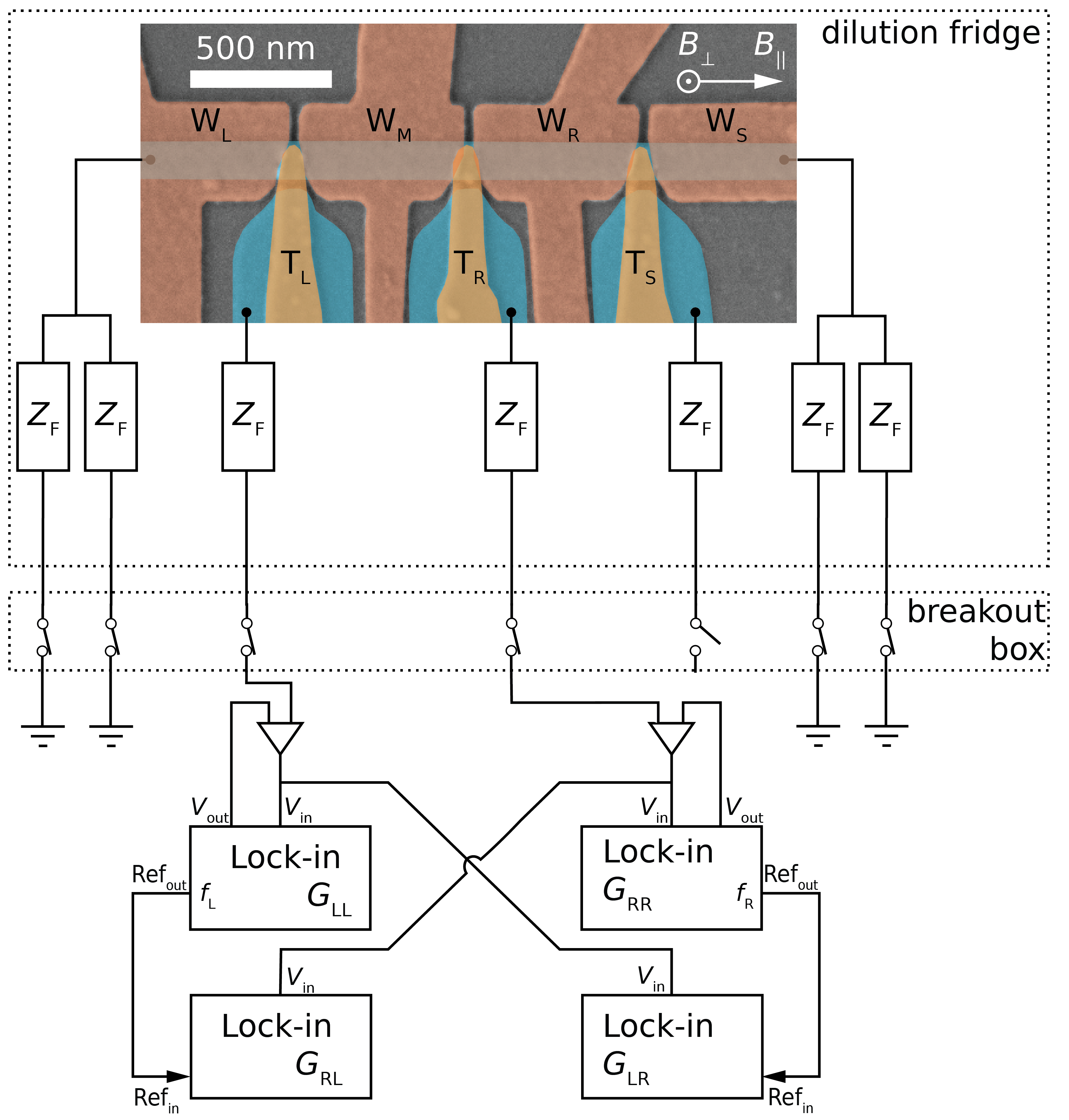}
\end{center}
\caption[Electrical measurement setup for nonlocal conductance spectroscopy]{\label{fig:detection} Schematic of the electrical measurement setup. The sample was cooled in a dilution refrigerator. Connections to measurement electronics were made by electrical lines of impedance $Z_{\mathrm{F}}$. Two current to voltage amplifiers were used to measure the tunneling current across the tunnel barriers under the gates $\mathrm{T_L}$ and $\mathrm{T_R}$. Four lock-in amplifiers in the shown configuration were used to measure the four elements of the conductance matrix. The lead under the gate $\mathrm{T_S}$ was terminated by an open circuit at the breakout box.}
\end{figure}

The devices were wire bonded with Al wire and cooled in a cryo-free dilution refrigerator (Oxford Instruments, Triton 400) equipped with a 1-1-\SI{6}{\tesla} vector magnet. The mixing chamber temperature was $\approx\SI{15}{\milli\kelvin}$ as measured by a $\mathrm{RuO_2}$ thermometer.\\
A schematic of the measurement setup is depicted in Fig.~\ref{fig:detection}. All electrical lines in the cryostat were equipped with in-house built, multi-stage cryogenic RF and RC filters (cutoff frequency \SI{80}{\mega\hertz} and \SI{0.7}{\kilo\hertz}). The resulting total line impedance $Z_{\mathrm{F}}$ at lock-in frequencies ($<\SI{100}{\hertz}$) is dominated by its resistance \SI{0.88}{\kilo\ohm} stemming from \SI{0.180}{\kilo\ohm} line resistance and \SI{0.70}{\kilo\ohm} filter resistance.  In order to mitigate circuit-effects that can occur for three terminal devices \cite{voltage_divider} a low resistance to ground of the parent superconductor is desirable. This was achieved by bonding each of the ground planes at the respective nanowire (NW) end to two electrical lines. These four lines were connected to ground at the breakout box at room temperature, creating a resistance to ground that is four times smaller compared to a single line. \\

For nonlocal conductance measurements, the detection scheme as depicted in the bottom half of Fig.~\ref{fig:detection} was used. The two leads under the gates $\mathrm{T_L}$ and  $\mathrm{T_R}$ were connected to individual low-noise high-stability current to voltage converting amplifiers (Basel Precision Instruments, SP983c) with a gain of $10^8\: \SI{}{\volt\per\ampere}$. In addition to the two tunnel probes under the gates $\mathrm{T_L}$ and $\mathrm{T_R}$, the device is equipped with a third lead under the gate $\mathrm{T_S}$ which was terminated with an open circuit at the breakout box for the measurements presented in the Section \ref{sec:2DEG_3terminal}. Voltage offsets $V_{\mathrm{SD}}^\mathrm{L}$, $V_{\mathrm{SD}}^\mathrm{R}$ at the current input of the amplifiers were applied by using the offset voltage input of the amplifier. The DC component of $V_{\mathrm{SD}}^\mathrm{L}$, $V_{\mathrm{SD}}^\mathrm{R}$ were supplied by an in-house built digital to analog converter while additional AC modulations $\mathrm{d}V_{\mathrm{SD}}^\mathrm{L}$, $\mathrm{d}V_{\mathrm{SD}}^\mathrm{R}$ at reference frequencies $f_\mathrm{L}$,  $f_\mathrm{R}$ were supplied by the outputs of lock-in amplifiers (Stanford Research Systems, SR830). The nonlocal conductance was detected by measuring the output of each amplifier with two lock-in amplifiers. One of the lock-in amplifiers was locked to  $f_\mathrm{L}$ while the other was locked to  $f_\mathrm{R}$. The same detection technique has been applied in previous works \cite{denise_nl_gapclosing, gerbold_nonlocal}.

\subsection{Symmetrization and anti-symmetrization of data}
In order to investigate the relation between the anti-symmetric parts of the different conductances with respect to $V_{\mathrm{SD}}$, the anti-symmetric parts were extracted from the measured data. The symmetric components with respect to $V_{\mathrm{SD}}$ were further extracted to determine $Q_\mathrm{j}$. \\

(Anti-)symmetrizing the data is sensitive to small offsets of $V_{\mathrm{SD}}$ from zero, which can occur in experiments despite careful calibration prior to measurements. In order to compensate for voltage offsets, that are smaller than the spacing of measured points in $V_{\mathrm{SD}}$, the data was up-sampled along the $V_{\mathrm{SD}}$ dimension to roughly three times the resolution. The up-sampled data was then (anti-)symmetrized around the symmetry point which is typically offset from zero by two to four pixels in the new, up-sampled dimension. After forming the (anti-)symmetric parts of the data, the data was down-sampled again to restore the resolution of the underlying raw data.

\subsection{Extraction of values at peak positions}

For the extraction of the quantities $Q_\mathrm{j}$ of a state at energy $E$, it is necessary to extract the values of $G^\mathrm{sym}_\mathrm{ij}$ and $G^\mathrm{anti}_\mathrm{ij}$ at the voltage $e V_\mathrm{SD}^\mathrm{j}=E$. Typically there is a local extremum in at least one of the measured conductance matrix elements around this voltage value. The routine used to determine this position was to search for local maxima in $|G^\mathrm{anti}_\mathrm{ij}(V_\mathrm{SD}^\mathrm{j})|$, $|G^\mathrm{sym}_\mathrm{ij}(V_\mathrm{SD}^\mathrm{j})|$, or $|G_\mathrm{sum}(V_\mathrm{SD})|$ for every value of gate voltage. Only one of the three quantities was used depending on which one yielded the best initial guess for the evolution of the state. Points at which the state is not located precisely due to the absence of a clear local maximum were excluded by hand. The number of such erroneously detected peak locations was below 5 out of 75 in every case reported here.

\subsection{Additional data on device 1}
\begin{figure}[h!]
\begin{center}
\includegraphics[width=\textwidth]{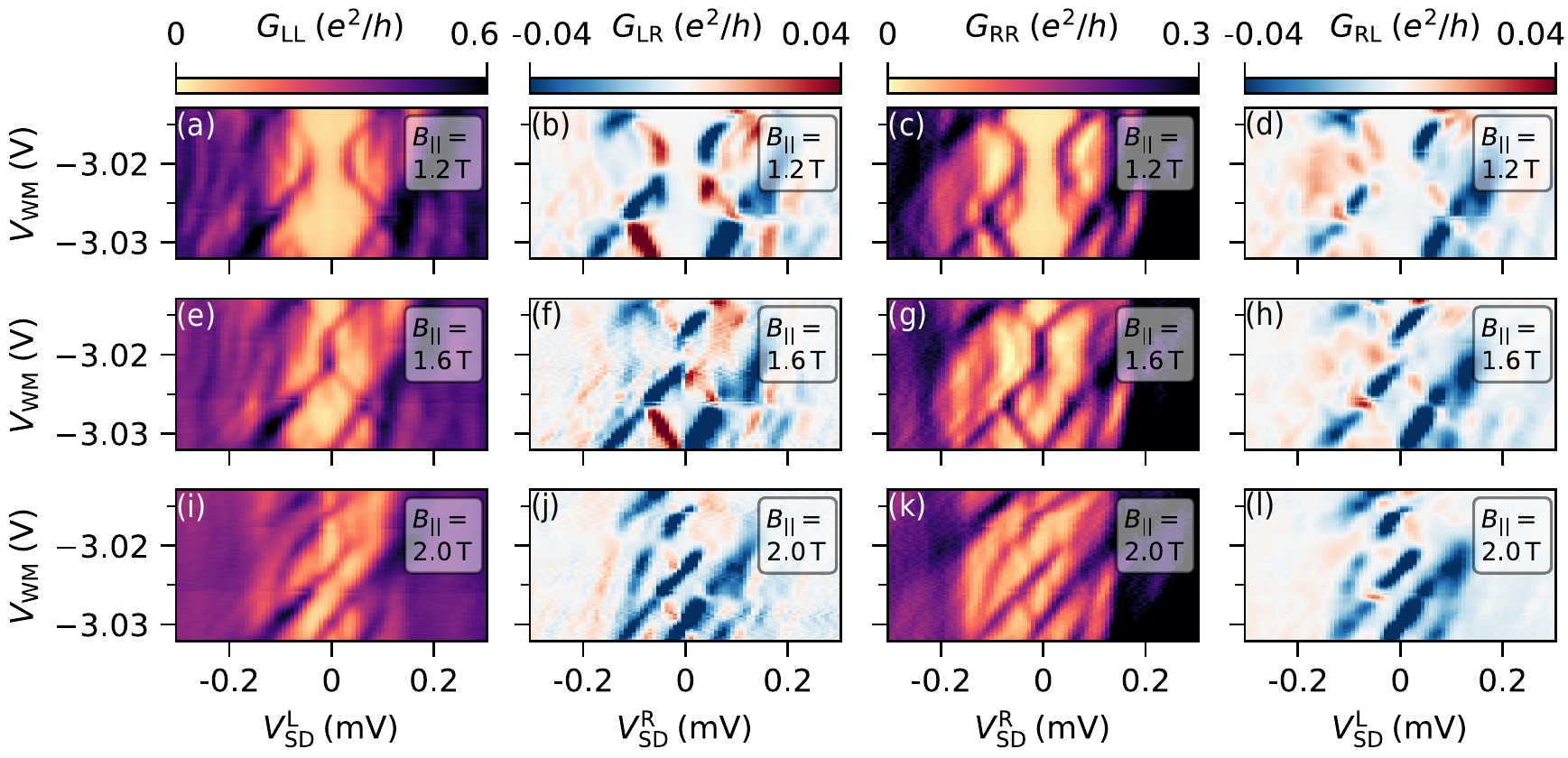}
\end{center}
\caption[]{\label{fig:full_plunger_dev_1}Each row is a measurements of all four elements of the conductance matrix as a function of gate voltage $V_\mathrm{WM}$ at a specific value of $B_{||}$. The two leftmost columns are the data presented in the main text.}
\end{figure}

The data in the following supplement the discussion in Section \ref{sec:2DEG_nl_gate}. In Fig.~\ref{fig:full_plunger_dev_1} we show all four conductances as a function of gate voltage $V_\mathrm{WM}$ at different magnetic field values $B_{||}$. The data for $G_\mathrm{LL}$ and $G_\mathrm{LR}$ are shown in Section \ref{sec:2DEG_nl_gate}. ABSs appear as subgap states in both nonlocal conductances, and are visible in both local conductances $G_\mathrm{LL}$ and $G_\mathrm{RR}$. 

\subsubsection{Relations between anti-symmetric parts of conductance matrix elements}
\label{sec:antisymmetries_dev1}

It was shown in Ref. \cite{karsten_nl_spectroscopy} within a linear transport theory that the anti-symmetric parts of the conductances $G_\mathrm{ij}$ and $G_\mathrm{ii}$ fulfill the relation
\begin{equation}
\label{eq:anti-symmetry}
    G_\mathrm{ij}^\mathrm{anti}(V_\mathrm{SD}^\mathrm{j}) = -G_\mathrm{ii}^\mathrm{anti}(V_\mathrm{SD}^\mathrm{i})
\end{equation}
as a consequence of current conservation and particle-hole symmetry at voltages below the energy gap of the parent superconductor $e V_\mathrm{SD}^\mathrm{i,j}<\Delta$. Note that this relation compares two quantities that are measured independently in the experiment, as the left hand side is measured as a function of $V_\mathrm{SD}^\mathrm{j}$ while $V_\mathrm{SD}^\mathrm{i} = \SI{0}{\volt}$ and the right hand side is measured as a function of $V_\mathrm{SD}^\mathrm{i}$ while $V_\mathrm{SD}^\mathrm{j} = \SI{0}{\volt}$.\\

\begin{figure}[h!]
\begin{center}
\includegraphics[scale=0.9]{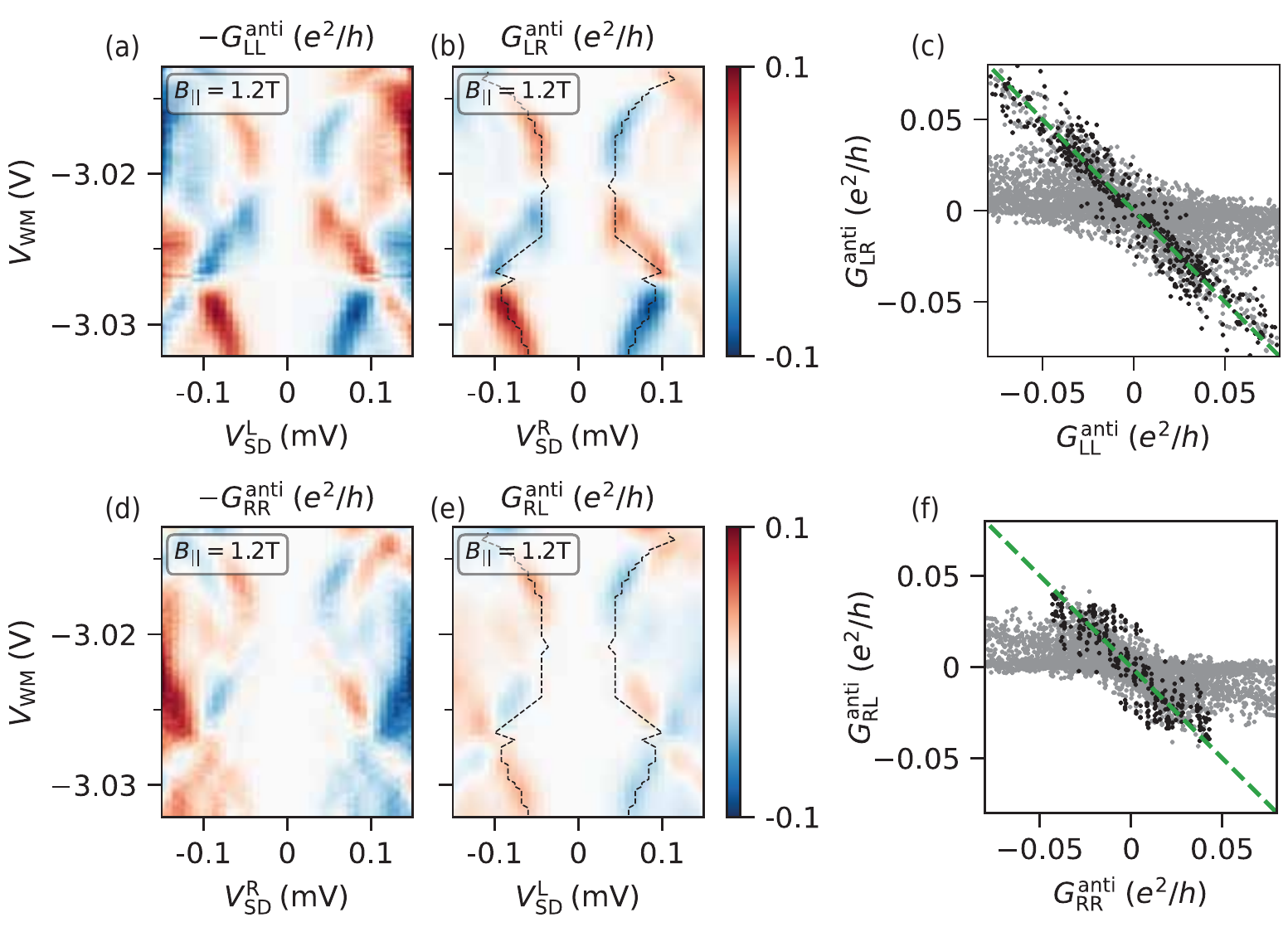}
\end{center}
\caption[]{\label{fig:antisymmetries_1.2T}(a, b) plots of $-G_\mathrm{LL}^\mathrm{anti}$ and $G_\mathrm{LR}^\mathrm{anti}$ which are expected to be identical according to a linear transport theory. (c) parametric plot of the data points in (a) and (b). The black data points correspond to the data in (a, b) taken in a three pixel window around the black dashed line in (b). The green dashed line in (c) denotes the relation expected from theory. (d, e) comparison of $-G_\mathrm{RR}^\mathrm{anti}$ and $G_\mathrm{RL}^\mathrm{anti}$. (f) shows a parametric plot of data points from (d, e) with the data points taken from a three pixel window around the dashed line in (e) shown in black.}
\end{figure}

Figure \ref{fig:antisymmetries_1.2T} shows the anti-symmetric parts of the local and nonlocal conductances for comparison. The underlying data is the same as in Fig.~\ref{fig:full_plunger_dev_1}(a-d) taken at $B_{||}=\SI{1.2}{\tesla}$. The data in Fig.~\ref{fig:full_plunger_dev_1}(a, b) show $-G_\mathrm{LL}^\mathrm{anti}$ and $G_\mathrm{LR}^\mathrm{anti}$. The two quantities are expected to be the same. The lowest excited state shows up with the same sign and similar strength in both plots. For energy values above the first excited state, the anti-symmetric part of the local conductance is larger than the anti-symmetric part of the nonlocal conductance. The data points from (a) and (b) are plotted parametrically in (c). The points that correspond to three pixels around the lowest excited state shown by the dashed line in (b) are shown as black dots in (c). They lie close to the green dashed line which corresponds to the relation predicted by theory. All other data points are plotted in grey. A large number of these points have a relatively small value in $|G_\mathrm{LR}^\mathrm{anti}|$ compared to their relatively large value in $|G_\mathrm{LL}^\mathrm{anti}|$. These data points originate not only from energies above $\Delta$, where a deviation from the relation in equation \ref{eq:anti-symmetry} is expected, but also from energies below $\Delta$ and above $\Delta_\mathrm{ind}$.

The quantities $-G_\mathrm{RR}^\mathrm{anti}$ and $G_\mathrm{RL}^\mathrm{anti}$ are plotted in Fig.~\ref{fig:antisymmetries_1.2T}(d, e). A parametric plot of the same data is shown in (f). Figure \ref{fig:antisymmetries_1.2T}(f) confirms the observation of the lowest energy state fulfilling the expected relation from Eq.~\ref{eq:anti-symmetry}.\\

\begin{figure}[h!]
\begin{center}
\includegraphics[scale=0.9]{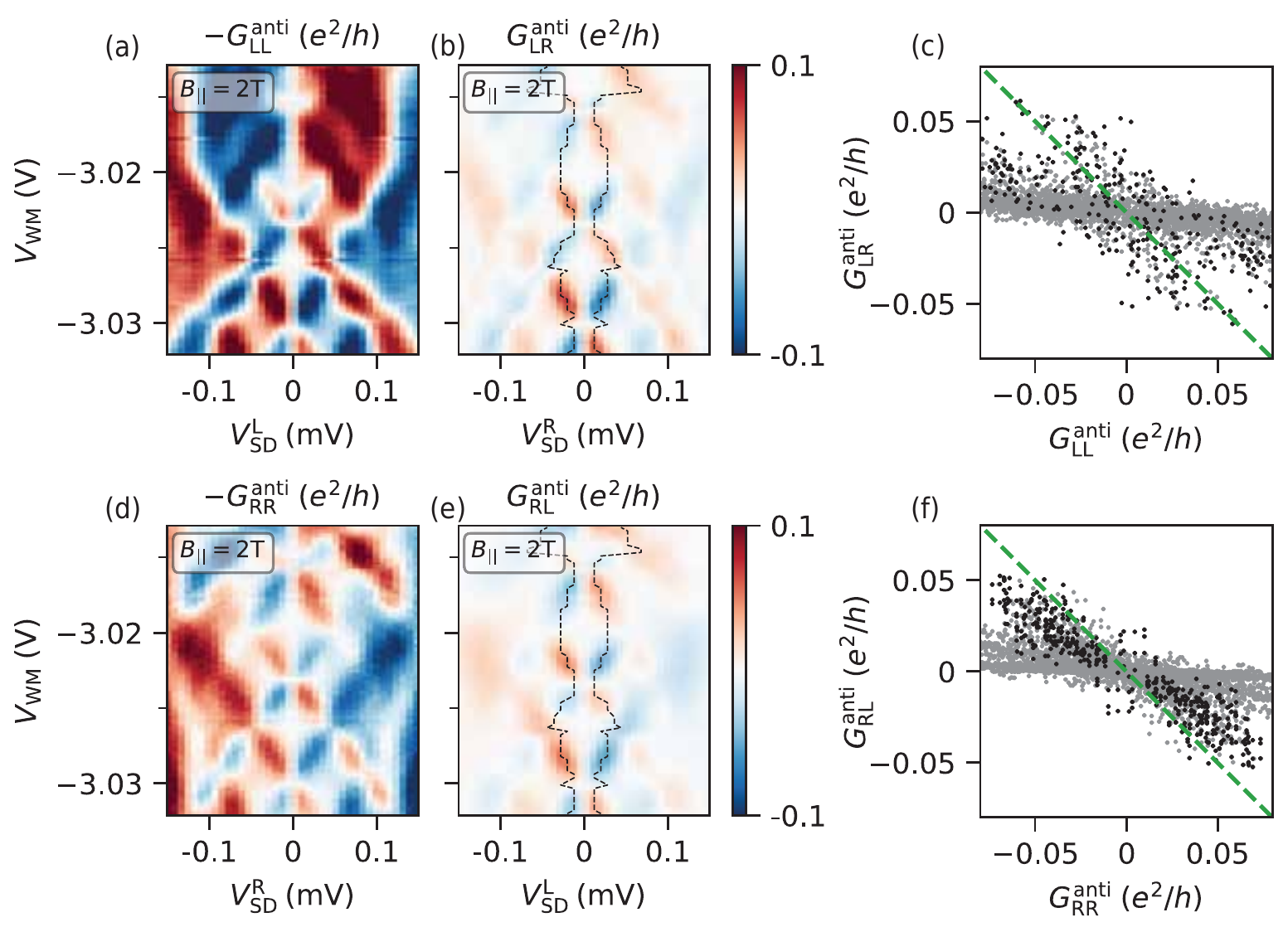}
\end{center}
\caption[]{\label{fig:antisymmetries_2T}(a, b) plots of  $-G_\mathrm{LL}^\mathrm{anti}$ and $G_\mathrm{LR}^\mathrm{anti}$ which are expected to be identical according to a linear transport theory. (c) parametric plot of the data points in (a) and (b). The black data points correspond to the data in (a, b) taken in a three pixel window around the black dashed line in (b). The green dashed line in (c) denotes the relation expected from theory. (d, e) comparison of $-G_\mathrm{RR}^\mathrm{anti}$ and $G_\mathrm{RL}^\mathrm{anti}$. (f) shows a parametric plot of data points from (d, e) with the data points taken from a three pixel window around the dashed line in (e) shown in black.}
\end{figure}

For the data taken at $B_{||}=\SI{2}{\tesla}$ in Fig.~\ref{fig:full_plunger_dev_1}(e-h) the same comparison of anti-symmetric components of the conductance matrix elements is shown in Fig.~\ref{fig:antisymmetries_2T}. In Fig.~\ref{fig:antisymmetries_2T}(a, b), $-G_\mathrm{LL}^\mathrm{anti}$ and $G_\mathrm{LR}^\mathrm{anti}$ are shown. A parametric plot of the same data is shown in Fig.~\ref{fig:antisymmetries_2T}(c), with the black data points taken in a three pixel window around the lowest energy state denoted by the dashed line in (b). In Fig.~\ref{fig:antisymmetries_2T}(d) and (e), $-G_\mathrm{RR}^\mathrm{anti}$ and $G_\mathrm{RL}^\mathrm{anti}$ are shown for comparison, together with a parametric plot of the same data in Fig.~\ref{fig:antisymmetries_2T}(f). The  low energy state shows up with the same sign in $-G^\mathrm{anti}_\mathrm{LL}$ and $G_\mathrm{LR}^\mathrm{anti}$. The values of $-G^\mathrm{anti}_\mathrm{RR}$ and $G_\mathrm{RL}^\mathrm{anti}$ for the low energy state are similar. The parametric plots in Fig.~\ref{fig:antisymmetries_2T}(f) demonstrate, that the data points taken around the first excited state are the ones that lie close to dashed green line, which is the theoretically expected relation at source drain voltages below the superconducting energy gap $eV_\mathrm{SD}<\Delta$. The black points in Fig.~\ref{fig:antisymmetries_2T}(c) stemming from the low energy state appear less correlated according to the theory expectation compared to the lower field value $B_{||}=\SI{1.2}{\tesla}$.\\

The relation between the anti-symmetric parts of the conductance matrix elements in Eq.~\ref{eq:anti-symmetry} has as a consequence that the sum of all conductance matrix elements 
\begin{equation}
    G_\mathrm{sum} (V_\mathrm{SD})= G_\mathrm{LL} (V_\mathrm{SD}^\mathrm{L})+G_\mathrm{RL}(V_\mathrm{SD}^\mathrm{L})+G_\mathrm{LR}(V_\mathrm{SD}^\mathrm{R})+G_\mathrm{RR}(V_\mathrm{SD}^\mathrm{R})
\end{equation}
is a symmetric function of $V_\mathrm{SD}$ \cite{karsten_nl_spectroscopy}. 

\begin{figure}[h!]
\begin{center}
\includegraphics[scale=0.9]{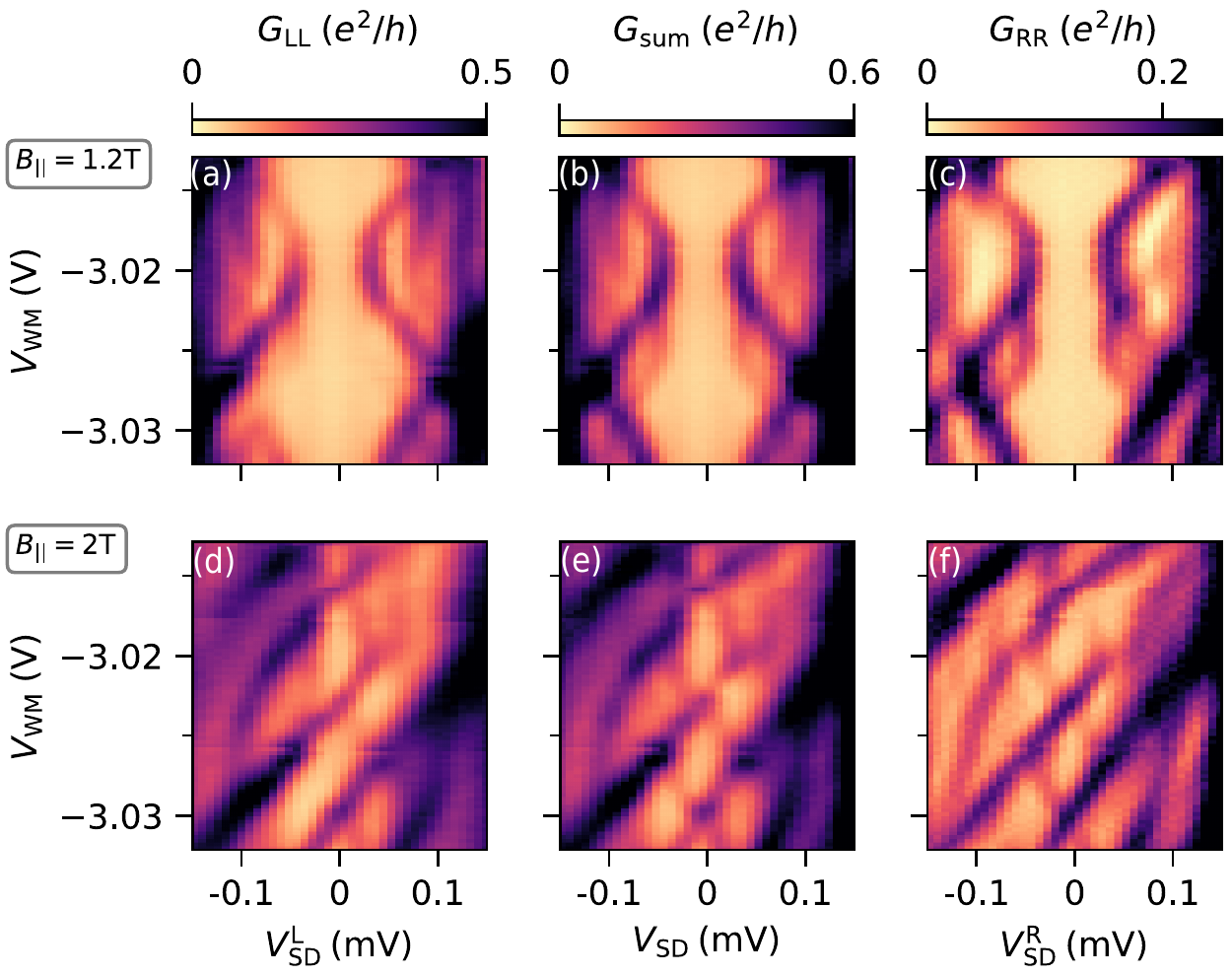}
\end{center}
\caption[]{\label{fig:symmetries_1.2T}(a) tunneling conductance $G_\mathrm{LL}$, (b) the sum of all conductance matrix elements $G_\mathrm{sum}$, (c) tunneling conductance $G_\mathrm{RR}$ measured at $B_{||}=\SI{1.2}{\tesla}$. Subgap states around $V_\mathrm{WM} = \SI{-3.03}{\volt}$ and $V_\mathrm{WM}=\SI{-3.025}{\volt}$ appear symmetric with respect to $V_\mathrm{SD}$ in $G_\mathrm{sum}$, but show a notable asymmetry in $G_\mathrm{LL}$ and $G_\mathrm{RR}$. (d) tunneling conductance $G_\mathrm{LL}$, (e) the sum of all conductance matrix elements $G_\mathrm{sum}$, (f) tunneling conductance $G_\mathrm{RR}$ measured at $B_{||}=\SI{2}{\tesla}$. The low energy state appears symmetric with respect to $V_\mathrm{SD}$ in $G_\mathrm{sum}$.}
\end{figure}

The quantity $G_\mathrm{sum}$ calculated from the data in Fig.~\ref{fig:full_plunger_dev_1}(a-d) measured at $B_{||}=\SI{1.2}{\tesla}$ is shown in Fig.~\ref{fig:symmetries_1.2T}(b). The two corresponding local conductances $G_\mathrm{LL}$ and $G_\mathrm{RR}$ are plotted for comparison in  Fig.~\ref{fig:symmetries_1.2T}(a, c). Qualitatively, $G_\mathrm{sum}$ is more symmetric in comparison to $G_\mathrm{LL}$  and $G_\mathrm{RR}$ for some of the subgap states.\\

$G_\mathrm{sum}$ calculated for data taken at $B_{||}=\SI{2}{\tesla}$ in Fig.~\ref{fig:full_plunger_dev_1}(i-l) is shown in Fig.~\ref{fig:symmetries_1.2T}(e). The two local conductances $G_\mathrm{LL}$  and $G_\mathrm{RR}$ are plotted in Fig.~\ref{fig:symmetries_1.2T}(d, f) for comparison. $G_\mathrm{sum}$ appears more symmetric with respect to source drain voltage $V_\mathrm{SD}$, in particular the low energy state, that oscillates around zero bias appears more symmetric in $G_\mathrm{sum}$ compared to $G_\mathrm{LL}$ and $G_\mathrm{RR}$. States at larger source drain voltages appear with a strong anti-symmetry in $G_\mathrm{sum}$ which is a result of the relation in Eq.~\ref{eq:anti-symmetry} only being fulfilled for the low energy state in the experiment.\\

\subsubsection{Extracted value of Q\textsubscript{L} and Q\textsubscript{R}}

For the nonlocal conductances measured at \SI{1.2}{\tesla} shown in Fig.~\ref{fig:full_plunger_dev_1}(b, d) we extracted
\begin{equation}
\label{eq:Q}
Q_\mathrm{j}=\mathrm{sign}(V_\mathrm{SD}^\mathrm{j}) \frac{G_\mathrm{ij}^\mathrm{sym}(V_\mathrm{SD}^\mathrm{j})}{G_\mathrm{ij}^\mathrm{anti}(V_\mathrm{SD}^\mathrm{j})}
\end{equation}
for the lowest energy ABS at $E=e V_\mathrm{SD}^\mathrm{j}$. This quantity is expected to reflect the local charge character of the state. 

\begin{figure}[h!]
\begin{center}
\includegraphics[scale=1]{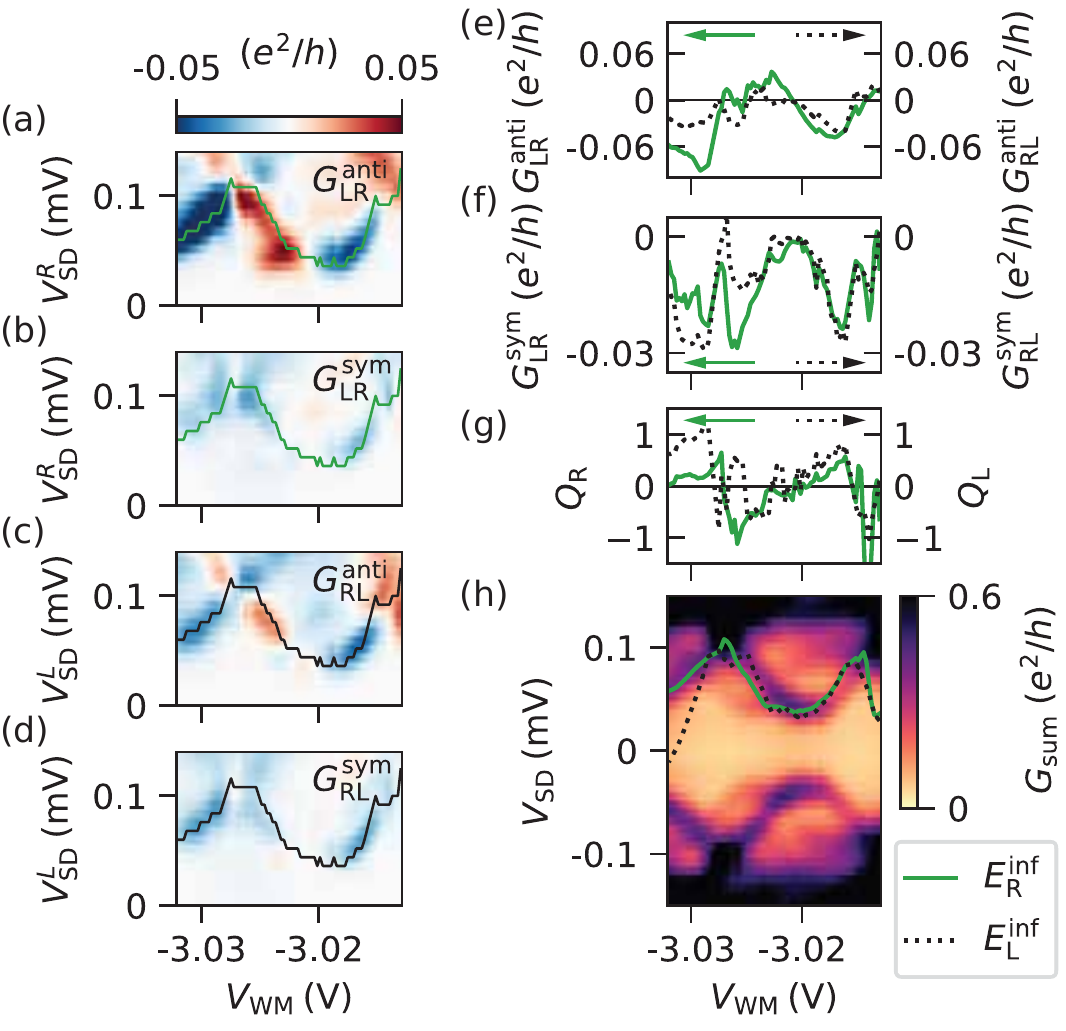}
\end{center}
\caption[]{\label{fig:Q_extraction_left_1.2T}(a-d) symmetric and anti-symmetric components of the nonlocal conductances measured at $B_{||}=\SI{1.2}{\tesla}$. The values at the position of the green and black line are plotted in (e, f) and were used to extract $Q_\mathrm{L}$, $Q_\mathrm{R}$, which are shown in (g). The two evolve similarly with $V_\mathrm{WM}$. An abrupt change from positive to negative values is seen where neighbouring ABS cross. (h) shows the sum of all conductances $G_\mathrm{sum}$. The dotted black, solid green lines display the energies $E_\mathrm{L,\:R}^{\mathrm{inf}}$ inferred from the integrated $Q_\mathrm{L}$, $Q_\mathrm{R}$. They roughly retrace the bound state energy.}
\end{figure}

In Fig.~\ref{fig:Q_extraction_left_1.2T}(a-d) we show the symmetric and anti-symmetric components of the nonlocal conductances. The values at the position of the lowest excited state are extracted at the position given by the green and black lines. The resulting values are plotted in Fig.~\ref{fig:Q_extraction_left_1.2T}(e, f). The resulting values for $Q_\mathrm{L}$, $Q_\mathrm{R}$ are shown in Fig.~\ref{fig:Q_extraction_left_1.2T}(g). The two quantities show roughly the same evolution as a function of gate voltage $V_\mathrm{WM}$. At the two points where ABS cross around $V_\mathrm{SD}=\SI{0.1}{\milli\volt}$ an abrupt change from positive to negative $Q_\mathrm{j}$ is observed. Around the gate voltage, where the ABS goes through a minimum in energy a smooth change from negative to positive $Q_\mathrm{j}$ can be seen. Regions of positive (negative) valued $Q_\mathrm{j}$ coincide with ranges where the subgap state has a positive (negative) slope with respect to $V_\mathrm{WM}$.  Integrating $Q_\mathrm{j}$, rescaling, and taking into account a linear background leads to the inferred energies
\begin{equation}
    \begin{aligned}
    E_\mathrm{j}^\mathrm{inf}=a\int Q_{j} \mathrm{d}V_\mathrm{WM} + b V_\mathrm{WM} + c.
    \end{aligned}
\end{equation} 
These lines are plotted in Fig.~\ref{fig:Q_extraction_left_1.2T}(h). They approximately retrace the energy of the lowest lying subgap state.
This evolution of $Q_\mathrm{L}\approx Q_\mathrm{R}$ in accordance with the energy of the subgap state suggests, that $Q_\mathrm{L}$ and $Q_\mathrm{R}$ measure the total charge of the subgap state. 

\subsubsection{Local and nonlocal conductance as a function of out of plane magnetic field}

\begin{figure}[h!]
\begin{center}
\includegraphics[width=\textwidth]{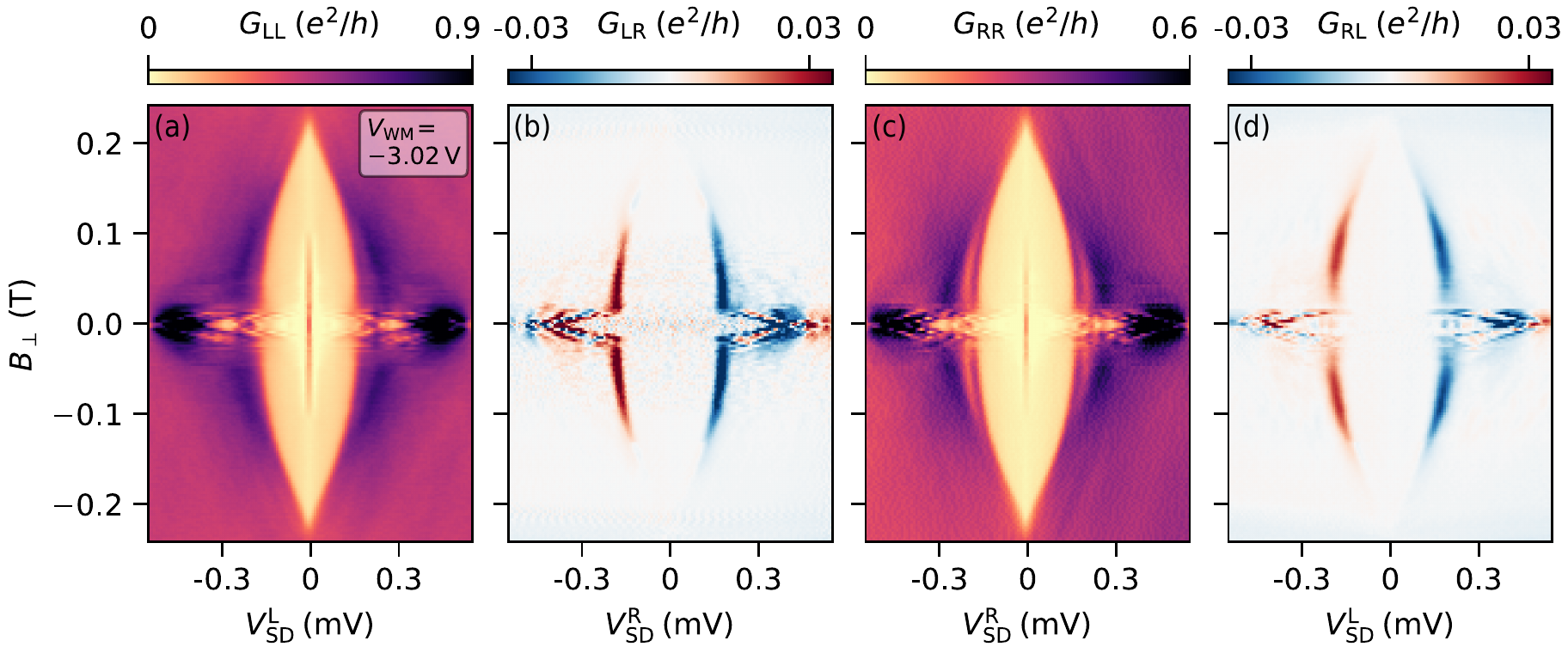}
\end{center}
\caption[]{\label{fig:perp_fieldscan_left}Measurement of all four conductances as a function of magnetic field $B_{\perp}$ pointing out of the 2DEG plane. (a, c) The local conductances show a clear supercurrent as a peak at zero source drain bias. The supercurrent does not appear in the nonlocal conductances in (b, d).}
\end{figure}

Figure \ref{fig:perp_fieldscan_left} shows a measurement of local and nonlocal conductance as a function of magnetic field $B_\perp$ oriented out of the 2DEG plane. The direction of the magnetic field is denoted in Fig.~\ref{fig:detection}. The critical field value of the Al film for this magnetic field direction is lowered in comparison to the in plane magnetic field $B_{||}$. The superconducting gap therefore closes at a magnetic field value around \SI{0.22}{\tesla} before subgap states enter the superconducting energy gap. The nonlocal conductance is restricted to a small region close to the gap of the superconductor. 

A supercurrent is visible at $V_\mathrm{SD}=\SI{0}{\volt}$ for field values $|B_{\perp}|<\SI{0.1}{\tesla}$. This can be explained by the wire and the leads being both in the superconducting state. The supercurrent appears clearly in the local conductances $G_\mathrm{LL}$ and $G_\mathrm{RR}$, but does not appear in the nonlocal conductances $G_\mathrm{LR}$ and $G_\mathrm{RL}$. This is a result of the supercurrent being mediated by Cooper pairs that go from the superconducting leads directly to the superconducting condensate of the nanowire. Nonlocal conductance, on the other hand, is mediated by $1e$ quasiparticles that are transmitted and Andreev reflected from one lead to the other.

\subsection{Nonlocal conductance spectroscopy of Andreev bound state localized under gate W\textsubscript{R}}

{Device} 1 shown in Fig.~\ref{fig:detection} allows for both nonlocal spectroscopy of an ABS under the gate $\mathrm{W_M}$ (as demonstrated in Section \ref{sec:2DEG_3terminal}) and measurements of an ABS under the gate $\mathrm{W_R}$. For this purpose, current to voltage converting amplifiers were connected to the leads under the gates $\mathrm{T_R}$ and $\mathrm{T_S}$, while the lead under the gate $\mathrm{T_L}$ was terminated with an open circuit at the breakout box. 

In the following, data on ABSs under the gate $\mathrm{W_R}$ are shown. The the lead under gate $\mathrm{T_R}$ ($\mathrm{T_S}$) and the respective current is labeled $j=L$ ($j=R$). This follows the commonly applied naming convention for nonlocal conductance, i.e., $G_\mathrm{LL}$ and $G_\mathrm{LR}$ are a result of a tunneling current at the left side of the NW segment of interest and $G_\mathrm{RR}$ and $G_\mathrm{RL}$ are a result of a tunneling current at the right side of the NW segment of interest.

\subsubsection{Measurements of the conductance matrix}

In order to measure an ABS that is confined to the NW segment under the gate $\mathrm{W_{R}}$, the gate voltages $V_\mathrm{WL},\; V_\mathrm{WM},\; V_\mathrm{WS}$ were set to $\SI{-7}{\volt}$. A measurement of the full conductance matrix as a function of magnetic field $B_{||}$ is shown in Fig.~\ref{fig:fieldscan_dev_2} with the gate voltage $V_\mathrm{WR}=\SI{-3.09}{\volt}$ - significantly more positive than the neighbouring gates. The ABS appears as subgap states in both local conductances $G_\mathrm{LL}$ and $G_\mathrm{RR}$. The states emerge from the continuum of QP at an energy $\Delta$ at low magnetic field $B_{||}\approx\SI{0.3}{\tesla}$ and cross zero bias at $B_{||}\approx\SI{1.6}{\tesla}$. The nonlocal conductances $G_\mathrm{LR}$ and $G_\mathrm{RL}$ are appreciably large at the energies of the low energy ABS and above. The signal is furthermore strongly suppressed for source drain voltages above the parent gap $eV_\mathrm{SD}>\Delta$. The nonlocal conductance shows a strong anti-symmetric component and the subgap states originating from the low energy ABS show a characteristic pattern as they cross zero energy. Specifically, the states cross without a change in sign of the nonlocal conductance. Similar behavior has been observed previously in numerical studies and experiments \cite{karsten_nl_spectroscopy,gerbold_nonlocal,SDS_nl_conductance}.\\

\begin{figure}[h!]
\begin{center}
\includegraphics[width=\textwidth]{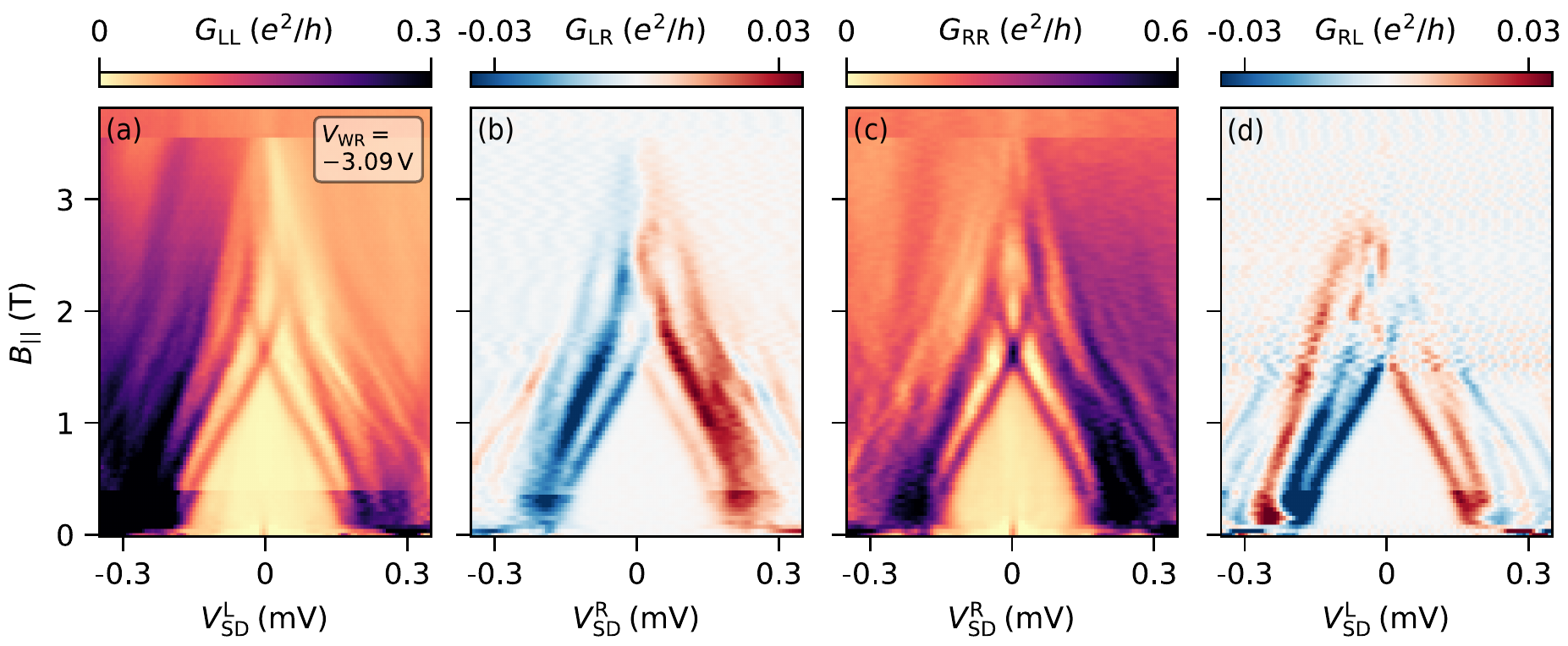}
\end{center}
\caption[]{\label{fig:fieldscan _dev_1_right}Measurement of all four conductances as a function of magnetic field $B_{||}$. (a, c) Local conductances reveal an ABS in the form of subgap states that cross zero source drain bias. (b, d) ABSs and higher excited states appear with sizeable nonlocla conductance.}
\end{figure}

\begin{figure}[h!]
\begin{center}
\includegraphics[width=\textwidth]{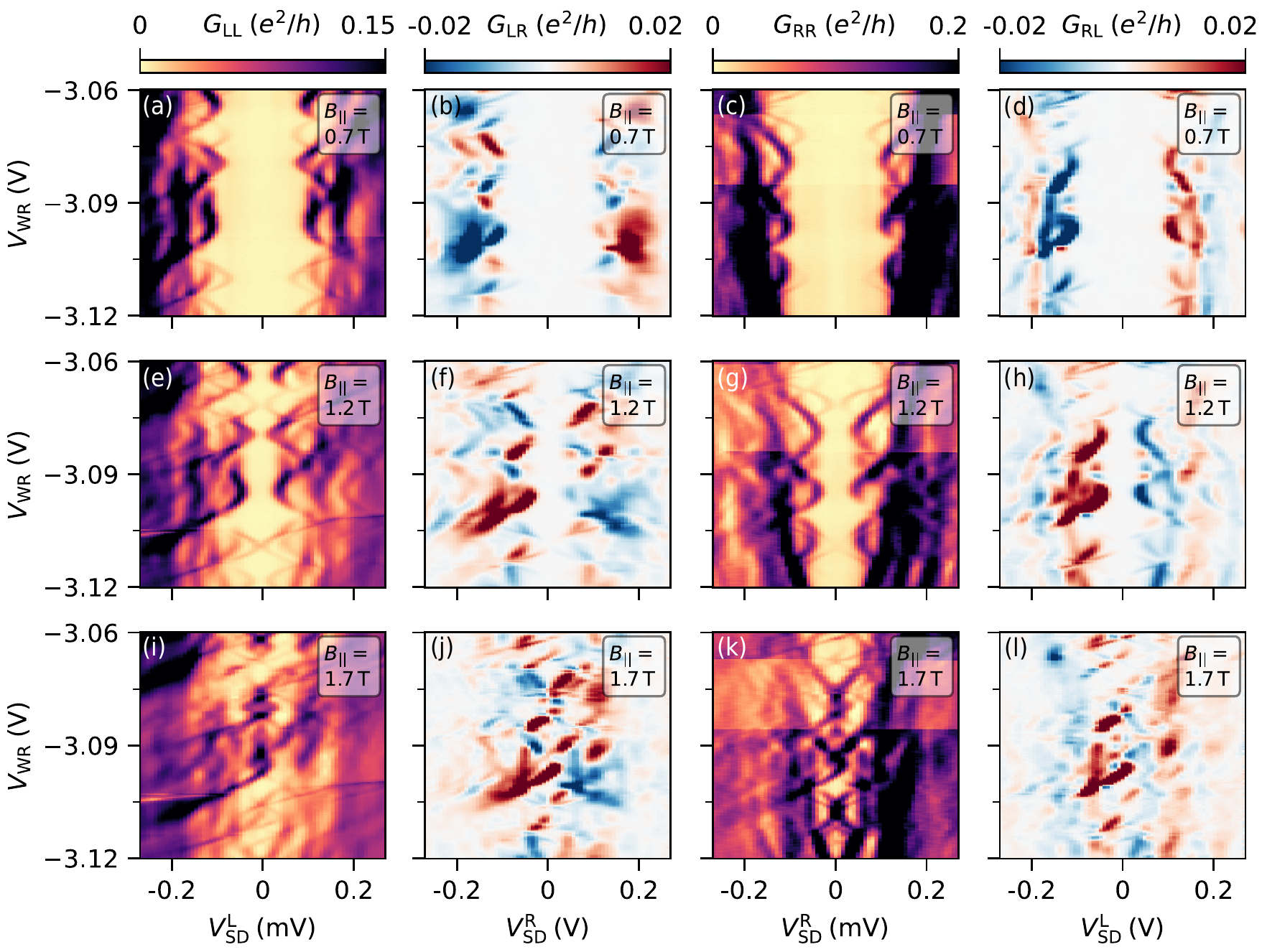}
\end{center}
\caption[]{\label{fig:plungerscans_dev_1_right}Each row shows all four conductances as a function of gate voltage $V_\mathrm{WR}$ at a specific value of $B_{||}$. (a-d) ABSs in the NW appear as lobe shaped subgap states that appear in both local conductances and nonlocal conductances at low magnetic fields. (e-h) With increasing filed $B_{||}$ the ABS are lowered in energy. (i-l) The subgap states merged and cross at zero bias.}
\end{figure}

The evolution of conductances as a function of gate voltage $V_\mathrm{WR}$ is depicted in Fig.~\ref{fig:plungerscans_dev_1_right}. Note that the voltage on the gates $\mathrm{T_R}$ and $\mathrm{T_R}$ were compensated according to the equations


in order to keep the transparency of the tunnel barriers constant throughout the measurement. To denote that more than one gate voltage was changed, the vertical axis in \ref{fig:plungerscans_dev_1_right} is denoted $\tilde{V}_\mathrm{WR}$ instead of ${V}_\mathrm{WR}$.

Data taken at a magnetic field $B_{||}=\SI{0.7}{\tesla}$ in \ref{fig:plungerscans_dev_1_right}(a-d) show subgap states as lobes close to the edge of the parent gap $\Delta$. The states appear in both local conductances $G_\mathrm{LL}$ and $G_\mathrm{RR}$. In the nonlocal conductances $G_\mathrm{LR}$ and $G_\mathrm{RL}$ subgap states can be seen with the charactereistic change in sign at points of minimal energy of the ABS or at intersection points of ABSs.

At a magnetic field value $B_{||}=\SI{1.2}{\tesla}$ in \ref{fig:plungerscans_dev_1_right}(e-h) the ABSs are lowered in energy due to an increased Zeeman energy. Between the low energy state and the parent gap $\Delta\approx\SI{0.2}{\milli\eV}$, higher excited states appear both in the local and nonlocal conductances. Characteristic sign changes are observed in nonlocal conductance at crossing points of ABSs at finite bias $V_\mathrm{SD}\approx\SI{0.1}{\milli\volt} $and at gate voltages at which the ABSs reach a minimum in energy.

The ABSs reach zero bias at a magnetic field value of $B_{||}=\SI{1.7}{\tesla}$ (see Fig.~\ref{fig:fieldscan _dev_1_right}). This results in a dense spectrum of subgap states that fill the superconducting gap in both the local and nonlocal conductances measured in \ref{fig:plungerscans_dev_1_right}(i-l). \\

\subsubsection{Extracted value of Q\textsubscript{L} and Q\textsubscript{R}}

\begin{figure}[h!]
\begin{center}
\includegraphics[scale=1.0]{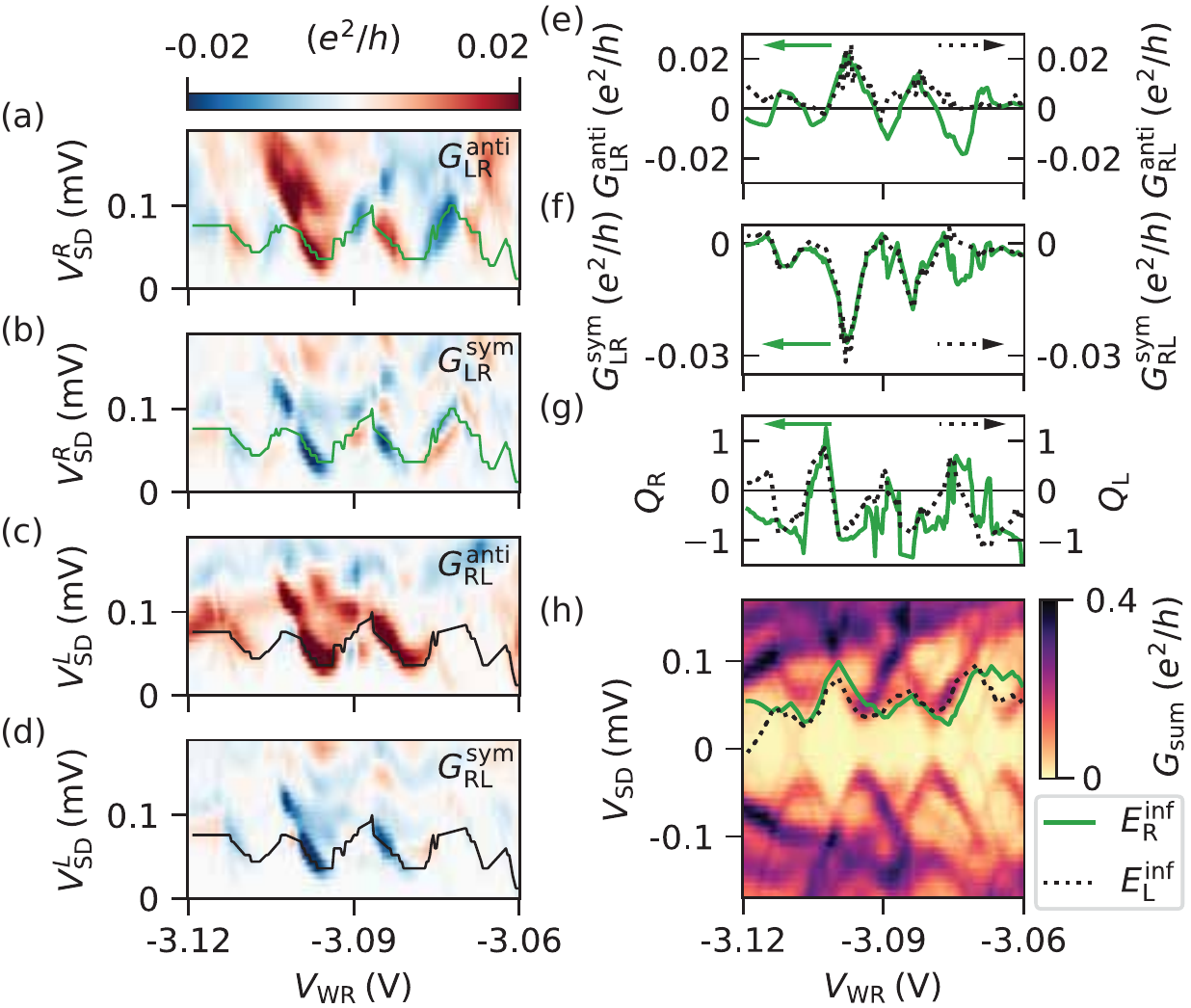}
\end{center}
\caption[]{\label{fig:Q_extr_right_1.2T}(a-d) symmetric and anti-symmetric components of the nonlocal conductances measured at $B_{||}=\SI{1.2}{\tesla}$. The values at the position of the green and black line are plotted in (e, f) and were used to calculate $Q_\mathrm{L}$, $Q_\mathrm{R}$, which are plotted in (g). The two quantities evolve similarly with $V_\mathrm{WM}$. (h) shows the sum of all conductances $G_\mathrm{sum}$. The dotted black, solid green lines display the energies $E_\mathrm{L,\:R}^{\mathrm{inf}}$ inferred from the integrated $Q_\mathrm{L}$, $Q_\mathrm{R}$.}
\end{figure}

Figure \ref{fig:Q_extr_right_1.2T}(a-d) shows the symmetric and anti-symmetric components of $G_\mathrm{LR}$ and $G_\mathrm{RL}$ measured at $B_{||}=\SI{1.2}{\tesla}$. The values at the position of the lowest lying state [shown by the green and black line in Fig.~\ref{fig:Q_extr_right_1.2T}(a-d)] are plotted in Fig.~\ref{fig:Q_extr_right_1.2T}(e, f). The values of $Q_\mathrm{L}$ and $Q_\mathrm{R}$ plotted in Fig.~\ref{fig:Q_extr_right_1.2T}(g) are similar. The energies $E_\mathrm{L,\:R}^\mathrm{inf}$ inferred from the integration of $Q_\mathrm{L,R}$ track the energy of the lowest lying state for a finite range of $V_\mathrm{WM}$ as seen from Fig.~\ref{fig:Q_extr_right_1.2T}(h). \\

\section{Supplementary information on device 2}
\label{sec:dev2_nl}
\begin{figure}[h!]
\begin{center}
\includegraphics[width=0.7\textwidth]{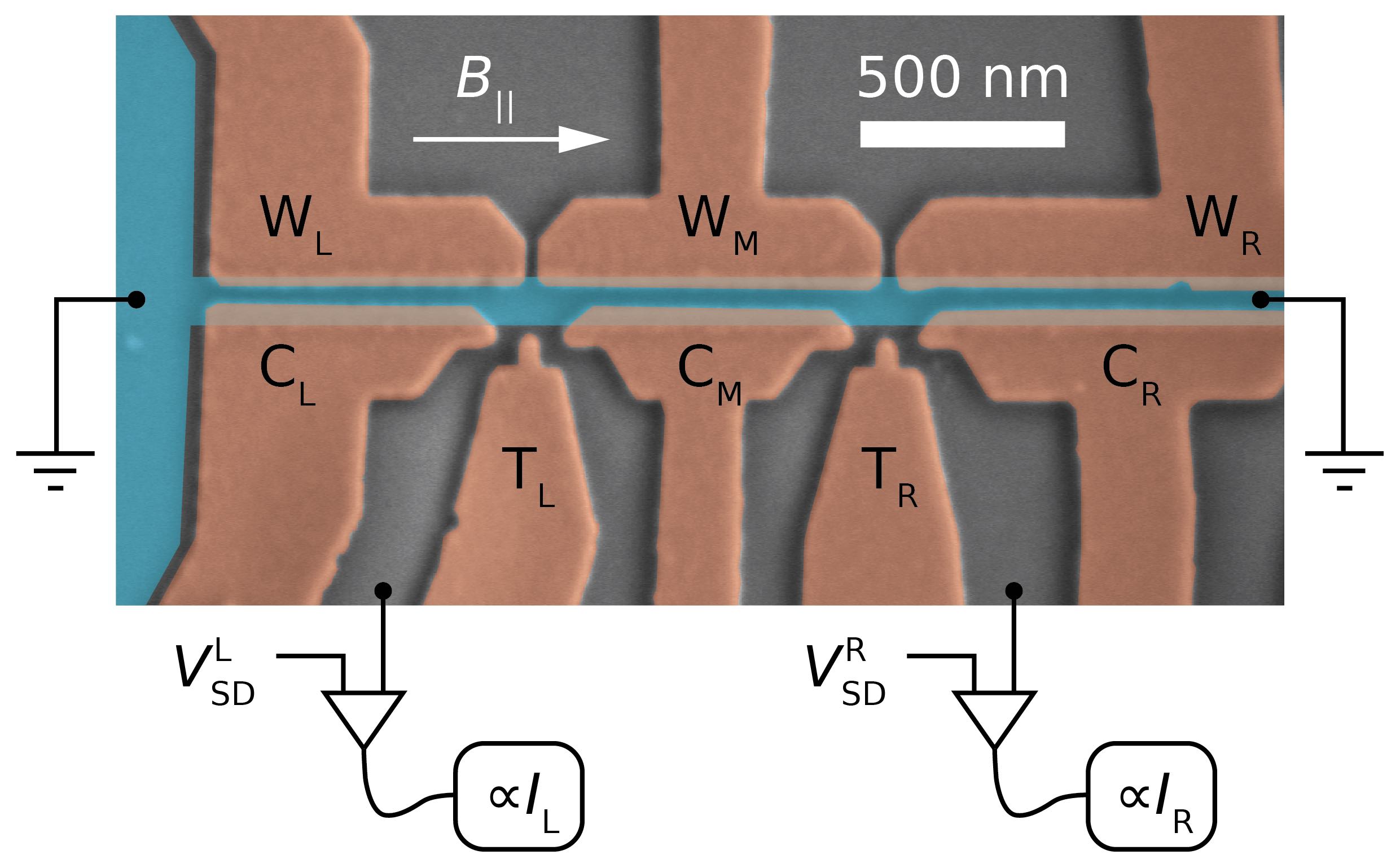}
\end{center}
\caption[]{\label{fig:dev_2_sem}False-color micrograph of {device} 2. A wire connected to electrical ground planes at its ends is formed by the Al (blue) on top of the InAs quantum well (gray). Electrostatic gates (red) control the electron density in the NW and control the quantum point contacts adjacent to the NW.}
\end{figure}

In the following, we present data on {device} 2, which is shown in the false-color electron micrograph in Fig.~\ref{fig:dev_2_sem}. The device consists of an Al film (shown in blue) that is shaped into a NW connected to ground planes at both of its ends. A single layer of electrostatic gates made from Ti/Au (shown in red) on top of $\mathrm{HfO_x}$ gate dielectric are used to control the device and shape the electron density in the 2DEG (shown in gray). Fabrication details for device 2 can be found in Appendix \ref{sec:fab_semi_sideprobe}. The gates labeled $\mathrm{W_L}$, $\mathrm{W_M}$, $\mathrm{W_R}$ are used to electrostatically confine the NW and tune the electron density in the semiconductor under the Al. The gates labeled $\mathrm{C_L}$, $\mathrm{C_M}$, $\mathrm{C_R}$ also electrostatically confine the NW under the Al, and in addition form quantum point contacts adjacent to the NW. The gates $\mathrm{T_L}$ and $\mathrm{T_R}$ offer additional control over the tunnel barrier formed by the point contact. The gates $\mathrm{T_L}$ and $\mathrm{T_R}$ screen the 2DEG region underneath them, which serve as semiconducting leads for tunneling measurements. {Device} 2 is based on an $\mathrm{In_{1-y}Al_{y}As-InAs-In_{1-x}Ga_{x}As}$ quantum well. Due to the larger band gap of $\mathrm{In_{1-y}Al_{y}As}$ compared to $\mathrm{In_{1-x}Ga_{x}As}$, the interface transparency between the InAs and superconducting Al is decreased for {device} 2 in comparison to {device} 1. \\

The same measurement setup as depicted in Fig.~\ref{fig:detection} for {device} 1 was applied for {device} 2. Each of the ground planes of Al at the NW ends are connected by two electrical lines and grounded at the breakout box. Each of the semiconducting leads under the gates $\mathrm{T_L}$ and $\mathrm{T_R}$ are connected to a current to voltage converting amplifier in order to measure the tunneling currents $I_\mathrm{L}$ and $I_\mathrm{R}$. Using the lock-in detection scheme as shown in Fig.~\ref{fig:detection} allows for the measurement of all four conductances $G_\mathrm{ij}=\mathrm{d}I_\mathrm{i}/\mathrm{d}V_\mathrm{SD}^\mathrm{j}$ ($i,j\in \{\mathrm{L}, \mathrm{R}\}$).\\

\subsection{Measurement of local and nonlocal conductance: Magnetic field dependence}

\begin{figure}[h!]
\begin{center}
\includegraphics[width=\textwidth]{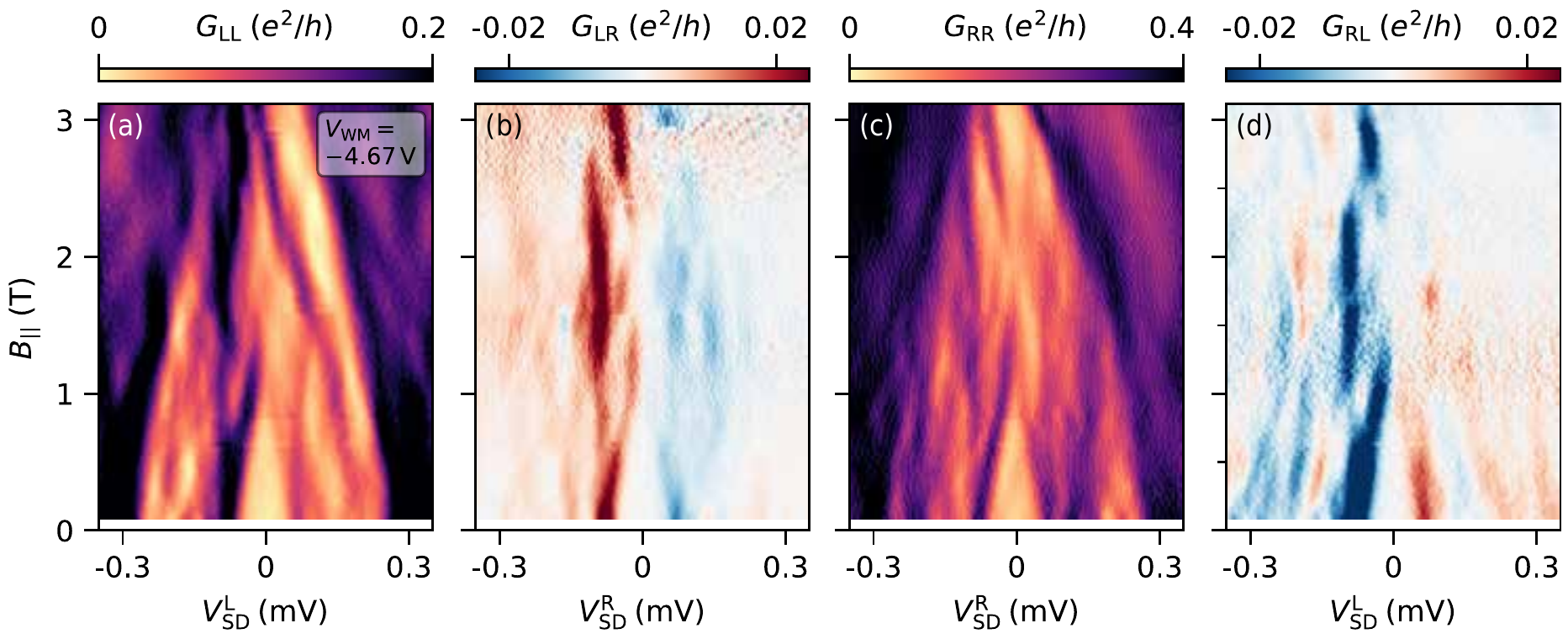}
\end{center}
\caption[]{\label{fig:fieldscan_dev_2}Local and nonlocal conductances as a function of magnetic field measured on {device} 2. The parent gap $\Delta$ is visible at high source drain bias in (a, c). The strongly reduced induced gap $\Delta_\mathrm{ind}$ is visible in all four conductance matrix elements and closes around $B_{||}\approx\SI{1}{\tesla}$. }
\end{figure}

A measurement of the local and nonlocal conductances a function of magnetic field $B_{||}$ is shown in Fig.~\ref{fig:fieldscan_dev_2}. The gate $\mathrm{W_M}$ was set to \SI{-4.67}{\volt}, while $V_\mathrm{WL}=V_\mathrm{WR}=\SI{-6}{\volt}$. This creates a modulation in the electron density along the elongated NW axis. Both local conductances in Fig.~\ref{fig:fieldscan_dev_2}(a,c) show a parent superconducting gap of \SI{0.25}{\milli\volt} at zero magnetic field which decreases with increasing magnetic field. Several subgap states are visible already at zero magnetic field. The induced gap $\Delta_\mathrm{ind}$ of the proximitized system given by the energy of the lowest ABSs is $\approx \SI{80}{\micro\eV}$ at zero magnetic field. The strongly reduced $\Delta_\mathrm{ind}$ can be interpreted as a result of the lower interface transparency between semiconductor and superconductor of the heterostructure used. With increasing magnetic field, the induced gap $\Delta_{\mathrm{ind}}$ closes at a value of $B_{||}\approx\SI{1}{\tesla}$ due to the ABSs moving to zero energy.\\

In the nonlocal conductances $G_\mathrm{LR}$ and $G_\mathrm{RL}$  [Fig.~\ref{fig:fieldscan_dev_2} (b, d)], the closing of the induced gap is visible at $B_{||}\approx\SI{1}{\tesla}$ where the low energy ABSs merge at zero bias. Higher excited subgap states are resolved in the nonlocal conductance. At voltages above the parent gap $|eV_\mathrm{SD}|>\Delta$ finite nonlocal conductance is visible.\\

\subsection{Measurement of local and nonlocal conductance: gate voltage dependence}

\begin{figure}[h!]
\begin{center}
\includegraphics[width=\textwidth]{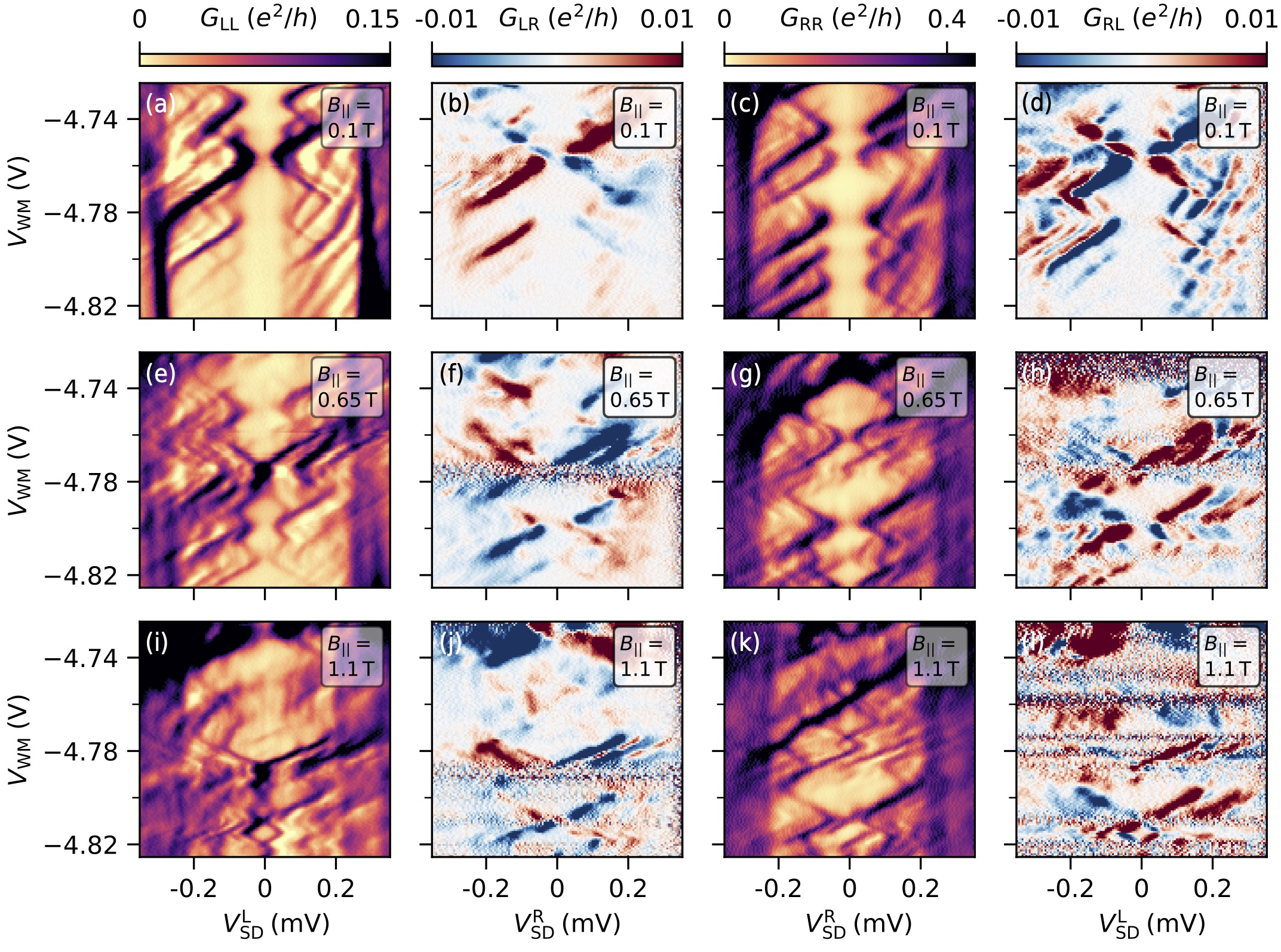}
\end{center}
\caption[]{\label{fig:plungerscans_1_dev_2}Measured conductance matrix elements at different magnetic field values. (a-d) the induced gap is visible in local and nonlocal conductances. The excited state spectrum consists of discrete states due to the finite size of the NW segment. (i-l) states have crossed zero bias and a low energy state that oscillates in energy is visible.}
\end{figure}

The different tunneling conductances as a function of gate voltage $V_\mathrm{WM}$ at four distinct magnetic field values are depicted in Fig.~\ref{fig:plungerscans_1_dev_2}. At low magnetic field $B_{||}=\SI{0.1}{\tesla}$, there is a finite induced gap $\Delta_\mathrm{ind}$ in the proximitized system. For an infinitely long wire, one expects a continuum of states at energies above the induced gap $\Delta_\mathrm{ind}$. Here, states are confined to a NW segment of length \SI{0.8}{\micro\meter}, which leads to a spectrum of discrete ABSs due to finite size effects instead of a continuum. These discrete states appear clearly as a dense spectrum of intersecting lobes at voltages $\Delta_\mathrm{ind}\leq e V_\mathrm{SD}<\Delta$ in Fig.~\ref{fig:plungerscans_1_dev_2}(a, c). At the voltage $eV_\mathrm{SD}=\Delta_\mathrm{ind}\approx\SI{80}{\micro\eV}$ the states go through a minimum in energy. Similar excited state spectra have been observed in numerical simulations \cite{Mishmash2016, dassarma_goodbadugly}. The presence of disorder, localized bound states in the NW, and multiple transverse modes may further affect the excited state spectrum. The lobe shaped ABSs lead to nonlocal conductance signal as seen from Fig.~\ref{fig:plungerscans_1_dev_2}(b, d) with changes in sign around their minimal energy $\approx\SI{80}{\micro\eV}$.\\

With increasing magnetic field, the ABSs lower their energy. At a magnetic field value $B_{||}=\SI{0.65}{\tesla}$ some of the ABSs merge at zero bias [Fig.~\ref{fig:plungerscans_1_dev_2} (e, g)]. At $B_{||}=\SI{1.1}{\tesla}$ the ABSs have crossed zero bias [see Fig.~\ref{fig:plungerscans_1_dev_2}(i, k)]. Around $V_\mathrm{WM}=\SI{-4.8}{\volt}$, subgap states oscillate around zero bias in Fig.~\ref{fig:plungerscans_1_dev_2}(i, k). Most subgap states in Fig.~\ref{fig:plungerscans_1_dev_2} appear both in the local and nonlocal conductances, which is a strong indication that they stem from extended bound states that couple to the normal leads at both ends of the NW segment.\\

\begin{figure}[h!]
\begin{center}
\includegraphics[width=\textwidth]{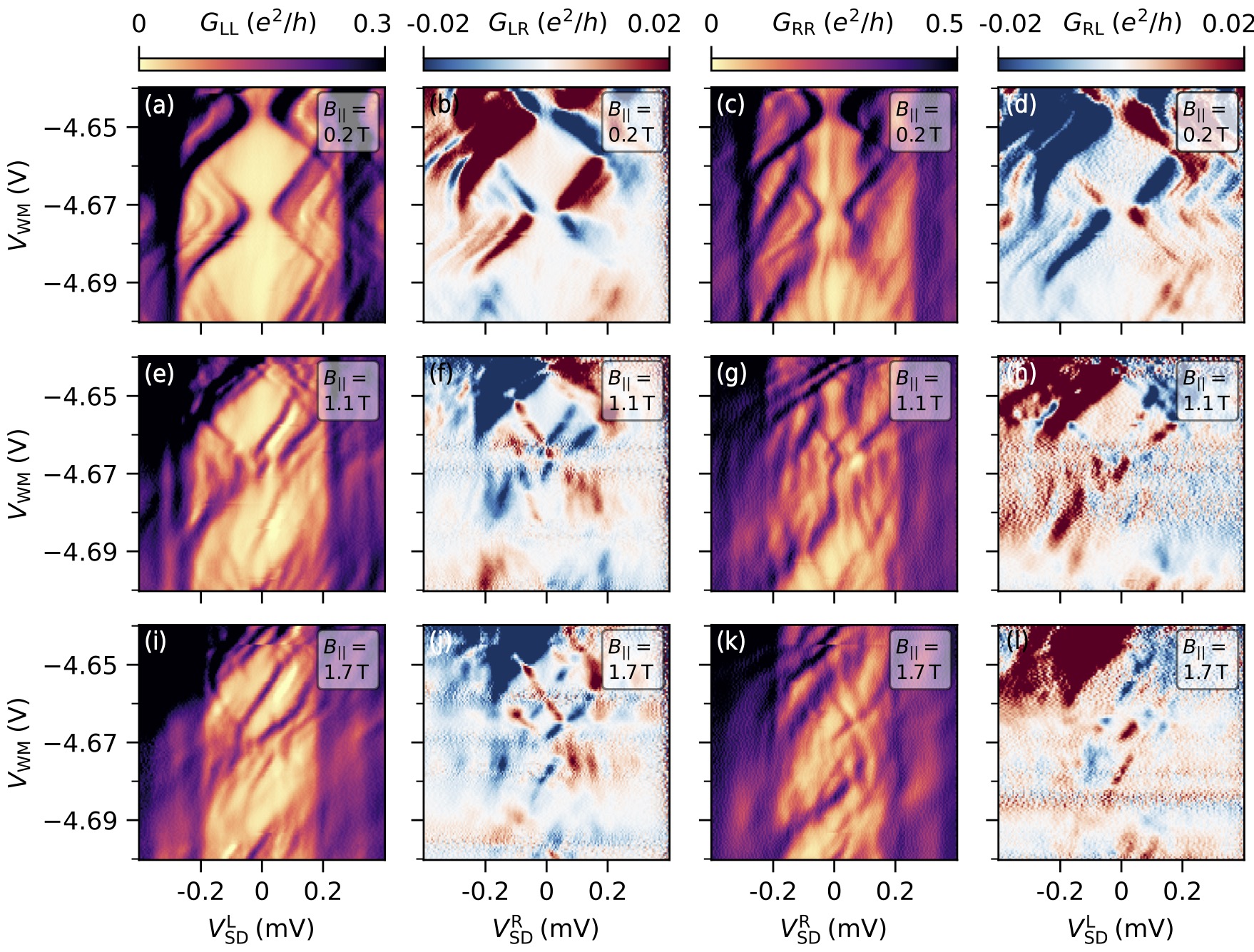}
\end{center}
\caption[]{\label{fig:plungerscans_2_dev_2}Local and nonlocal conductances at different values of $B_{||}$. The subgap state close to zero bias in (c) is not visible in the three other conductances measured at the same field value. Some of the lobe shaped ABSs appear in both local conductances (a, c) and in the nonlocal conductances (b, d) with the characteristic sign changes. (e-h) subgap state around $V_\mathrm{WM}=\SI{-4.67}{\volt}$ oscillates around zero bias and is visible in all local and nonlocal conductances.}
\end{figure}

Local and nonlocal conductances at more positive gate voltage $V_\mathrm{WM}$ are depicted in Fig.~\ref{fig:plungerscans_2_dev_2}. At low magnetic field $B_{||}=\SI{0.2}{\tesla}$ subgap states due to finite size confinement appear in the local and nonlocal conductances [Fig.~\ref{fig:plungerscans_2_dev_2}(a-d)]. The nonlocal conductance signal of individual subgap states furthermore undergoes the characteristic sign changes at points where the state reaches a minimum in energy ($\approx\SI{80}{\micro\eV}$) or where subgap states cross ($\approx \SI{0.2}{\milli\eV}$). The measurement of $G_\mathrm{RR}$ furthermore shows a subgap state around zero bias. This state does not appear in the measurement of $G_\mathrm{LL}$ suggesting that it does not couple to both leads at this magnetic field value. The state consequently is not visible in the nonlocal conductances [Fig.~\ref{fig:plungerscans_2_dev_2}(b, d)]. At a higher magnetic field value of $B_{||}=\SI{1.1}{\tesla}$ two subgap states merge forming a zero bias peak in the measurement of $G_\mathrm{RR}$ [Fig.~\ref{fig:plungerscans_2_dev_2}(b)]. This state appears only for a small range of gate voltage around $V_\mathrm{WM}=\SI{-4.67}{\volt}$ in $G_\mathrm{LL}$ in Fig.~\ref{fig:plungerscans_2_dev_2}. In this range, the two states cross zero bias and overshoot. This leads to a characteristic signature in nonlocal conductance with a sign change at the point of local maximum in energy of the bound states.
At an even higher magnetic field value $B_{||}=\SI{1.7}{\tesla}$, the conductance acquires a strong asymmetry, making it difficult to track the evolution of individual states in the local conductances.

\subsection{Comparison between anti-symmetric parts of local and nonlocal conductances}
\begin{figure}[h!]
\begin{center}
\includegraphics[width=\textwidth]{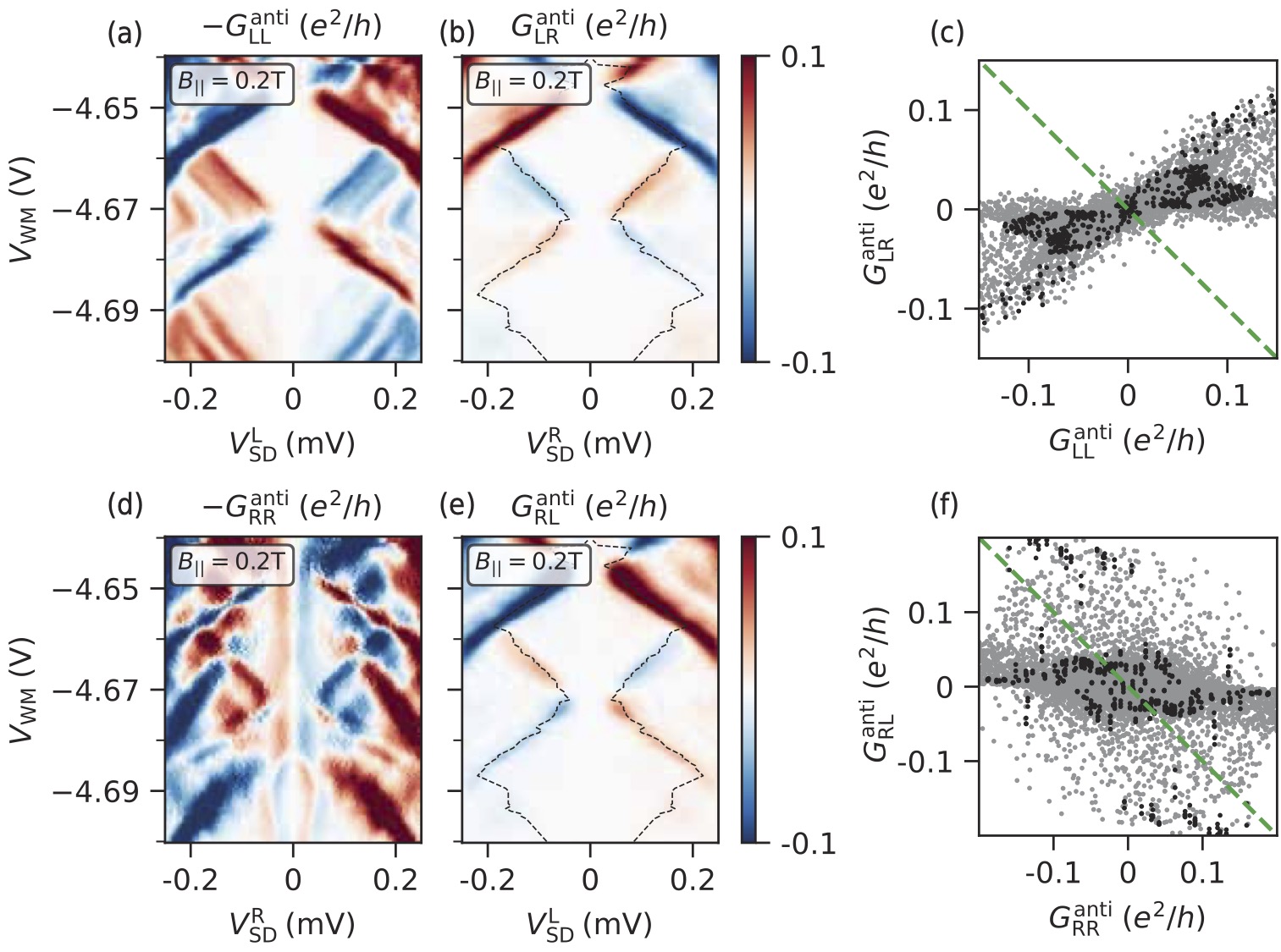}
\end{center}
\caption[]{\label{fig:antisymmetries_lowB_dev_2}(a, b) Comparison of $-G_\mathrm{LL}^\mathrm{anti}$ and $G_\mathrm{LR}^\mathrm{anti}$. (c) parametric plot of the data in (a, b). The black data points originate from a three pixel wide window around the black dashed line in (b). (d, e) shows a comparison of $-G_\mathrm{RR}^\mathrm{anti}$ and $G_\mathrm{RL}^\mathrm{anti}$. (f) Parametric plot of the data in (d, e). The black data points originate from a three pixel wide window around the black dashed line in (e).}
\end{figure}

Similar to the the analysis in Section \ref{sec:antisymmetries_dev1}, the anti-symmetric parts of the conductance matrix elements are calculated from the data in Fig.~\ref{fig:plungerscans_2_dev_2}. The quantities $-G_\mathrm{LL}^\mathrm{anti}$ and $G_\mathrm{LR}^\mathrm{anti}$ at $B_{||}=\SI{0.2}{\tesla}$ are shown in Fig.~\ref{fig:antisymmetries_lowB_dev_2}(a, b). While the two quantities are expected to be equal according to Eq.~\ref{eq:anti-symmetry} predicted by theory \cite{karsten_nl_spectroscopy}, the experimental data shows that $-G_\mathrm{LL}^\mathrm{anti}$ has a larger magnitude and different sign than $G_\mathrm{LR}^\mathrm{anti}$. This lack of similarity between the data is also visible in the parametric plot in Fig.~\ref{fig:antisymmetries_lowB_dev_2}(c), where all data points from panels (a) and (b) are plotted with the data points taken in a three pixel window around the state shown as a dashed black line in (b).

Between $-G_\mathrm{RR}^\mathrm{anti}$ and $G_\mathrm{RL}^\mathrm{anti}$ there is a large discrepancy visible in Fig.~\ref{fig:antisymmetries_lowB_dev_2}(d, e). A particularly large anti-symmetric component of the local conductance $G_\mathrm{RR}$ is visible at low values of source-drain bias $|V_\mathrm{SD}^\mathrm{R}|\leq\SI{80}{\micro\V}$ due to the presence of the low energy subgap state that only appears in $G_\mathrm{RR}$ as discussed earlier. A parametric plot of the data in (d, e) is shown in (f) and confirms the lack of points close to the expected theory relation given by the green dashed line.

\begin{figure}[h!]
\begin{center}
\includegraphics[width=\textwidth]{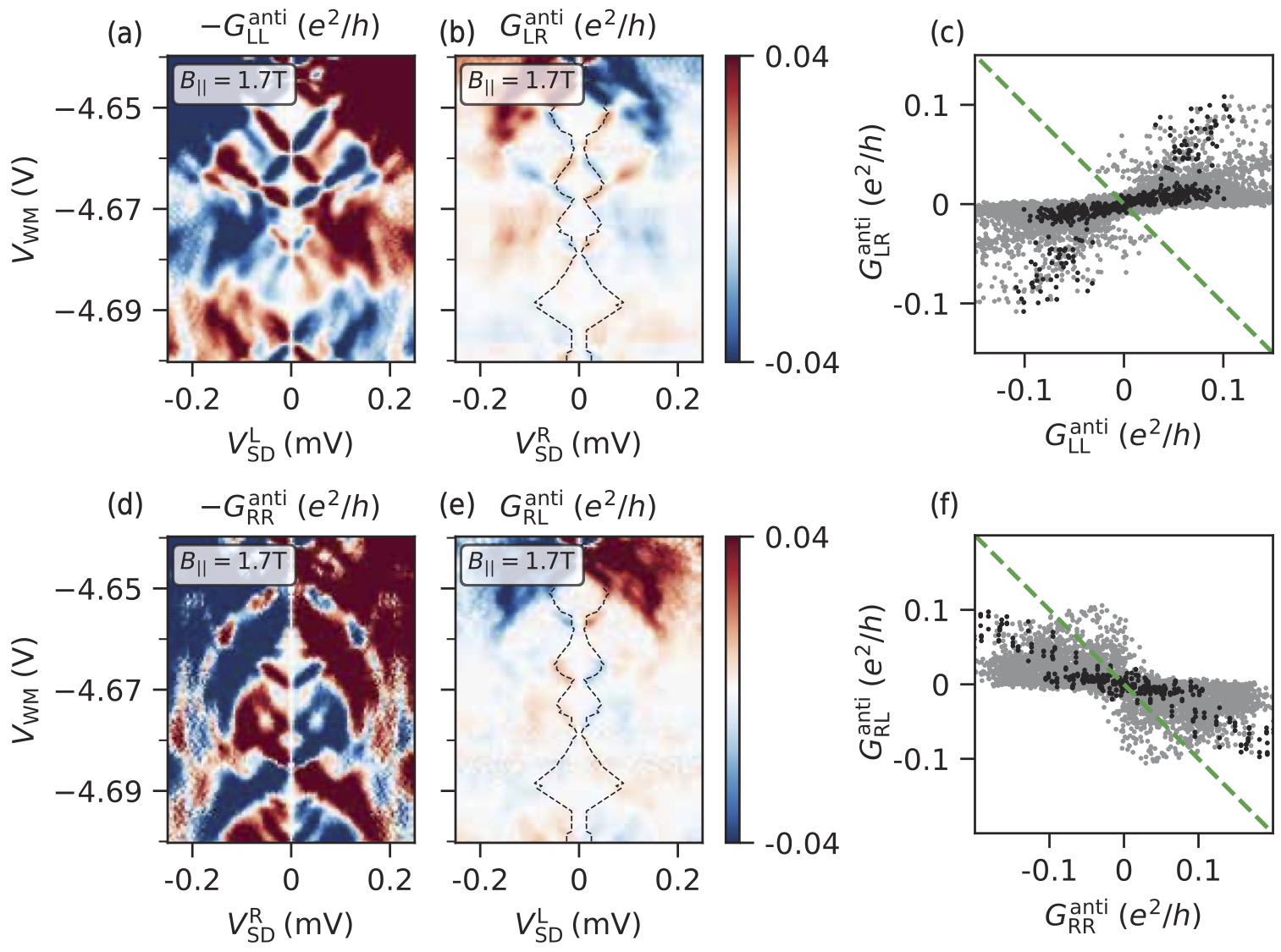}
\end{center}
\caption[]{\label{fig:antisymmetries_highB_dev_2}(a,b) Comparison of $-G_\mathrm{LL}^\mathrm{anti}$ and $G_\mathrm{LR}^\mathrm{anti}$. (c) Parametric plot of the data in (a,b). The green dashed line indicates the relation between data points expected from theory and the black data points originate from three pixel wide window around the black dashed line in (b). (d, e) comparison of $-G_\mathrm{RR}^\mathrm{anti}$ and $G_\mathrm{RL}^\mathrm{anti}$. (f) Parametric plot of the data in (d, e).  The green dashed line indicates the relation between data points expected from theory and the black data points originate from three pixel wide window around the black dashed line in (e).}
\end{figure}

In Fig.~\ref{fig:antisymmetries_highB_dev_2} the anti-symmetric component of the conductances measured at $B_{||}=\SI{1.7}{\tesla}$ is shown. There is no equality of $-G_\mathrm{LL}^\mathrm{anti}$ and $G_\mathrm{LR}^\mathrm{anti}$. The same holds true for $-G_\mathrm{RR}^\mathrm{anti}$ and $G_\mathrm{RL}^\mathrm{anti}$.

\begin{figure}[h!]
\begin{center}
\includegraphics[scale=0.95]{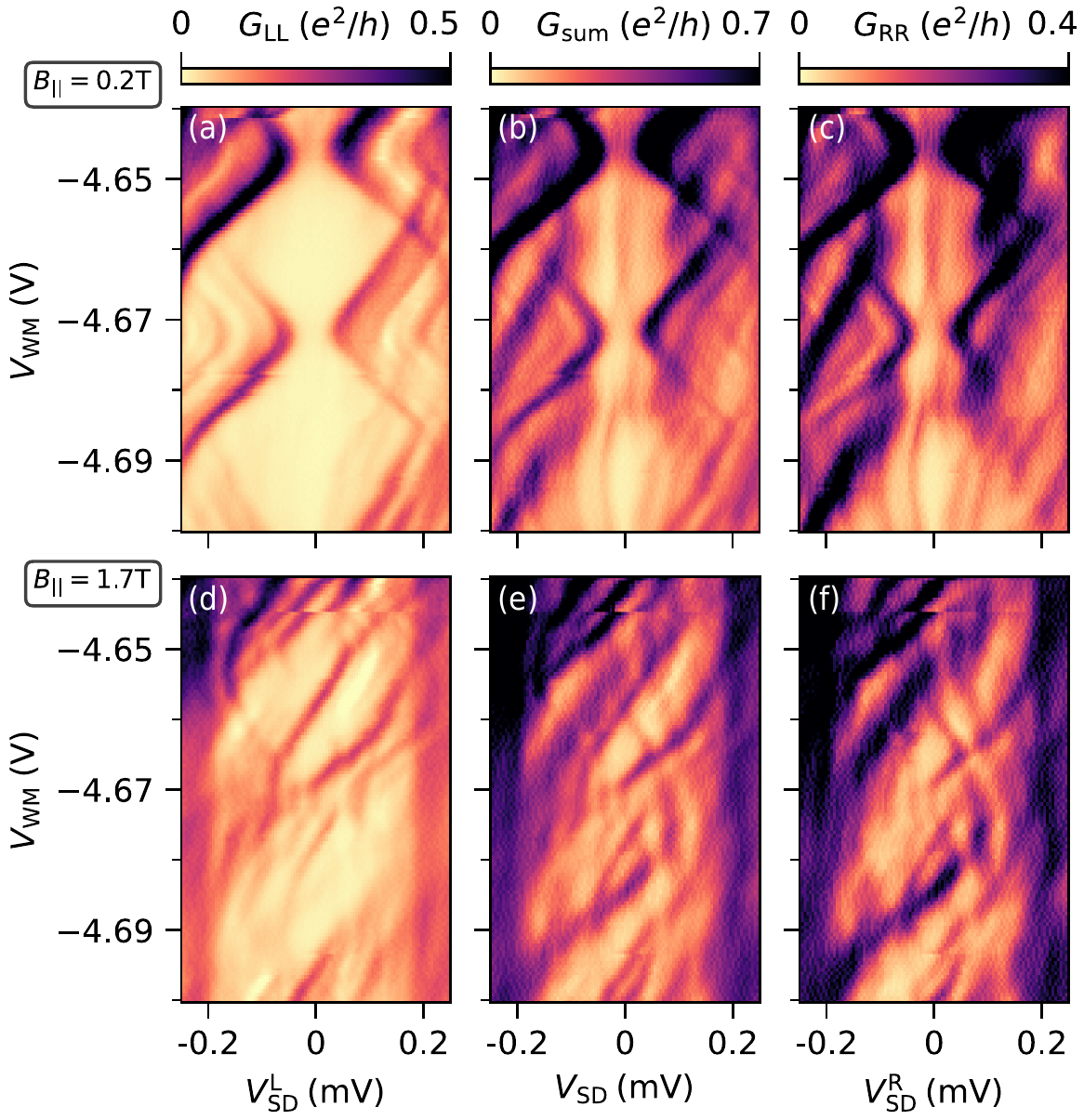}
\end{center}
\caption[]{\label{fig:symmetries_lowB_dev_2}(a) Tunneling conductance $G_\mathrm{LL}$, (b) the sum of all conductance matrix elements $G_\mathrm{sum}$, (c) tunneling conductance $G_\mathrm{RR}$ measured at $B_{||}=\SI{0.2}{\tesla}$. A notable asymmetry of $G_\mathrm{sum}$ with respect to $V_\mathrm{SD}$ is visible. (d) tunneling conductance $G_\mathrm{LL}$, (e) the sum of all conductance matrix elements $G_\mathrm{sum}$, (f) tunneling conductance $G_\mathrm{RR}$ measured at $B_{||}=\SI{1.7}{\tesla}$.}
\end{figure}

As a consequence of the experimental data not fulfilling Eq.~\ref{eq:anti-symmetry}, forming the sum $G_\mathrm{sum}$ of all conductance matrix elements does not recover a symmetric function with respect to $V_\mathrm{SD}$ compared to the local conductances. This can be clearly seen for $B_{||}=\SI{0.2}{\tesla}$ in Fig.~\ref{fig:symmetries_lowB_dev_2}(a-c) and for $B_{||}=\SI{1.7}{\tesla}$ in Fig.~\ref{fig:symmetries_lowB_dev_2}(d-f) where $G_\mathrm{LL}$, $G_\mathrm{sum}$, and $G_\mathrm{RR}$ are plotted side-by-side.\\

While we found that the lowest energy subgap state measured on {device} 1 fulfills the relation given by Eq. \ref{eq:anti-symmetry}, the data for {device} 2 do not follow this relation. Several reasons for Eq. \ref{eq:anti-symmetry} being violated have been given in previous works \cite{gerbold_nonlocal,melo_asym}. The two devices presented here also differ in the composition of their underlying hybrid heterostructure and in the NW length.

\subsection{Extracted values of Q\textsubscript{L} and Q\textsubscript{R}}

\begin{figure}[h!]
\begin{center}
\includegraphics[scale=0.95]{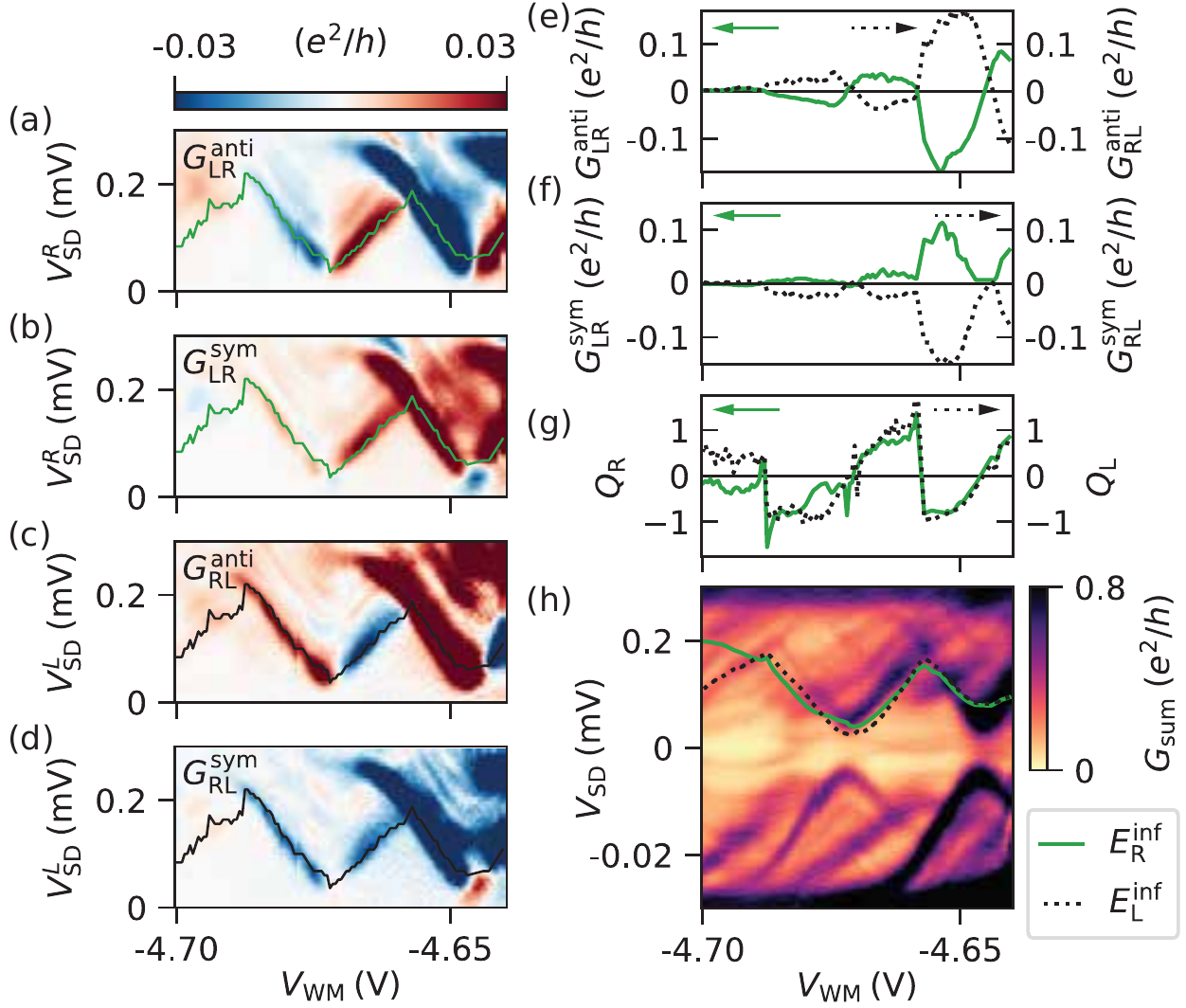}
\end{center}
\caption[]{\label{fig:Q_extr_dev_2_0.2T}(a-d) Symmetric and anti-symmetric components of the nonlocal conductances measured at $B_{||}=\SI{0.2}{\tesla}$. The values at the position of the green and black line are plotted in (e, f) and were used to extract $Q_\mathrm{L}$, $Q_\mathrm{R}$, which are shown in (g). The two quantities show a continuous transition from -1 to +1 where the ABS reaches a minimum in energy. Sharp transitions are visible at the crossing of two states at $V_\mathrm{SD}=\SI{0.2}{\milli\volt}$. (h) shows the sum of all conductances $G_\mathrm{sum}$. The dotted black, solid green display the energies $E_\mathrm{L ,\:R}^{\mathrm{inf}}$ inferred from the integrated $Q_\mathrm{L}$, $Q_\mathrm{R}$.}
\end{figure}

Figure \ref{fig:Q_extr_dev_2_0.2T}(a-d) shows the symmetric and anti-symmetric component $G_\mathrm{LR,\:RL}^\mathrm{sym(anti)}$ of the nonlocal conductance in Fig.~\ref{fig:plungerscans_2_dev_2}(b, d) measured at $B_{||}=\SI{0.2}{\tesla}$. The value for the lowest excited state is shown in \ref{fig:Q_extr_dev_2_0.2T}(e, f). The resulting values of $Q_\mathrm{L}$ and $Q_\mathrm{R}$ are depicted in Fig.~\ref{fig:Q_extr_dev_2_0.2T}(g). The curves for $Q_\mathrm{L}$ and $Q_\mathrm{R}$ are similar. At the value where the ABS goes through a minimum in energy, $Q_\mathrm{j}$ undergoes a continuous crossover from negative to positive values. At points where the ABS intersects with a neighboring ABS around $V_\mathrm{SD}\approx\SI{0.2}{\milli\volt}$ an abrupt change from $Q_\mathrm{j}>0$ to $Q_\mathrm{j}<0$ is visible. Regions of positive (negative) $Q_\mathrm{j}$ coincide with regions where the state energy has a positive (negative) slope with respect to the gate voltage $V_\mathrm{WM}$. This is in agreement with the interpretation of $Q_\mathrm{j}$ reflecting the charge character of the bound state. 

Integrating $Q_\mathrm{L}$ and  $Q_\mathrm{R}$, taking into account a linear background and rescaling by a constant lever arm leads to the curves shown in Fig.~\ref{fig:Q_extr_dev_2_0.2T}(h). The curve matches the state energy over a large range of gate voltage $V_\mathrm{WM}$, which suggests that $Q_\mathrm{L}$, $Q_\mathrm{R}$ are proportional to the total charge $Q$ \cite{karsten_nl_spectroscopy}. 

\begin{figure}[h!]
\begin{center}
\includegraphics[scale=0.95]{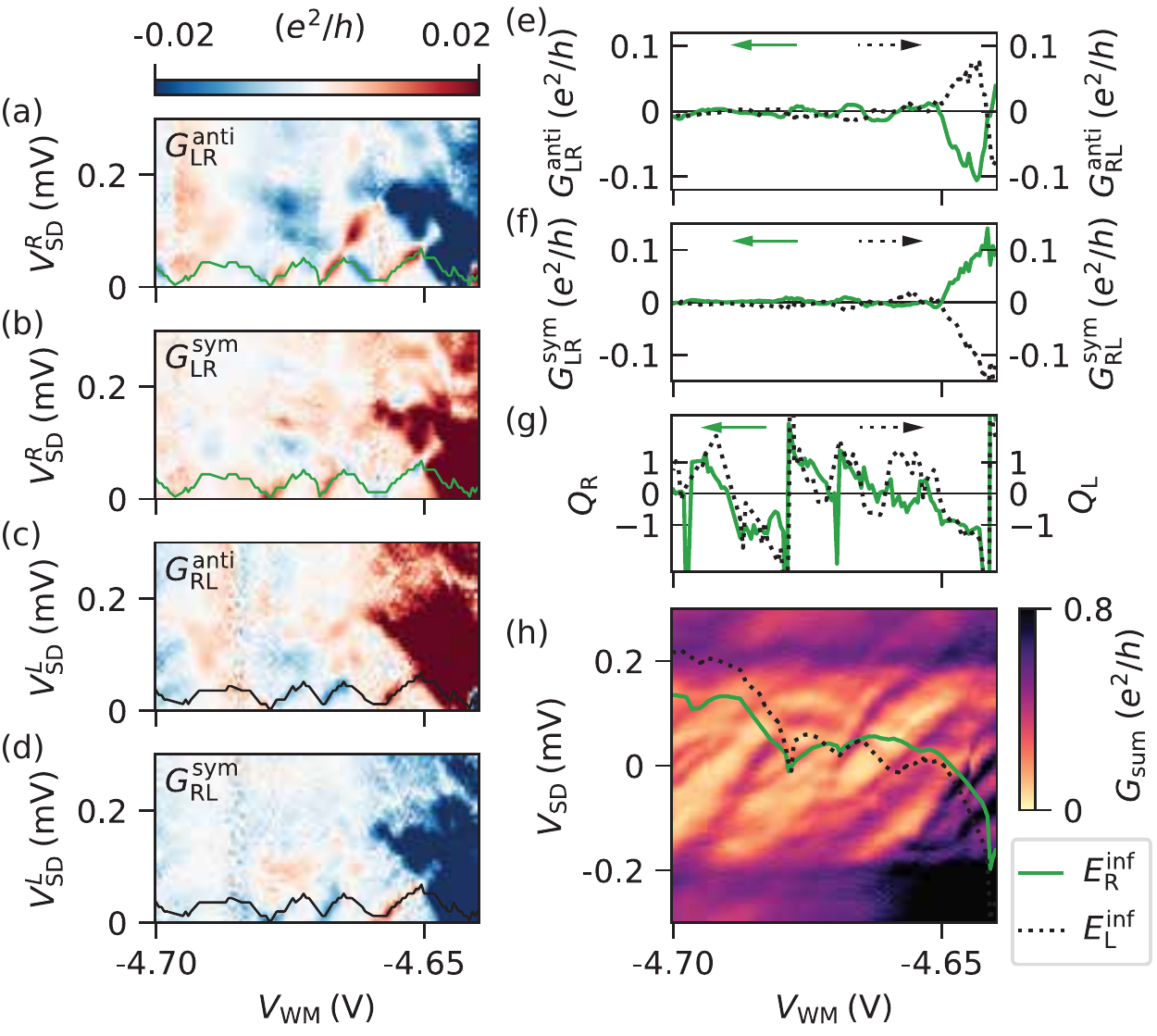}
\end{center}
\caption[]{\label{fig:Q_extr_dev_2_1.7T}(a-d) Symmetric and anti-symmetric components of the nonlocal conductances measured at $B_{||}=\SI{1.7}{\tesla}$. The values at the position of the green and black line are plotted in (e, f) and were used to extract $Q_\mathrm{L}$, $Q_\mathrm{R}$, which are plotted in (g). (h) shows the sum of all conductances $G_\mathrm{sum}$. The dotted black, solid green lines in represent $E_\mathrm{L,\:R}^{\mathrm{inf}}$ inferred from the integrated $Q_\mathrm{L}$, $Q_\mathrm{R}$.}
\end{figure}

For the nonlocal conductance data in Fig.~\ref{fig:plungerscans_2_dev_2}(i-l) at $B_{||}=\SI{1.7}{\tesla}$, $Q_\mathrm{L}$ and $Q_\mathrm{R}$ were extracted. The underlying (anti-)symmetric components are plotted in Fig.~\ref{fig:Q_extr_dev_2_1.7T}(a-f). The resulting $Q_\mathrm{L}$ and $Q_\mathrm{R}$ are plotted in Fig.~\ref{fig:Q_extr_dev_2_1.7T}(g). The two curves show a similar evolution with $V_\mathrm{WM}$. The inferred energy $E_\mathrm{L,\:R}^\mathrm{inf}$ in Fig.~\ref{fig:Q_extr_dev_2_1.7T}(h) does not match the energy of the low energy subgap state. The extracted values $Q_\mathrm{j}$ are the quotient of two experimentally measured quantities subjected to noise. The anti-symmetric part enters this quotient as denominator and is zero at $V_\mathrm{SD}=\SI{0}{\volt}$ by construction. Consequently, the extraction of $Q_\mathrm{j}$ is very sensitive for states close or at $V_\mathrm{SD}=\SI{0}{\volt}$. This and the overall small signal may lead to a deviation between the energy of the state and $E_\mathrm{L,\:R}^\mathrm{inf}$ in the case shown in Fig.~\ref{fig:Q_extr_dev_2_1.7T}(h).

\chapter{Fingerprint of localized bound states}\label{ch:sideprobe}

In the following chapter, we present tunneling spectroscopy at the end of long ($\gtrsim\SI{1.5}{\micro\meter}$) nanowires (NWs) based on gated two-dimensional electron gas (2DEG) semiconductor-superconductor hybrid heterostructures. Subgap states stemming from Andreev bound states (ABSs) were found in tunneling spectroscopy at finite magnetic field. The occurrences of ABSs at the two NW ends were in general uncorrelated with respect to the gate voltage that tunes the electron density in the NW. A similar evolution of the subgap states at both NW ends with respect to magnetic field or gate voltage was only achieved by fine-tuning local gates. The data for uncorrelated states at both NW ends and correlated states due to fine-tuning are compared with the behavior of extended ABSs in shorter NWs ($\lesssim \SI{0.8}{\micro\meter}$). The described device geometry opens a path for tunneling spectroscopy at discrete points along a NW. 

\section{Introduction}

Proximitized NWs allow for the investigation of bound states that form in the presence of spin-orbit coupling, Zeeman energy, and superconducting pairing. Multiple experimental techniques exist to date that allow for the investigation of these bound states \cite{Anders_leaded_cqed, DS_DR_rfsensing, gerbold_nonlocal, henri_lead}. Among the most widespread are tunneling spectroscopy experiments, where electronic transport through a potential barrier between a normal lead and a hybrid system reveals the energy-resolved spectrum of excitations in the hybrid system \cite{chang_YSR,chang_hardgap,morten_sqpcn,henri_lead}. This has been used to investigate various bound states in conventional VLS NWs and gate-defined NWs based on 2DEGs. Some of the reported results are suggestive of the existence of a topological phase of matter that manifests itself in a zero-bias conductance peak in tunneling spectroscopy experiments \cite{nichele_scaling,mingtang_science,mourik,sole_fullshell}. However, multiple follow-up experiments and numerical investigations were able to explain the same signature by the presence of ABSs due to the presence of a quantum dot (QD) \cite{lee_singletdoublet, lee_scaling, dassarma_against_mingtang, dassarma_goodbadugly, trivial_fullshell, elsa_trivial_fullshell} at the NW end, quasi-Majoranas \cite{vuik_quasiMBS,frolov_ubiquitoustheory, San-Jose2016}, or disorder in the material \cite{dassarma_goodbadugly}.\\

A promising device geometry that allows for an easier distinction of states that are localized at the NW end and states that extend in the NW is the three-terminal geometry. It consists of a semiconductor NW proximitized by a superconductor which is electrically grounded. Each of the ends of the semiconductor NW are coupled to a normal lead. This allows for tunneling spectroscopy at both NW ends. First devices of this type based on selective-area-grown NWs were used to demonstrate nonlocal conductance spectroscopy \cite{gerbold_nonlocal, karsten_nl_spectroscopy} and the disappearance of end-to-end correlations in NWs longer than \SI{300}{\nano\meter} \cite{gian_correlations}. Similar experiments applied the same device geometry for VLS NWs to rule out the presence of extended states \cite{frolov_quantized}. In the proceeding chapters we showed how a three-terminal device can be realized in InAs 2DEGs proximitized by superconducting Al. In particular, we showed in Chapter \ref{ch:pradaclarke}  how extended ABSs appear in NWs as end-to-end correlated subgap resonances and how they transform when hybridized at one end with a localized QD. In Chapter \ref{ch:2DEG_nonlocal}, we investigated the signatures of these ABSs in nonlocal conductance. The devices used for these experiments were gate-defined NWs of length \SI{0.6}{\micro\meter} and \SI{0.8}{\micro\meter}.

The natural question that arises is what happens when longer NWs are used. It is expected theoretical that a topological phase arises if the material is clean enough and the length $L$ of the device exceeds the decay length of MZMs. ABSs that are found in short NWs may cross over to well isolated MZMs at the NW ends with strong end-to-end correlations in this scenario \cite{Mishmash2016, dassama_meta,das_sarma_correlation}. If the material is disordered such that the localization length is smaller than the device length $L$, the topological phase is broken and only localized states can be found \cite{Brouwer2000,Gruzberg2005,Brouwer2011,Lobos2012}. Soft confinement of the NW at its ends furthermore promotes the formation of quasi-Majoranas \cite{vuik_quasiMBS}. Here we investigate devices that allow for tunneling at the ends of gate-defined NWs in two-dimensional InAs electron gas proximitized by superconducting Al. The devices have a length $L=\SI{1.5}{\micro\meter}$ and allow for tunneling spectroscopy at the ends of the NW and for a tunnel probe in the middle of the NW. Only localized states are found, pointing to the scenario where the crossover to a topological phase in long NWs is prohibited. 

\section{Breakdown of end-to-end correlations}
\label{sec:uncorrelated}
\begin{figure}[h!]
\begin{center}
\includegraphics[width=\textwidth]{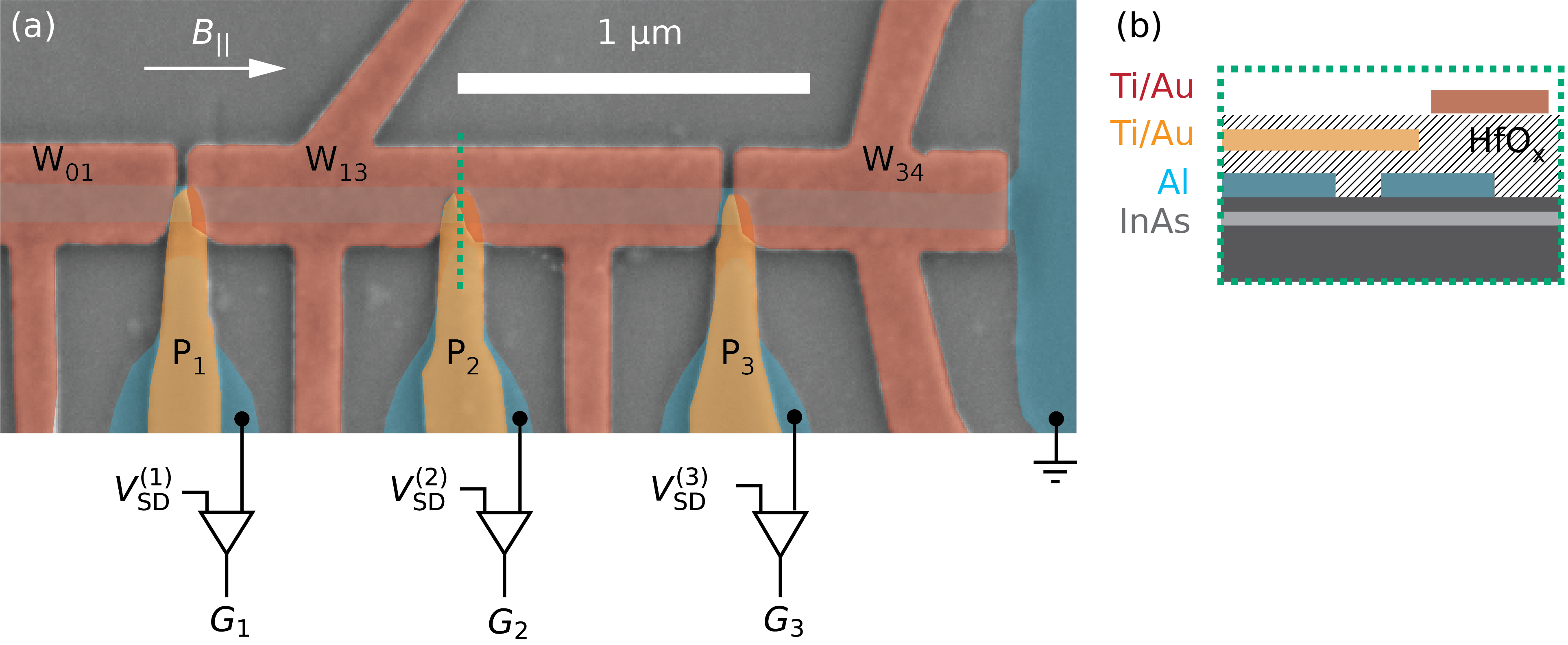}
\end{center}
\caption[Micrograph of device with superconducting sideprobes]{\label{fig:super_sideprobe_SEM}(a) False-color electron micrograph of sideprobe device with tunnel probes made from Al (blue). The gate $\mathrm{W_{13}}$ controls the electron density of the \SI{1.5}{\micro\meter} long NW formed under a strip of Al. The probes $j\in\{1,2,3\}$ are tunnel-coupled to the ends and the middle of the NW. (b) Illustration of the heterostructure along the green dotted line in (a).}
\end{figure}

Figure \ref{fig:super_sideprobe_SEM} shows the device under investigation (device 4). It consists of a superconducting strip of Al on top of semiconducting InAs 2DEG. The gates $\mathrm{W}_\mathrm{kl}$, $kl\in \{ 01, 13,34\}$ are used to deplete the 2DEG self-aligned with a strip of epitaxial Al and tune the density in individual segments of the NW. Superconducting probes are coupled laterally to the NW and a tunnel barrier between each probe and the NW can be tuned by the gates $\mathrm{P}_\mathrm{j}$ with $j\in \{ 1, 2,3\}$. Individual current-to-voltage converting amplifiers are connected to each of the tunnel probes. This allows for tunneling spectroscopy at each probe position, equivalent to the electrical setup described in Chapter \ref{ch:pradaclarke}. Note that tunnel probes 1 and 3 are coupled the end of the NW controlled by the gate $\mathrm{W}_\mathrm{13}$ while probe 2 is weakly coupled to the center of the NW. Details on the fabrication of device 4 is given in Appendix \ref{sec:fab_super_sideprobe}.\\

\begin{figure}[h!]
\begin{center}
\includegraphics[scale=0.9]{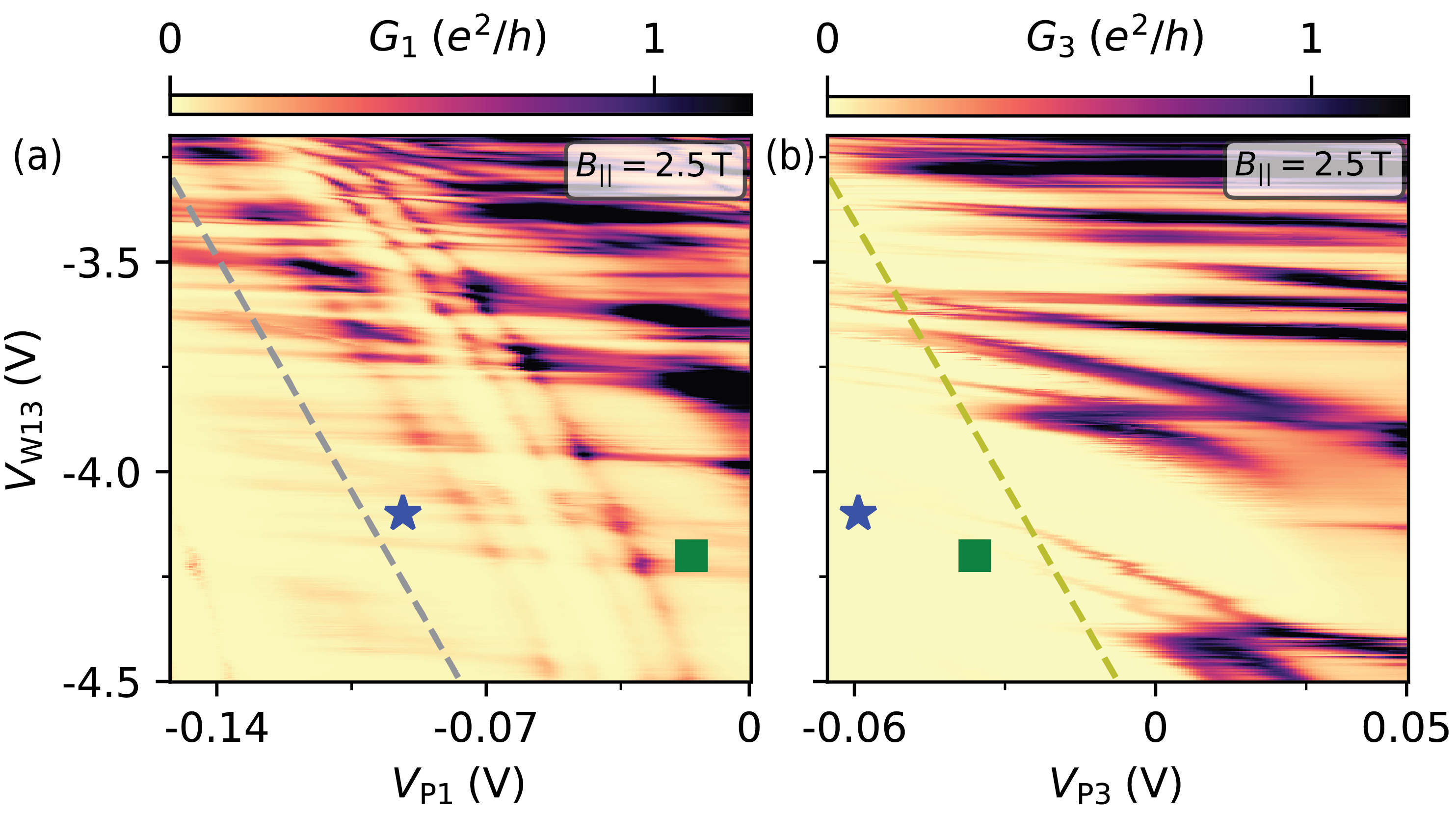}
\end{center}
\caption[Gate dependence of localized states at the ends of a \SI{1.5}{\micro\meter} long nanowire]{\label{fig:sideprobe_gatemap_dev1}(a, b) Simultaneously measured zero-bias differential conductance at the ends of the NW segment as a function of the voltages on gates $\mathrm{W13}$, $\mathrm{P1}$, and $\mathrm{P3}$. The resonances are largely independent of the gates $\mathrm{P1}$, $\mathrm{P3}$ that tune the tunnel barriers. The states at gate voltage values marked by $\bigstar$  and \ding{110} are investigated as a function of magnetic field in the following.}
\end{figure}

\begin{figure}[h!]
\begin{center}
\includegraphics[scale=0.9]{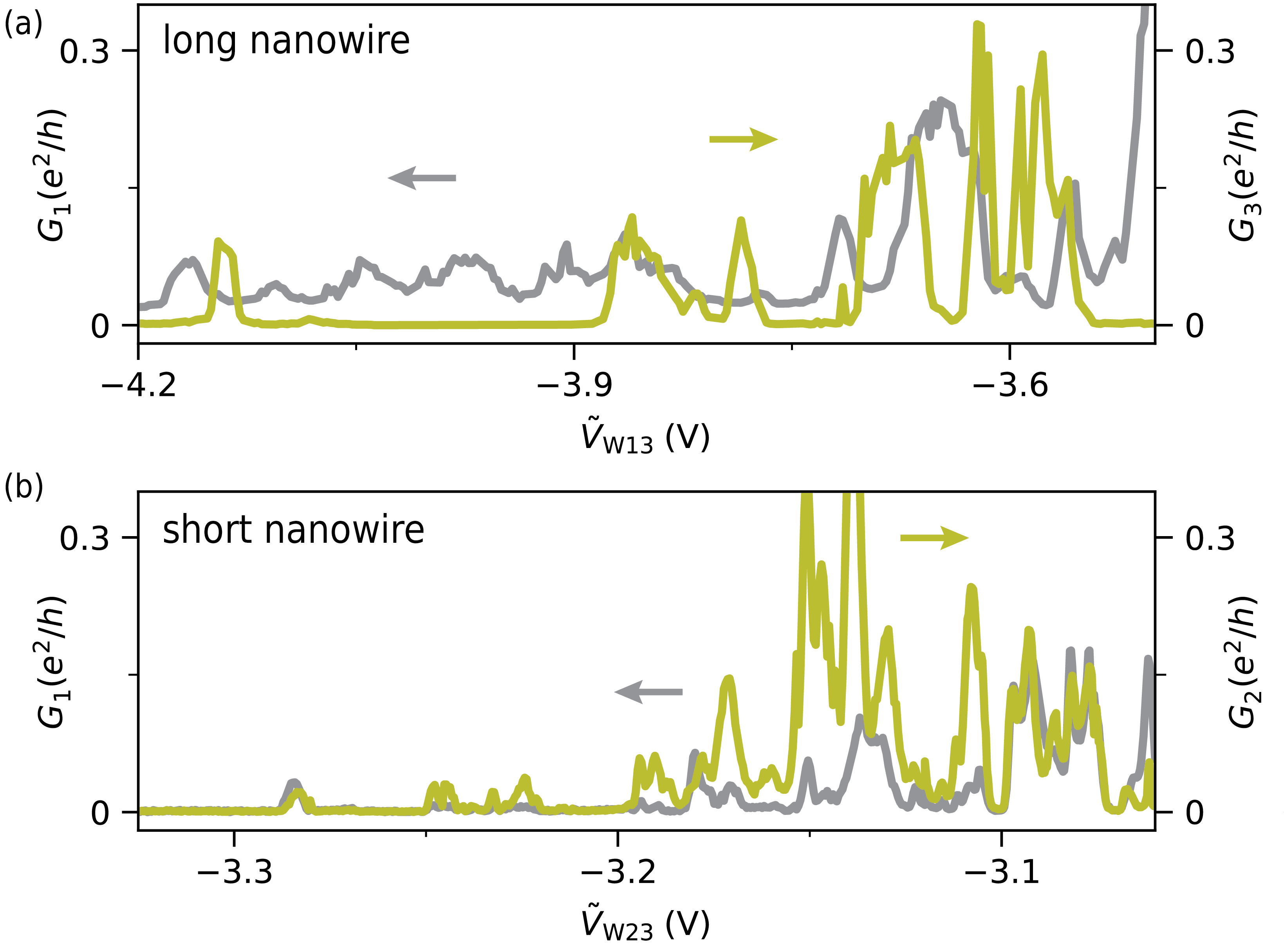}
\end{center}
\caption[Comparison: Unocrrelated subgap states from localized states and correlated subgap states from extended states]{\label{fig:sideprobe_pcorrelation_dev1}Differential conductance traces for a NW with localized states at its ends and a NW with extended states. (a) Tunneling conductance at the ends of a long (\SI{1.5}{\micro\meter}) NW segment of device 4. Most of the peaks measured in $G_1$ have no counterpart in the measured trace of $G_3$. (b) Tunneling conductance at the ends of a short (\SI{0.6}{\micro\meter}) NW segment of device 1 at zero source-drain bias from Fig.~\ref{fig:evolution}(c) for comparison. The majority of conductance peaks measured in $G_{1}$ has a counterpart in $G_{2}$ at the same gate voltage, but with differing conductance value. }
\end{figure}

We simultaneously mapped out subgap resonances at the locations of probe 1 and 3 at $V_\mathrm{SD}^{(j)}=\SI{0}{\volt}$ and magnetic field parallel to the NW $B_{||}=\SI{2.5}{\tesla}$ as a function of gate voltages $V_\mathrm{W13}$, which tunes the density in the wire, and $V_\mathrm{P1}, \: V_\mathrm{P3}$, which tune the tunnel barriers. A large number of resonances can be observed in the resulting measurements of local conductances $G_1$ and $G_3$ in Fig.~\ref{fig:sideprobe_gatemap_dev1}. Both show numerous resonances that are strongly coupled to $V_\mathrm{W13}$ and therefore point towards states that are located in the NW \cite{frolov_ubiquitousexp, frolov_ubiquitoustheory}. In order to evaluate if there are extended states, we plot the traces along the dashed lines in Fig.~\ref{fig:sideprobe_gatemap_dev1} together in Fig.~\ref{fig:sideprobe_pcorrelation_dev1}(a). The observed resonances in $G_1$ and $G_3$ appear largely uncorrelated. This is in stark contrast to the data that arises due to extended ABS in short NWs. Corresponding line cuts of tunneling conductance taken at $V_\mathrm{SD}^\mathrm{(j)}=\SI{0}{\volt}$ at the ends of a \SI{0.6}{\micro\meter} long NW are shown for comparison in Fig.~\ref{fig:sideprobe_pcorrelation_dev1}(b), where almost every resonance has a counterpart that peaks at the same $\tilde{V}_\mathrm{W23}$ value [Note that this is the same data shown in Fig.~\ref{fig:evolution}(c)].\\

\begin{figure}[h!]
\begin{center}
\includegraphics[scale=0.9]{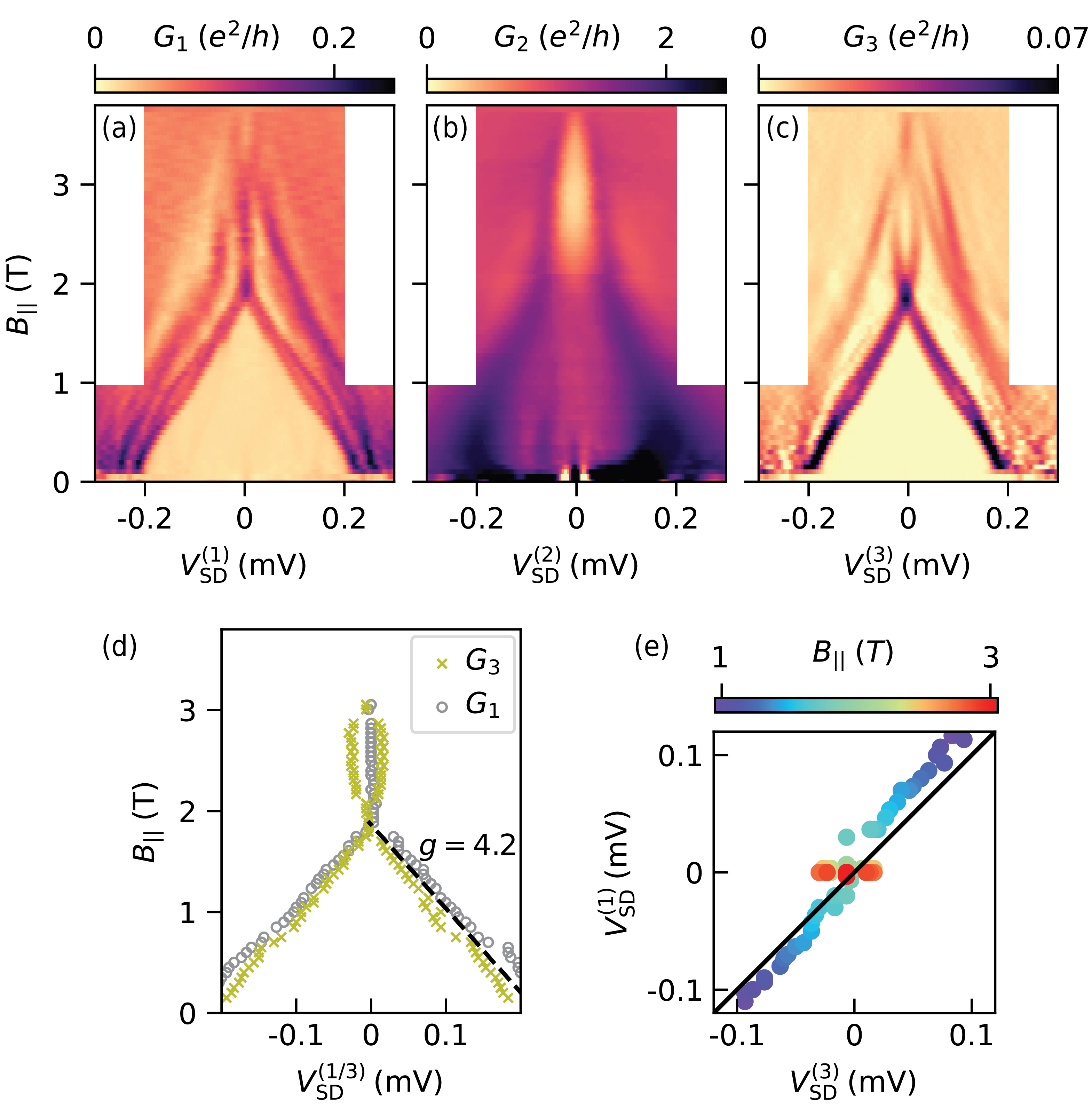}
\end{center}
\caption[Localized bound states at the nanowire ends show no correlated evolution with magnetic field]{\label{fig:sideprobe_fieldscan1_dev1}Bias spectroscopy at the ends and in the middle of the NW as a function of magnetic field parallel to the NW. The gate voltages were set to the values given by the blue $\bigstar$ in Fig.\ref{fig:sideprobe_gatemap_dev1}. (a) Conductance at the left end of the NW segment shows a persisting zero-bias peak at high magnetic fields. (b) spectroscopy in the open transport regime shows no discrete states, but a finite gap at high magnetic fields is visible. (c) Spectroscopy at the right NW end shows states crossing and merging at zero bias with increasing field. (d) Peak locations of the low energy subgap states extracted from (a) and (c). Linear fit yielding $g$ factor of $4.2$ shown as dashed line. (e) Parametric plot of the peak locations from (d). Identity line shown in solid black.
}
\end{figure}

The subgap resonances observed here lack end-to-end correlation in the parameter space spanned by gate voltages. The lack of correlation is also apparent when investigating particular subgap states as function of magnetic field $B_{||}$, an example of which is depicted in Fig.~\ref{fig:sideprobe_fieldscan1_dev1}. The subgap resonances at both ends peal off from the continuum of states at low magnetic fields and stick to zero bias at one end and cross zero bias followed by an overshoot at the other end. We extracted the peak positions from the localized states at the NW ends using the method described in Section \ref{sec:peakextraction} and compared them in a parametric plot [see Fig.~\ref{fig:sideprobe_fieldscan1_dev1}(e)]. The difference in energy at low magnetic fields and the overshoot versus sticking behavior at the two ends in $G_1$ and $G_3$ are in disagreement with a single extended bound state. The observed subgap states are consistent with two localized ABS at each NW end. Note that tunneling spectroscopy from the probe in the middle of the segment is plotted in Fig.~\ref{fig:sideprobe_fieldscan1_dev1}(b). While the tunneling barrier was tuned to a rather open regime, a dip in conductance at high magnetic fields is observed. This feature may easily be mistaken for a closing and reopening of the bulk gap of the NW \cite{das_sarma_qpgaps}.

\begin{figure}[h!]
\begin{center}
\includegraphics[scale=0.9]{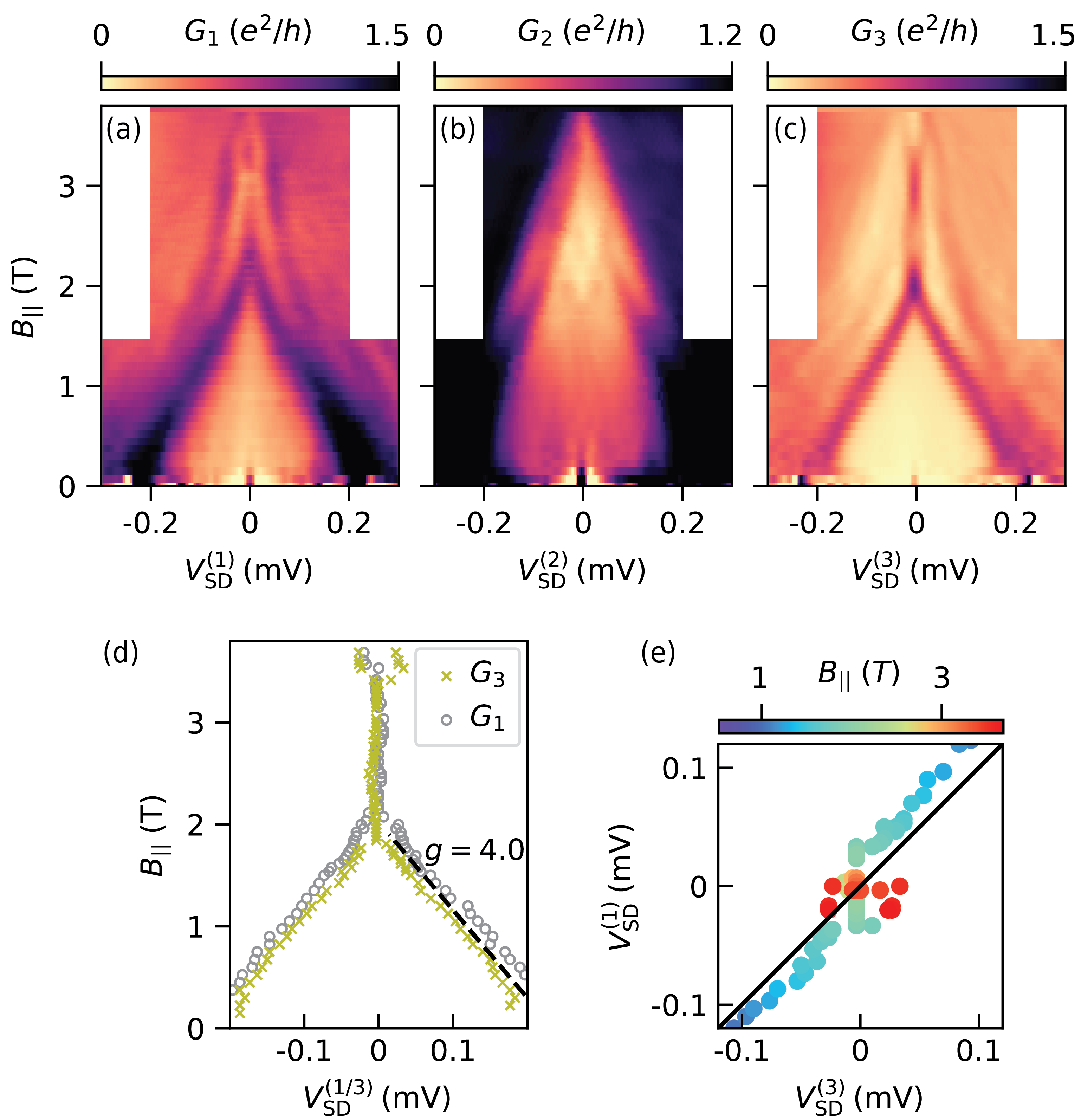}
\end{center}
\caption[Localized states at the nanowire ends appear seemingly correlated]{\label{fig:sideprobe_fieldscan2_dev1}Bias spectroscopy at the ends and in the middle of the NW as a function of magnetic field parallel to the NW. The gates were set to the values marked by the green \ding{110} in Fig.\ref{fig:sideprobe_gatemap_dev1}. (a, c) Conductance spectroscopy at both NW ends show states that merge at high magnetic field. They are pinned around zero bias for an extended region. (b) The probe in the middle of the NW shows subgap states overshooting around $B_{||}=\SI{2}{\tesla}$. (d) Extracted peak locations of the low energy subgap state extracted from (a) and (c). Linear fit yielding $g$ factor $g=4.0$ shown as dashed line. (e) Parametric plot of the peak locations from (d). Identity line shown in solid black.}
\end{figure}

The evolution of subgap states - in particular the presence or absence of an overshoot at high bias - is strongly dependent on the gate voltages. To illustrate this, we show the evolution of subgap states at the three probe locations as a function of magnetic field $B_{||}$ at a slightly different gate voltage configuration in Fig.~\ref{fig:sideprobe_fieldscan2_dev1}. In this example, zero-bias states form above $B_{||}=\SI{2}{\tesla}$. A signature that is expected for Majorana bound states \cite{SDS_substrateRenormalization,frolov_ubiquitoustheory,elsa_pinning}. Oscillations around zero bias may be present at higher magnetic fields, but are only resolved in the form of a peak broadening. A direct comparison of the subgap state energies from $G_1$ and $G_3$ in Fig.~\ref{fig:sideprobe_fieldscan2_dev1}(d) shows that the states merge at zero bias at different magnetic fields and have different energies at low magnetic fields. The oscillations about zero bias are furthermore uncorrelated between $G_1$ and $G_3$, which rules out MZMs as the origin of the zero-bias peaks. The middle probe conductance $G_2$ in Fig.~\ref{fig:sideprobe_fieldscan2_dev1}(b) shows faint subgap states with a zero crossing around $B_{||}=\SI{2}{\tesla}$ followed by an overshoot, which may be mistaken for spectroscopy of the bulk-gap reopening \cite{das_sarma_qpgaps}. Tunneling spectroscopy as a function of gate voltage $V_\mathrm{W13}$ (not shown) around the same point in gate voltage directly shows that there is no end-to-end correlations in the evolution of subgap states in $G_1$ and $G_3$ as observed for shorter NW in Chapter \ref{ch:pradaclarke}. \\

In Section \ref{sec:semi_sideprobe} we show an alternative device geometry that does not use superconducting sideprobes but the bare semiconductor coupled to the gate-defined NW through a quantum point contact. In total 15 devices were investigated and confirmed the absence of end-to-end correlated subgap resonances in tunneling spectroscopy measured at the ends of $\gtrsim\SI{1.5}{\micro\meter}$ long NW segments. Large parts of parameter space were explored by a guided search with the help of maps as the one in Fig.~\ref{fig:sideprobe_gatemap_dev1}, magnetic field scans as shown in Fig.~\ref{fig:sideprobe_fieldscan1_dev1}, and tunneling spectroscopy as a function of the gate voltage that tunes the electron density in the NW.

\section{End-to-end correlated subgap states due to fine-tuning}
\label{sec:fine_tuning}
Uncorrelated ABSs can appear at the ends of NWs of sufficient length. These ABSs typically lack end-to-end correlations over a finite region in parameter space. Fine-tuning of gate voltages can be used to make these uncorrelated ABSs appear correlated at both NW ends for certain cuts in parameter space spanned by gate voltages and magnetic field. Here we show a particular example, how localized ABSs can be fine-tuned to imitate the presence of an extended bound state.\\

\begin{figure}[h!]
\begin{center}
\includegraphics[width=\textwidth]{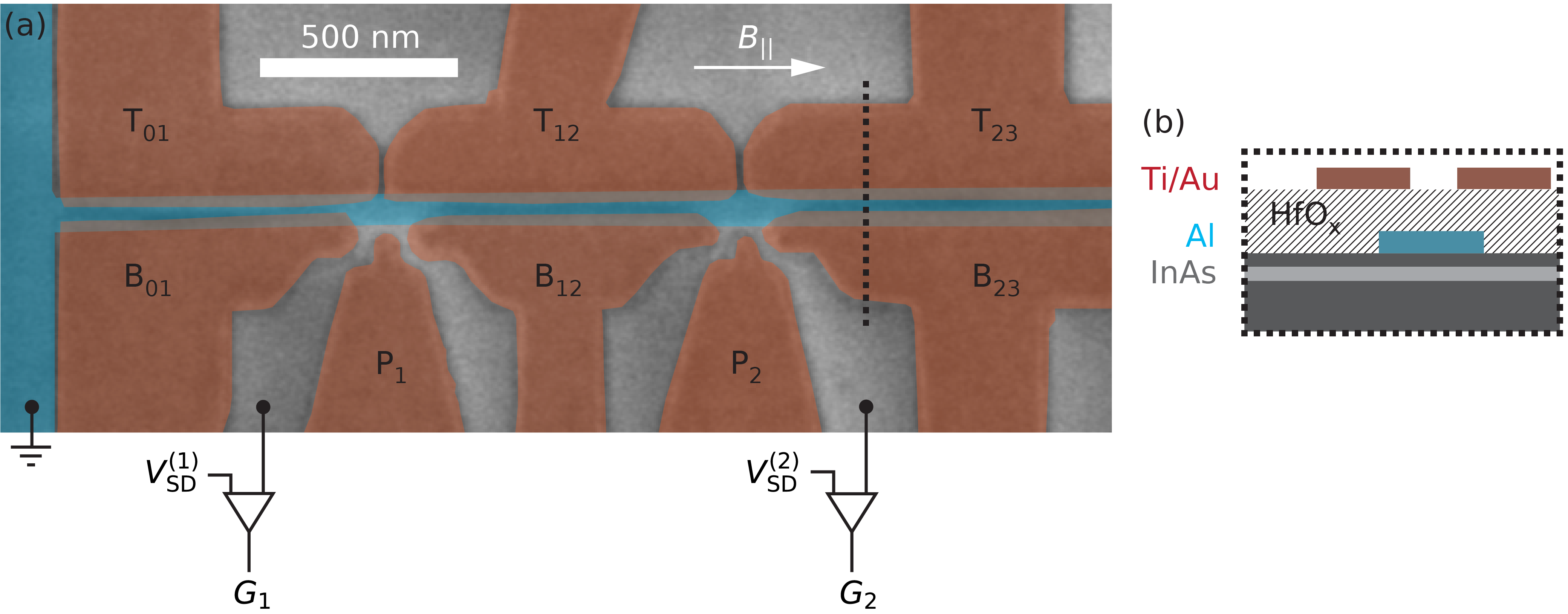}
\end{center}
\caption[Micrograph of short nanowire segment with semiconducting sideprobes]{\label{fig:semi_sideprobe_SEM}(a) Micrograph of a sideprobe device with \SI{0.8}{\micro\meter} long NW segment. The NW is formed in the InAs 2DEG by self-aligned depletion using the gates $\mathrm{T_{kl}}$, $\mathrm{B_{kl}}$ ($kl\in\{01, \: 12, \: 23 \}$) on top of the Al strip (blue). The two quantum point contacts at the side of the wire an be tuned by the gates $\mathrm{P_{j}}$, ($j\in \{1,\,2\}$). (b) Schematic of the heterostructure along the dotted line in (a).}
\end{figure}

\begin{figure}[h!]
\begin{center}
\includegraphics[scale=0.9]{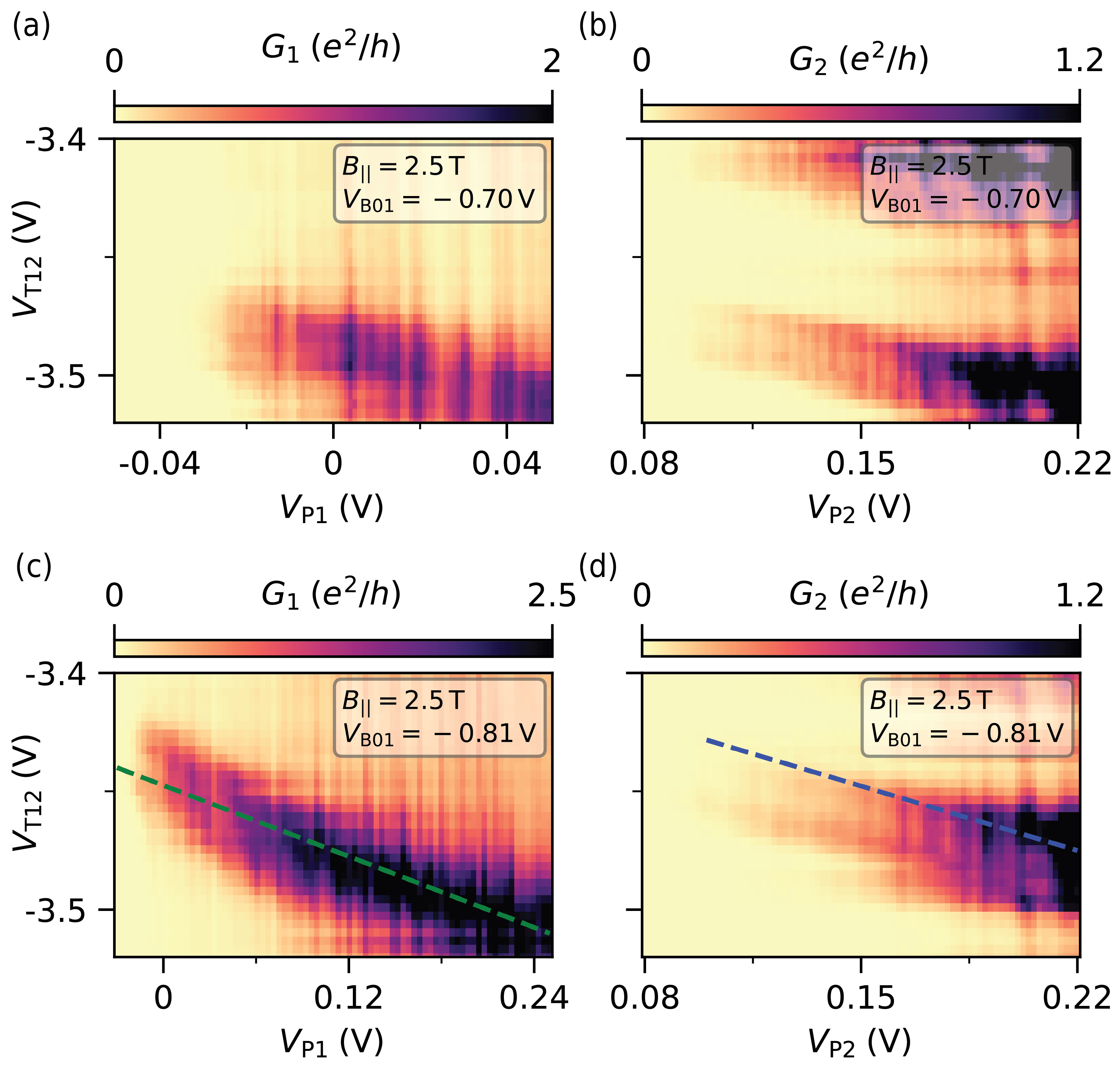}
\end{center}
\caption[Localized states appear at different gate voltages after fine-tuning of nearby gates]{\label{fig:fine_tuned_gatemap}Zero-bias conductance as a function of the gate voltage that tunes the electron density in the NW and the gate voltage that controls the tunnel barrier. (a, b) A resonance that depends weakly on $V_\mathrm{P1}$ and $V_\mathrm{P2}$ is visible in tunneling conductance at both ends of the NW.  The states show a similar slope and shape. (c, d) The same measurement repeated for a different value of the voltage on gate $\mathrm{B01}$ which is adjacent to the quantum point contact on the left. The state in the tunneling conductance at the left NW end in (c) is located at more positive gate voltage $V_\mathrm{T12}$ now in comparison to (a), while the state measured on the right end in (d) is only shifted slightly in comparison to (b). Tunneling spectroscopy along the dashed lines is shown in supplementary figure \ref{fig:fine_tuned_hscan}. }
\end{figure}

A device (device 3) of the geometry shown in Fig.~\ref{fig:semi_sideprobe_SEM} that allows for tunneling spectroscopy at the ends of a \SI{0.8}{\micro\meter} long NW segment was used.  Details on the fabrication of device 3 is given in Appendix \ref{sec:fab_semi_sideprobe}. The device did not show any extended states presumably due to disorder. Conductance resonances from two distinct ABSs at the two NW ends as a function of the gate voltages which tune the tunnel barriers and the overall electron density in the NW segment are depicted in Fig.~\ref{fig:fine_tuned_gatemap}(a, b). The two resonances in $G_1(V_\mathrm{T12}, V_\mathrm{P1})$ and $G_2(V_\mathrm{T12}, V_\mathrm{P2})$ belong to two distinct states despite their similar shape and slope. The simultaneous measurements of conductance at the other NW end $G_2(V_\mathrm{T12}, V_\mathrm{P1})$ and $G_1(V_\mathrm{T12}, V_\mathrm{P2})$ are furthermore shown in the supplementary Fig.~\ref{fig:fine_tuned_gatemap_2}. These maps show that the resonances measured simultaneously at the other NW end do not depend on $V_\mathrm{P1}$ and $V_\mathrm{P2}$ respectively. The gate voltage $V_\mathrm{B01}$ has a strong influence on the localized state at the left NW end, while it has only a small influence on the state localized at the right end. Changing the gate voltage $V_\mathrm{B01}$ from  $\SI{-0.70}{\volt}$ to $\SI{-0.81}{\volt}$ [see Fig.~\ref{fig:fine_tuned_gatemap}(c, d)] leaves the resonances observed in $G_2$ almost unchanged, the resonance in $G_1$ is displaced significantly.

\begin{figure}[h!]
\begin{center}
\includegraphics[scale=0.95]{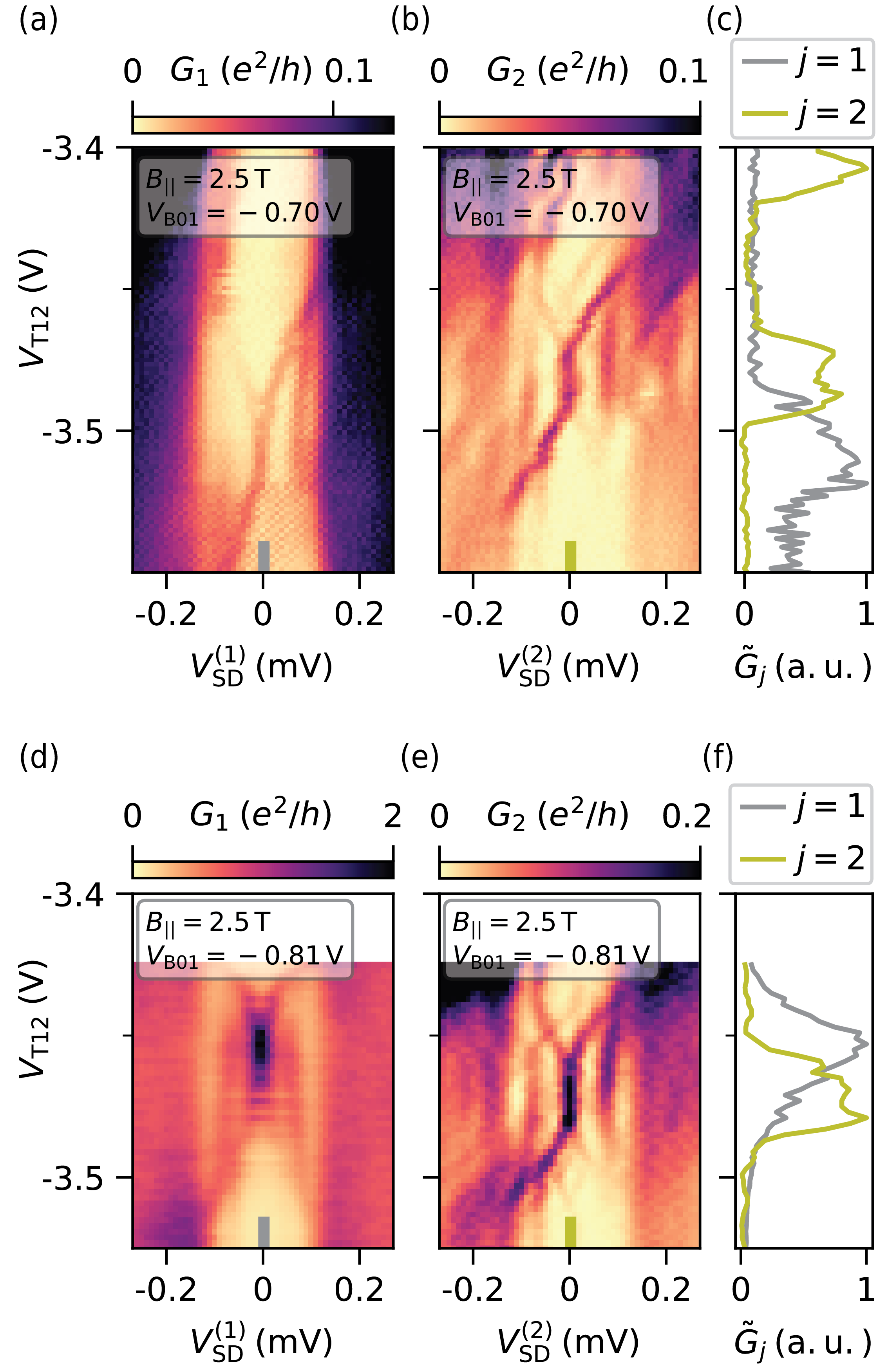}
\end{center}
\caption[Localized states appear correlated due to fine-tuning]{\label{fig:fine_tuned_plunger} Tunneling spectroscopy as a function of gate voltage $V_\mathrm{T12}$ which tunes the electron density in the NW segment. (a) Tunneling conductance on the left NW end shows ABSs emerging from the continuum and forming a zero-bias conductance peak. (b) Tunneling conductance on the rigth NW end shows very similar behavior. The states do not appear at identical values of $V_\mathrm{T12}$ which is also seen in the zero-bias cuts shown in (c).  (d) For a more negative gate voltage $V_\mathrm{B01} =\SI{-0.81}{\volt}$, the subgap states measured on the left NW end appears at more positive values of $V_\mathrm{T12}$. (e) The evolution of the state measured on the right is unaffected by the change in $V_\mathrm{B01}$. There is now a finite range of $V_\mathrm{T12}$ where the states appear correlated as seen in the zero-bias line cuts in (f). }
\end{figure}

The gate voltage $V_\mathrm{B01}$ can make distinct ABSs at the two NW ends appear correlated over a small range of $V_\mathrm{T12}$. This is best seen in tunneling spectroscopy at two distinct values of gate voltage $V_\mathrm{B01}$. For the case $V_\mathrm{B01}=\SI{-0.70}{\volt}$ shown in Fig.~\ref{fig:fine_tuned_plunger}(a, b), the ABSs appear as subgap states in both $G_1$ and $G_2$. While the states show a similar behavior as a function of $V_\mathrm{T12}$, they do not coincide as the line cut in Fig.~\ref{fig:fine_tuned_plunger}(c) illustrates. For the case $V_\mathrm{B01}=\SI{-0.81}{\volt}$ shown in Fig.~\ref{fig:fine_tuned_plunger}(d, e) the ABSs are moved with respect to each other and the regions where they are close to zero bias overlap [see Fig.~\ref{fig:fine_tuned_plunger}(f)]. This illustrates how end-to-end correlations as a function of one parameter can be produced by fine-tuning a second parameter. \\

\begin{figure}[h!]
\begin{center}
\includegraphics[scale=0.95]{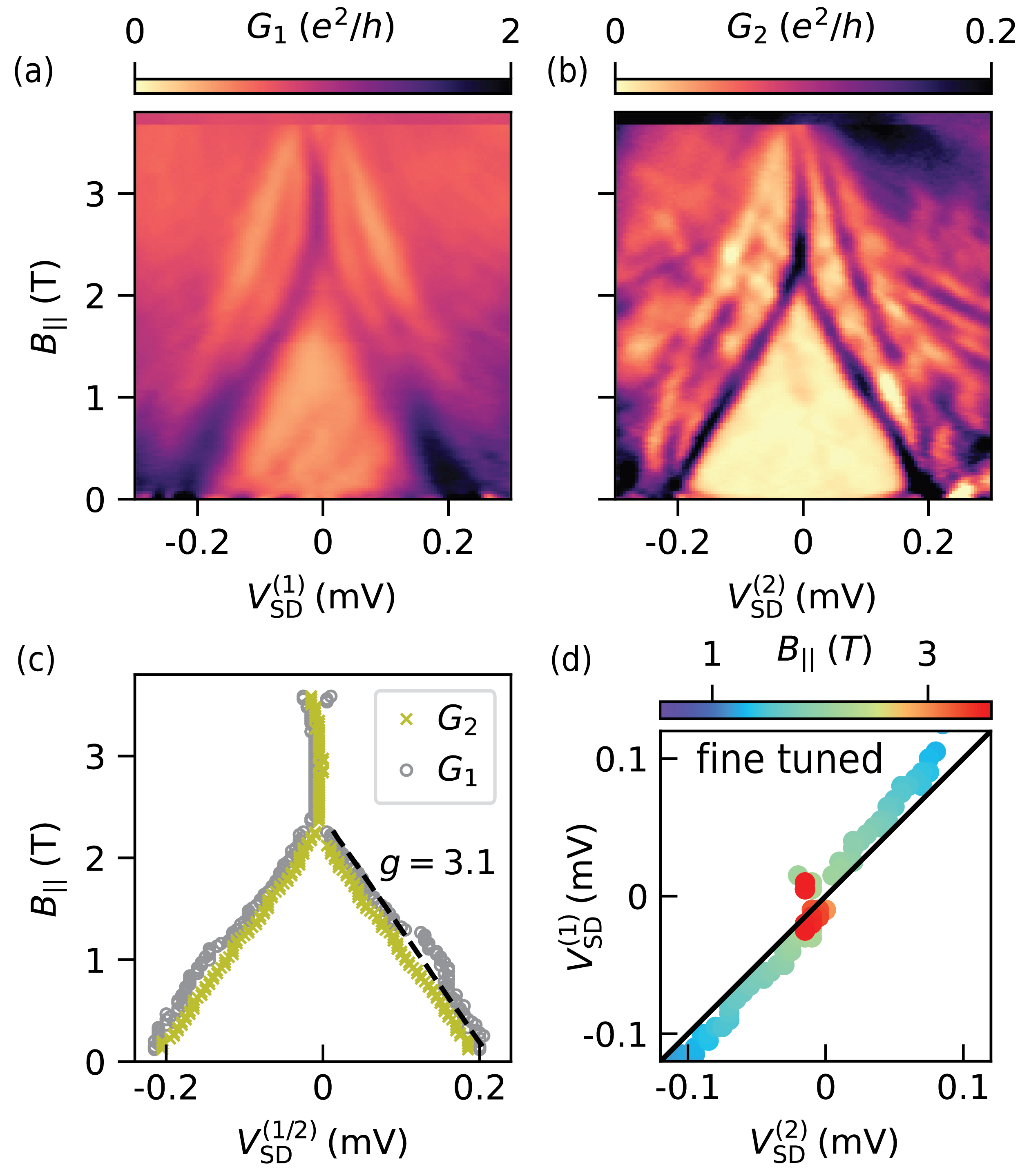}
\end{center}
\caption[Generic field dependence of localized Andreev bound states leads to correlation at sweet spot in gate voltage]{\label{fig:fine_tuned_fieldscan}Bias spectroscopy of two distinct ABSs at fine-tuned gate voltage value as a function of magnetic field parallel to the NW. (a) Tunneling conductance on the left shows states merging and forming a zero-bias peak at high magnetic field. (b) Tunneling conductance on the right NW end shows subgap states with identical evolution to the one in (a). (c) The extracted peak positions from (a) and (b) lie on top of each other. The dashed line shows a linear fit with $g$ factor 3.1. (d) Parametric plot of extracted peaks from (c) with the identity line shown in solid black.}
\end{figure}

The same subgap states as a function of magnetic field $B_{||}$ parallel to the NW is shown in Fig.~\ref{fig:fine_tuned_fieldscan}(a, b). The two ABSs show almost identical evolution with magnetic field, as seen from the extracted peak positions in Fig.~\ref{fig:fine_tuned_fieldscan}(c). When plotted parametrically, the peak positions fall very close to the line of perfect correlation. 

\begin{figure}[h!]
\begin{center}
\includegraphics[scale=0.9]{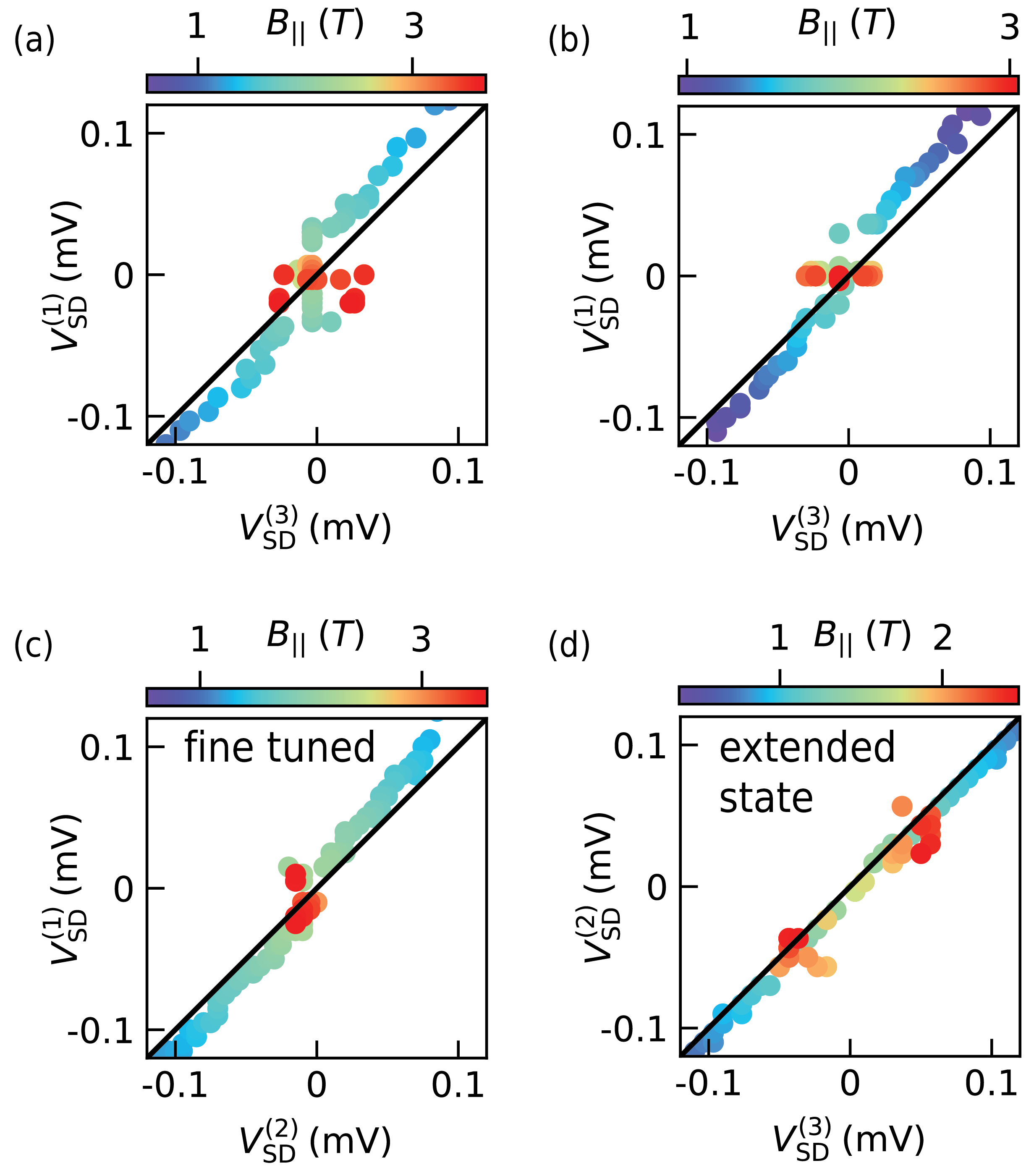}
\end{center}
\caption[Comparison: correlation of magnetic field dependence for localized states, localized states after fine-tuning, and extended bound states]{\label{fig:sideprobe_fieldcorrelation_dev1}Correlation analysis of localized low energy states at the two NW ends in comparison to state extending throughout the NW. (a, b) Parametric plot of peak positions of states localized at the ends of a long (\SI{1.5}{\micro\meter}) NW. The peak positions are from Fig.~\ref{fig:sideprobe_fieldscan1_dev1} and Fig.~\ref{fig:sideprobe_fieldscan2_dev1} respectively. (c) Parametric plot of fine-tuned, localized states from Fig.~\ref{fig:fine_tuned_fieldscan}. (d) Parametric plot from Fig.~\ref{fig:fieldscan_pc} of the peak positions for extended states in a short (\SI{0.6}{\micro\meter}) NW. The points lie close to the identity line (solid black) which indicates perfect correlations. The difference between correlations due to fine-tuned localized states and due to extended bound states may be small and subtle to recognize. }
\end{figure}

In Fig.~\ref{fig:sideprobe_fieldcorrelation_dev1}, we compare the parametric plots of peak positions for three cases: the uncorrelated ABSs at the ends of long NWs from Section \ref{sec:uncorrelated}, the fine-tuned states from Fig.~\ref{fig:fine_tuned_fieldscan}, and the extended ABSs from Chapter \ref{ch:pradaclarke}. Note that the data in Fig.~\ref{fig:sideprobe_fieldcorrelation_dev1}(a-c) all show a deviation of the data points from the identity line at low magnetic field values, whereas the data of extended bound states in Fig.~\ref{fig:sideprobe_fieldcorrelation_dev1}(d) closely follows the identity line for all field values. Different evolutions of distinct ABSs localized at the two ends at high magnetic fields lead to the points that are scattered around the origin in  Fig.~\ref{fig:sideprobe_fieldcorrelation_dev1}(a, b). However, the spread of these points can be reduced in some cases by an appropriate choice of gate voltages leading to the fine-tuned case shown in Fig.~\ref{fig:sideprobe_fieldcorrelation_dev1}(c). This case is hard to distinguish from the data in Fig.~\ref{fig:sideprobe_fieldcorrelation_dev1}(d), which shows data from extended bound states. Note that finite level broadening, spectroscopic resolution, and visibility of conductance peaks lead to points that deviate from the ideal correlations shown by the black identity line even in the case of the extended bound states. The data in Fig.~\ref{fig:sideprobe_fieldcorrelation_dev1}(c, d) may suggest that one can barely distinguish extended states from fine-tuned localized bound states at the NW ends. The difference between the two cases lies in the stability of these parametric plots. The fine-tuned case results in a plot that suggest correlated states only in a limited region, sometimes just a single point of parameter space spanned by the gate voltages. In contrast, extended bound states produce almost perfect correlated states at both NW ends over almost the full parameter space in which the ABSs can be observed in tunneling spectroscopy.\\

\begin{figure}[h!]
\begin{center}
\includegraphics[scale=0.95]{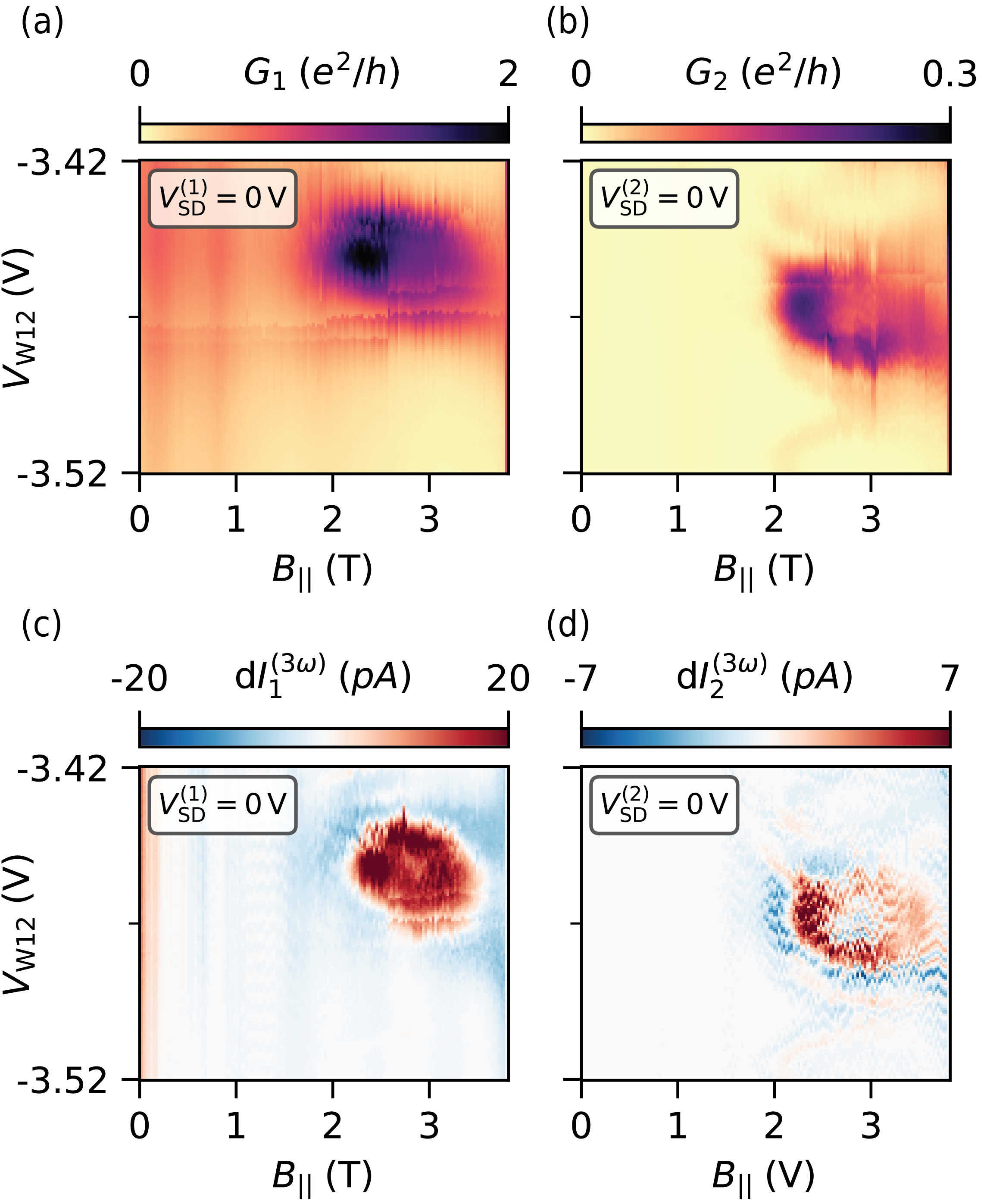}
\end{center}
\caption[Ellipsoidal area with zero-bias state in the magnetic field - gate voltage plane mimicking a phase diagram]{\label{fig:fine_tuned_phasediagram}Zero-bias conductance as a function of the gate $\mathrm{T_{12}}$ that tunes the electron density in the NW and magnetic field parallel to the NW. (a,b) An ellipse shape of conductance can be seen in conductance measured on both the left and the right NW end as a result of the evolution with $V_\mathrm{T12}$ in Fig. \ref{fig:fine_tuned_pscans}. The ellipses appear roughly at the gate voltage due to fine-tuning of $V_{\mathrm{B01}}$ and $V_\mathrm{H1}$. (c,d) the third harmonic of the tunneling current as a measure for peaks around zero bias confirm the ellipse shaped region hosts zero-bias conductance peaks. }
\end{figure}

The pitfall of mistaking fine-tuned localized ABSs at the two NW ends for an extended bound state is particularly relevant in the presence of ABSs that show a rather generic dependence on gate voltage and magnetic field, such as the states shown in Fig.~\ref{fig:fine_tuned_plunger} and Fig.~\ref{fig:fine_tuned_fieldscan}. Subgap states that show a parabolic lobe shape as a function of gate voltage and merge at high magnetic fields to form a zero bias peak can be found frequently. If there are bound states with a more complicated spectrum as a function of gate voltage, e.g., the states investigated in Chapter \ref{ch:pradaclarke}, one is less prone to mistake correlations due to fine-tuning for correlations due to extended bound states.

The rather simple dependence of the localized ABSs on gate voltage has a further drawback. When the conductance resonances at zero bias due to these ABSs are investigated as a function of magnetic field and gate voltage $V_\mathrm{T12}$ they appear in a typical ellipse shaped area, such as shown in Fig.~\ref{fig:fine_tuned_phasediagram}. The differential conductance measurements $G_1$ and $G_2$ measured on both NW ends show ellipse-shaped regions of approximately same size which can be made to appear correlated using the voltage $V_\mathrm{B01}$ on the gate adjacent to the left NW end. In Fig.~\ref{fig:fine_tuned_phasediagram}(c, d) we furthermore show the third harmonic of the tunneling current measured with the lock-in amplifier. It shows positive signal where a peak at zero bias is present \cite{AlexThesis}. This characteristic appearance of zero-bias peaks at high magnetic fields may easily be misinterpreted as the presence of MZMs which should appear with a parabolic phase boundary according to the Lutchyn-Oreg model \cite{frolov_phasediagram}. Bias spectroscopy at selected magnetic field values is shown in the supplementary Fig.~\ref{fig:fine_tuned_pscans} to illustrate how the ellipse-shaped regions with a zero-bias peak result from the subgap states. In the supplementary Fig.~\ref{fig:fine_tuned_hscan}, we furthermore show tunneling spectroscopy of the ABSs at $B_{||}=\SI{2.5}{\tesla}$ as the tunnel barriers are opened by moving the gates along the green (blue) dashed lines in Fig.~\ref{fig:fine_tuned_gatemap}. The state in measured in $G_1$ ($G_2$) sticks to zero bias while the state measured at the other end in $G_2$ ($G_1$) moves away from zero bias. This is a further indication, that we are not dealing with a single extended ABS, but two localized ABSs at each NW end. The ABS at the left NW end shows a value of conductance close to the conductance quantum $G_1\approx \SI{2}{e^2/h}$, which may be mistaken for the conductance quantization of MZMs \cite{frolov_quantized, SDS_against_hao}. 

\section{Conclusion}
Using devices that allow for tunneling spectroscopy at discrete points along a gate-defined NW in proximitized 2DEG, we showed subgap states due to localized ABSs at the two NW ends. No end-to-end correlations between subgap sepctra are observed over extended regions in gate space for devices longer than $\SI{0.8}{\micro\meter}$. Fine-tuning of gate voltages can be used to manipulate ABSs localized at each NW end, such that they appear with almost indentical evolution in selected cuts of parameter space imitating an ABS that extends through the NW. These ABSs may have signatures similar to the ones of MZMs at the end of a topological non-trivial superconducting phase. End-to-end correlations are therefore only a good indicator for extended bound states in the NW if they persist over extended regions of gate space, preferably over large parts of parameter space in which subgap states can be observed in tunneling spectroscopy. Additional tools, like nonlocal conductance spectroscopy (Chapter \ref{ch:2DEG_nonlocal}) and nonlocal signatures due to hybridization of a localized quantum dot resonance with the bound state (Chapter \ref{ch:pradaclarke}) can be used to distinguish extended states that couple to the probes at both NW ends from states that are localized at one NW end. 

The absence of bound states that extend in NWs longer than $\SI{0.8}{\micro\meter}$ suggests that the semiconductor-superconductor hybrid material of the current state-of-the-art is too disordered to show a clean topological phase with localized MZMs at its ends \cite{SDS_mobility, SDS_impurity}. A different scenario, that is in agreement with the presented data, is that the nonlocal fermion formed by two MZMs cannot be detected by tunneling spectroscopy as easily as trivial, localized ABSs. The localized bound states at the NW ends may be due to disorder, quantum dots forming in the NW, or quasi-Majoranas. Further improvements of the quality in terms of material growth and device fabrication have to be undertaken to realize a topological phase of matter in this material system. The devices and techniques developed within this thesis form a foundation for material characterization in future works.

\section{Supplementary information}


\subsection{\label{app:wafer}Wafer information and fabrication}
The material used for device 3 and device 4 was an InGaAs/InAs/InAlAs heterostructure covered with an \emph{in-situ} epitaxially grown Al top layer. Electron mobility in the InAs quantum well after removal of the Al by wet etching was measured using a Hall bar to be $\SI{25000}{\centi\meter\squared\per\volt\per\second}$. Fabrication details for device 3 can be found in Appendix \ref{sec:fab_semi_sideprobe} and for device 4 in Appendix \ref{sec:fab_super_sideprobe}.

\subsection{Sideprobe devices with semiconducting leads}

\begin{figure}[h!]
\begin{center}
\includegraphics[width=\textwidth]{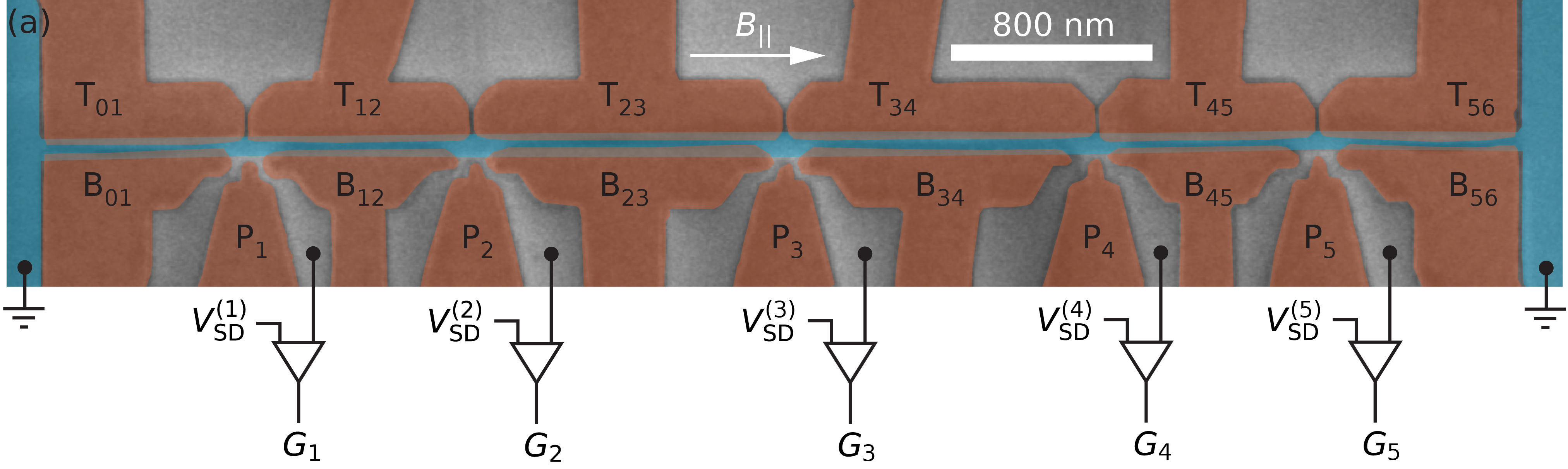}
\end{center}
\caption[Micrograph of device with long nanowire segments and semiconducting sideprobes]{\label{fig:device_semi_long}False-color micrograph of electrostatically defined NW on InAs 2DEG proximitized by a strip of superconducting Al (blue). The gate electrodes (red) allow for tuning of the electron density in individual segments of the NW. Quantum point contacts at the side of the NW tunnel couple semiconducting leads to the NW.}
\end{figure}

Here we present an alternative to the device geomtery with superconducting tunnel probes adjacent to the gate-defined NW. The geometry is shown in Fig.~\ref{fig:device_semi_long}. It consists of a strip of superconducting Al on top of a InAs 2DEG. Gates labeled $\mathrm{T_{kl}}$, $\mathrm{B_{kl}}$ ($kl\in \{01,\,12,\,23,\,34,\,45,\,56\}$) are used to deplete the electrons in the 2DEG self-aligned with the Al strip to form a quasi-one-dimensional NW. The gates labeled $\mathrm{B_{kl}}$ furthermore form quantum point contacts next to the NW. The tip of the gates labeled $\mathrm{P_{j}}$ ($j\in \{1,\,2,\,3,\,4,\,5\}$) allow for fine-tuning of the tunnel barrier. The extruded body of the $\mathrm{P_{j}}$ gates screens the 2DEG from the electric fields of the nearby $\mathrm{B_{kl}}$ gates.

\label{sec:semi_sideprobe}

\begin{figure}[h!]
\begin{center}
\includegraphics[scale=0.95]{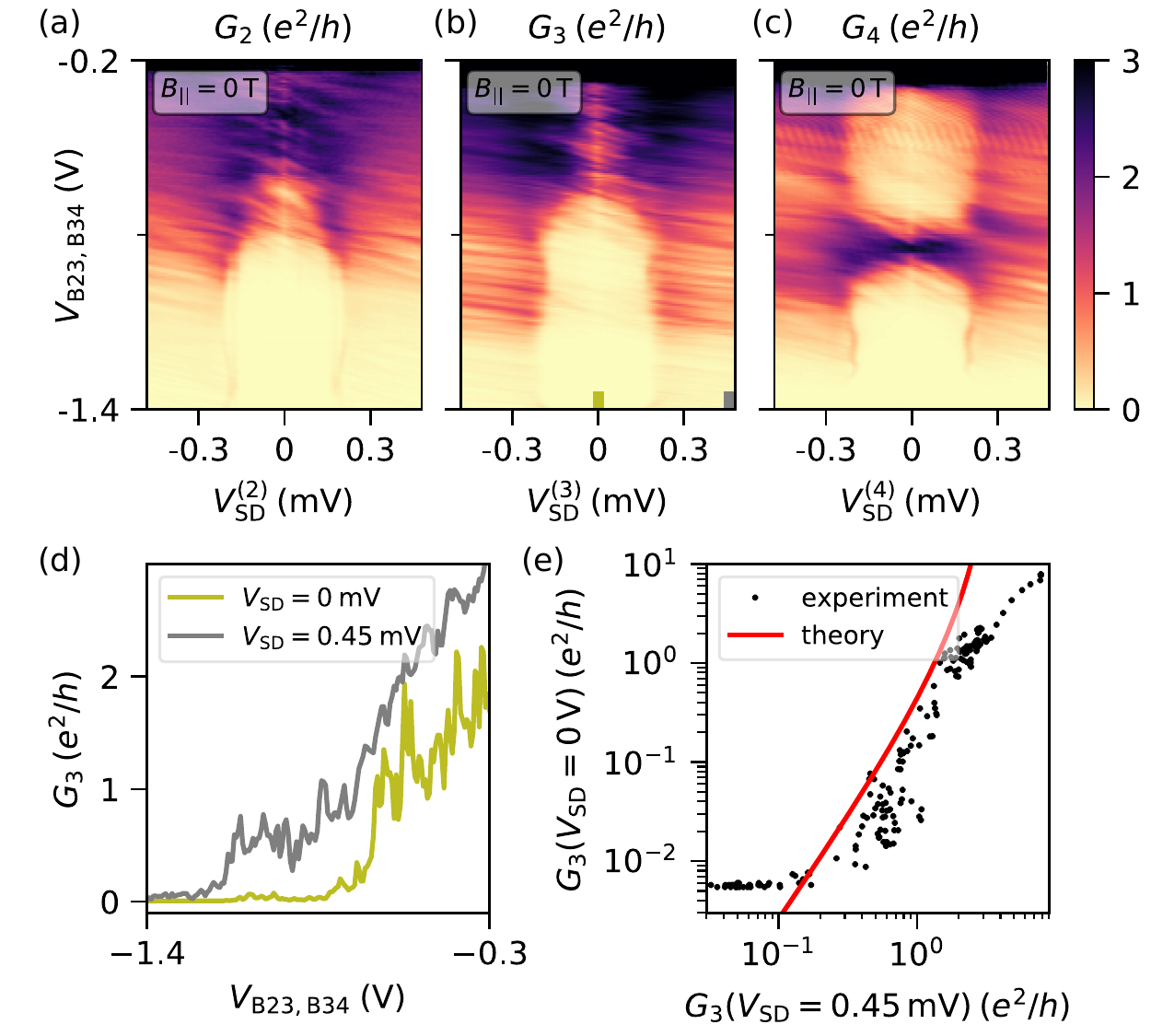}
\end{center}
\caption[Tunable point contact for sideprobe device with semiconducting probes]{\label{fig:SQPCN_sideprobe}Spectroscopy of the superconducting gap of a NW through adjacent quantum point contacts at zero magnetic field measured on device 3 which has the geometry shown in \ref{fig:device_semi_long}. (a-c) Tunneling spectroscopy at the three locations of the quantum point contacts as a function of the voltage $V_\mathrm{B23,B34}$ applied to the gates $\mathrm{B23}$ and $\mathrm{B34}$. All three conductances show a gap in the tunneling regime over an extended range of gate voltage. (d) Comparison of conductance at zero bias and \SI{0.45}{\milli\volt} bias from the measurement in (b). (e) Parametric plot of the data in (d) with the curve for an ideal single-mode point contact as predicted by theory \cite{beenakker_microjunctions}. }
\end{figure}

Device 3 is of this type and the pinch-off behavior for its quantum point contacts is illustrated in Fig.~\ref{fig:SQPCN_sideprobe}(a-c). All three point contacts show a hard superconducting gap. The conductance is not always a monotonic function of gate voltage [see \ref{fig:SQPCN_sideprobe}(c)]. For an ideal quantum point contact at a normal-superconductor interface conductance plateaus are expected. The lowest plateau is expected at the value $4e^2/h$ in contrast to the value $2e^2/h$ for the normal-normal interface (i.e., at high bias) \cite{morten_sqpcn}. The line traces shown in Fig.~\ref{fig:SQPCN_sideprobe}(d) are not able to show these plateaus, because the point contact cannot be opened enough. Figure \ref{fig:SQPCN_sideprobe}(e) compares a parametric plot of the line cuts at zero bias and high bias with the theory curve expected for an ideal single-mode quantum point contact \cite{beenakker_microjunctions}. Deviations at small conductance values arise were the signal is dominated by noise. Other deviations may be a result of multiple modes or a remnant normal scattering probability.

\subsection{Additionl data on fine-tuned subgap states}

\begin{figure}[h!]
\begin{center}
\includegraphics[scale=0.9]{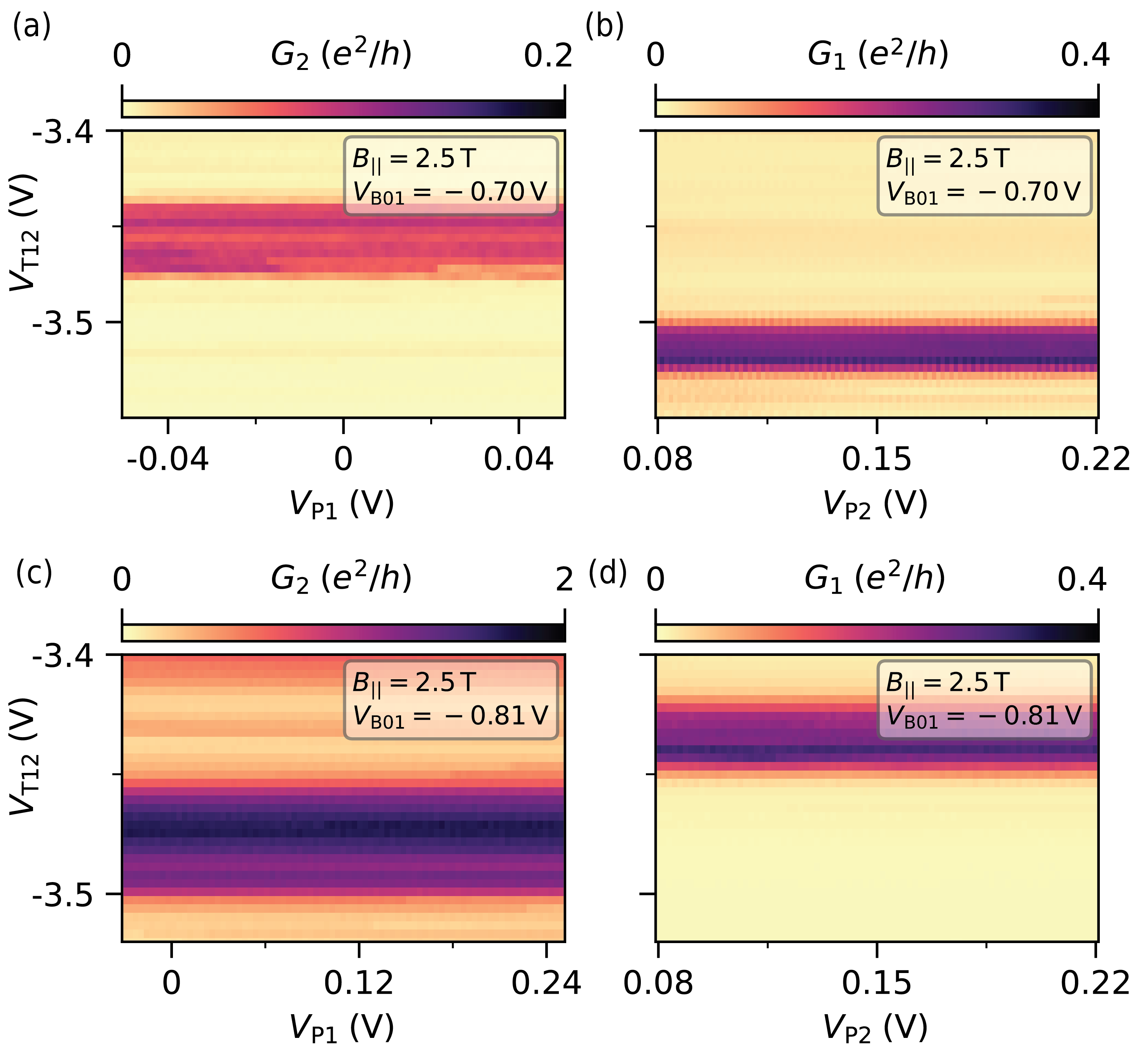}
\end{center}
\caption[Gate response of fine-tuned zero-bias states as function of local gates at the neighboring prob]{\label{fig:fine_tuned_gatemap_2}Conductances $G_1$ ($G_2$) measured simultaneously with the data of $G_2$ ($G_1$) in Fig.~\ref{fig:fine_tuned_gatemap}. Conductance resonances appear as straight lines whereas the conductance resonance on the other end in Fig.~\ref{fig:fine_tuned_gatemap} appear with a finite slope in $G_1$ ($G_2$) as a function of the gate voltage $V_\mathrm{T12}$ and the local gate voltages $V_\mathrm{P1}$ ($V_\mathrm{P2}$). }
\end{figure}

Here we show additional data on the fine-tuned localized ABS at the two ends of the NW device 3. Figure \ref{fig:fine_tuned_gatemap_2} shows the respective conductance measurements on the respective other end taken during the measurements in Fig.~\ref{fig:fine_tuned_gatemap}. While all the states in Fig.~\ref{fig:fine_tuned_gatemap} appeared with a finite slope, none of the states measured simultaneously at the other end in Fig.~\ref{fig:fine_tuned_gatemap_2} show similar behavior. This strongly suggests, that the subgap states measured in $G_1$ and $G_2$ are a result of distinct ABSs at the two NW ends.

\begin{figure}[h!]
\begin{center}
\includegraphics[scale=0.9]{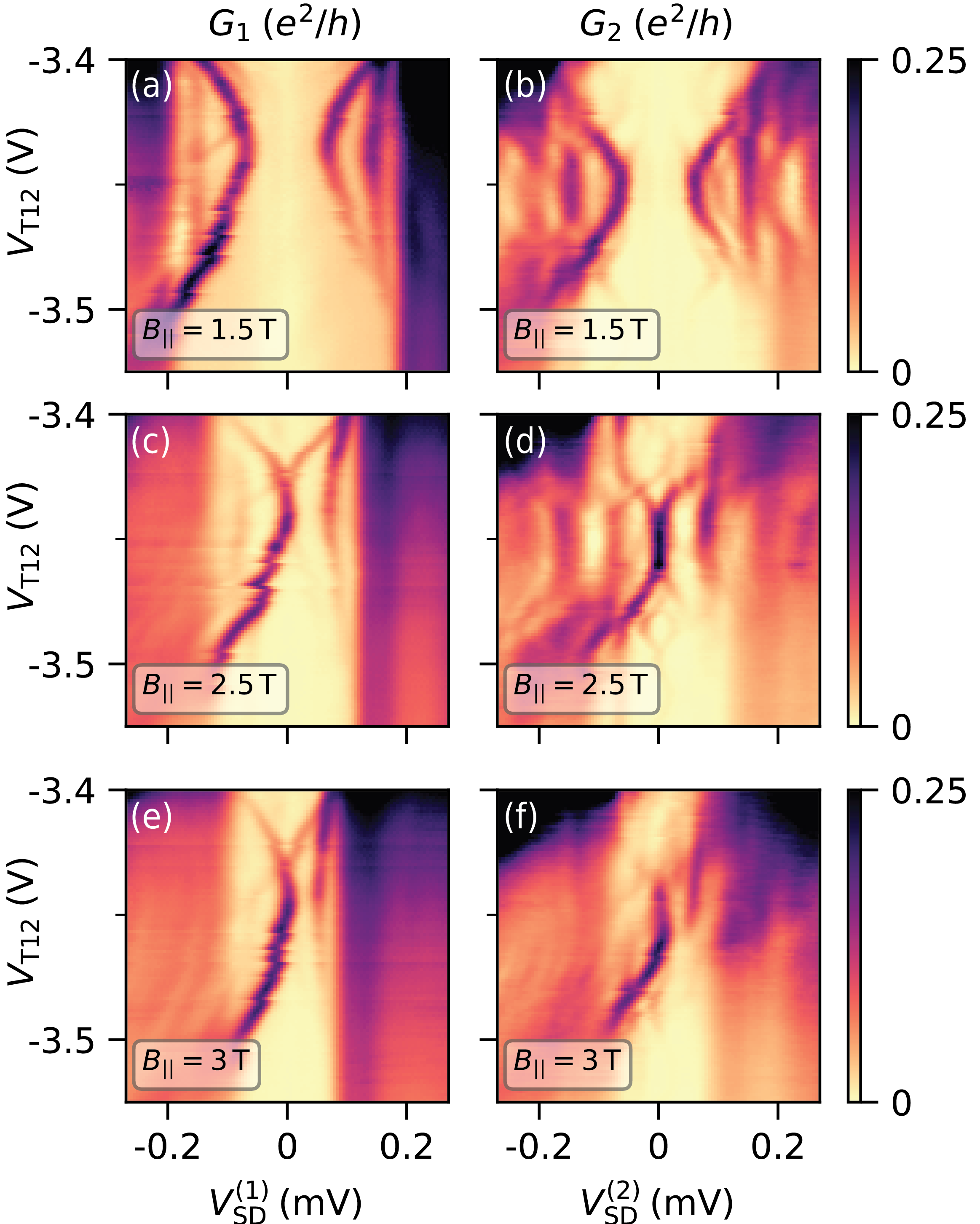}
\end{center}
\caption[Evolution of generic, fine-tuned, localized bound states]{\label{fig:fine_tuned_pscans}Bias spectroscopy as a function of gate voltage $V_\mathrm{T12}$ that tunes the electron density in the NW segment at different magnetic field values. (a, b) Tunneling conductance at both NW ends reveals resonances that have a parabolic shape with respect to gate voltage. (c, d) at higher magnetic fields, the resonances merge at zero bias. (e, f) at higher magnetic field the parent gap reduces and the states form a state close to zero bias over a small range. }
\end{figure}

Localized ABS frequently appear as parabolic lobes at subgap energies. The evolution of the fine-tuned ABSs is shown in \ref{fig:fine_tuned_pscans} and illustrates how the ellipsoidal area with a zero-bias conductance peak in \ref{fig:fine_tuned_phasediagram} emerges.

\begin{figure}[h!]
\begin{center}
\includegraphics[scale=0.95]{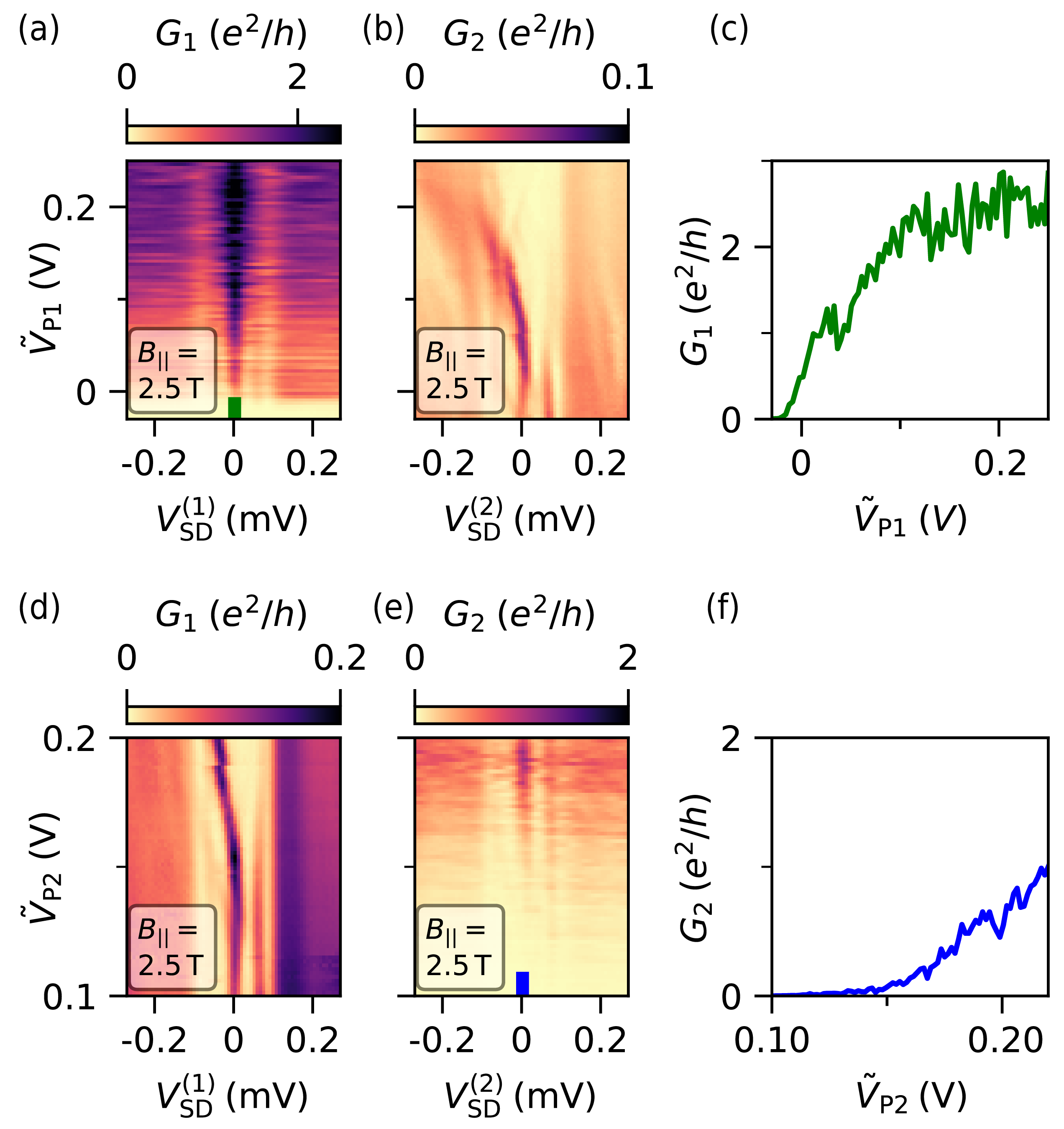}
\end{center}
\caption[Fine-tuned states at different tunneling rates]{\label{fig:fine_tuned_hscan}Bias spectroscopy as a function of varying tunnel barrier transparency. (a, b) tunneling conductance while changing the gate voltages $V_\mathrm{P1}$ and $V_\mathrm{T12}$ given by the green dashed line in Fig. \ref{fig:fine_tuned_gatemap}(c). The state measured on the left end of the NW is pinned to zero, while the state at the right end moves away from zero. (c) The line cut taken at zero bias in (a) reveals that the peak saturates at a conductance value $\approx 2 \, e^2/h$. (d, e) Tunneling conductance at both NW ends, while changing the gate voltages $V_\mathrm{P2}$ and $V_\mathrm{T12}$ along the blue dashed line in \ref{fig:fine_tuned_gatemap}(d). The state measured on the left end splits, while the state measured on the right end is pinned at zero source-drain bias. (f) Line cut of the data in (e) taken at zero bias. }
\end{figure}

\chapter{Three-terminal device based on full-shell nanowires}\label{ch:fullshell_nonlocal} 
A different material system that potentially hosts Majorana zero modes (MZMs) are vapor-liquid-solid grown InAs nanowires with a shell of superconducting Al that covers the full nanowire (NW) surface. In this chapter, the realization of a three terminal device based on full-shell Al-InAs NWs is presented. The device geometry allows for tunneling spectroscopy at both ends of a NW and measurements of nonlocal conductance. A rich structure of subgap states is observed. The spectra of conductance resonances on both NW ends bear similarity but are not identical. When hybridizing subgap states with local quantum dot states in the tunnel barrier on one NW end, the spectrum is unaffected  on the other NW end. This is in disagreement with states that are extending through the whole NW. No signatures of a gap reopening that would be expected for a topological phase have been observed in nonlocal conductance measurements. Quantum dot states can furthermore dominate the local conductance spectra and lead to strong zero-bias conductance enhancements mimicking the signatures of a topological phase. 

The NWs for the study presented in the following were provided by the Krogstrup group at the University of Copenhagen. Device Fabrication was done by M. E. Kloster, and A. Pöschl. E. B. Hansen, and A. Pöschl contributed equally to measurements and data analysis.

\section{Superconducting cylinders and a topological phase with a twist}
The superconducting phase gives rise to a macroscopic quantum state characterized by a complex wavefunction $\Psi\propto e^{i\theta}$ with phase ${\theta}$. Superconductors support a dissipationless supercurrent density $\mathbf{j}_\mathrm{S}$ that is connected to the vector potential $\mathbf{A}$ and the phase of the wavefunction through the expression

\begin{equation}
\Lambda \mathbf{j}_\mathrm{S} = - \left( \mathbf{A}(\mathbf{r},t)-\frac{\hbar}{2e}\nabla \theta(\mathbf{r},t)\right).
\end{equation}

$\Lambda$ is the London-coefficient. Integrating this equation along a closed path $C$ that encloses a surface $F$ yields 

\begin{equation}\label{eq:floxoid}
\oint_C \left( \Lambda \mathbf{j}_S\right)\cdot \mathrm{d}\mathbf{\boldsymbol\ell} + \int_F \mathbf{B}\cdot \mathrm{d}\mathbf{F} = n \frac{h}{2e}.
\end{equation}

The left-hand side of eq. \ref{eq:floxoid} is the so-called fluxoid. It is the sum of the magnetic flux $\Phi$ supplied by an external magnetic field $\mathbf{B}$ and the flux created by the supercurrent. The righ-hand side denotes multiples of the magnetic flux quantum $\Phi_0=h/2e$. $n$ is required to be an integer by the condition of a single-valued wavefunction $\Psi$. For the case of a non-simply connected superconductor that can be threaded by a magnetic field - like a hollow cylinder - equation \ref{eq:floxoid} predicts finite, quantized values of the fluxoid.\\

In pioneering experiments, metallic hollow cylinders were used to demonstrate the quantization of trapped flux in superconducting cylinders \cite{fairbank_fluxquantization, doll_naebauer_fluxquant}. The supercurrent through superconducting cylinders and their transition temperature are furthermore found to oscillate with varying magnetic flux piercing the cylinder with a period of $\Phi_0=h/2e$ \cite{littleparks}. This is known as the Little-Parks effect. In the cases of hollow cylinders with a radius smaller than their coherence length, the Little-Parks effect can be strong enough to completely destroy the superconducting state in an extended interval of finite magnetic flux near odd integer multiples of $\Phi_0/2$. A re-entrant superconducting phase around integer multiples of $\Phi_0$ occurs in this case where a superconducting state is recovered \cite{destructive_littleparks, sole_little_parks, oreg_destructive, oreg_destructive_2}. In a Ginzburg-Landau theory, the superconducting order parameter $\Delta$ can be determined for specific values of $\Phi$. For the re-entrant phase of superconductivity around a value of $\Phi=n\Phi_0$ the phase of the superconducting order parameter winds by $2\pi n$ for paths encircling the axis of the superconducting cylinder.\\ Semiconducting InAs nanowires fully covered by an Al shell can be MBE grown using the vapour-liquid-solid method \cite{krogstrup_epitaxially}. An Al shell can be deposited in the same growth chamber while rotating the sample holder in order to cover the full NW surface. This combines a non-simply connected superconductor with a semiconducting NW. The Little-Parks effect aswell as a destruction of superconductivity with re-entrant superconductivity at integer multiple values of magnetic flux $\Phi_0$ have been studied for full-shell Al-InAs NWs \cite{sole_little_parks}.\\

For the realization of a topological phase, there are several benefits resulting from the full-shell geometry. The metal film of Al covers the full surface of the semiconducting InAs and protects it during processing. When cooled below its critical temperature, the superconducting shell screens the semiconducting wire from electric fields and induces superconductivity in the semiconductor via the proximity effect. For the realization of a topological phase, the cylindrical geometry of the superconductor is important. It allows for a controled winding of the superconducting order parameter by $2\pi n$ in the presence of $n$ magnetic flux quanta $\Phi_0$ threading the NW core. Twisting the phase $n$ times around the NW axis can be seen as $n$ vortices penetrating the superconducting cylinder. This is analogous to vortices penetrating type II superconductors. The winding of the phase may allow for a topological phase even in the absence of a Zeeman energy at moderate field scales. A proposed Hamiltonian governing the behavior of the electrons inside the proximitized semiconductor in cylindrical coordinates reads \cite{elsa_topo_fullshell, sole_fullshell}
\begin{equation}
\label{eq:fullshell_hamiltonian}
H= \left[ \frac{\left(\bm{p}+e\bm{A}\right)^2}{2m^*} - \mu(r)+\bm{\alpha}(r)\cdot\bm{\sigma}\times[\bm{p}+e\bm{A}(r)] \right]\tau_z+\sigma_y\tau_y|\Delta(r)|e^{in\phi}
\end{equation}

The effect of the magnetic field is incorporated in the vector potential $\bm{A}$ and the $n$ winds of the superconducting order parameter. For a magnetic field applied along the NW axis, the vector potential reads $\bm{A}(\bm{r})=\frac{r\Phi}{2\pi R^2}\bm{\hat{\Phi}}$ with a tangential unitvector $\bm{\hat{\Phi}}$ and magnetic flux $\Phi=\pi B R^2$. A Zeeman energy term was omitted, as it is negligible compared to the last term in the Hamiltonian in eq. \ref{eq:fullshell_hamiltonian}. Note the that the Hamiltonian is invariant under rotations and can be diagonalized for individual values of the total angular momentum quantum number $m_j$.
For the Al-InAs interface a band offset between the superconductor and the semiconductor leads to band bending in the semiconducting region. This leads to a charge accumulation at the interface between semiconductor and superconductor. It was therefore suggested that full-shell Al-InAs NWs can be modeled as a thin walled hollow cylinder \cite{sole_fullshell}. Using this approximation  the Hamiltonian reduces to
\begin{equation}
\label{eq:fullshell_effective}
H_{m_j}= \left[ \frac{p_z^2}{2m^*} - \mu_{m_j}\right]\tau_z+V_z \sigma_z + A_{m_j} + C_{m_j}\sigma_z\tau_z+\alpha p_z \sigma_y \tau_z +\Delta \tau_x.
\end{equation}
This Hamiltonian is equivalent to the Lutchyn-Oreg model \cite{OREG, LUTCHYN} with the effective parameters $\mu_{m_j}$, $V_z$, $A_{m_j}$, and $C_{m_j}$ depending on $\Phi$, $m_j$, $m^*$, $n$, the radius of the cylinder, and the spin-orbit strength. Following this elegant mapping of the hollow cylinder model $H_{m_j}$ onto the Lutchyn-Oreg NW model it is not suprising that full-shell NWs can host MZMs in theory, in particular for an odd number of winding numbers $n$. Numerical simulations beyond the hollow-cylinder approximation confirmed this also for the more realistic hexagonal nanowire cross section \cite{sole_fullshell}.\\ 

Despite this promising theoretical result and first experimental sightings of zero-bias conductance peaks (ZBPs) in agreement with the existence of MZMs \cite{sole_fullshell}, several drawbacks of the full-shell geometry have been identified. The superconducting shell covering the full NW surface prohibits any control of the electron density via electrostatic gating. Consequently, the electron density is fixed by the bulk properties of InAs and the band bending at the interface to the Al. The only experimental parameters to achieve a topological phase are therefore the radius of the semiconducting NW and the thickness of the superconducting shell. Several numerical studies confirmed that fine tuning of these parameters is necessary to realize a topological phase in this system \cite{elsa_topo_fullshell, dassarma_fullshell}. The superconducting full-shell furthermore decreases the spin-orbit strength, because it screens the InAs NW core from electric fields. This leads to a decrease in the topological energy-gap that can be reached \cite{dassarma_fullshell}.\\

The aforementioned difficulties are specific to the full-shell NW geometry. In addition to those, the general problems of trivial ABS, quasi-Majoranas, and quantum dot states at the ends of the NW arise. This makes the unambiguous identification of MZMs in tunneling spectroscopy at a single NW end difficult. Spacings of conductance peaks of Coulomb blockaded full-shell nanowires were found to oscillate \cite{sole_fullshell}. The amplitude of these oscillations decays exponentially with NW length - a behavior that is expected for MZMs but can also be mimicked by trivial ABSs \cite{dassarma_against_sven}. \\

There is a need for more sophisticated experiments, which may be able to shine light on the nature of the ZBP in full-shell Al-InAs nanowires and potentially enable a measurement of a topological gap. One possible experiment is a three-terminal device based on full-shell NWs \cite{andreev_rectifier,karsten_nl_spectroscopy,SDS_nl_conductance}.

\section{Three-terminal devices based on full-shell nanowires}

\begin{figure}
\begin{center}
\includegraphics[scale=0.85]{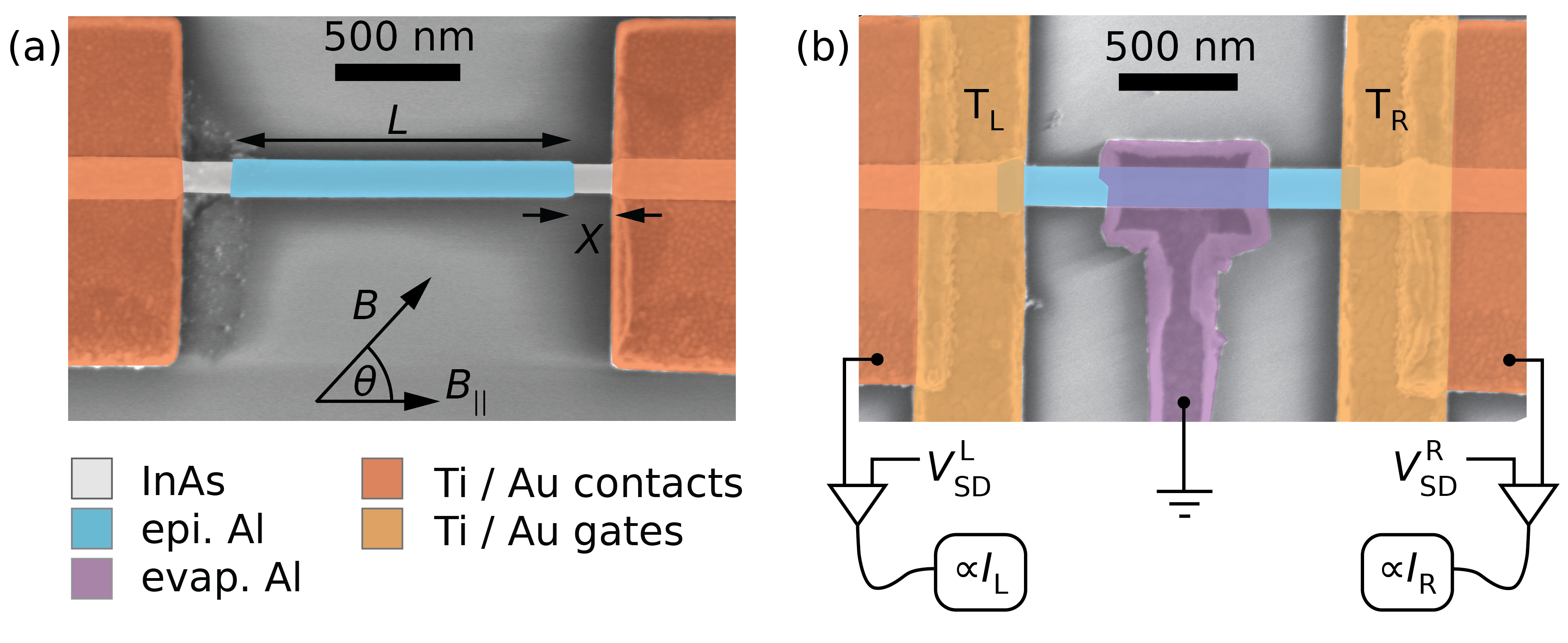}
\end{center}
\caption[Micrograph of three-terminal device based on InAs nanowire with full-shell of Al]{\label{fig:fullshell_SEM}False-color scanning electron micrograph of an example three-terminal device based on full-shell nanowires. (a) InAs nanowire (white) after selective etch of the Al shell (blue) and deposition of ohmic contacts (orange). (b) completed device with superconducting contact (purple) and topgates $\mathrm{T_L}$ and $\mathrm{T_R}$ (yellow).}
\end{figure}

In order to do tunneling spectroscopy at both ends of a full-shell nanowire and measure nonlocal conductance, it is necessary to tunnel couple both NW ends by a gatable barrier to normal leads and ground the superconducting shell with a superconductor \cite{gerbold_malinowski_qpfilter}. Figure \ref{fig:fullshell_SEM}(a) shows a full-shell Al-InAs NW after removing the Al at the NW ends by wet etching. Both ends are contacted by metal contacts, leaving a short segment of semiconducting NW between the metal contact and the Al covered part of the NW. A superconducting contact can be used to electrically ground the superconducting shell. A technique that uses a ramp fabricated from cross-linked resist next to the NW, removal of the native oxide of the Al shell by ion milling, and deposition of a thin layer of Al was developed in Ref. \cite{Razmadze2020} and was adapted for the present work. A completed device can be seen in \ref{fig:fullshell_SEM}(b). The NW ends are covered by gate dielectric and top gates $\mathrm{T_L}$ and $\mathrm{T_R}$. A back gate is provided by the doped silicon/silicon substrate. The top gates combined with the back gate allow for gating of the InAs NW from all sides to form a tunnel barrier.
 
In the following, measurements of local and nonlocal tunneling conductance for two devices are presented. The length of the Al covered part of the NW was $L\approx\SI{2.1}{\micro\meter}$ for device 5 and \nolinebreak{$L\approx\SI{2.6}{\micro\meter}$} for device 6. Both NW had a semiconducting core diameter of $\approx \SI{80}{\nano\meter}$ and $\approx \SI{5}{\nano\meter}$ thick Al shell. The lock-in detection technique as described in Ref. \cite{gerbold_nonlocal} was used. The experimental setup is sketched in Fig.~\ref{fig:fullshell_SEM}(b).

\section{Local and nonlocal conductance measurements}
\begin{figure}
\begin{center}
\includegraphics[width=\textwidth]{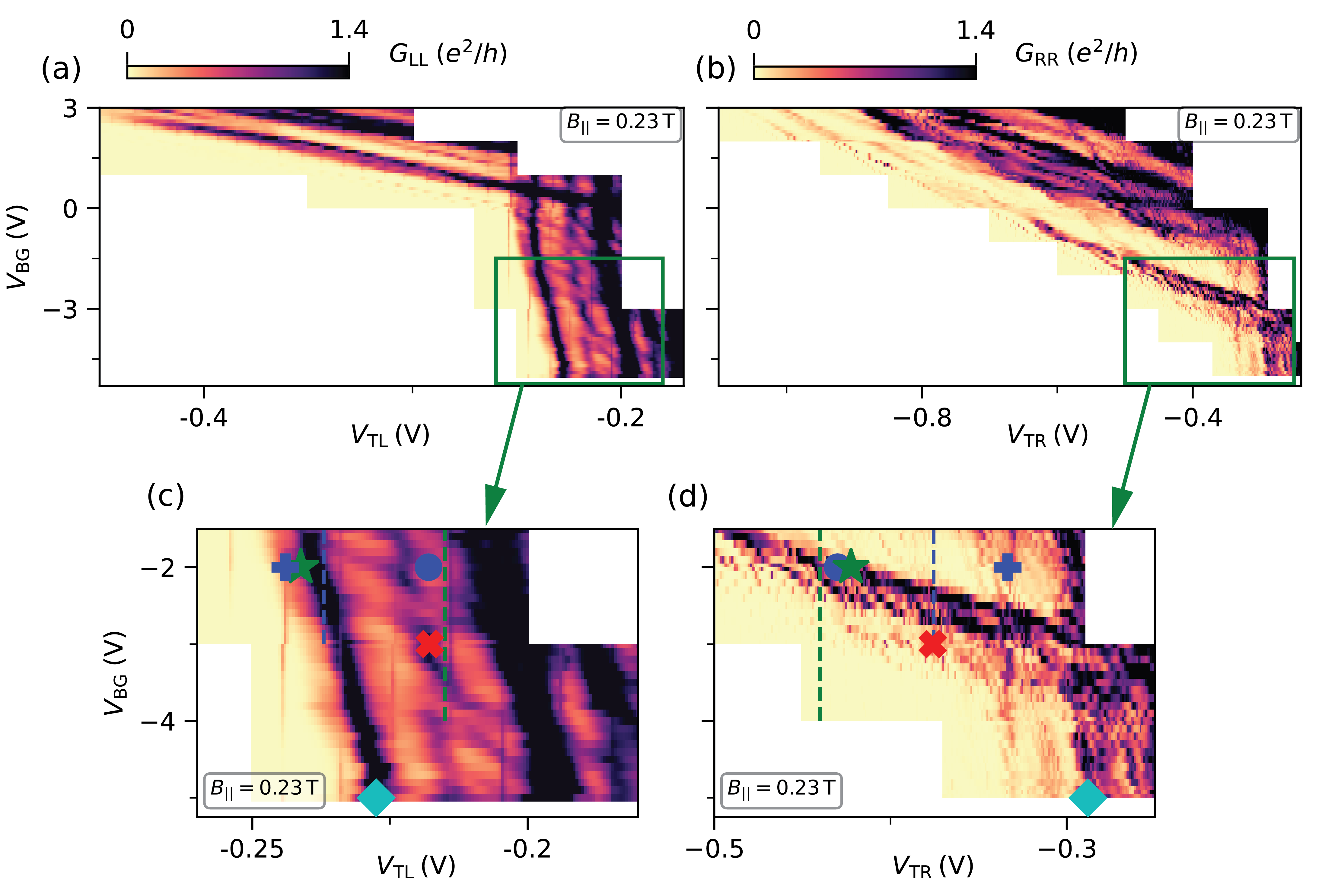}
\end{center}
\caption[Tuning of zero-bias resonances by gate voltages]{\label{fig:fullshell_gatemap}Gate dependence of conductance resonances at zero source-drain voltage in the first lobe. (a, b) Conductance resonances as a function of both backgate and respective topgate voltage at both nanowire ends. (c, d) Zoom-in on the data inside the green rectangles in (a, b). Markers denote the gate configuration of measurements shown below.}
\end{figure}

A magnetic field $B_{||}=\SI{0.23}{\tesla}$ parallel to the NW was applied to reach the re-entrant superconducting phase of the Al shell around $\Phi=\Phi_0$ for device 5. Differential conductance at zero source-drain bias $V_\mathrm{SD}^\mathrm{L}=V_\mathrm{SD}^\mathrm{R}=\SI{0}{\volt}$ reveals a multitude of resonances that are strongly dependent on either the back gate voltage $V_\mathrm{BG}$ or the respective top gate voltage $V_\mathrm{TL}$, $V_\mathrm{TR}$ [see Fig.~\ref{fig:fullshell_gatemap}(a,b)]. The resonances can be due to localized states in the tunnel barrier. Another cause of origin for the conductance resonances may be states inside the Al covered part of the NW. Measurements of the conductance spectra as a function of magnetic field parallel to the NW $B_{||}$ are presented for the gate voltage configurations given by the different markers in Fig.~\ref{fig:fullshell_gatemap}(c, d). \\

\begin{figure}
\begin{center}
\includegraphics[width=\textwidth]{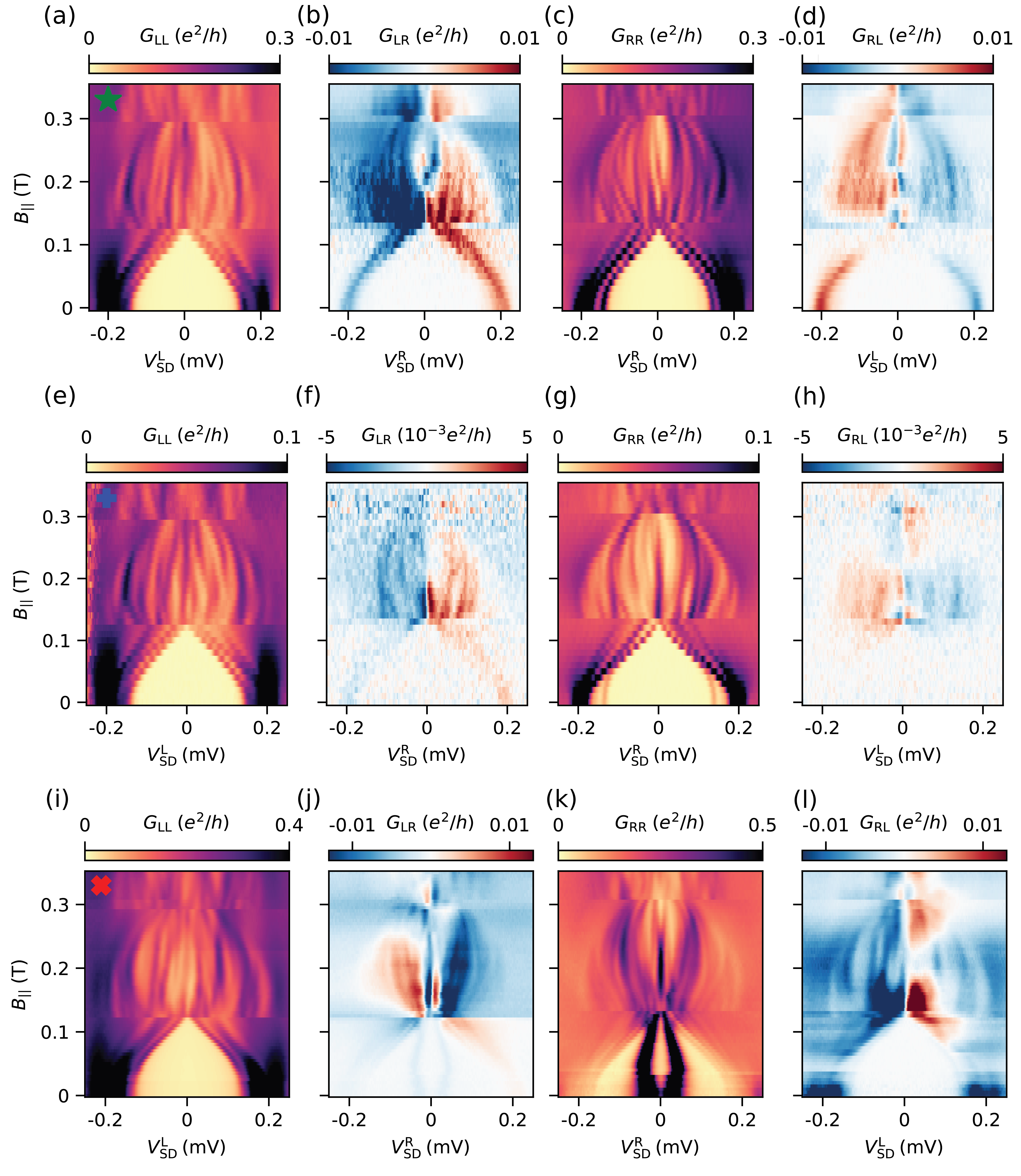}
\end{center}
\caption[Tunneling spectroscopy as function of magnetic field at selected gate voltages]{\label{fig:fullshell_fieldscans}Bias spectroscopy as function of parallel magnetic field. (a-d) All four conductance matrix elements with the gate voltages at the values denoted by $\bigstar$ in Fig.~\ref{fig:fullshell_gatemap}. While a gap is visible in the zeroth lobe in all four measurements, no gap is seen in the first lobe. (e-h) the full conductance matrix at gate voltages denoted by \ding{58} in Fig.~\ref{fig:fullshell_gatemap}. $G_\mathrm{RR}$ in (g) shows conductance peaks $\approx$ \SI{0.02}{\milli\eV} below the superconducting gap at zero magnetic field and a strong ZBP in the first lobe. (i-l) All four conductance matrix elements with gate voltages at the values given by \ding{54} in Fig.~\ref{fig:fullshell_gatemap}. $G_\mathrm{RR}$ in (k) shows two strong conductance resonances close to zero bias in the zeroth lobe that evolve into a ZBP in the first lobe.}
\end{figure}

A measurement of all four conductance matrix elements, consisting of the two local conductances $G_\mathrm{LL}$ and $G_\mathrm{RR}$ and the two nonlocal conductances $G_\mathrm{LR}$ and $G_\mathrm{RL}$, is shown in Fig.~\ref{fig:fullshell_fieldscans}(a-d). The gates were set to the values marked by $\bigstar$ in Fig.~\ref{fig:fullshell_gatemap}. The conductance spectrum shows a hard superconducting gap $\Delta\approx \SI{150}{\micro\eV}$ at zero magnetic field in both the local and nonlocal conductances. The gap closes with increasing magnetic field and vanishes at a magnetic field value $B_{||}=\SI{0.12}{\tesla}$. An abrupt change in conductance measurement is visible before the gap of the re-entrant superconducting phase reopens around $n=1$ quanta of magnetic flux. The superconducting phases around flux values $\Phi=n\Phi_0$ create a structure of consecutive lobes in tunneling conductance. In the first lobe around $n=1$ flux quanta, multiple subgap states fill the gap. A conductance peak at zero bias is visible only in a small portion of the first lobe in the local conductances. The subgap states measured as peaks in the local conductance may originate due to strongly localized states in the tunnel barriers or due to extended states in the Al covered part of the NW. The nonlocal conductances show a hard gap closing in the zeroth lobe and a gap filled with conductance signal in the first lobe. The signal has a strong anti-symmetric component with respect to zero bias. No finite gap is visible around zero bias in the nonlocal conductance as one would expect for a NW that hosts a topological phase and is longer than the topological coherence length. Possible explanations for the absence of a gap in the nonlocal conductance may be that the NW are not in the right parameter regime that is required to be in the topological phase. This is determined by the core radius of the NW \cite{dassarma_fullshell, elsa_topo_fullshell}. The results are also consistent with a topological gap $\lesssim \SI{10}{\micro\eV}$, which is on the order of the electron temperature and unlikely to be detectable by this technique. The topological gap can be reduced to a minigap due to the occupation of higher orbital momentum bands which interact with the topological band and renormalize the topological gap \cite{elsa_nonlocality}. The background of conductance due to the additional bands may give rise to finite nonlocal conductance. 

Figure \ref{fig:fullshell_fieldscans}(e-h) shows a conductance matrix measurement with a different value of $V_\mathrm{TR}$. The gate configuration is marked by \ding{58} in Fig.~\ref{fig:fullshell_gatemap}. The local conductance $G_\mathrm{RR}$ shows two discrete subgap states that are slightly separated from the continuum of states at zero magnetic field. With increasing magnetic field they merge into the continuum of states as the superconducting gap closes due to the Little-Parks effect around $\Phi=\Phi_0/2$. Throughout the whole first lobe there is a strong, sharp ZBP visible in $G_{LL}$. The subgap states at low field together with a state at zero bias most likely originate from a quantum dot formed in the tunnel barrier region. Experimental studies found that similar signatures occur due to quantum dots formed in tunnel barriers longer than $\SI{150}{\nano\meter}$ \cite{trivial_fullshell}. The ones used in this experiment measure $\approx\SI{100}{\nano\meter}$. Numerical studies found that quantum dot states may be pushed towards zero bias in the first lobe as a result of the circular symmetric coupling of the quantum dot to the superconducting shell hosting a vortex \cite{elsa_trivial_fullshell}. The example shown in Fig.~\ref{fig:fullshell_fieldscans}(g) may be an experimental example of this phenomenon and shows that QD states in the tunnel barrier can sometimes be hard to disentangle from states inside the NW based on their evolution with magnetic field $B_{||}$ alone. The local states around the ZBP as in Fig.~\ref{fig:fullshell_fieldscans}(g) furthermore mimic the behavior of the first excited state of a topological phase. However, none of the nonlocal conductances in Figs~\ref{fig:fullshell_fieldscans}(f, h) show a gap. This illustrates the potential pitfall when performing only local tunneling spectroscopy, where localized states mimic the behavior of a topological phase despite the bulk of the NW being trivial. 

Another example of local and nonlocal conductance measurements is shown in Figs.~ \ref{fig:fullshell_fieldscans}(i-l), where the gates were set to the values marked by \ding{54} in Fig.~\ref{fig:fullshell_gatemap}. Here $G_\mathrm{RR}$ shows a more obvious case of a quantum dot, visible as two strong resonances far away from the quasiparticle continuum in the zeroth lobe and a strong ZBP in the first lobe \cite{trivial_fullshell}. This example shows that the exact spectrum of subgap states strongly depends on the precise gate settings. The bulk of the NW is screened from the electric fields by the superconducting shell and should therefore show a generic behavior independent of gate voltage settings. The tunnel barriers in the experiment still require tuning in the full-shell geometry which gives rise to multiple phenomena that mask both local and nonlocal conductance measurements, potentially obscuring the generic bulk properties of the NW.

\begin{figure}
\begin{center}
\includegraphics[width=\textwidth]{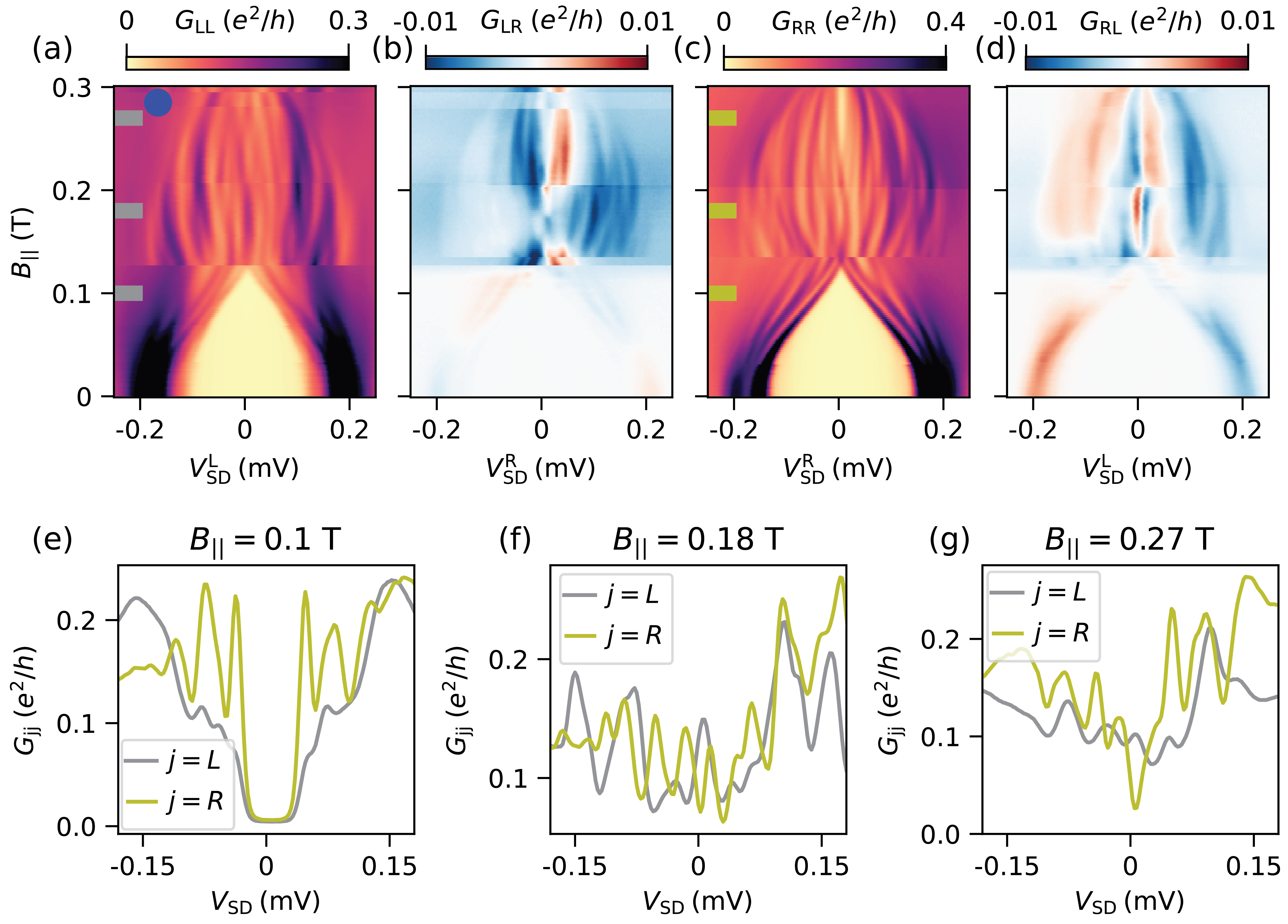}
\end{center}
\caption[Absence of end-to-end correlations of subgap states in first lobe]{\label{fig:fullshell_linecuts}Bias spectroscopy as a function of magnetic field and comparison of the spectra measured in $G_\mathrm{LL}$ and $G_\mathrm{RR}$. (a-d) Measurement of all four conductance matrix elements at gate voltage values denoted by \ding{108} in Fig.~\ref{fig:fullshell_gatemap}. (e-g) comparison of line cuts of $G_\mathrm{LL}$ and $G_\mathrm{RR}$ at the three magnetic field values. The spectra are similar in the number and density of conductance peaks. The position of the peaks in $G_\mathrm{LL}$ and $G_\mathrm{RR}$ do not occur at the exact same value of $V_\mathrm{SD}$, in particular the low-energy states in the first lobe shown in (f,g). }
\end{figure}

Most notable, however, is the close similarity between spectra that is present in local spectroscopy on the left and right end for most gate settings. The conductance spectroscopy data shown in Figs.~\ref{fig:fullshell_linecuts}(a-d) is an example that was taken at the gate configuration marked by \ding{108} in Fig.~\ref{fig:fullshell_gatemap}. No obvious QD states are visible in the local conductances $G_\mathrm{LL}$ and $G_\mathrm{RR}$. The number and density of subgap states is similar, however. This is confirmed by the line cuts at different values of magnetic field [see Figs.~\ref{fig:fullshell_linecuts}(e-g)]. The line cuts taken at magnetic field values $B_{||}=\SI{0.18}{\tesla}$ and $B_{||}=\SI{0.27}{\tesla}$ in the first lobe show that the peaks in the local conductances measured at the two NW ends are not aligned. This is in particular true for states close to zero bias. This is in disagreement with the same extended states being probed at the two ends of the NW because this would require a precise alignment of the states in $V_\mathrm{SD}$ \cite{gian_correlations}.

\begin{figure}
\begin{center}
\includegraphics[width=\textwidth]{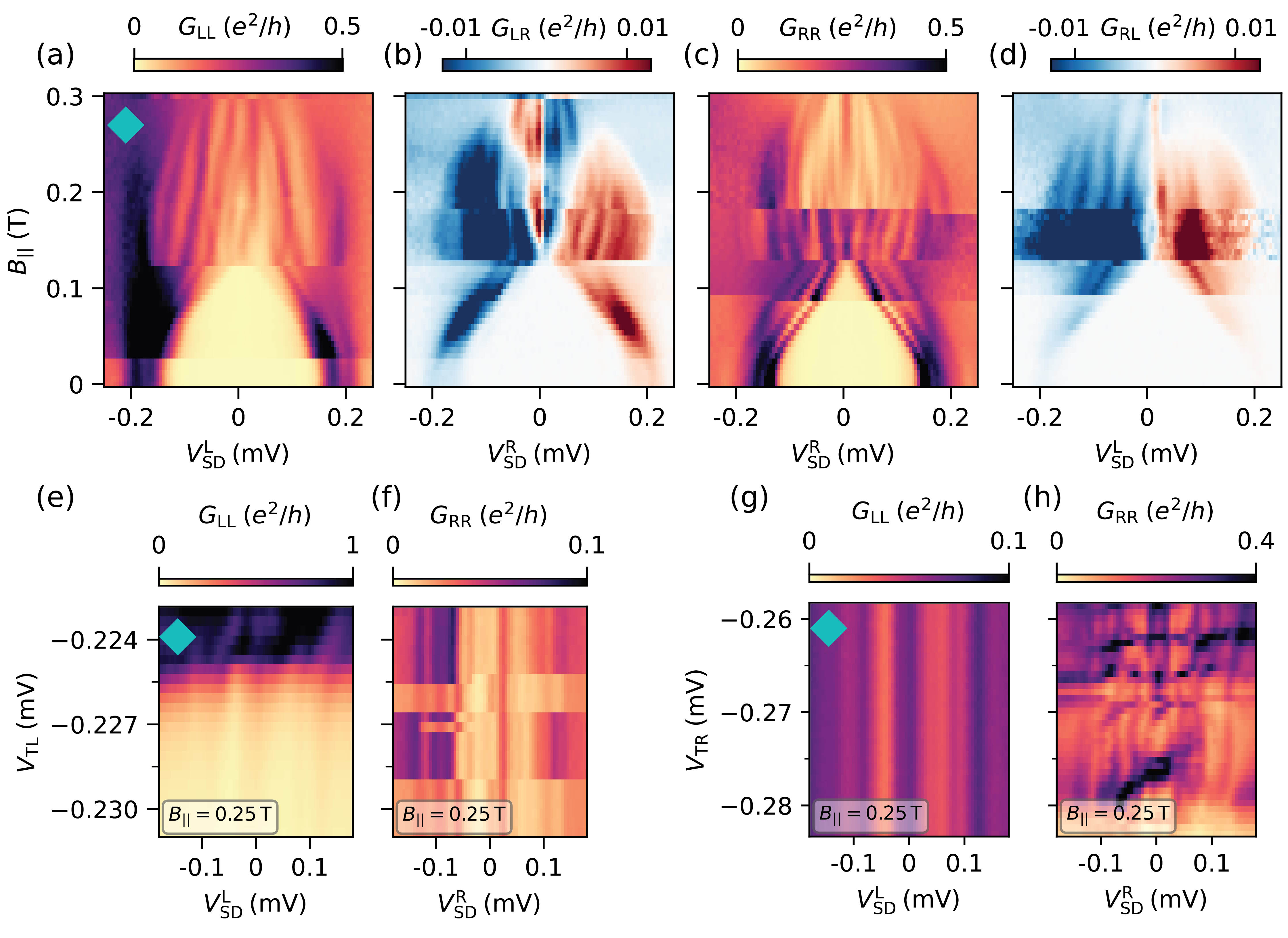}
\end{center}
\caption[Absence of nonlocal signature due to quantum dot-bound state hybridization]{\label{fig:fullshell_pradaclarke}Bias spectroscopy as function of magnetic field and topgate voltages. (a-d) Measurement of the all four conductance matrix elements as a function of magnetic field at gate voltages denoted by \ding{117} in Fig.~\ref{fig:fullshell_gatemap}. (e, f) Local tunneling conductances as a function of left topgate voltage around the gate voltage values denoted by \ding{117} in Fig.~\ref{fig:fullshell_gatemap}. There is no change in the local conductance measured on the right, except for electrostatic switches. (g,h) Measurement of local tunneling conductances as a function of right topgate voltage. The spectrum of subgap states in $G_\mathrm{RR}$ changes due to energy level hybridization with quantum dot resonances. The spectrum in $G_\mathrm{LL}$ stays constant. }
\end{figure}

Tunneling spectroscopy as a function of magnetic field with the gates voltages set to the values marked by \ding{117} in Fig.~\ref{fig:fullshell_gatemap}  is shown in Fig.~\ref{fig:fullshell_pradaclarke}. The top gate voltages were scanned in a small range at a magnetic field value $B_{||}=\SI{0.25}{\tesla}$ within the first lobe. The subgap spectrum in the local conductance $G_\mathrm{LL}$ as a function of $V_\mathrm{TL}$ shows only changes at the point were the overall conductance enhances strongly as seen from Fig.~\ref{fig:fullshell_pradaclarke}(e). $G_\mathrm{RR}$ stays constant during the change of $V_\mathrm{TL}$ [see Fig.~\ref{fig:fullshell_pradaclarke}(f)]. The measured local conductance spectra as a function of $V_\mathrm{TR}$ in Fig.\ref{fig:fullshell_pradaclarke}(g, h) reveal multiple QD resonances in the right tunnel barrier. The subgap states visible in $G_\mathrm{RR}$ undergo a change in energy that is correlated with the states of the QD crossing zero bias. This change in energy of the subgap states can be seen as a result of the QD and states in the NW being tunnel coupled, such that the energy levels hybridize \cite{elsa_nonlocality, clarke_quality, elsa_quantifying, mingtang_nonlocality}. The spectrum of subgap states in $G_\mathrm{LL}$ measured at the other end shows no change, however. This indicates that the subgap resonances in the spectra measured at the left and the right end of the NW are not stemming from the same extended state in the NW. 

\begin{figure}
\begin{center}
\includegraphics[width=\textwidth]{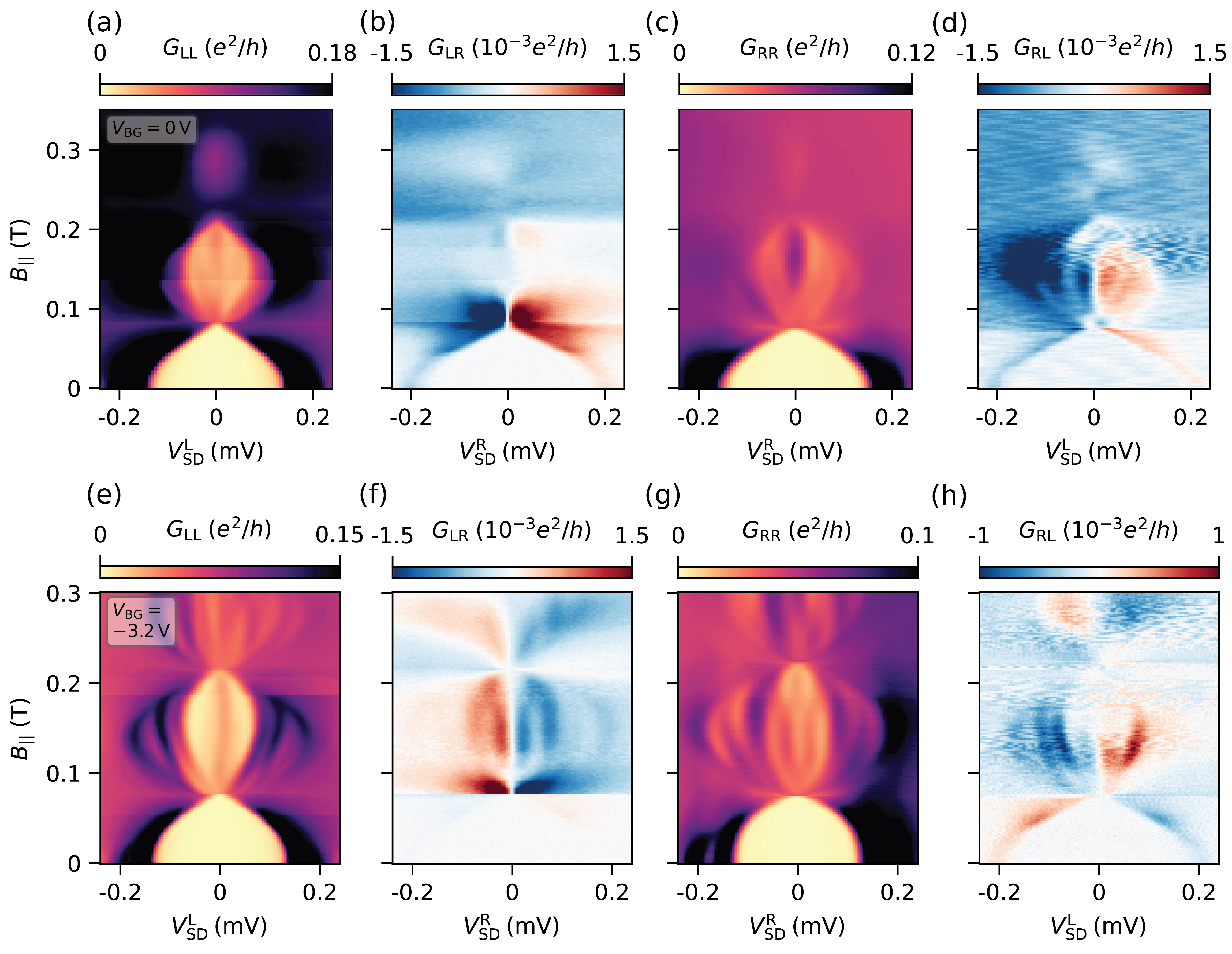}
\end{center}
\caption[Spectrum at both NW ends for second three-terminal device]{\label{fig:fullshell_2nd_dev_fieldscan}Bias spectroscopy as a function of magnetic field measured on a different device. (a-d) Full conductance matrix at zero backgate voltage. An enhancement of zero bias conductance is visible in both local conductances, no gap is visible in the nonlocal conductances. (e-h) All four conductance matrix elements at more negative backgate voltage. The spectra on the two local conductances show close similarity, with a difference in the lowest excited states. }
\end{figure}

Conductance spectroscopy from device 6 is shown in Fig.~\ref{fig:fullshell_2nd_dev_fieldscan}. Results for two distinct values of back gate voltage $V_\mathrm{BG}$ are presented. Both measurements confirm the findings made on device 5. A hard gap is found in the zeroth lobe around zero magnetic field. In the first lobe qualitatively similar spectra are found in the local conductances. Only for small parts of the first lobe a ZBP is visible in general and the conductance peaks of the low-energy states appear at different values of $V_\mathrm{SD}$ in $G_\mathrm{LL}$ and $G_\mathrm{RR}$. The nonlocal conductances show a hard gap for the zeroth lobe and no clear gap reopening in the first lobe, even in parts of the first lobe where ZBP are visible in both local conductances.

\section{Conclusion}

The realization of a full-shell NW device with both NW ends coupled to normal leads and a grounded superconducting shell was outlined. The local tunneling spectroscopy shows spectra that are similar at both NW ends but do not stem from the same extended states. This is revealed by a difference in $V_\mathrm{SD}$ positions of conductance peaks at both ends. A change in energy of the state should furthermore occur on both ends of the NW for extended states that are hybridized with a QD state at one end, which could not be observed. A plausible explanation of the states observed are states that are localized at one end of the NW, for example due to disorder. The exact spectrum is determined by the precise geometry of the NW. This includes cross-sectional size and shape. For the case of full-shell NW these two factors determine the electron density, number of occupied radial modes, and consequently the measured spectrum in tunneling conductance  \cite{sole_fullshell, elsa_topo_fullshell, dassarma_fullshell}. A constant cross section throughout the NW leads to similar conductance spectra at both NW ends. The cross section shape between wires can differ, which explains different conductance spectra for different devices.

A difficulty that arises in tunneling spectroscopy experiments are quantum dot states forming in the tunnel barrier region and ABS localized at the ends of the NW, which are sometimes called quasi-Majoranas. These can have a large influence on the observed conductance spectrum and imitate the signatures of a topological phase in the form of a ZBP in the $n=1$ lobe, examples of which were presented here. In the nonlocal conductance, no gap reopening was measured in the presence of a ZBP at both ends in all measured cases. This suggests that the NW did not host a topological phase that had an extent larger than the topological coherence length or that the topological gap is very small. A potential systematic issue of this experiment may arise additionally due to the third lead that is used to contact the NW and ground the superconducting shell. The fabrication process can potentially introduce defects in the Al. The deposited Al only covers half of the NW which breaks the symmetry of the full-shell geometry. This may result in unwanted subgap states or a disruption of the topological phase at the  point were the full-shell is contacted \cite{dassarma_circuits}.


\cleardoublepage
\part{Summary and Outlook}\label{pt:outlook}
\chapter{Summary and outlook}
Within the scope of my PhD, devices were developed for the study of bound states in nanowires (NWs) based on two-dimensional electron gas (2DEG) hybrid-heterostructures and conventional NWs with a full-shell of Al. The experiments focused in particular on devices that make use of more than one tunneling probe in order to study bound states that extend over the NW and distinguish them from states localized at one end of the NW. Extended Andreev bound states (ABSs) were studied in local and nonlocal conductance measurements. \\

\begin{figure}[h!]
\begin{center}
\includegraphics[width=\textwidth]{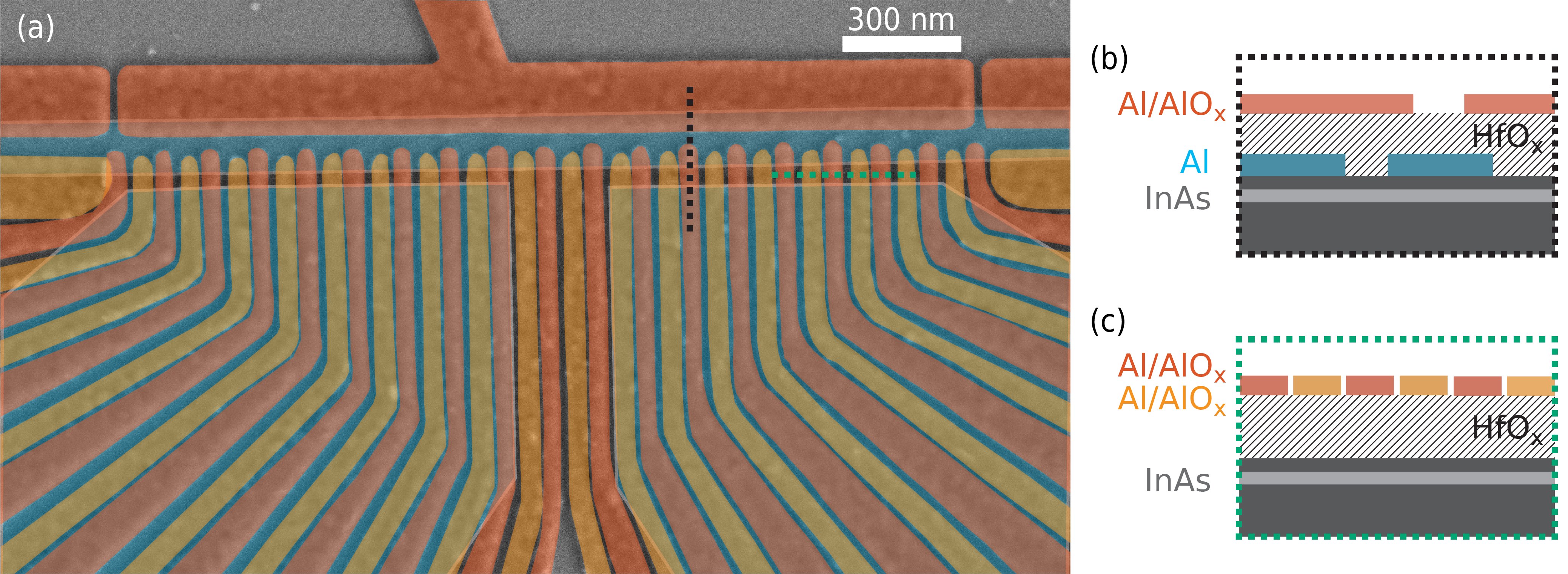}
\end{center}
\caption[Gate-defined nanowire with a gate array]{\label{fig:finger_piano}(a) False-color micrograph of NW device with an array of finger-gate electrodes. An Al strip (blue) defines the NW and can be gate controlled by wide gates (red) on its upper edge and by an array of finger gates (orange and red) on its lower edge. Putting a single finger-gate to a voltage around zero volt allows for tunneling between the NW and the elongated leads (blue) next to the NW. (b) Schematic of the heterostructure along the black dotted line in (a). (c) Schematic of the heterostructure along the green dotted line in (a).}
\end{figure}

Local tunneling conductance at both ends of the ABSs were used to investigate signatures of the hybridization of the ABS with a local quantum dot. Nonlocal conductance measurements were applied to identify the charge character of the ABSs which evolves in accordance with the change in state energy as a function of gate voltage. These techniques and device geometries may inspire future work, that uses gate-defined quantum dots as tool for the investigation of bound states in 2DEG hybrid-heterostructures. The demonstration of a three-terminal device geometry that allows for the measurement of nonlocal conductance is also relevant for research concerned with Majorana zero modes. While the proof of principle devices were demonstrated in this work unambiguously for ABSs, they can be applied directly in the form of longer NWs for the research concerning Majoranas. Here a topological phase would manifest itself in end-to-end correlated zero-bias states concomitant with a reopening of the bulk gap visible in nonlocal conductance. The current theoretical knowledge and experimental insights using longer NWs gained within this thesis suggest, however, that this also requires improved material quality. 

A progression of the sideprobe devices with superconducting leads is a device with an array of finger gates. Such a device was fabricated according to Appendix \ref{sec:fab_finger_piano} and is shown in Fig.~\ref{fig:finger_piano}. It consists of a NW in the form of a strip of Al on top of InAs 2DEG. Gate electrodes deplete the electron density to form a NW under the Al. One edge of the NW is confined by a single, extended gate. The opposite edge is depleted by an array of narrow, elongated finger-gates each of which can be set to an individual voltage value. This allows for a high control of the electrostatic confining potential along the NW. Two elongated leads made from superconducting planes of Al under the array of finger-gates can be tunnel coupled to the NW by putting a selected finger-gate to a positive voltage. This allows for a three-terminal measurement of bound states at arbitrary points along the NW with high control of the chemical potential. \\

Experiments with devices like the ones presented in this thesis will further help to shine light on the quantum states that emerge in superconductor-semiconductor hybrid devices. Unambiguous experimental results, better theoretical understanding of the underlying physics, and improved material quality may naturally lead to the realization of a topological phase in the future.

\appendix
\cleardoublepage
\part{Appendix}
\chapter{Fabrication protocols}
\label{ch:fab}
The following gives detailed fabrication protocols of the devices presented in this work.

\section{\label{sec:fab_super_sideprobe}Fabrication of sideprobe device with superconducting probes}
Device 1 and device 4 were fabricated according to this protocol.
\begin{itemize}
\item Cleaving
\begin{itemize}
\item Wafer is coated with post-baked A4 PMMA resist.
\item Scribe and cleave chips of suitable size (e.g., $\SI{3}{\milli\meter}$) using manual scriber.
\end{itemize}

\item Fabrication of registration marks

	\begin{itemize}
   \item Use resist that was applied before cleaving the chip (avoids etch beads) or spin coat A4 PMMA resist for \SI{45}{\second}, post-bake \SI{2}{\min} at \SI{185}{\celsius}.
\item  EBL with \SI{150}{\micro\meter} write field, 60000 dots, pitch 1. Beam current \SI{0.5}{\nano\ampere}, dose time = \SI{0.085}{\micro\second\per dot} [proximity effect correction for InAs substrate, \SI{200}{\nano\meter} PMMA, \SI{100}{\kilo\volt}, optimal contrast].
   \item Develop \SI{45}{\second} in MIPA:IPA (1:3), \SI{5}{\second} IPA, blowdry. 
   \item Ash for \SI{60}{\second}.
   \item Metal deposition in AJA
   \begin{itemize}
   \item Deposit \SI{5}{\nano\meter} Ti.
   \item Deposit \SI{55}{\nano\meter} Au.
   \end{itemize}
   \item Clean and liftoff in 1,3-dioxolane for \SI{5}{\min} under ultrasonication at \SI{80}{\kilo\hertz}. Rinse with squirt of aceton, rinse with squirt of IPA, blowdry.
	\end{itemize}

\item Mesa etch

\begin{itemize}
\item Spin coat with A4 PMMA for \SI{45}{\second} at 4000 rpm, post-bake \SI{120}{\second} at \SI{185}{\celsius}.
\item  EBL of fine features with \SI{150}{\micro\meter} write field, 60000 dots, pitch 1. Beam current \SI{0.5}{\nano\ampere}, dose time = \SI{0.085}{\micro\second\per dot} [proximity effect correction for InAs substrate, \SI{200}{\nano\meter} PMMA, \SI{100}{\kilo\volt}, optimal contrast].
\item  EBL of coarse features with \SI{600}{\micro\meter} write field, 20000 dots, pitch 1. Beam current \SI{0.5}{\nano\ampere}, dose time = \SI{0.34}{\micro\second\per dot} [proximity effect correction for InAs substrate, \SI{200}{\nano\meter} PMMA, \SI{100}{\kilo\volt}, optimal contrast].
\item Develop \SI{60}{\second} in MIPA:IPA (1:3), \SI{5}{\second} IPA, blowdry. 
\item Ash for \SI{60}{\second}.
\item Reflow \SI{120}{\second} at \SI{120}{\celsius}
\item Al etch
\begin{itemize}
\item Prepare two beakers wtith MQ water, and one beaker with Transene D Al etchant in a hot bath. The level of liquid should be the same in all three beakers and it should be aligned with the water level in the hot bath.
\item Check that the temperature inside the beakers is  \SI{50\pm0.1}{\celsius} by measuring the temperature of one of the MQ beakers.
\item Etch for \SI{5}{\second} in the Transene D Al etchant. 
\item Rinse for \SI{20}{\second} in the beaker with MQ that was not used to check the temperature.
\item Rinse for \SI{40}{\second} in MQ water at room temperature.
\item Blowdry.
\end{itemize}

\item Mesa etch
\begin{itemize}
\item Prepare a solution of solution of $\mathrm{H_2O:C_6H_8O7:H_3PO_4:H_2O_2}$ (220:55:3:3).
\item Stir with magnetic stir.
\item Etch for \SI{9}{\min} while rotating the chip every \SI{30}{\second} by \SI{90}{\degree}.
\item Rinse in MQ water for \SI{60}{\second}.
\end{itemize}

     \item Strip resist for \SI{15}{\min} in 1,3-dioxolane, \SI{10}{\second} in aceton, \SI{10}{\second} in IPA, blowdry.
\end{itemize}

\item Selective removal of Al

\begin{itemize}
\item Spin coat with A4 PMMA for \SI{45}{\second} at 4000 rpm, post-bake \SI{120}{\second} at \SI{185}{\celsius}.
\item  EBL with \SI{150}{\micro\meter} write field, 60000 dots, pitch 1. Beam current \SI{0.1}{\nano\ampere}, dose time = \SI{0.45}{\micro\second\per dot} [proximity effect correction for InAs substrate, \SI{200}{\nano\meter} PMMA, \SI{100}{\kilo\volt}, optimal contrast].
\item Develop \SI{45}{\second} in MIPA:IPA (1:3), \SI{5}{\second} IPA, blowdry. 
\item Ash for \SI{60}{\second}.
\item Reflow \SI{120}{\second} at \SI{120}{\celsius}
\item Al etch
\begin{itemize}
\item Prepare two beakers wtith MQ water, and one beaker with Transene D Al etchant in a hot bath. The level of liquid should be the same in all three beakers and it should be aligned with the water level in the hot bath.
\item Check that the temperature inside the beakers is  \SI{50\pm0.1}{\celsius} by measuring the temperature of one of the MQ beakers.
\item Etch for \SI{5}{\second} in the Transene D Al etchant. 
\item Rinse for \SI{20}{\second} in the beaker with MQ that was not used to check the temperature.
\item Rinse for \SI{40}{\second} in MQ water at room temperature.
\item Blowdry.
\end{itemize}

\item Strip resist for \SI{30}{\min} in 1,3-dioxolane, \SI{10}{\second} in aceton, \SI{10}{\second} in IPA, blowdry.
\end{itemize}

\item Atomic layer deposition of first gate dielectric
 \begin{itemize}
 \item Pump down for \SI{10}{\hour} at \SI{90}{\celsius}.
 \item 150 cycles of TDMAH precursor pulse, \SI{150}{\second} waiting time, and $\mathrm{H_2O}$ pulse at \SI{90}{\celsius}. 
 \item Results in $\approx\SI{15}{\nano\meter}$ $\mathrm{HfO_x}$.
 \end{itemize}

\item Fine parts of first metallic gate layer on top of the mesa.

	\begin{itemize}
   \item Spin coat A4 PMMA resist for \SI{45}{\second} at 4000 rpm, post-bake \SI{120}{\second} at \SI{185}{\celsius}. 
      \item  EBL with \SI{150}{\micro\meter} write field, 60000 dots, pitch 1. Beam current \SI{0.1}{\nano\ampere}, dose time = \SI{0.38}{\micro\second\per dot} [proximity effect correction for InP substrate, \SI{200}{\nano\meter} PMMA, \SI{100}{\kilo\volt}, optimal contrast].
   \item Develop \SI{45}{\second} MIPA:IPA (1:3), \SI{5}{\second} IPA, blowdry. 
   \item Ash for \SI{60}{\second}.
   \item Metal deposition in AJA
   \begin{itemize}
   \item Evaporate \SI{5}{\nano\meter} Ti .
   \item Evaporate \SI{20}{\nano\meter} Au. 
   \item Evaporate \SI{5}{\nano\meter} Ti .
   \end{itemize}
   \item Liftoff in 1,3-dioxolane overinight, rinse with aceton squirt, rinse with IPA squirt, blowdry.
\end{itemize}

\item  Coarse parts of first metallic gate layer: lines crawling up the mesa, bondpads.

	\begin{itemize}
   \item Spin coat with EL9 Copolymer resist for \SI{45}{\second} at 4000 rpm, post-bake \SI{120}{\second} at \SI{185}{\celsius}. 
   \item Spin coat EL9 Copolymer resist for \SI{45}{\second} at 4000 rpm, post-bake \SI{120}{\second} at \SI{185}{\celsius}. 
   \item Spin coat A4 PMMA resist for \SI{45}{\second} at 4000 rpm, post-bake \SI{120}{\second} at \SI{185}{\celsius}. 
      \item  EBL of fine features with \SI{300}{\micro\meter} write field, 60000 dots, pitch 1. Beam current \SI{0.5}{\nano\ampere}, dose time = \SI{0.32}{\micro\second\per dot} [proximity effect correction for InP substrate, \SI{200}{\nano\meter} PMMA, \SI{100}{\kilo\volt}, optimal contrast].
            \item  EBL of coarse features with \SI{600}{\micro\meter} write field, 20000 dots, pitch 1. Beam current \SI{20}{\nano\ampere}, dose time = \SI{0.3}{\micro\second\per dot} [proximity effect correction for InP substrate, \SI{200}{\nano\meter} PMMA, \SI{100}{\kilo\volt}, optimal contrast].
   \item Develop \SI{45}{\second} MIPA:IPA (1:3), \SI{5}{\second} IPA, blowdry. 
   \item Ash for \SI{60}{\second}.
   \item Metal deposition in AJA
   \begin{itemize}
   \item Evaporate \SI{10}{\nano\meter} Ti at \SI{5}{\degree} tilt and medium rotation.
   \item Evaporate \SI{30}{\nano\meter} Au at \SI{5}{\degree} tilt and medium rotation. 
   \item Evaporate \SI{300}{\nano\meter} Au at \SI{0}{\degree} tilt and medium rotation. 
   \item Evaporate \SI{50}{\nano\meter} Au at \SI{10}{\degree} tilt and medium rotation. 
   \end{itemize}
   \item Liftoff in 1,3-dioxolane overinight, rinse with aceton squirt, rinse with IPA squirt, blowdry.
\end{itemize}

\item Atomic layer deposition of second gate dielectric
 \begin{itemize}
 \item Pump down for \SI{10}{\hour} at \SI{90}{\celsius}.
 \item 120 cycles of TDMAH precursor pulse, \SI{120}{\second} waiting time, and $\mathrm{H_2O}$ pulse at \SI{90}{\celsius}. 
 \item Results in $\approx\SI{12}{\nano\meter}$ $\mathrm{HfO_x}$.
 \end{itemize}

\item Fine parts of second metallic gate layer on top of the mesa.

	\begin{itemize}
   \item Spin coat A4 PMMA resist for \SI{45}{\second} at 4000 rpm, post-bake \SI{120}{\second} at \SI{185}{\celsius}. 
      \item  EBL with \SI{150}{\micro\meter} write field, 60000 dots, pitch 1. Beam current \SI{0.1}{\nano\ampere}, dose time = \SI{0.38}{\micro\second\per dot} [proximity effect correction for InP substrate, \SI{200}{\nano\meter} PMMA, \SI{100}{\kilo\volt}, optimal contrast].
   \item Develop \SI{45}{\second} MIPA:IPA (1:3), \SI{5}{\second} IPA, blowdry. 
   \item Ash for \SI{60}{\second}.
   \item Metal deposition in AJA
   \begin{itemize}
   \item Evaporate \SI{5}{\nano\meter} Ti .
   \item Evaporate \SI{20}{\nano\meter} Au.
   \end{itemize}
   \item Liftoff in 1,3-dioxolane overinight, rinse with aceton squirt, rinse with IPA squirt, blowdry.
\end{itemize}
\item  Coarse parts of second metallic gate layer: lines crawling up the mesa, bondpads.

Same as coarse parts of second metallic gate layer.

\end{itemize}

\section{\label{sec:fab_semi_sideprobe}Fabrication of sideprobe device with semiconducting leads}
Device 2 and device 3 were fabricated according to this protocol.
\begin{itemize}
\item Cleaving
\begin{itemize}
\item Wafer is coated with post-baked A4 PMMA resist.
\item Scribe and cleave chips of suitable size (e.g., $\SI{3}{\milli\meter}$) using manual scriber.
\end{itemize}

\item Fabrication of registration marks

	\begin{itemize}
   \item Use resist that was applied before cleaving the chip (avoids etch beads) or spin coat A4 PMMA resist for \SI{45}{\second}, post-bake \SI{2}{\min} at \SI{185}{\celsius}.
\item  EBL with \SI{150}{\micro\meter} write field, 60000 dots, pitch 1. Beam current \SI{0.5}{\nano\ampere}, dose time = \SI{0.085}{\micro\second\per dot} [proximity effect correction for InAs substrate, \SI{200}{\nano\meter} PMMA, \SI{100}{\kilo\volt}, optimal contrast].
   \item Develop \SI{45}{\second} in MIPA:IPA (1:3), \SI{5}{\second} IPA, blowdry. 
   \item Ash for \SI{60}{\second}.
   \item Metal deposition in AJA
   \begin{itemize}
   \item Deposit \SI{5}{\nano\meter} Ti.
   \item Deposit \SI{55}{\nano\meter} Au.
   \end{itemize}
   \item Clean and liftoff in 1,3-dioxolane for \SI{5}{\min} under ultrasonication at \SI{80}{\kilo\hertz}. Rinse with squirt of aceton, rinse with squirt of IPA, blowdry.
	\end{itemize}

\item Mesa etch

\begin{itemize}
\item Spin coat with A4 PMMA for \SI{45}{\second} at 4000 rpm, post-bake \SI{120}{\second} at \SI{185}{\celsius}.
\item  EBL of fine features with \SI{150}{\micro\meter} write field, 60000 dots, pitch 1. Beam current \SI{0.5}{\nano\ampere}, dose time = \SI{0.085}{\micro\second\per dot} [proximity effect correction for InAs substrate, \SI{200}{\nano\meter} PMMA, \SI{100}{\kilo\volt}, optimal contrast].
\item  EBL of coarse features with \SI{600}{\micro\meter} write field, 20000 dots, pitch 1. Beam current \SI{0.5}{\nano\ampere}, dose time = \SI{0.34}{\micro\second\per dot} [proximity effect correction for InAs substrate, \SI{200}{\nano\meter} PMMA, \SI{100}{\kilo\volt}, optimal contrast].
\item Develop \SI{60}{\second} in MIPA:IPA (1:3), \SI{5}{\second} IPA, blowdry. 
\item Ash for \SI{60}{\second}.
\item Reflow \SI{120}{\second} at \SI{120}{\celsius}
\item Al etch
\begin{itemize}
\item Prepare two beakers wtith MQ water, and one beaker with Transene D Al etchant in a hot bath. The level of liquid should be the same in all three beakers and it should be aligned with the water level in the hot bath.
\item Check that the temperature inside the beakers is  \SI{50\pm0.1}{\celsius} by measuring the temperature of one of the MQ beakers.
\item Etch for \SI{5}{\second} in the Transene D Al etchant. 
\item Rinse for \SI{20}{\second} in the beaker with MQ that was not used to check the temperature.
\item Rinse for \SI{40}{\second} in MQ water at room temperature.
\item Blowdry.
\end{itemize}

\item Mesa etch
\begin{itemize}
\item Prepare a solution of solution of $\mathrm{H_2O:C_6H_8O7:H_3PO_4:H_2O_2}$ (220:55:3:3).
\item Stir with magnetic stir.
\item Etch for \SI{9}{\min} while rotating the chip every \SI{30}{\second} by \SI{90}{\degree}.
\item Rinse in MQ water for \SI{60}{\second}.
\end{itemize}

     \item Strip resist for \SI{15}{\min} in 1,3-dioxolane, \SI{10}{\second} in aceton, \SI{10}{\second} in IPA, blowdry.
\end{itemize}

\item Selective removal of Al

\begin{itemize}
\item Spin coat with A4 PMMA for \SI{45}{\second} at 4000 rpm, post-bake \SI{120}{\second} at \SI{185}{\celsius}.
\item  EBL with \SI{150}{\micro\meter} write field, 60000 dots, pitch 1. Beam current \SI{0.1}{\nano\ampere}, dose time = \SI{0.45}{\micro\second\per dot} [proximity effect correction for InAs substrate, \SI{200}{\nano\meter} PMMA, \SI{100}{\kilo\volt}, optimal contrast].
\item Develop \SI{45}{\second} in MIPA:IPA (1:3), \SI{5}{\second} IPA, blowdry. 
\item Ash for \SI{60}{\second}.
\item Reflow \SI{120}{\second} at \SI{120}{\celsius}
\item Al etch
\begin{itemize}
\item Prepare two beakers wtith MQ water, and one beaker with Transene D Al etchant in a hot bath. The level of liquid should be the same in all three beakers and it should be aligned with the water level in the hot bath.
\item Check that the temperature inside the beakers is  \SI{50\pm0.1}{\celsius} by measuring the temperature of one of the MQ beakers.
\item Etch for \SI{5}{\second} in the Transene D Al etchant. 
\item Rinse for \SI{20}{\second} in the beaker with MQ that was not used to check the temperature.
\item Rinse for \SI{40}{\second} in MQ water at room temperature.
\item Blowdry.
\end{itemize}

\item Strip resist for \SI{30}{\min} in 1,3-dioxolane, \SI{10}{\second} in aceton, \SI{10}{\second} in IPA, blowdry.
\end{itemize}

\item Atomic layer deposition of gate dielectric
 \begin{itemize}
 \item Pump down for \SI{10}{\hour} at \SI{90}{\celsius}.
 \item 150 cycles of TDMAH precursor pulse, \SI{150}{\second} waiting time, and $\mathrm{H_2O}$ pulse at \SI{90}{\celsius}. 
 \item Results in $\approx\SI{15}{\nano\meter}$ $\mathrm{HfO_x}$.
 \end{itemize}

\item Fine parts of metallic gate layer on top of the mesa.

	\begin{itemize}
   \item Spin coat A4 PMMA resist for \SI{45}{\second} at 4000 rpm, post-bake \SI{120}{\second} at \SI{185}{\celsius}. 
      \item  EBL with \SI{150}{\micro\meter} write field, 60000 dots, pitch 1. Beam current \SI{0.1}{\nano\ampere}, dose time = \SI{0.38}{\micro\second\per dot} [proximity effect correction for InP substrate, \SI{200}{\nano\meter} PMMA, \SI{100}{\kilo\volt}, optimal contrast].
   \item Develop \SI{45}{\second} MIPA:IPA (1:3), \SI{5}{\second} IPA, blowdry. 
   \item Ash for \SI{60}{\second}.
   \item Metal deposition in AJA
   \begin{itemize}
   \item Evaporate \SI{5}{\nano\meter} Ti .
   \item Evaporate \SI{20}{\nano\meter} Au. 
   \item Evaporate \SI{5}{\nano\meter} Ti .
   \end{itemize}
   \item Liftoff in 1,3-dioxolane overinight, rinse with aceton squirt, rinse with IPA squirt, blowdry.
\end{itemize}

\item  Coarse parts of metallic gate layer: lines crawling up the mesa, bondpads.

	\begin{itemize}
   \item Spin coat with EL9 Copolymer resist for \SI{45}{\second} at 4000 rpm, post-bake \SI{120}{\second} at \SI{185}{\celsius}. 
   \item Spin coat EL9 Copolymer resist for \SI{45}{\second} at 4000 rpm, post-bake \SI{120}{\second} at \SI{185}{\celsius}. 
   \item Spin coat A4 PMMA resist for \SI{45}{\second} at 4000 rpm, post-bake \SI{120}{\second} at \SI{185}{\celsius}. 
      \item  EBL of fine features with \SI{300}{\micro\meter} write field, 60000 dots, pitch 1. Beam current \SI{0.5}{\nano\ampere}, dose time = \SI{0.32}{\micro\second\per dot} [proximity effect correction for InP substrate, \SI{200}{\nano\meter} PMMA, \SI{100}{\kilo\volt}, optimal contrast].
            \item  EBL of coarse features with \SI{600}{\micro\meter} write field, 20000 dots, pitch 1. Beam current \SI{20}{\nano\ampere}, dose time = \SI{0.3}{\micro\second\per dot} [proximity effect correction for InP substrate, \SI{200}{\nano\meter} PMMA, \SI{100}{\kilo\volt}, optimal contrast].
   \item Develop \SI{45}{\second} MIPA:IPA (1:3), \SI{5}{\second} IPA, blowdry. 
   \item Ash for \SI{60}{\second}.
   \item Metal deposition in AJA
   \begin{itemize}
   \item Evaporate \SI{10}{\nano\meter} Ti at \SI{5}{\degree} tilt and medium rotation.
   \item Evaporate \SI{30}{\nano\meter} Au at \SI{5}{\degree} tilt and medium rotation. 
   \item Evaporate \SI{300}{\nano\meter} Au at \SI{0}{\degree} tilt and medium rotation. 
   \item Evaporate \SI{50}{\nano\meter} Au at \SI{10}{\degree} tilt and medium rotation. 
   \end{itemize}
   \item Liftoff in 1,3-dioxolane overinight, rinse with aceton squirt, rinse with IPA squirt, blowdry.
\end{itemize}

\end{itemize}

\section{\label{sec:fab_finger_piano}Fabrication of sideprobe device with array of gate electrodes}

The device shown in Fig.~\ref{fig:finger_piano} was fabricated according to this protocol.

\begin{itemize}
\item Cleaving
\begin{itemize}
\item Wafer is coated with post-baked A4 PMMA resist.
\item Scribe and cleave chips of suitable size (e.g., $\SI{3}{\milli\meter}$) using manual scriber.
\end{itemize}

\item Fabrication of registration marks

	\begin{itemize}
   \item Use resist that was applied before cleaving the chip (avoids etch beads) or spin coat A4 PMMA resist for \SI{45}{\second}, post-bake \SI{2}{\min} at \SI{185}{\celsius}.
\item  EBL with \SI{150}{\micro\meter} write field, 60000 dots, pitch 1. Beam current \SI{0.5}{\nano\ampere}, dose time = \SI{0.085}{\micro\second\per dot} [proximity effect correction for InAs substrate, \SI{200}{\nano\meter} PMMA, \SI{100}{\kilo\volt}, optimal contrast].
   \item Develop \SI{45}{\second} in MIPA:IPA (1:3), \SI{5}{\second} IPA, blowdry. 
   \item Ash for \SI{60}{\second}.
   \item Metal deposition in AJA
   \begin{itemize}
   \item Deposit \SI{5}{\nano\meter} Ti.
   \item Deposit \SI{55}{\nano\meter} Au.
   \end{itemize}
   \item Clean and liftoff in 1,3-dioxolane for \SI{5}{\min} under ultrasonication at \SI{80}{\kilo\hertz}. Rinse with squirt of aceton, rinse with squirt of IPA, blowdry.
	\end{itemize}

\item Mesa etch

\begin{itemize}
\item Spin coat with A4 PMMA for \SI{45}{\second} at 4000 rpm, post-bake \SI{120}{\second} at \SI{185}{\celsius}.
\item  EBL of fine features with \SI{150}{\micro\meter} write field, 60000 dots, pitch 1. Beam current \SI{0.5}{\nano\ampere}, dose time = \SI{0.085}{\micro\second\per dot} [proximity effect correction for InAs substrate, \SI{200}{\nano\meter} PMMA, \SI{100}{\kilo\volt}, optimal contrast].
\item  EBL of coarse features with \SI{600}{\micro\meter} write field, 20000 dots, pitch 1. Beam current \SI{0.5}{\nano\ampere}, dose time = \SI{0.34}{\micro\second\per dot} [proximity effect correction for InAs substrate, \SI{200}{\nano\meter} PMMA, \SI{100}{\kilo\volt}, optimal contrast].
\item Develop \SI{60}{\second} in MIPA:IPA (1:3), \SI{5}{\second} IPA, blowdry. 
\item Ash for \SI{60}{\second}.
\item Reflow \SI{120}{\second} at \SI{120}{\celsius}
\item Al etch
\begin{itemize}
\item Prepare two beakers wtith MQ water, and one beaker with Transene D Al etchant in a hot bath. The level of liquid should be the same in all three beakers and it should be aligned with the water level in the hot bath.
\item Check that the temperature inside the beakers is  \SI{50\pm0.1}{\celsius} by measuring the temperature of one of the MQ beakers.
\item Etch for \SI{5}{\second} in the Transene D Al etchant. 
\item Rinse for \SI{20}{\second} in the beaker with MQ that was not used to check the temperature.
\item Rinse for \SI{40}{\second} in MQ water at room temperature.
\item Blowdry.
\end{itemize}

\item Mesa etch
\begin{itemize}
\item Prepare a solution of solution of $\mathrm{H_2O:C_6H_8O7:H_3PO_4:H_2O_2}$ (220:55:3:3).
\item Stir with magnetic stir.
\item Etch for \SI{9}{\min} while rotating the chip every \SI{30}{\second} by \SI{90}{\degree}.
\item Rinse in MQ water for \SI{60}{\second}.
\end{itemize}

     \item Strip resist for \SI{15}{\min} in 1,3-dioxolane, \SI{10}{\second} in aceton, \SI{10}{\second} in IPA, blowdry.
\end{itemize}

\item Selective removal of Al

\begin{itemize}
\item Spin coat with A4 PMMA for \SI{45}{\second} at 4000 rpm, post-bake \SI{120}{\second} at \SI{185}{\celsius}.
\item  EBL with \SI{150}{\micro\meter} write field, 60000 dots, pitch 1. Beam current \SI{0.1}{\nano\ampere}, dose time = \SI{0.45}{\micro\second\per dot} [proximity effect correction for InAs substrate, \SI{200}{\nano\meter} PMMA, \SI{100}{\kilo\volt}, optimal contrast].
\item Develop \SI{45}{\second} in MIPA:IPA (1:3), \SI{5}{\second} IPA, blowdry. 
\item Ash for \SI{60}{\second}.
\item Reflow \SI{120}{\second} at \SI{120}{\celsius}
\item Al etch
\begin{itemize}
\item Prepare two beakers wtith MQ water, and one beaker with Transene D Al etchant in a hot bath. The level of liquid should be the same in all three beakers and it should be aligned with the water level in the hot bath.
\item Check that the temperature inside the beakers is  \SI{50\pm0.1}{\celsius} by measuring the temperature of one of the MQ beakers.
\item Etch for \SI{5}{\second} in the Transene D Al etchant. 
\item Rinse for \SI{20}{\second} in the beaker with MQ that was not used to check the temperature.
\item Rinse for \SI{40}{\second} in MQ water at room temperature.
\item Blowdry.
\end{itemize}

\item Strip resist for \SI{30}{\min} in 1,3-dioxolane, \SI{10}{\second} in aceton, \SI{10}{\second} in IPA, blowdry.
\end{itemize}

\item Atomic layer deposition of first gate dielectric
 \begin{itemize}
 \item Pump down for \SI{10}{\hour} at \SI{90}{\celsius}.
 \item 150 cycles of TDMAH precursor pulse, \SI{150}{\second} waiting time, and $\mathrm{H_2O}$ pulse at \SI{90}{\celsius}. 
 \item Results in $\approx\SI{15}{\nano\meter}$ $\mathrm{HfO_x}$.
 \end{itemize}

\item Metallic gate connectors on top of the mesa that make connection between fine Al gates and connecting lines to the bond pads.

	\begin{itemize}
   \item Spin coat A4 PMMA resist for \SI{45}{\second} at 4000 rpm, post-bake \SI{120}{\second} at \SI{185}{\celsius}. 
      \item  EBL with \SI{150}{\micro\meter} write field, 60000 dots, pitch 1. Beam current \SI{0.1}{\nano\ampere}, dose time = \SI{0.38}{\micro\second\per dot} [proximity effect correction for InP substrate, \SI{200}{\nano\meter} PMMA, \SI{100}{\kilo\volt}, optimal contrast].
   \item Develop \SI{45}{\second} MIPA:IPA (1:3), \SI{5}{\second} IPA, blowdry. 
   \item Ash for \SI{60}{\second}.
   \item Metal deposition in AJA
   \begin{itemize}
   \item Evaporate \SI{3}{\nano\meter} Ti .
   \item Evaporate \SI{15}{\nano\meter} Au.
   \end{itemize}
   \item Liftoff in 1,3-dioxolane overinight, rinse with aceton squirt, rinse with IPA squirt, blowdry.
\end{itemize}

\item First layer of fine inner gates made from Al, capped with native oxide.

	\begin{itemize}
   \item Spin coat A3 PMMA resist for \SI{45}{\second} at 4000 rpm, post-bake \SI{120}{\second} at \SI{185}{\celsius}. 
      \item  EBL with \SI{150}{\micro\meter} write field, 60000 dots, pitch 1. Beam current \SI{0.1}{\nano\ampere}, dose time = \SI{0.11}{\micro\second\per dot} [proximity effect correction for InP substrate, \SI{200}{\nano\meter} PMMA, \SI{100}{\kilo\volt}, optimal contrast, multi-pass 4].
   \item Develop \SI{45}{\second} MIPA:IPA (1:3), \SI{5}{\second} IPA, blowdry. 
   \item Ash for \SI{45}{\second}.
   \item Metal deposition in AJA
   \begin{itemize}
   \item Evaporate \SI{3}{\nano\meter} Ti .
   \item Evaporate \SI{27}{\nano\meter} Al.
   \end{itemize}
   \item Liftoff in 1,3-dioxolane overinight, rinse with aceton squirt, rinse with IPA squirt, blowdry.
\item Oxidize Al on the hotplate at \SI{185}{\celsius} for \SI{10}{\min} in ambient conditions. Spin coat with resist of next fabrication step directly afterwards.
\end{itemize}

\item Second layer of fine inner gates made from Al, capped with native oxide (alternatively Au without oxide capping may be used here).

	\begin{itemize}
   \item Spin coat A3 PMMA resist for \SI{45}{\second} at 4000 rpm, post-bake \SI{120}{\second} at \SI{185}{\celsius}. 
      \item  EBL with \SI{150}{\micro\meter} write field, 60000 dots, pitch 1. Beam current \SI{0.1}{\nano\ampere}, dose time = \SI{0.11}{\micro\second\per dot} [proximity effect correction for InP substrate, \SI{200}{\nano\meter} PMMA, \SI{100}{\kilo\volt}, optimal contrast, multi-pass 4].
   \item Develop \SI{45}{\second} MIPA:IPA (1:3), \SI{5}{\second} IPA, blowdry. 
   \item Ash for \SI{45}{\second}.
   \item Metal deposition in AJA
   \begin{itemize}
   \item Evaporate \SI{3}{\nano\meter} Ti .
   \item Evaporate \SI{27}{\nano\meter} Al.
   \end{itemize}
   \item Liftoff in 1,3-dioxolane overinight, rinse with aceton squirt, rinse with IPA squirt, blowdry.
\item Oxidize Al on the hotplate at \SI{185}{\celsius} for \SI{10}{\min} in ambient conditions. Spin coat with resist of next fabrication step directly afterwards.
\end{itemize}

\item  Coarse parts of metallic gate layer: lines crawling up the mesa and overlapping the connectors, bondpads.

	\begin{itemize}
   \item Spin coat with EL9 Copolymer resist for \SI{45}{\second} at 4000 rpm, post-bake \SI{120}{\second} at \SI{185}{\celsius}. 
   \item Spin coat EL9 Copolymer resist for \SI{45}{\second} at 4000 rpm, post-bake \SI{120}{\second} at \SI{185}{\celsius}. 
   \item Spin coat A4 PMMA resist for \SI{45}{\second} at 4000 rpm, post-bake \SI{120}{\second} at \SI{185}{\celsius}. 
      \item  EBL of fine features with \SI{300}{\micro\meter} write field, 60000 dots, pitch 1. Beam current \SI{0.5}{\nano\ampere}, dose time = \SI{0.32}{\micro\second\per dot} [proximity effect correction for InP substrate, \SI{200}{\nano\meter} PMMA, \SI{100}{\kilo\volt}, optimal contrast].
            \item  EBL of coarse features with \SI{600}{\micro\meter} write field, 20000 dots, pitch 1. Beam current \SI{20}{\nano\ampere}, dose time = \SI{0.3}{\micro\second\per dot} [proximity effect correction for InP substrate, \SI{200}{\nano\meter} PMMA, \SI{100}{\kilo\volt}, optimal contrast].
   \item Develop \SI{45}{\second} MIPA:IPA (1:3), \SI{5}{\second} IPA, blowdry. 
   \item Ash for \SI{60}{\second}.
   \item Metal deposition in AJA
   \begin{itemize}
   \item Evaporate \SI{10}{\nano\meter} Ti at \SI{5}{\degree} tilt and medium rotation.
   \item Evaporate \SI{30}{\nano\meter} Au at \SI{5}{\degree} tilt and medium rotation. 
   \item Evaporate \SI{300}{\nano\meter} Au at \SI{0}{\degree} tilt and medium rotation. 
   \item Evaporate \SI{50}{\nano\meter} Au at \SI{10}{\degree} tilt and medium rotation. 
   \end{itemize}
   \item Liftoff in 1,3-dioxolane overinight, rinse with aceton squirt, rinse with IPA squirt, blowdry.
\end{itemize}

\end{itemize}

\section{\label{sec:fab_fullshell}Fabrication of fullshell nanowire three-terminal device}
Device 5 and device 6 were fabricated according to this protocol.
\begin{itemize}
\item Nanowire deposition 
\begin{itemize}
\item Clean carrier chip for \SI{5}{\min} in aceton, \SI{5}{\min} in IPA, blowdry.
\item Ash for \SI{120}{\second}
\item Deposit nanowires using micromanipulator.
\end{itemize}

\item Aluminum etch

	\begin{itemize}
   \item Ash for \SI{60}{\second}.
   \item Spin coat with adhesion promoter AR 300-80 NEW for \SI{45}{\second} at 4000 rpm, post-bake \SI{120}{\second} at \SI{115}{\celsius}. 
   \item Strip adhesion promoter for \SI{120}{\second} in aceton, \SI{120}{\second} in 1,3-dioxolane, \SI{30}{\second} in IPA. Blowdry.
   \item Spin coat with EL9 for \SI{45}{\second} at 4000 rpm, post-bake \SI{180}{\second} at \SI{185}{\celsius}.
   \item EBL with \SI{300}{\micro\meter} write field, 60000 dots, pitch 1. Beam current \SI{0.5}{\nano\ampere}, dose time = \SI{0.22}{\micro\second\per dot} [proximity effect correction for Si substrate, \SI{200}{\nano\meter} PMMA, \SI{100}{\kilo\volt}, optimal contrast].
   \item Develop \SI{22}{\second} in MIPA:IPA (1:3), \SI{20}{\second} IPA, blowdry. 
   \item At room temperature, etch for \SI{33}{\second} in 0.17 N TMAH, rinse \SI{15}{\second} in mq water, rinse \SI{15}{\second} in mq water, blowdry.
     \item Strip resist for \SI{120}{\second} in 1,3-dioxolane, \SI{120}{\second} in aceton, \SI{30}{\second} in IPA, blowdry.
      \end{itemize}

\item Ohmic contacts to semiconductor

	\begin{itemize}
   \item Spin coat A6 PMMA resist for \SI{45}{\second} at 4000 rpm, post-bake \SI{120}{\second} at \SI{115}{\celsius}. 
      \item Spin coat A4 PMMA resist for \SI{45}{\second} at 4000 rpm, post-bake \SI{120}{\second} at \SI{115}{\celsius}. 
       \item EBL with \SI{300}{\micro\meter} write field, 60000 dots, pitch 1. Beam current \SI{0.5}{\nano\ampere}, dose time = \SI{0.47}{\micro\second\per dot} [proximity effect correction for Si substrate, \SI{200}{\nano\meter} PMMA, \SI{100}{\kilo\volt}, optimal contrast].
   \item Develop \SI{75}{\second} MIPA:IPA (1:3), \SI{10}{\second} IPA, blowdry. 
   \item Ash for \SI{120}{\second}.
   \item Metal deposition in AJA
   \begin{itemize}
   \item Kaufmann milling at medium rotation, 15 sccm Ar, 1 mTorr, 100 V, 5 min discharge, 2 min warm up, 8 min milling.
   \item Let sample cool down for 5 min. 
   \item Evaporate at medium rotation \SI{5}{\nano\meter} Ti, \SI{200}{\nano\meter} Au. 
   \end{itemize}
   \item Liftoff for \SI{5}{min} at \SI{65}{\celsius} in NMP, \SI{1}{\min} at room temperature in aceton, \SI{1}{\min} at room temperature in IPA, blowdry. 
   \item Ash for \SI{120}{\second}.
\end{itemize}

\item Al contact to Al shell

	\begin{itemize}
	\item Ash \SI{120}{\second}.
   \item Spin coat A6 PMMA resist for \SI{45}{\second} at 4000 rpm, post-bake \SI{120}{\second} at \SI{115}{\celsius}. 
       \item EBL with \SI{300}{\micro\meter} write field, 60000 dots, pitch 1. Beam current \SI{0.5}{\nano\ampere}, dose time = \SI{0.48}{\micro\second\per dot} [proximity effect correction for Si substrate, \SI{200}{\nano\meter} PMMA, \SI{100}{\kilo\volt}, optimal contrast].
   \item Develop \SI{40}{\second} MIPA:IPA (1:3), \SI{10}{\second} IPA, blowdry. 
   \item Ash for \SI{60}{\second}.
    \item Reflow for \SI{30}{\second} at \SI{185}{\celsius}.
     \item Crosslink resist by EBL with \SI{500}{\micro\meter} write field, 50000 dots, pitch 1. Beam current \SI{3}{\nano\ampere}, dose time = \SI{13}{\micro\second\per dot} [proximity effect correction for Si substrate, \SI{200}{\nano\meter} PMMA, \SI{100}{\kilo\volt}, optimal contrast].
        \item Strip resist for \SI{5}{min} in aceton, \SI{10}{\second} in IPA, blowdry. 
       \item Spin coat A6 PMMA resist for \SI{45}{\second} at 4000 rpm, post-bake \SI{60}{\second} at \SI{115}{\celsius}. 
       \item EBL with  \SI{300}{\micro\meter} write field, 60000 dots, pitch 1. Beam current \SI{0.5}{\nano\ampere}, dose time = \SI{0.48}{\micro\second\per dot} [proximity effect correction for Si substrate, \SI{200}{\nano\meter} PMMA, \SI{100}{\kilo\volt}, optimal contrast].
     \item Develop \SI{40}{\second} MIPA:IPA (1:3), \SI{10}{\second} IPA, blowdry. 
   \item Ash for \SI{60}{\second}.
   \item Metal deposition in AJA
   \begin{itemize}
   \item Kaufmann milling at medium rotation, 15 sccm Ar, 1 mTorr, 100 V, 5 min discharge, 2 min warm up, \SI{4.5}{\min} milling.
   \item Let sample cool down for 5 min. 
   \item Evaporate \SI{25}{\nano\meter} Al.
   \end{itemize}
   \item Liftoff for \SI{5}{min} at \SI{80}{\celsius} in NMP, \SI{1}{\min} at room temperature in aceton, \SI{1}{\min} at room temperature in IPA, blowdry. 
	\end{itemize}
	
\item Atomic layer deposition of gate dielectric
 \begin{itemize}
 \item Pump down for \SI{10}{\hour} at \SI{90}{\celsius}.
 \item 70 cycles of TDMAH precursor pulse, \SI{150}{\second} waiting time, and $\mathrm{H_2O}$ pulse at \SI{90}{\celsius}. 
 \item Results in $\approx\SI{7}{\nano\meter}$ $\mathrm{HfO_x}$.
 \end{itemize}
	
\item Metallic gates

	\begin{itemize}
   \item Spin coat A6 PMMA resist for \SI{45}{\second} at 4000 rpm, post-bake \SI{60}{\second} at \SI{115}{\celsius}. 
       \item EBL with  \SI{300}{\micro\meter} write field, 60000 dots, pitch 1. Beam current \SI{0.5}{\nano\ampere}, dose time = \SI{0.47}{\micro\second\per dot} [proximity effect correction for Si substrate, \SI{200}{\nano\meter} PMMA, \SI{100}{\kilo\volt}, optimal contrast].
   \item Develop \SI{60}{\second} MIPA:IPA (1:3), \SI{10}{\second} IPA, blowdry. 
   \item Ash for \SI{60}{\second}.
   \item Metal deposition in AJA
   \begin{itemize}
   \item Evaporate at medium rotation \SI{5}{\nano\meter} Ti, \SI{100}{\nano\meter} Au with no tilt. 
      \item Evaporate at medium rotation \SI{30}{\nano\meter} Au at \SI{10}{\degree} tilt. 
            \item Evaporate at medium rotation \SI{80}{\nano\meter} Au with no tilt. 
   \end{itemize}
   \item Liftoff for \SI{7}{min} at \SI{80}{\celsius} in NMP, \SI{1}{\min} at room temperature in aceton, \SI{1}{\min} at room temperature in IPA, blowdry.
\end{itemize}

\end{itemize}
 
\cleardoublepage
\defbibheading{bibintoc}[\bibname]{%
  \phantomsection
  \manualmark
  \markboth{\spacedlowsmallcaps{#1}}{\spacedlowsmallcaps{#1}}%
  \addtocontents{toc}{\protect\vspace{\beforebibskip}}%
  \addcontentsline{toc}{chapter}{\tocEntry{#1}}%
  \chapter*{#1}%
}
\printbibliography[heading=bibintoc]

\end{document}